%% file: paper.tex
\documentclass[final,nopageno]{jpaper}

\usepackage[normalem]{ulem}

\input{font}
\usepackage{mystyle}
\usepackage{macros}
\usepackage{pifont}
\usepackage{indentfirst}
\usepackage{setspace}
\usepackage{etoolbox}
\usepackage[bookmarks=false,hidelinks]{hyperref}
\usepackage{enumitem} 
\usepackage{fancyhdr}
\usepackage{cite}
\usepackage{url}
\usepackage{hhline}
\usepackage[titletoc]{appendix}

\usepackage{subcaption}
\usepackage{dblfloatfix}

\usepackage{titlesec}
\DeclareTextFontCommand{\var}{\normalfont\ttfamily}

\makeatletter
\patchcmd{\ttlh@hang}{\parindent\z@}{\parindent\z@\leavevmode}{}{}
\patchcmd{\ttlh@hang}{\noindent}{}{}{}
\makeatother

\newcommand{\LDescriptor}[0]{Locality Descriptor\xspace}
\newcommand{\X}[0]{XMem\xspace}

\newcommand{\Xlib}[0]{\texttt{XMemLib}\xspace}
\newcommand{\Xfull}[0]{Expressive Memory\xspace}
\newcommand{\revision}[1]{{\leavevmode\color{black}{ #1}}}
\newcommand{\shepherd}[1]{{\leavevmode\color{black}{#1}}}
\newcommand{\newII}[1]{{\leavevmode\color{black}{#1}}}
\newcommand{\highlight}[1]{{\leavevmode\color{black}{#1}}}
\newcommand{\tech}[1]{{\leavevmode\color{black}{#1}}}

\newcommand{\techreport}[1]{}
\newcommand{\One}{\emph{(i)}~}
\newcommand{\Two}{\emph{(ii)}~}
\newcommand{\Three}{\emph{(iii)}~}
\newcommand{\Four}{\emph{(iv)}~}

\newcommand{\intrathread}[0]{\texttt{INTRA-THREAD}\xspace}
\newcommand{\interthread}[0]{\texttt{INTER-THREAD}\xspace}
\newcommand{\noreuse}[0]{\texttt{NO-REUSE}\xspace}
\newcommand{\histo}[0]{\texttt{histo}\xspace} 
\newcommand{\dtile}[0]{\mbox{D-Tile}\xspace} 
\newcommand{\ctile}[0]{\mbox{C-Tile}\xspace} 
\newcommand{\dtiles}[0]{\mbox{D-Tiles}\xspace} 
\newcommand{\ctiles}[0]{\mbox{C-Tiles}\xspace} 
\newcommand{\remove}[1]{}
\newcommand{\ignore}[1]{}
\newcommand{\thoughts}[1]{}
\newcommand{\secref}[1]{\S\ref{#1}}
\newcommand{\xref}[1]{\S\ref{#1}}
\newcommand{\xmem}[0]{(Chapter~\ref{sec:xmem})\xspace}
\newcommand{\ldesc}[0]{(Chapter~\ref{sec:ldesc})\xspace}
\newcommand{\zorua}[0]{(Chapter~\ref{sec:zorua})\xspace}
\newcommand{\caba}[0]{(Chapter~\ref{sec:caba})\xspace}
\newcommand{\all}[0]{(Chapter~\ref{sec:xmem}-\ref{sec:caba})\xspace}

\newcounter{rowcntr}[table]
\renewcommand{\therowcntr}{\arabic{rowcntr}}
\newcolumntype{N}{>{\refstepcounter{rowcntr}\therowcntr}c}
\AtBeginEnvironment{tabular}{\setcounter{rowcntr}{0}}
\begin{document}

\title{
\vspace{2cm}
{\fontsize{19}{10}\selectfont Enhancing Programmability, Portability, and Performance\\ with Rich
Cross-Layer Abstractions} 
\vspace{1cm}
	}
\date{}
\maketitle
\begin{center}
\Large\emph{Submitted in partial fulfillment of the requirements\\ for the degree
of Doctor of Philosophy\\ in Electrical and Computer Engineering\\}
\vspace{2.5cm}
\LARGE\textbf{Nandita Vijaykumar} \\
\vspace{1cm}
\large{M.S., Electrical and Computer Engineering, Carnegie Mellon University}\\ 
\large{B.E., Electrical and Electronics Engineering, P.E.S. Institute of
Technology}\\
\vspace{2.5cm}
\Large{Carnegie Mellon
University \\}
\Large{Pittsburgh, PA\\}
\vspace{3cm}
\Large\textbf{December 2019}
\\[\baselineskip]
\end{center}
 
\newgeometry{lmargin=1in,rmargin=1in,tmargin=1in,bmargin=1in}
\newpage
\vspace*{\fill}
\begin{center}
\large{\textcopyright~~Nandita Vijaykumar, 2019\\All Rights Reserved}
\end{center}
\newpage
\onehalfspacing
\thispagestyle{empty}
\pagestyle{plain}
\input{sections/acknowledgments}

\input{sections/abstract}
\newpage
\tableofcontents
\newpage
\listoffigures
\listoftables
\input{sections/introduction}

\input{sections/xmem}

\input{sections/xmem_impl}

\input{sections/ldesc}

\input{sections/zorua}

\input{sections/caba}

\input{sections/misc}
\titleformat{\section}[display]{\sffamily\huge\bfseries}{Appendix \thesection}%
  {0.5ex}{\vspace{1ex}}[\vspace{2ex}]
\input{sections/appendix}
\fancyfoot[C]{\thepage} 
\newpage

\bibliographystyle{plain}
\singlespacing
\bibliography{paper}

\end{document}

%% file: font.tex
\usepackage[T1]{fontenc}
\usepackage{libertine}
\usepackage{pifont}
\usepackage{inconsolata}

%% file: sections/acknowledgments.tex
\vspace*{\fill}
\begin{center}
\emph{To my loving parents, Porul and Vijay.}
\end{center}
\vspace*{\fill}

\section*{\hfill Acknowledgments \hfill}

\vspace{0.8cm}

There have been many many people, more than I can name, who have contributed in
very different ways to this thesis and to my academic journey over the last 6
years. This is but a brief and humble reflection on their invaluable contributions.

First and foremost, I would like to thank my advisors, Prof. Onur Mutlu and
Prof. Phillip B. Gibbons. Onur generously gave me all the resources,
opportunities, guidance, and freedom I needed to grow and succeed. His
unrelenting emphasis on clarity of thinking and writing and his indomitable
work ethic was a constant source of inspiration and learning. I am
immensely grateful for his unwavering support and confidence in me throughout
the ups and downs of my PhD.  I thank Phil for his tremendous generosity with
his time and faith in me. His technical perspectives played a big role in
widening the scope of my
research and challenging my thinking. I am very grateful for the safe and
stimulating  
environment Phil provided during my PhD journey.  My advisors' constant encouragement and nurturing
shaped me into the researcher that I am today. 

I am grateful to my thesis committee members, Prof. James Hoe, Prof. Mattan
Erez, and Prof. Wen-Mei Hwu, for their valuable feedback in making this thesis
stronger and their support during my academic  job search.  I am very grateful
to Prof. Hoe for all his advice and for providing the stimulating research
environment in CALCM at CMU. I sincerely thank Prof. Todd Mowry and Prof. Chita Das for the research
collaborations that strengthened this thesis and their invaluable support during
my academic job search. 

Graduate school was a long and lonely journey, and I am immensely grateful to
the members of the SAFARI research group for all their companionship: I found my
family away from home in SAFARI.  I will always be grateful to Kevin Hsieh for
the many long hours of brainstorming, his endless support and positivity, and
our invaluable research synergy.
Gennady Pekhimenko was an irreplaceable mentor, friend, and big brother, who
stood by me right from the beginning. I am very thankful to Samira Khan for
being a great friend, mentor, and confidante. Her unrelenting high standards
pushed me to succeed and her support kept me sane during my hardest times. I
thank Hongyi Xin for being a wonderful friend and an unending source of laughter
and support. I thank Vivek Seshadri for being a great mentor and a role model to
aspire to. I thank Lavanya Subramanian for her support, kindness, and warmth
throughout my PhD. These lifelong friendships are one of the most valuable
outcomes of my PhD and I cannot express in words my gratitude to them.  

I am very grateful to all the other members of SAFARI at CMU and ETH for being
great friends and colleagues and providing a stimulating research environment:
Yoongu Kim for setting high standards in the group and the many intellectual
discussions; Donghyuk Lee for his kindness, technical expertise, and all his support during my Nvidia internship; Saugata Ghose for his support,
positivity, encouragement, and all the enjoyable dinners; Rachata Ausavarungnirun for the technical
discussions and for being a kind friend; and Juan Gomez Luna for helping me
settle into ETH. I also thank Justin Meza, Kevin Chang, Yixin Luo, Amirali
Boroumand, Hasan Hassan, Minesh Patel, Damla Senol, Jeremie Kim, Giray Yaglikci,
Nika Mansouri, Mohammed Alser, and the other SAFARI members at CMU and ETH. 

I thank the members of CALCM and PDL at CMU and the Systems Group at ETH for all
the technical feedback and for creating a rich stimulating research environment.
I also thank my collaborators at Penn State, Prof. Mahmut Kandemir, Adwait Jog,
and Onur Kayiran for the research collaborations. 

I am grateful to all the students and mentees who worked closely with me over the
years: Abhilasha Jain, Ashish Shrestha, Konstantinos Kanellopoulos, Mehrshad
Lotfi, Diptesh Majumdar, Abhishek Bhowmick, Nisa Bostanci, and Ataberk Olgun.
This thesis would not be possible without them.

I would like to thank all my internship mentors and managers for their guidance
during my many internships:  Eiman Ebrahimi, Steve Keckler, Dave Nellans, Chris
Wilkerson, Kingsum Chow, Alaa Alameldeen, Zeshan Chisti, Olatunji Ruwase, and
Trishul Chilimbi. I sincerely thank Intel, Nvidia, and Microsoft for these
opportunities.

I gratefully acknowledge the generous support from Facebook, Google, IBM, Intel,
Microsoft, Qualcomm, VMware, Nvidia, Samsung, Alibaba, the Intel Science and
Technology Center for Cloud Computing, the Semiconductor Research Corporation
and NSF (grants 0953246, 1065112, 1205618, 1212962, 1213052, 1302225, 1302557,
1317560, 1320478, 1320531, 1409095, 1409723, 1423172, 1439021, and 1439057). I
am also grateful for the support from the Benjamin Garver Lamme/Westinghouse
fellowship at CMU.

I sincerely thank my mentors and colleagues at AMD for introducing me to
computer architecture and PhD research:  Kanishka Lahiri, Hrishikesh
Murukkathampoondi, Anasua Bhowmick, Ramkumar Jayaseelan, and Subhash Kunnoth. 

I thank my family and friends for all their love and support during this
journey. I am tremendously blessed to have a big and wonderful family and
amazing friends. This thesis would in no way be possible without them and I hope
I have made them proud. I especially thank Santosh Harish and Manzil Zaheer for their
support and companionship. My heartfelt gratitude is to Neha Singh for our decades-long
friendship and her constant support
and care. I will forever be grateful to my aunt, Akila Suresh, for her
unending love and care and for making my travel-filled PhD so much easier.  I
thank my sister, Dipti, for being a constant source of love, patience, and
support throughout graduate school. 

Finally, I thank my parents, Porul and Vijay. I will forever be indebted to them for their unconditional love,
unwavering support, and the inspiration to pursue my academic dreams. They
cannot know the immense role they played over the last few years and words
cannot express my gratitude to them. I will forever be grateful for everything
they have given me and enabled me to do, and this thesis is dedicated to them.

%% file: sections/abstract.tex
\newpage
\begin{abstract}
\thispagestyle{plain}
\vspace{0.5in}
Programmability, performance portability, and resource efficiency have emerged as
critical challenges in harnessing complex and diverse architectures today to
obtain high performance and energy efficiency. While there is abundant research,
and thus significant improvements, 
at different levels of the stack that address these very challenges, in this
thesis, we observe that we are fundamentally limited by the \emph{interfaces}
and abstractions between the application and the underlying
system/hardware{\textemdash}specifically, the hardware-software interface. 
The existing narrow interfaces poses two critical challenges. First, significant effort and expertise are required to write high-performance
code to harness the full potential of today’s diverse and sophisticated
hardware. Second, as a hardware/system designer, architecting faster and more
efficient systems is challenging as the vast majority of the program’s semantic
content gets lost in translation with today’s hardware-software interface.
Moving towards the future, these challenges in programmability and efficiency
will be even more intractable as we architect increasingly heterogeneous and
sophisticated systems.

This thesis makes the case for rich low-overhead cross-layer abstractions as a
highly effective means to address the above challenges. These abstractions are
designed to communicate higher-level program information from the application to
the underlying system and hardware in a highly efficient manner, requiring only
minor additions to the existing interfaces. In doing so, they enable a rich
space of hardware-software cooperative mechanisms to optimize for performance.  
We propose 4 different approaches to designing
richer abstractions between the application, system software, and hardware
architecture in different contexts to significantly improve programmability,
portability, and performance in CPUs and GPUs:
\One Expressive Memory: A unifying cross-layer abstraction to express and
communicate higher-level program semantics from the application to the
underlying system/architecture to enhance memory optimization; 
\Two The Locality Descriptor: A cross-layer abstraction to express and exploit
data locality in GPUs; 
\Three Zorua: A framework to decouple the programming model from management of on-chip
resources and parallelism in GPUs; 
\Four Assist Warps: A helper-thread abstraction to dynamically leverage
underutilized
compute/memory bandwidth in GPUs to perform useful work. In this thesis, we
present each
concept and describe how communicating higher-level program information from the
application can enable more intelligent resource management by the architecture
and system software to significantly improve programmability, portability, and
performance in CPUs and GPUs.
\end{abstract}

%% file: sections/introduction.tex
\section{Introduction}
\label{sec:intro}
\pagenumbering{arabic}
Efficient management of compute and memory resources, today and in the future, is as critical
as ever to
maximize system performance and energy efficiency. Important goals when it comes to effectively managing system resources include:
\textbf{\textit{programmability}}, to minimize programmer effort in optimizing for
performance; \textit{\textbf{portability}} of software 
optimizations across architectures with different resources/characteristics and
in the presence of co-running applications that share resources; and
\textit{\textbf{resource efficiency}}, to maximize utilization of all available
resources and effectively leverage features of a given architecture. 

The importance of managing the diverse compute/memory resources in an
easy-to-program, portable, and efficient manner has inspired a large body of
research in programming languages, software frameworks,
compilers, and architectures. We, however, argue that ever-growing complexity at each
level of the stack \emph{cannot}, in isolation, fully achieve the three-fold goal of
programmability, portability, and resource efficiency: we are
fundamentally constrained by current cross-layer abstractions that are
\textit{not} designed to optimize for these goals. 
The levels of the computing 
stack{\textemdash}\textit{\textbf{application} (the application/programming
model), \textbf{system} (OS/runtime/compiler),
and \textbf{architecture} (the hardware architecture)}{\textemdash}still interact with
the traditional interfaces and abstractions (e.g., virtual memory,
instruction set architecture (ISA)), which were primarily designed to convey
\textit{functionality}, rather than for the efficient management of resources which
is critical for \textit{performance}. 

%

\subsection{Motivation: Narrow Hardware-Software Interfaces Constrain Performance,
Programmability, and Portability.}
The existing interfaces have two important implications that make achieving
programmability, portability, and resource efficiency significantly 
challenging: 

\textbf{Implication 1: \textit{Diminishing returns from hardware-only
approaches.}} 
The existing interfaces between layers of the computing stack, e.g., the
instruction-set architecture (ISA) and virtual memory, strip any
application down to the basics of what is required to execute code
correctly: a sequence of instructions and memory accesses. 
Higher-level information{\textemdash}even the simple notion of data structures,
their
access semantics, data types, and properties{\textemdash}are all lost in
translation. In other words, there is a large \textit{semantic gap} between the
application and the underlying system/hardware. Today, system designers \emph{work around}
this gap and the vast majority of optimizations in the hardware architecture (optimizing caches, memory,
coherence, computation, and so on) try to predict/infer program behavior.
However, these approaches are fundamentally
limited by how much application information is visible. Hence, we are seeing
diminishing returns from hardware-only approaches in general-purpose computing. 
There have been \emph{numerous}
proposals for cross-layer optimizations and hardware-software cooperative
mechanisms, but since they require full-stack
changes for a \emph{single} optimization, they are challenging to adopt.


\textbf{Implication 2: \textit{Application/system 
manages 
low-level hardware resources with limited visibility and access.}} 
With the existing narrow hardware-software interface, the architecture is heavily
constrained in resource management. Thus, we rely on the application software to
do much of the heavy lifting in optimizing code to the specifics of each
architecture. The application and system software need to be aware of
low-level system resources, and manage them appropriately to tune for
performance. GPU programming is challenging task today, as many hardware
resources that are critical for performance need to be directly allocated and
managed by the programmer. This causes challenges in
programmability, portability, and resource efficiency. The software may \emph{not}
always have visibility into available resources such as available cache space (e.g., in virtualized
environments) and even if it does, software has little access to many hardware
features that are critical when optimizing for performance (e.g., caching
policies, memory mapping). Furthermore, software \emph{cannot} easily adapt to changes
in the runtime environment (e.g, co-running applications, input data).

\subsection{Our Approach: \textit{Rich Cross-Layer Abstractions
to Enable Hardware-Software Cooperative Mechanisms}}

In this thesis, we propose \emph{unifying} cross-layer abstractions to
bridge the semantic gap between the application and underlying system/hardware.
These abstractions directly communicate higher-level program information, such
as data structure semantics, parallelism, and data access properties, to the
lower levels of the stack: compiler, OS, and hardware. This information is
conveyed by the programmer using our programming abstractions or automatically
inferred using software tools. These abstractions enable a rich space of
hardware-software cooperative mechanisms to improve performance. The abstractions are expressive enough to convey
a wide range of program information. At the same time, they are designed to be
highly portable and low overhead, requiring only small additions to existing
interfaces. This makes them \emph{highly practical} and easy to adopt. We look
at 4 different contexts in CPUs and GPUs where new rich cross-layer abstractions
enable new hardware-software cooperative mechanisms that address challenges in
performance, programmability, and portability. 

Cross-layer cooperative mechanisms are highly effective because providing the compiler, OS, and hardware a \emph{global} view of program
behavior, ahead of time, enables these components to actively optimize
resources accordingly. For example, we demonstrated that knowledge of data
structures and their access semantics enables the OS to place data intelligently
in memory to maximize locality and parallelism~\xmem. Similarly, knowledge
of the locality semantics of GPU programs enables the hardware thread scheduler
to co-locate threads that share data at the same core to enhance data
locality~\ldesc. These are just two simple examples in a large space of
hardware-software codesigns in general-purpose
processors, including numerous
previously-proposed cross-layer optimizations.

 
In this thesis, we propose cross-layer abstractions, along with the
new hardware-software mechanisms that they enable in
different contexts in CPUs and GPUs, including: (1) rich
 programming abstractions that enable expression of application-level
 information and programmer intent, completely agnostic to the underlying system
 and hardware; (2) a cross-layer system that efficiently integrates the
 OS/runtime system and compiler, enabling these components to flexibly tap into
 a rich reservoir of application information; and (3) a low-overhead
 implementation in the hardware architecture.

This thesis, hence, provides evidence for the following thesis
statement:

\textbf{\textit{A rich low-overhead cross-layer interface that
communicates higher-level application information to hardware enables many
hardware-software cooperative mechanisms that significantly
improve performance, portability, and programmability.}} 

\subsection{Key Benefits}
\label{sec:benefits}
While there is a wide space of research opportunities into what a rich
cross-layer abstraction enables, the benefits we demonstrated are detailed
below.

\textbf{\textit{(1) Enabling intelligent  and application-specific cross-layer
optimizations in the system/hardware:}} In addition to the aforementioned
examples, more generally, communicating program information enables
more intelligent cross-layer optimizations to manage caches, memory, coherence,
computation, and so on. Examples of this information include semantics of 
how a program accesses its data structures and properties of the data itself.
The system/hardware can now effectively \emph{adapt} to
the application at runtime  
to improve overall system performance (Chapter~\ref{sec:xmem},~\ref{sec:ldesc}): For example, the
Locality Descriptor~\ldesc leverages knowledge of an application's data
access properties to enable coordinated thread scheduling and data placement in
NUMA (non-uniform memory access) architectures. A richer cross-layer abstraction
also enables \emph{customized optimizations}: Expressive Memory~\xmem in
CPUs enables adding specialization in the
\emph{memory hierarchy} to accelerate specific applications/code segments. For
example, the abstraction enables transparently integrating a customized
prefetcher for different types of data structures and their access semantics
(e.g, hash table, linked list, tensor). Similarly, with Assist
Warps~\caba, we demonstrate customized hardware data compression in GPUs,
specific to the data layout of any data structure.

\textbf{\textit{(2) Enabling hardware to do more for productivity and
portability:}} 
In addition to performance, access to program semantics and programmer intent
enables hardware to improve productivity and portability. We
demonstrate~\all that architectural techniques
can improve the performance portability of optimized code and reduce the effort
required to write performant code when privy to the semantics of software
optimizations (e.g., cache tiling, graph locality optimizations). For example,
by communicating to hardware the tile size and access pattern when using a
cache-blocking optimization (e.g., in stencils and linear algebra), the cache
can intelligently coordinate eviction policies and hardware prefetchers to
retain as much as possible of the tile in the cache, irrespective of available
cache space~\xmem. This avoids cache thrashing and the resulting
performance cliffs that may occur when the available cache is less than what the
program was optimized for{\textemdash}thus improving \emph{performance
portability}. 
Similarly, performance cliffs are rampant in GPU programs, 
requiring great precision in tuning code to efficiently use resources such as
scratchpad memory, registers, and the available parallelism. We demonstrate how
enabling more intelligent hardware can significantly alleviate performance
cliffs and enhance productivity and portability in GPUs via careful management
of resources at runtime (Chapter~\ref{sec:zorua},~\ref{sec:caba}).

\subsection{Overview of Research}
We propose 4 different approaches to designing
richer abstractions between the application, system software, and hardware
architecture, which we briefly describe next. 
\subsubsection{Expressive Memory~\cite{xmem}: A rich and low-overhead cross-layer
abstraction in CPUs to enable hardware-software cooperative mechanisms in
memory~\xmem}
In this work, we proposed a new cross-layer interface in CPUs,
Expressive Memory (XMem), to communicate higher-level
program semantics from the application to the operating
system and architecture. To retain the existing abstraction as is, we instead
associate higher-level program semantics with \emph{data} in memory. Examples of
this information include:
\emph{(i)} data structure semantics and access properties; \emph{(ii)}
properties of the data values
contained in the data structures, e.g., data  sparsity,
data type; \emph{(iii)} data locality. The OS and any component in hardware
(e.g., cache, memory controller) can simply query the XMem system with any
memory address to retrieve the associated program information. 
XMem was architected such that the entire interface is simply defined by three
key operators that are recognized by all levels of the
stack{\textemdash}\texttt{CREATE}, \texttt{MAP}, and \texttt{ACTIVATE}. These
operators enable rich expressiveness to describe more complex semantics in any
programming language. These operators can be demonstrably implemented with low
overhead and are portable across different architectures. We demonstrated
significant benefits in enabling the system/hardware to do more for performance,
productivity, and portability and specializing for different applications. The
abstraction was designed to flexibly support many cross-layer optimizations. To
demonstrate its utility and generality, we presented 9 different use cases.

\subsubsection{The Locality Descriptor~\cite{ldesc}: Expressing and leveraging data locality
in GPUs with a rich cross-layer abstraction~\ldesc}
While modern GPU programming models are designed to explicitly express
\emph{parallelism}, there is no clear way to express semantics of the program
itself: i.e.,
how the thousands of concurrent threads access its data structures. 
We designed a rich cross-layer abstraction that describes how the hierarchy of
threads in the GPU programming model access each data structure and the access
semantics/properties of the data structures themselves. We then leverage this
abstraction to significantly improve the efficacy and ease with which we can
exploit \emph{data locality} in modern GPUs{\textemdash}both \emph{reuse-based}
locality, to make
efficient use of the caches, and  \emph{NUMA} locality, to place data and
computation in near proximity in a non-uniform memory access (NUMA) system. 

Exploiting data locality in GPUs today is a challenging but elusive feat both to
the programmer and the architect. Software has no access to key components (such
as the thread scheduler) and hardware misses key information such as: which
threads share data?
We designed a powerful abstraction (named by its use case: the Locality
Descriptor) that communicates the locality semantics of any application to the
hardware. This enables hardware to transparently coordinate many locality
optimizations such as co-scheduling threads that share data at the same core and
placing data close to the threads that use it. The programming interface is
designed to be seamlessly integrated into modern GPU programming models (e.g.,
CUDA and OpenCL). The abstraction's semantics are defined such that it can be
automatically generated via software tools and is highly portable, making no
assumptions about the underlying architecture. 

\subsubsection{Zorua~\cite{zorua,zorua-bc}: Decoupling the GPU programming model from hardware resource
management~\zorua}
In accelerators, such as modern day GPUs, the available parallelism as well as
the memory resources need to be \textit{explicitly} managed by the
programmer. There exists no powerful abstraction between the
architecture and the programming
model, and the management of many hardware resources
is tied to the programming model itself. This leads to underutilization of
resources (and hence, significant loss in performance) when the application is
not well tuned for a given GPU.
Even when an application is perfectly tuned for one GPU, there can still be a
significant degradation in performance when running the same program on a
different GPU. Furthermore, it is unclear how to write such architecture-specific
programs in virtualized environments, where the same resources are being shared
by multiple programs. 

To achieve the three-fold goal of enhanced programmability, portability, and
resource efficiency, we designed a new hardware-software cooperative framework,
\textit{\textbf{Zorua}}, to enhance the interface between the
programming
model and architecture for the management of several critical compute and memory
resources. Zorua \textit{decouples} the resource management as
specified by the programming model and the actual utilization in the system
hardware by effectively
\textit{virtualizing} each of the resources (register file, scratchpad memory,
and thread slots) in hardware. Zorua also communicates fine-grained information
regarding the application's future resource requirements at regular intervals to
the hardware with a new hardware-software interface. This virtualization, along with the rich interface, 
enables the hardware to intelligently and dynamically manage these resources depending on the
application behavior and makes the
performance of any program far
less sensitive to the software-provided resource specification. High performance
is thus made far easier to
attain and performance is portable across generations of the architecture which
may have varying amounts of each resource. 

\subsubsection{Assist Warps~\cite{caba,caba-bc}: A cross-layer abstraction and hardware-software
framework to leverage idle
memory and compute resources \caba}
In modern throughput-oriented processors, \textit{even with highly optimized
code}, imbalances between the compute/memory resources
requirements of an application
and the resources available in hardware can lead to significant idling of compute
units and available memory bandwidth. The current programming models and
interface to the architecture provide \textit{no simple abstraction} to manage the
utilization of critical resources such as memory/compute bandwidth and on-chip
memory. To \textit{leverage} this undesirable
wastage to perform useful work,
we propose a new hardware-software abstraction{\textemdash}\textit{the assist
warp}{\textemdash}in the GPU programming model and architecture. Assist
warps enable
light-weight execution of helper-thread code alongside the primary application to
perform optimizations to accelerate the program (such as
data compression or prefetching) and perform background tasks, system-level tasks,
etc. Assist warps automatically adapt to the availability of resources and
unbalances in the primary application's execution to increase overall efficiency and
throughput.  

\subsection{Related Work}
\label{sec:relatedwork}
In this section, we provide a brief overview of prior work that address similar
challenges in enhancing programmability, portability, and resource efficiency
and works that propose cross-layer interfaces.
We then contrast the \textit{general} approaches taken by these works with
the approaches
taken in this thesis. We discuss related work specific to each of the proposed
works at the
end of each chapter. 

\subsubsection{Expressive Programming Models and Runtime Systems} 
\highlight{Numerous software-only approaches tackle the 
disconnect between an application, the OS, and the underlying resources
via
programming models
and runtime systems
that allow explicit expression of data
locality and independence~\cite{x10-charles-oopsla05,chapel-chamberlain-ijhpca,sequoia-fatahalian-sc06,legiondependent-treichler-oopsla16,legionstructure-bauer-sc14,
realm-treichler-pact14,regent-slaughter-sc15,hpt-yan-lcpc09,hta-bikshandi-ppopp06,
legion-bauer-sc12, programming-guo-ppopp08, tida-unat-hpc16} in
the programming model.} This explicit expression enables the programmer and/or
runtime system to make effective 
memory placement decisions in a NUMA system or produce code that is
optimized to effectively leverage the cache hierarchy. For example, the Legion
programming system~\cite{legion-bauer-sc12} provides software abstractions to
describe properties of data such as locality and independence. The programmer
and runtime system can then explicitly place data in the memory hierarchy to
maximize parallelism and memory efficiency.  
These approaches in general have several shortcomings. First, they are
entirely software-based and are hence limited to using the \emph{existing} interfaces
to the architectural resources. Second, programming model-based approaches
\emph{require} rewriting applications to suit the model. For example, programs
need to adapted to the Legion programming model to expose parallelism and
locality using the model's abstractions. This requires explicit
programmer effort to ensure correctness is retained. 
Third, these
systems are very specific to an application type (e.g., operations on
tiles~\cite{tida-unat-hpc16},
arrays~\cite{sequoia-fatahalian-sc06}). Only those programs that can be expressed
with Legion's task-based programming model can leverage its benefits. 

In contrast, in this thesis, we proposed abstractions~\all,
that are, first, \textit{cross-layer},
and are hence \emph{not} limited to existing interfaces between hardware and
software and enable hardware-software cooperative mechanisms. As we demonstrate, this enables significant performance,
programmability, and portability benefits. Second, all the abstractions proposed in this
thesis \emph{retain existing programming models and execution paradigms} to
minimize programmer and developer effort. Third, each approach taken in this thesis
is \emph{general} and is \emph{not} limited to any programming
language or application.  Programming model/runtime system approaches are
orthogonal to our proposed approaches, and can
\textit{leverage} the abstractions we propose to enable a wider range of
optimizations.

\subsubsection{Leveraging Hints, Annotations, and Software Management for
Performance Optimization}
A large body of prior work aims to leverage the benefits of static program
information in the form of hints, annotations, or directives in 
performance optimization. For example, Cooperative Cache
Scrubbing~\cite{cooperativescrubbing-sartor-pact14} is a hardware-software
mechanism that communicates to the cache which data blocks will not be used any
longer in the program (dead blocks), so they can be evicted from the cache to
make space for more useful data. More generally, these include \One hint-based
approaches, such as 
software prefetch instructions~\cite{intel-prefetching} and cache
bypass/insertion/eviction hints~\cite{swcache-jain-iccad01,compilerpartitioned-ravindran-lctes07,popt-gu-lcpc08,pacman-brock-ismm13,evictme-wang-pact02,generatinghints-beyls-jsysarch05,keepme-sartor-interact05,compilerassisted-yang-lcpc04,cooperativescrubbing-sartor-pact14,runtimellc-pan-sc15,prefetchtasklifetimes-papaefstathiou-ics13,radar-manivannan-hpca16,compile-beyls-epic2,modified-tyson-micro95}; \Two hardware-software
cooperative prefetch techniques~\cite{effective-chen-micro95,compilerassisted-vanderwiel-vlsi99,chiueh-sunder-sc94,integrated-gornish-icpp94,guided-wang-isca03,controlling-jain-csg01,hybrid-skeppstedt-icpp97,prefetchingtechnique-karlsson-hpca00,loopaware-fuchs-micro14,techniques-ebrahimi-hpca09}
that use compiler
analysis or annotations
to inform a hardware prefetch engine; and \Three program annotations to place
data in heterogeneous memories
(e.g.,~\cite{pageplacement-agarwal-asplos15,characterizing-luo-dsn14,flikker-liu-asplos11}).

The approaches taken in this thesis, XMem~\xmem, Locality
Descriptor~\ldesc, Zorua~\zorua, and CABA~\caba,
differ from these works in several ways. First, many of these
prior works seek to inform hardware components with \emph{specific directives}
that override dynamic
policies by enforcing \emph{static} policies. This loss in dynamism introduces
challenges when the workload behavior changes, the underlying architecture
changes or is unknown (portability), or in the presence of co-running
applications~\cite{comparing-leverich-isca07,whirlpool-mukkara-asplos16,zorua}.
Our approaches
do \emph{not} \emph{direct} policy at any component but only provide
higher-level program semantics. System/hardware components can use this information
to \emph{supplement}
their dynamic management policies. 
Second, these prior works are \emph{specific} to an optimization (e.g., prefetching,
cache insertion/eviction). For example, Cooperative Cache Scrubbing is only
applicable to dead block eviction from the cache. Our approaches, XMem~\xmem and Locality
Descriptor~\ldesc, however, provide a \emph{general} interface to communicate
program semantics that can be leveraged by \textit{many} system/architectural
components.

\subsubsection{Enhancing Programming Ease and Portability in GPUs}
There is a large body of work that aims to improve programmability and
portability of modern GPU applications using software tools, such as
auto-tuners~\cite{toward-davidson-iwapc10,atune,maestro,autotuner1,autotuner-fft},
optimizing
compilers~\cite{g-adapt,optimizing-compiler1,parameter-selection,porple,optimizing-compiler2,sponge},
and high-level programming languages and
runtimes~\cite{cuda-lite,halide,hmpp,hicuda}. For example, Porple~\cite{porple}
is an optimizing compiler that automatically selects the memory allocation in
registers and scratchpad memory for GPU programs. hiCUDA~\cite{hicuda} is a
directive-based programming language that automatically chooses the lower-level
specifications required in GPU programs (i.e., registers, scratchpad, threads
per block), based on the directives provided by the programmer. These tools tackle a multitude of
optimization challenges, and have been
demonstrated to be very effective in generating high-performance portable
code. 

However, there are several shortcomings in these works in contrast with our works, Locality Descriptor~\ldesc and Zorua~\zorua. First,
these prior works
often require profiling
runs~\cite{toward-davidson-iwapc10,atune,maestro,porple,optimizing-compiler1,optimizing-compiler2}
on the GPU to determine the best performing resource specifications. Porple
requires runtime profiling to determine the program's access patterns before
selecting how much register space or scratchpad memory should be allocated to
the program. These runs have to
be repeated for each new input set and GPU generation. Second, software-based approaches still require significant
programmer effort to write code in a manner that can be exploited by software
to optimize the resource utilization. hiCUDA requires rewriting any program with
its directive-based language. Zorua~\zorua is software transparent and the
Locality Descriptor~\ldesc does not require rewriting the application, but
only requires provide hint-based annotations that do not affect program
correctness.
Third, selecting the best-performing resource specifications statically using
software tools is a challenging task in virtualized
environments (e.g., clouds), where it is unclear which kernels
may be run together on the same Streaming Multiprocessor (SM) or where it is not known, a priori, which GPU
generation the application 
may execute on. Finally, software tools assume a fixed amount of available
resources. This leads to \textit{dynamic} underutilization due to static allocation of
resources, which cannot be easily addressed by these tools. None of the above
tools, including Porple and hiCUDA, can handle dynamic recompilation in the
presence of co-running applications or address dynamic underutilization of
hardware resources.
Furthermore, these
prior works are largely orthogonal can be used in conjunction with our proposed approaches to
further improve performance.

\subsubsection{Tagged Architectures} Prior work proposes to associate
software-defined metadata with each memory
location in the form of tagged/typed
memory~\cite{architectural-dhawan-asplos15,tagged-feustel-taco73,hardware-zeldovich-osdi08,mondrian-witchel-asplos02\remove{,raksha-dalton-isca07}}.
These proposals are typically used for fine-grained memory access protection,
debugging, etc., and usually incur non-trivial
performance/storage overhead. For example,
PUMP~\cite{architectural-dhawan-asplos15} associates every word in the memory
system with a software-defined metadata tag. These tags are then used to enforce
security policies in hardware, e.g., avoiding buffer overflows. In contrast,
XMem~\xmem aims to deliver \emph{general} program
semantics to many system/hardware components to aid in
\emph{performance optimization}. This necessitates a low overhead implementation
that is also general enough to enable a wide range of cross-layer optimizations,
while not sacrificing programmability. While a fundamental \emph{component}
of XMem is the metadata tracking system similar to tagged memories, to achieve
the above goal in performance optimization requires several other key
components: a hardware-software \emph{translator} that enables a many
cross-layer optimizations with a common set of information provided by the
programmer and alleviates challenges in portability and programmability; and
a full hardware-software system design that partitions work between the compiler,
OS, and hardware to minimize system complexity.
Furthermore, \X is designed to enable a number of features and benefits that
cannot be obtained from tagged/typed architectures: \One a flexible and
extensible abstraction to \emph{dynamically} describe program behavior with
\Xlib; and \Two low-overhead interfaces to many hardware components to easily
access the expressed semantics, including the prefetcher, caches, memory
controller, etc. 
PARD~\cite{pard-ma-asplos15} and Labeled RISC-V~\cite{labeled-yu} are tagged
architectures that enable labeling memory requests with tags to
applications, VMs, etc. These tags are used to convey an
application's QoS,
security requirements, etc., to hardware. \X is similar in that it provides an
interface to hardware to convey information from software. However, unlike these
works~\cite{pard-ma-asplos15,labeled-yu}, we design a new abstraction (the atom) to flexibly
express program semantics that can be seamlessly integrated into programming
languages, runtime systems, and modern ISAs. The atom lends itself to a
low-overhead implementation to convey software semantics to hardware
components \emph{dynamically} and at flexible granularities. \X can potentially
\emph{leverage} tagged
architectures to communicate atom IDs to
different hardware components. Hence, PARD and Labeled RISC-V are complementary
to XMem.

%% file: sections/xmem.tex
\section{Expressive Memory}
\label{sec:xmem}
This chapter proposes a rich low-overhead interface to enable the operating system and
hardware architecture to leverage key higher-level program information regarding
how an application accesses its data to optimize memory system performance in
CPUs. We
first motivate the programmability, portability, and resource efficiency
challenges that exist in modern CPUs when optimizing for memory system
performance. We then describe our proposed cross-layer abstraction, Expressive
Memory, and detail its design and effectiveness in addressing these challenges.  
\subsection{Overview}

As discussed in Section~\ref{sec:intro}, traditionally, the key interfaces between the software stack and the architecture (the
ISA and virtual memory) have been primarily designed to convey
program \emph{functionality} to ensure the program is executed as required by
software. An application is converted into 
ISA instructions and a 
series of accesses to virtual memory for execution in hardware. The
application is, hence, stripped down to the basics of what is necessary to execute the
program correctly, and the \emph{higher-level semantics} of the program are lost. For example, even the simple higher-level notion of different \emph{data structures} in a
program is \emph{not} available to the OS or hardware architecture, which
deal only with virtual/physical
pages and addresses. While the
higher-level semantics of data structures may be
irrelevant for correct execution, these semantics could prove very useful to the system
for \emph{performance optimization}. 

\remove{Additional knowledge in terms of
classifying data that is \emph{similar} in terms of the data that is stored and how it
is accessed based on the higher-level semantics of a data structure could prove
very useful in different system-level or architectural memory optimizations such
as intelligent data placement. }

\remove{In this work, we narrow the scope of interest to the program semantics that are
primarily useful for \emph{memory optimization}, i.e., software and
hardware techniques that aim to improve the performance and efficiency of the
memory system. Another example of such key program semantics relevant to the
memory system is the use of
software optimizations such as \emph{cache tiling}
~\cite{mehta-tile-taco13, automatic-yuki-cgo10,
  spl-xiong-pldi01, automated-whaley-pc01,polly-bondhugula-pldi08,
  tile-coleman-pldi95, stencil-henretty-sc13} 
which increases reuse in the cache by tuning the working set to fit into a specific
level of cache. The caching hierarchy, however, is largely unaware of such
software optimizations and sophisticated generic caching policies may not
be able to capture the program behavior leading to \emph{cache thrashing} if
the tile is not appropriately optimized to fit the cache.}

There is, in other words, a \emph{disconnect} or
\emph{semantic gap} between the levels of the computing 
stack when it comes to conveying higher-level program semantics from the
application to the wide range of system-level and architectural
components that aim to improve performance. While the implications of the
disconnect are far-reaching, in this work, we narrow the focus to 
a critical component in determining the overall system efficiency, \emph{the memory
  subsystem}.
Modern systems employ a large variety of components to optimize memory
performance (e.g., prefetchers, caches, memory controllers).
The semantic gap has two important implications:

\textbf{Implication 1.}
The OS and hardware memory subsystem components are forced to
\emph{predict} or \emph{infer} program behavior when optimizing for
performance. This
is challenging because: \One each component (e.g., L1 cache, memory
controller) sees only a \emph{localized} view of the data accesses made by
the application and misses the bigger picture, \Two 
\emph{specialized} hardware may be required for each component optimizing for memory, and \Three the optimizations are typically only \emph{reactive}
as the program behavior is \emph{not} known a priori. 

\textbf{Implication 2.}
\highlight{Software is forced to optimize code to the specifics of the
underlying 
architecture (e.g., by tuning \emph{tile
size} to fit a specific cache size). 
Memory resource availability, however, can change or be unknown (e.g., in virtualized
environments or in the presence of co-running applications).
As a result, software optimizations are often unable to make accurate
assumptions regarding memory
resource availability, leading to significant challenges in \emph{performance
portability}.} 


The challenges of predicting program behavior and hence the benefits of
knowledge from software in memory system optimization are well known~\cite{swcache-jain-iccad01,compilerpartitioned-ravindran-lctes07,popt-gu-lcpc08,pacman-brock-ismm13,evictme-wang-pact02,generatinghints-beyls-jsysarch05,keepme-sartor-interact05,compilerassisted-yang-lcpc04,cooperativescrubbing-sartor-pact14,
runtimellc-pan-sc15,prefetchtasklifetimes-papaefstathiou-ics13,radar-manivannan-hpca16,modified-tyson-micro95,pageplacement-agarwal-asplos15,whirlpool-mukkara-asplos16,datatiering-dulloor-eurosys16,
flikker-liu-asplos11,effective-chen-micro95,compilerassisted-vanderwiel-vlsi99,chiueh-sunder-sc94,integrated-gornish-icpp94,guided-wang-isca03}.
There have been numerous hardware-software cooperative
techniques proposed in the form of fine-grain hints implemented as new ISA
instructions (to aid cache replacement, prefetching, etc.)~\cite{swcache-jain-iccad01,compilerpartitioned-ravindran-lctes07,popt-gu-lcpc08,pacman-brock-ismm13,evictme-wang-pact02,generatinghints-beyls-jsysarch05,keepme-sartor-interact05,compilerassisted-yang-lcpc04,cooperativescrubbing-sartor-pact14,
runtimellc-pan-sc15,prefetchtasklifetimes-papaefstathiou-ics13,radar-manivannan-hpca16,modified-tyson-micro95}, 
program annotations or directives to convey program
semantics and programmer
intent~\cite{pageplacement-agarwal-asplos15,whirlpool-mukkara-asplos16,datatiering-dulloor-eurosys16,
flikker-liu-asplos11,popt-gu-lcpc08}, or hardware-software co-designs for 
specific optimizations~\cite{effective-chen-micro95,compilerassisted-vanderwiel-vlsi99,chiueh-sunder-sc94,integrated-gornish-icpp94,guided-wang-isca03}.
These approaches, however, have  two significant
shortcomings. First, they are designed for a \emph{specific memory optimization}
and are 
limited in their implementation to address only challenges specific to that
optimization. As a result, they require changes across the stack for a
\emph{single optimization} (e.g., cache replacement, prefetching, or data
placement). Second, they are often very specific directives to instruct
a particular component to behave in a certain manner (e.g., instructions to
prefetch specific data or prioritize certain cache lines). These specific
directives create
portability and programmability concerns because these optimizations may \emph{not}
apply across different architectures and they require significant
effort to
understand the hardware architecture to ensure the directives are useful.

\textbf{Our Goal.}  
In this work, we ask the question: \emph{can we design a unifying general abstraction and
a cohesive set of interfaces
between the levels of the system stack to communicate key program
semantics from the application to all the system-level and architectural components?} In response, we
present \Xfull (\X), a rich cross-layer
interface that provides a new \emph{view} of the program data to the entire
system.
Designing \X in a \emph{low-overhead, extensible, and general} manner requires addressing several non-trivial 
challenges involving conflicting tradeoffs between generality and overhead,
programmability and effectiveness (\secref{sec:challenges}). In this work,
we provide a first attempt at designing a new end-to-end system to achieve our goal while addressing these challenges.

{\textbf{\Xfull} comprises two key components: 

\textbf{(1) The Atom.} We introduce a new hardware-software
abstraction, the \emph{atom}, which is a region of virtual memory with
a set of well-defined properties (\secref{sec:atom-abstraction}).
Each atom maps to data that is \emph{semantically similar}, e.g., a
data structure, a \emph{tile} in an array, or any pool of data with
similar properties.  Programs explicitly specify atoms that are
communicated to the OS and hardware. Atoms carry program information
such as: \One data properties (e.g., data type, sparsity,
approximability), \Two access properties (e.g., access pattern,
read-write characteristics), and \Three data locality (e.g., data
reuse, working set). The atom can also track properties of data that
\emph{change} during program execution.

\highlight{\textbf{(2) System and Cross-layer
Interfaces.\@} Figure~\ref{fig:xmem_overview} presents an overview of
this component:
\One The interface to the application
enables software to explicitly express atoms via
program annotation, static compiler analysis, or dynamic profiling \ding{182};
\Two The \X system enables summarizing, conveying, and storing the expressed
atoms \ding{183}; \Three The interface to the OS and 
architectural components (e.g., caches, prefetchers) provides key supplemental information to aid optimization \ding{184}. This
interface enables any system/architectural component to simply
\emph{query} the \X system for the higher-level semantics attached to a
memory address \ding{185}.} 

\begin{figure}[h]
  \centering
  \vspace{-2mm}
  \includegraphics[width=0.49\textwidth]{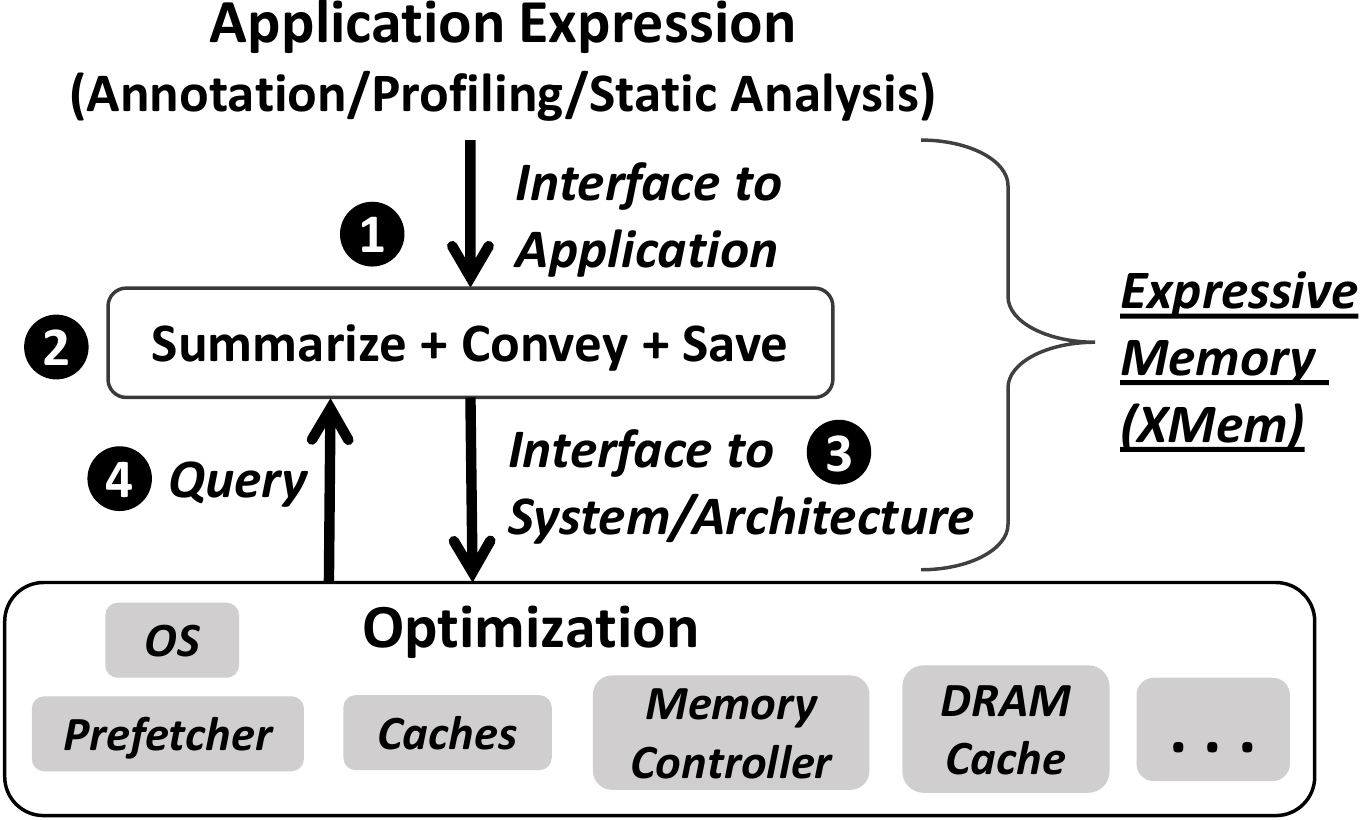}
  \caption{\X: the system and interfaces.}
  \label{fig:xmem_overview}
\vspace{-2mm}
\end{figure} 

\remove{\X
enables \emph{explicit} expression of the key program semantics by the software
(via
program annotation, static compiler analysis, or dynamic profiling)
, which
is then made available to the OS and architectural components (e.g.,
cache, prefetchers, memory controllers, etc.) to aid in a variety of memory 
optimization techniques. } 

\begin{table*}[p]\small
  \centering
  \caption{Summary of the example memory optimizations that \X aids.}
  \vspace{-1.5mm}
  \label{table:usecases}
  \input{tables/usecases}

\end{table*}

\textbf{Use Cases.}  \X is designed to be a \emph{general} interface
to aid a wide range of memory optimizations.  In this work, we first
demonstrate the benefits of \X in enhancing the portability of
\emph{software-based cache optimizations} (\secref{sec:cache}). The
effectiveness of
such optimizations (e.g., \emph{cache tiling}~\cite{mehta-tile-taco13, automatic-yuki-cgo10,
  spl-xiong-pldi01, automated-whaley-pc01,polly-bondhugula-pldi08,
  tile-coleman-pldi95, stencil-henretty-sc13}) is highly susceptible to
changes in cache space availability:\remove{ (e.g., due to co-running
applications) as these optimizations make assumptions regarding available cache
space.}
If the available cache space at runtime is \emph{less} than what the program was
optimized for, \emph{cache thrashing} often ensues. We demonstrate that by leveraging data locality
semantics (working set size and data reuse), we can enhance and coordinate the cache
management and prefetching policies to avoid cache thrashing and
ensure high hit rates are retained, thereby improving the portability of the
optimization. We demonstrate that when software optimizations inaccurately
assume available cache space, \X reduces the loss in
performance from 55\% in the baseline system to 6\% on average. 
Second, we demonstrate the performance benefits of \emph{intelligent
OS-based page placement in DRAM} by leveraging knowledge of data
structures and their access semantics (\secref{sec:dram}). \X improves performance by \One
isolating regular data structures with high row buffer locality in
separate banks and \two spreading out irregular data
structures across many banks/channels to maximize parallelism. \highlight{Our
experimental evaluation demonstrates an 8.5\% average performance improvement (up
to 31.9\%) over state-of-the-art techniques.}

More generally, Table~\ref{table:usecases} presents nine example
memory optimizations and the benefits \X can
provide over prior works that propose these optimizations in a specialized
manner.  \X's
benefits arise in three ways. First, it provides a \emph{unifying},
central interface for a wide range of optimizations that use many of
the \emph{same} semantics. Second, it partitions data into pools of
semantically-similar data.  This enables using different policies
(e.g., cache policies, compression algorithms) for different
pools of data. Third, it enhances optimizations by providing higher-level
semantics that \One are unavailable locally to each component
at runtime (e.g., distinguishing between data structures, data
properties), \Two are challenging to accurately infer (e.g., working set
size, data reuse) or \Three require profiling/monitoring to determine
(e.g., read-only/read-write, private/shared data characteristics).

This work makes the following \textbf{contributions}:
\begin{itemize}
\item This work is the first attempt to design a \emph{holistic and general} 
  cross-layer interface
  to communicate higher-level program semantics  
to the different system and architectural components in order to enable more
effective
memory optimizations in modern CPUs. 
\item To this end, we introduce \X, which comprises a new software-hardware
abstraction{\textemdash}the Atom{\textemdash}and a full end-to-end system
design.
\X 
\One is general and flexible enough to cover a wide range of program semantics
and use cases, \Two is completely decoupled from system functionality and
affects only performance not correctness, \Three can react to
phase changes in data properties during execution, and \Four has a low-overhead implementation.
\item We quantitatively demonstrate the benefits of using \X to
\One improve the portability of software-based cache optimizations
by leveraging \emph{data locality} semantics and 
\Two enhance OS-based DRAM placement by leveraging semantics of data
structures and their access properties. We highlight seven other
use cases (Table~\ref{table:usecases}).
\end{itemize}

\subsection{Goals and Challenges}
\subsubsection{Key Requirements}
There are several key requirements 
and invariants that drive the design of the proposed system: 

\textbf{\One Supplemental and hint-based.} The new interface should \emph{not} affect
functionality or correctness of the program in any way{\textemdash}it
provides only \emph{supplemental} information to help improve \emph{performance}. This reduces the necessity of obtaining precise or detailed hints, and system implementation can be \emph{simpler} as information can be conveyed/stored imprecisely.

\textbf{\Two Architecture agnosticism.} The abstraction for expressing semantics
must be based on the \emph{application} characteristics rather than the
specifics of the system, e.g., cache size, memory banks available.
This means that
the programmer/software need \emph{not} be aware of the precise workings of the
memory system resources, and it
significantly alleviates the portability challenges
when the programmer/software optimizes for performance. 

\textbf{\Three Generality and extensibility.} The interface should be general enough to flexibly
express a
wide range of program semantics that could be used to aid many
system-level and architectural
(memory) optimizations, and extensible to support more semantics and optimizations. 

\textbf{\Four Low overhead.} The interface must be amenable to an
implementation with low storage area and performance
overheads, while preserving the semantics of existing interfaces.

\subsubsection{Challenges}
\label{sec:challenges}
Current system and architectural components see only a description of the program's data 
in terms of \emph{virtual/physical addresses}.
To provide higher-level program-related semantics, we 
need to associate each address with much more information than is available
to the entire system today, addressing the following three challenges:

\textbf{Challenge 1: Granularity of expression.} The granularity of
associating program semantics with program data is challenging because the
best granularity for \emph{expressing} program semantics
is program dependent.
Semantics could be available at the granularity of an entire data structure, or
at much smaller granularities, such as a \emph{tile} in an array.
We cannot simply map program semantics to \emph{every} individual virtual address as that
would incur too 
much overhead, and the fixed granularity of a \emph{virtual page} may be too
coarse-grained, inflexible and challenging for
programmers to reason about.  

\textbf{Challenge 2: Generality vs. specialization.}  
Our archi\-tec\-ture-agnosticism requirement implies that we express higher-level information from the
application's or the programmer's point of view{\textemdash}without any
knowledge/assumptions of the memory
resources or specific directives to a
hardware component.
As a consequence, much of the conveyed
information may be either
irrelevant, too costly to manage effectively, or \emph{too
  complex} for different hardware components
to easily use. For example, 
hardware components like prefetchers are operated by simple hardware structures
and need only know prefetchable access patterns. Hence, the abstraction
must be \One \emph{high-level and architecture-agnostic}, so it can be easily expressed
by the programmer and \Two \emph{general}, in order to express a range of information useful
to many components. At the same time, it should be still amenable to translation into simple directives for each component.

\textbf{Challenge 3: Changing data semantics.} As the program executes, the
semantics of the data structures and the way they are 
accessed can change. Hence, we need to be able to 
express \emph{dynamic} data attributes in \emph{static} code, and these changing attributes
need to be conveyed to the running system at the appropriate time.
This ensures that the data attributes seen
by the memory components are accurate any time during execution.
Continual updates to data attributes at runtime can impose
significant overhead
that must be properly managed to make the approach practical.

\subsection{Our Approach: Expressive Memory}
We design \Xfull (\X), a new rich cross-layer
interface that enables 
\emph{explicit} expression and availability of key program semantics. \X comprises two
key components: \One a new hardware-software \emph{abstraction} with a
well-defined set of properties to convey program semantics and \Two a rich set of \emph{interfaces} to convey and store
that information at different system/memory components.
\subsubsection{The Atom Abstraction}
\label{sec:atom-abstraction} 
We define a new hardware-software abstraction, called an \emph{Atom}, that serves as the \emph{basic unit} 
of expressing and conveying program semantics to the system and architecture. 
An atom forms both an \emph{abstraction} for information expressed as well as a
\emph{handle} for communicating, storing, and retrieving the conveyed
information across different levels of the stack. Application programs can
dynamically create atoms in the program code,
each of which describes a specific range of program data at any given time
during execution. The OS and hardware architecture can then interpret atoms
specified in the program when the program is executed. There are three key components to an atom:
\One \emph{Attributes:} higher-level data semantics that it conveys;
\Two \emph{Mapping:} the virtual address range that it describes;
and \Three \emph{State:} whether the atom is currently active or inactive.


\subsubsection{Semantics and Invariants of an Atom}
\label{sec:atom-semantics} 
We define the invariants of the atom abstraction and
then describe the operators that realize 
the semantics of the atom.  
\begin{itemize}[wide, labelwidth=!, labelindent=0pt]
\item \textbf{Homogeneity:} All the data that maps to the same atom has the
  \emph{same} set of attributes.
\item \textbf{Many-to-One VA-Atom Mapping:} At any given time, any virtual address (VA)
can map to \emph{at most} one atom. 
\highlight{Hence, the
system/architecture can query for the atom (if any) associated with a VA and
thereby obtain any attributes associated with the VA, at any given time.}
\item \textbf{Immutable Attributes:} While atoms are dynamically created, the attributes of an atom 
\emph{cannot} be changed once created. To express different attributes for the same data, a new
atom should be created.
Atom attributes can, hence, be specified \emph{statically} in the program code.
Because any atom's attributes \emph{cannot} be updated during
execution, these attributes can be summarized and conveyed
to the system/architecture at compile/load time \emph{before execution}. This
minimizes expensive communication at runtime (Challenge 3). 
\item \textbf{Flexible mapping to data:} Any atom can be flexibly and
  dynamically mapped/unmapped to any set of data of any size.
  By selectively mapping and/or unmapping data to the same atom, an atom can be mapped to
  non-contiguous data (Challenge 1). 
\item \textbf{Activation/Deactivation:} While the attributes of an atom are
immutable and statically specified, an atom itself can be \emph{dynamically} activated and
deactivated in the program. The attributes of an atom are recognized by the
system \emph{only} when
the atom is currently active. This enables updating the attributes of any data region as the program
executes: when the atom no longer accurately describes the data,
it is deactivated and a new atom is mapped to the data. This ensures that
the system always sees the correct data attributes during runtime,
even though the attributes themselves are communicated earlier at load
time (Challenge 3). 
\end{itemize}
\begin{figure}[h]
  \centering
  \vspace{-1mm}
  \includegraphics[width=0.45\textwidth]{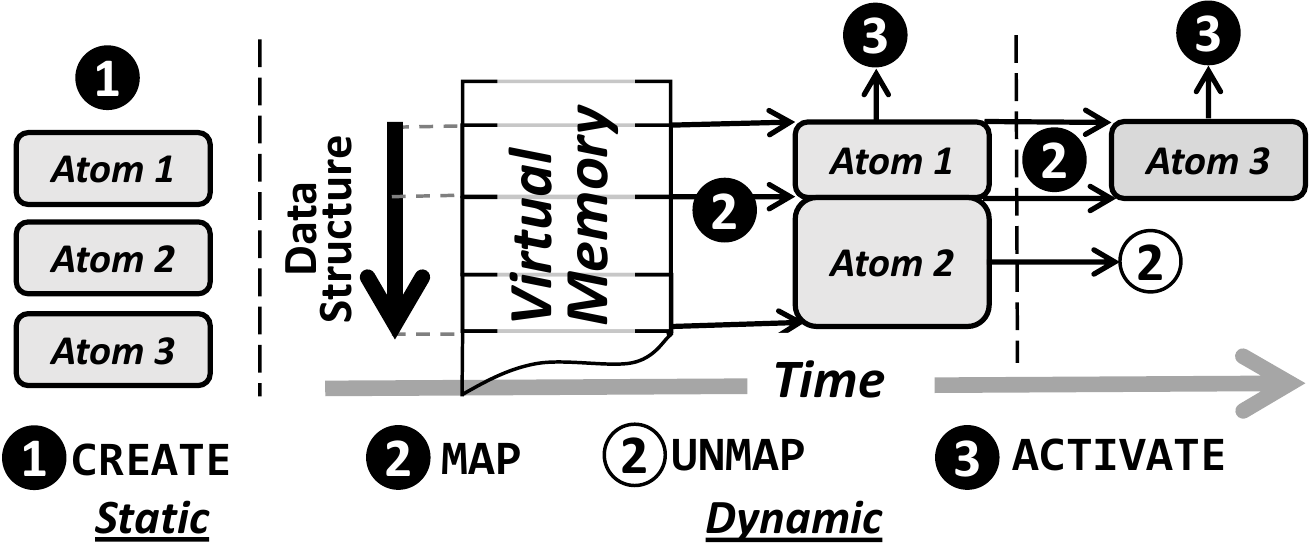}
  \vspace{-1mm}
  \caption{Overview of the three atom operators.}
  \label{fig:atom_overview}
 \vspace{-1mm}
\end{figure} 

To manipulate the three components (\emph{Attributes}, \emph{Mapping},
and \emph{State}), there are three corresponding operations that can be
performed on an atom via a corresponding library call: \One
\textbf{\texttt{CREATE}} an atom, providing it with immutable
statically-specified attributes; \Two
\textbf{\texttt{MAP/UNMAP}} an atom to/from
a range of data; and \Three \textbf{\texttt{ACTIVATE/DEACTIVATE}} an atom to dynamically tell the
memory system that the attributes of the atom are now (in)valid for
the data the atom is mapped to.

\highlight{Figure~\ref{fig:atom_overview} depicts an overview of the atom operators}. Atoms are first created in the program with statically-specified
attributes \ding{182}. 
During memory allocation (e.g., malloc), data structures are allocated ranges of
virtual memory. 
After allocation, atoms with the
appropriate attributes can be flexibly \emph{mapped} to the corresponding VA range
that each atom describes \ding{183}. Once the mapped atom is \emph{activated}, all the
system components recognize the attributes as valid \ding{184}. 
Data can be easily remapped to a different atom that describes it better as the program moves into a
different phase of execution (using the \texttt{MAP} operator \ding{183}), or just unmapped
\ding{173} when the atom no
longer accurately describes the data.
The \texttt{MAP/UNMAP} operator can be flexibly called to
selectively map/unmap multiple data ranges to/from the same atom at any granularity. The
\texttt{ACTIVATE/DEACTIVATE} operator dynamically validates/invalidates the atom attributes
relating to \emph{all} data the atom is mapped to.

\subsubsection{Attributes of an Atom and Use Cases}
\label{sec:attributes} 
Each atom contains an extensible set of attributes that convey the key program
semantics to the rest of the system. 
Table~\ref{table:usecases} lists example use cases for these attributes. 
The three classes of attributes (to date) are: 

\textbf{(1) Data Value Properties:} An expression of the attributes of the
data values contained in the data pool mapped to an atom. It is implemented as
an extensible list using a
single bit for each attribute. These attributes include data type
(e.g., \texttt{INT32}, \texttt{FLOAT32}, \texttt{CHAR8}) and data
properties (e.g., \texttt{SPARSE}, \texttt{APPROXIMABLE}, \texttt{POINTER},
\texttt{INDEX}). \remove{This class of attributes is useful for optimizations such
as cache/memory compression (\secref{sec:compression}), data approximation
(\secref{sec:approximation}), index/pointer-based prefetching
(\secref{sec:prefetching}), etc.}

\textbf{(2) Access Properties:} This describes three key characteristics of the
data the atom is mapped to:
\begin{sloppypar}
\begin{itemize}[wide, labelwidth=!, labelindent=0pt]
\item \textbf{\texttt{AccessPattern}}: This attribute 
defines the \texttt{PatternType}, currently either \texttt{REGULAR} (with a specific stride that is also
specified), \texttt{IRREGULAR} (when the access pattern is \emph{repeatable} within the data range, but with no repeated stride, e.g., graphs), or 
\texttt{NON\_DET}
(when there is no repeated pattern). \remove{The access pattern is
useful for accurate hardware prefetching (\secref{sec:prefetching}) and data
placement in DRAM (\secref{sec:dram}). }
\item \textbf{\texttt{RWChar}}: This attribute describes the read-write
characteristics of data at any given time, currently either \texttt{READ\_ONLY}, \texttt{READ\_WRITE}, or \texttt{WRITE\_ONLY}. It could also be extended to include varying degrees of read-write intensity,
and include shared/private information. \remove{This attribute is useful in managing
non-volatile memories (\secref{sec:nvm}), which have asymmetric read/write
characteristics; or in optimizations that leverage \texttt{READ\_ONLY} data to
reduce synchronization overhead or leverage data replication, as in management
schemes for NUMA (\secref{sec:numa}) or NUCA (\secref{sec:nuca}).} 
\item \textbf{\texttt{AccessIntensity}}: 
\highlight{This attribute conveys the 
access frequency or ``hotness'' of the data relative to other data at
any given time.} 
This attribute can be provided by the programmer, compiler, or profiler.
It is represented using an 8-bit integer, with 0 representing the
lowest frequency. Higher values imply 
an increasing amount of intensity
\emph{relative} to other data.
\highlight{Hence, this attribute conveys an access intensity ranking between different data mapped
to different atoms.} 
\remove{This attribute is key to several optimizations
including managing data placement in heterogeneous memories
(\secref{sec:hetero-systems}), managing DRAM caching (\secref{sec:dram-cache}), and NUCA
systems (\secref{sec:nuca}).} 
\end{itemize}
\end{sloppypar} 
\textbf{(3) Data Locality:} This attribute serves to express
software optimizations for cache locality explicitly (e.g., cache tiling,
stream buffers, partitions, etc.). The key attributes include
\emph{working set size} (which is inferred from the size of data the atom is
mapped to) and \emph{reuse},
for which we use a simple 8-bit integer (0 implying
no reuse and higher values implying a higher amount of reuse
\emph{relative} to other data).
\remove{This information is primarily useful in managing the cache hierarchy
(\secref{sec:cache}), including managing DRAM caches (\secref{sec:dram-cache}).
}

\shepherd{Note that the atom abstraction and its interface do \emph{not} a priori
  limit the program
  attributes that an atom can express. This makes the interface flexible and forward-compatible in terms of extending and changing the expressed program semantics.
  The above attributes
  have been selected for their memory optimization
  benefits (Table~\ref{table:usecases}) and ready translation into simple directives
for the OS and hardware components.}

\subsubsection{The \X System: Key Design Choices}
\label{sec:system-design} 
Before we describe \X's system implementation, we explain the rationale behind
the key design choices. 
\begin{itemize}[wide, labelwidth=!, labelindent=0pt]
\item \shepherd{\textbf{Leveraging Hardware Support:} 
For the \X design, we leverage hardware support for two major reasons. First,
a key design goal for \X is to minimize the runtime \emph{overhead} of
tracking and retrieving semantics at a fine granularity (even semantics that
change as the program executes). We hence leverage support in \emph{hardware} to efficiently perform several key
functionalities of the \X system{\textemdash}mapping/unmapping of semantics to
atoms and activating/deactivating atoms at runtime. This avoids the high
overhead of frequent system calls, memory updates, etc. Second, we aim to enable
the many hardware-managed components in the memory
hierarchy to leverage \X. To this end, we use hardware support to
efficiently transmit key semantics to the different hardware components.}
\item \textbf{Software Summarization and Hardware Tracking:} Because the
potential atoms and their attributes are known statically (by examining the
application program's \texttt{CREATE} calls), the compiler can summarize them
at compile time, and the OS can load them into kernel memory at load time.
The program directly communicates an atom's active status and address
mapping(s) at runtime (via \texttt{MAP} and \texttt{ACTIVATE} calls) with the help of new instructions in the ISA
(\secref{sec:design}) and hardware support. This 
minimizes expensive software intervention/overhead at runtime. In other words,
the static \texttt{CREATE} operator is handled in software before program
execution and the dynamic \texttt{MAP} and \texttt{ACTIVATE} operators
are handled
by hardware at runtime. 
\item \textbf{Centralized Global Tracking and Management:}
All the statically-defined atoms in the program are assigned a global \emph{Atom
ID} (within a process) that the \emph{entire} system recognizes. Tracking which atoms are active at runtime and the inverse mapping
between a VA and atom ID is also centralized at a \emph{global} hardware table, to minimize storage and communication cost (i.e., all architectural
components access the same table to identify the active atom for a VA). 
\item \textbf{Private Attributes and Attribute Translation:}  
The atom attributes provided by the application may be too complex and excessive
for easy interpretation by components like the cache or prefetcher.
To address this challenge (Challenge 2), when
the program is loaded for execution or after a context switch, the OS
invokes a \emph{hardware translator} that converts the
higher-level attributes to sets of
specific primitives relevant to each hardware component and the
optimization the component performs. Such specific primitives are then saved \emph{privately} at each
component, e.g., the
prefetcher saves only the access pattern for each atom.
\end{itemize}

\subsubsection{\X: Interfaces and Mechanisms}
\label{sec:system-interfaces} 
Figure~\ref{fig:components} depicts an overview of the system components to
implement the semantics of \X.
\begin{figure*}[h]
  \centering
  \includegraphics[width=0.99\textwidth]{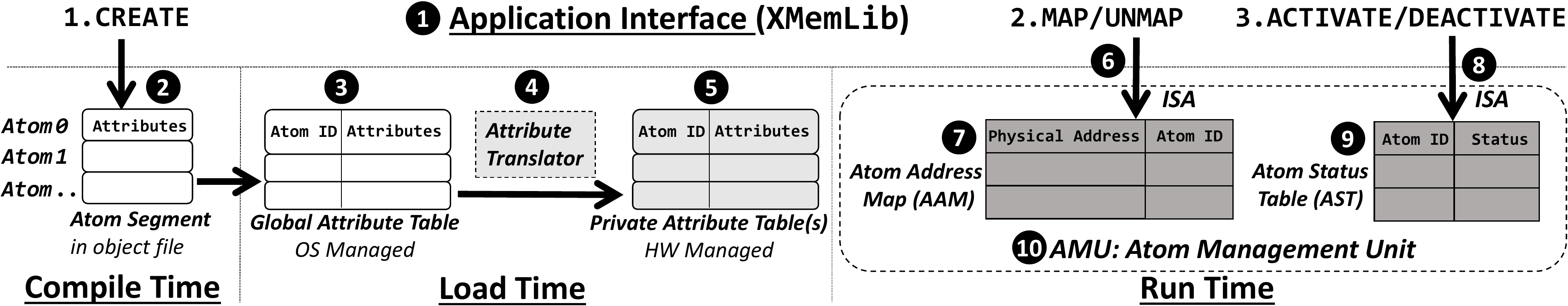}
  \vspace{1mm}
  \caption{XMem: Overview of the components.}
  \vspace{2mm}
  \label{fig:components}
\end{figure*} 

\textbf{Programmer/Application Interface.} 
The application interface to \X is via a library, \Xlib (\ding{182}).
An atom is defined by a class data
structure that defines the attributes of the atom and the operator functions 
(\texttt{CREATE}, \texttt{MAP/UNMAP}, and
\texttt{ACTIVATE/DEACTIVATE}).  An atom and its static attributes can be instantiated in the program code
(\texttt{CREATE}) by the programmer, autotuner, or compiler.

\textbf{System/Architecture Interface.}
\label{sec:system-overview} 
\Xlib communicates with the OS and architecture in the following two ways.

First, at compile time, the compiler summarizes all the atoms in the program
statically and creates a table for atom attributes, indexed by atom ID. During
run time, the same static atom can have many instances (e.g., 
within a \texttt{for} loop or across multiple function calls). All of the
calls to create the same atom will, however, be mapped to the same static atom
(and Atom ID). This is possible because atom attributes are immutable.
However, the address mapping of each atom is typically not known at compile time
because virtual address ranges are only resolved at runtime. The compiler
creates a table of all the atoms in the program
along with the atom attributes. This table is placed in the \emph{atom segment} of the program object file
(\ding{183}). When the program is loaded into memory for execution by the OS,
the OS also reads the atom segment and saves the attributes for each atom in
the \texttt{GLOBAL ATTRIBUTE TABLE (GAT\,\ding{184})}, which is managed by the OS in kernel space. The OS also 
invokes a \emph{hardware translator} (\ding{185}) that converts the
higher-level attributes saved in the \texttt{GAT} to sets of
specific hardware primitives relevant to each hardware component, and saves
them in a per-component \texttt{PRIVATE ATTRIBUTE
TABLE (PAT\,\ding{186})}, managed in hardware by each component. 

\begin{table*}[t]\small
  \centering
  \caption{\highlight{The \X operators and corresponding  \Xlib functions
  and ISA instructions (sizes/lengths in bytes).}}
  \input{tables/operators}

  \label{table:operators}
\end{table*}

Second, at run time, \X operators, in the form of function calls in \Xlib, are
translated into \emph{ISA instructions} that inform the system and
architecture of the atoms' activation/deactivation and mapping.
Conceptually, the \texttt{MAP/UNMAP} operator (\ding{187}) is converted into
ISA instructions that update the \texttt{ATOM ADDRESS MAP (AAM\,\ding{188})},
which enables looking up the atom ID associated with a physical address (PA). We use
the PA to index the \texttt{AAM} instead of the VA to simplify the table design
(\secref{sec:systemdesign}). The
\texttt{ACTIVATE/DEACTIVATE} operator (\ding{189}) is converted into ISA
instructions that update an atom's active status in the \texttt{ATOM STATUS TABLE (AST\,\ding{190})}.
The \texttt{AST} and \texttt{AAM} are managed
by the Atom Management Unit (AMU\,\ding{191}).
Because tables with entries for each PA are infeasible, we use
simple mechanisms to avoid them.  These mechanisms, and 
the functioning and
implementation of these hardware and software components, are described in
\secref{sec:systemdesign}.  

\shepherd{\textbf{Flexibility and Extensibility.} The system/architecture
interface ensures that the ISA and the microarchitecture need only implement the
three operators, but does \emph{not} dictate what application
attributes can be conveyed.  The attributes are stored in the binary as a
separate metadata segment with a version identifier to identify the information
format. The information format can be enhanced across architecture generations,
ensuring flexibility and extensibility, while  the version identifier ensures
forward/backward compatibility. Any future architecture can interpret
the semantics and older \X architectures can simply ignore unknown formats.
}

\subsection{\X: Detailed Design}
\label{sec:design}

We now detail the design and implementation of the interfaces
and components in \X. We describe the application, OS, and architecture
interfaces (\secref{sec:interfaces}), the key components of \X
(\secref{sec:systemdesign}), the use of \X in virtualized environments
(\secref{sec:virtualized}), and the overheads of our design (\secref{sec:overheads}).

\subsubsection{The Interfaces}
\label{sec:interfaces}
\textbf{Application Interface.}
The primary interface between \X and the application is \Xlib, a library
that provides type definitions and function calls for atoms.
\Xlib includes an atom class definition with the attributes described
in \secref{sec:attributes}. 
\Xlib provides three types of \X operations on atoms in the form of
function calls. These operations are the
interface to manipulate the attributes, mappings and state of an
atom. Table~\ref{table:operators} summarizes the definition of all the
functions (also discussed below): 

\textbf{(1) \texttt{CREATE}}: The function \texttt{CreateAtom} creates
an atom with the attributes specified by the input parameters, and
returns an \emph{Atom ID}. Multiple invocations of \texttt{CreateAtom}
at the same place in the program code always return the same Atom ID
(without reinvoking the function).

\textbf{(2) \texttt{MAP/UNMAP}}: 
These functions take
an Atom ID and an address range as parameters, and invoke
corresponding ISA instructions to tell the \texttt{Atom Management Unit (AMU)}
to update the \texttt{Atom Address Map} (\secref{sec:systemdesign}). We
create multiple functions so that we can easily map or unmap
multi-dimensional data structures (e.g., 2D/3D arrays). For example,
\texttt{Atom2DMap} maps/unmaps a 2D block of data of width
\texttt{sizeX} and height \texttt{sizeY}, in a 2D data structure that
has a row length \texttt{lenX}.

\textbf{(3) \texttt{ACTIVATE/DEACTIVATE}}: The functions
\texttt{AtomActivate} and \texttt{AtomDeactivate} serve to (de)activate the specified atom at any given time. They invoke
corresponding ISA instructions that update the Atom Status Table
(\secref{sec:systemdesign}) at run time.


\textbf{Operating System Interface.} \X interfaces with the OS in two ways. First, the OS manages the
\texttt{Global Attribute Table (GAT)} (\secref{sec:systemdesign}), which
holds the attributes of all the atoms in each application. Second,
the OS can optionally query for the \emph{static} mapping between VA ranges and
atoms through an interface to the memory allocator. This interface ensures that the OS knows the mapping
\emph{before} the virtual pages are mapped to physical pages, so that
the OS can perform static optimizations, such as memory placement based
on program semantics. Specifically, we augment the memory allocation APIs (e.g.,
\texttt{malloc}) to take Atom ID as a parameter. The memory allocator,
in turn, passes the Atom ID to the OS via augmented system calls that
request virtual pages.
The memory allocator maintains the static mapping between atoms and
virtual pages by returning virtual pages that match the requesting
Atom ID. The compiler converts the pair \texttt{A=malloc(size);
  AtomMap(atomID,A,size);} into this augmented API: \texttt{A=malloc(size,atomID);
  AtomMap(atomID,A,size);}. This interface enables the OS to
manipulate the virtual-to-physical address mapping \emph{without} extra system call overheads.

\textbf{Architecture Interface.}
\label{sec:isa} 
We add two new ISA instructions to enable \X to talk to the hardware at run
time: \One \texttt{ATOM\_MAP}/\texttt{ATOM\_UNMAP} tells the \texttt{Atom Management Unit (AMU)} to update the
address ranges of an atom. When this instruction is executed, the parameters required to convey the
address mapping for the different mapping types
(Table~\ref{table:operators}) are implicitly saved in \texttt{AMU}-specific
registers and accessed by the \texttt{AMU}. To map or unmap the address range
to/from the specified atom, the
\texttt{AMU} asks the Memory Management Unit (MMU) to translate the
virtual
address ranges specified by \texttt{ATOM\_MAP} to physical address ranges, and updates the \texttt{Atom
Address Map (AAM)} (\secref{sec:systemdesign}). \Two
\texttt{ATOM\_ACTIVATE}/\texttt{ATOM\_DEACTIVATE} causes the \texttt{AMU} to update the \texttt{Atom
Status Table (AST)} to
activate/deactivate the specified atom.

\subsubsection{System Design: Key Components}
\label{sec:systemdesign}
The system/architecture retrieves the data semantics associated with each 
memory address in three steps: \One determine to which atom (if any) a given
address maps; \Two determine whether
the atom is \emph{active}; and \Three retrieve the atom
attributes. \X enables this with four key components:

\textbf{(1) Atom Address Map (\texttt{AAM}):} This component determines the
latest atom (if any) associated with any PA. 
Because the storage overhead of maintaining a mapping table between \emph{each
  address} and Atom ID would be prohibitively large,
%
we employ an
\emph{approximate mapping} between atoms and address ranges at a
configurable granularity. The system decides the smallest
\emph{address range unit} the \texttt{AAM} stores for each
address-range-to-atom mapping. The default granularity is 8 cache
lines (512B), which means each consecutive 512B can map only to
one atom. This design significantly reduces the storage overhead as we
need only store one Atom ID for each 512B (0.2\% storage overhead assuming an
8-bit Atom ID). We can 
reduce this overhead further by increasing the granularity or limiting
the number of atoms in each application. For instance, if we
support only 6-bit Atom IDs with a 1KB address range unit, the storage
overhead becomes 0.07\%. Note that because \X provides only
\emph{hints} to the system, our approximate mapping may
cause optimization inaccuracy but it has no impact on functionality and
correctness.

To make it easy to look up the Atom ID for each address, the \texttt{AAM} stores the
Atom IDs \emph{consecutively} for \emph{all} the physical pages. The index of
the table is the physical page index and each
entry stores all Atoms IDs in each page. In the default configuration, each of the 
Atom IDs require 8B of storage per page (8 bits
times 8 subpages). With this design, the OS or the hardware
architecture can simply use the physical address that is queried as the table index to
find the Atom ID. 

We use the PA instead of the VA to index this table because \One there
are far fewer PAs compared to VAs and \Two this
enables the simplified lookup scheme discussed above.

\textbf{(2) Atom Status Table (\texttt{AST}):} We use a bitmap to
store the status (active or inactive) of all atoms in each
application. Because \texttt{CreateAtom} assigns atom IDs
consecutively starting at 0,
this table is efficiently accessed using the atom ID as index.
Assuming up to 256 atoms per application (all benchmarks in our
experiments had under 10 atoms, all in performance-critical sections),
the AST is only 32B per application.  The \texttt{Atom Management Unit
  (AMU)} updates the bitmap when an \texttt{ATOM\_(DE)ACTIVATE}
instruction is executed.

\textbf{(3) Attribute Tables (\texttt{GAT} and \texttt{PAT}) and the
  Attribute Translator:} As discussed in \secref{sec:system-design}, we
store the attributes of atoms in a \texttt{Global Attribute Table
(GAT)} and multiple \texttt{Private Attribute Tables
(PAT)}. \texttt{GAT} is managed by the OS in kernel space.\remove{ It contains \emph{all} the attributes
expressed in the program for each atom. On the other hand,} Each
hardware component that benefits from \X maintains its own
\texttt{PAT}, which stores a \emph{translated} version of the
attributes\remove{The translated version contains only the attributes that
the component needs for optimization, and is also converted into lower
level primitives specific to the hardware component} (an example of
this is in~\secref{sec:cache}). This translation is done by the
\emph{Attribute Translator}, a
hardware runtime system that translates attributes for each component at
program load time and during a context switch.

\textbf{(4) Atom Management Unit (\texttt{AMU}):} This is a hardware
unit that is responsible for \One managing the \texttt{AAM} and
\texttt{AST} and \Two looking up the Atom ID given a physical
address. When the CPU executes an \X ISA instruction, the CPU sends the
associated command to the \texttt{AMU} to update the \texttt{AAM} (for
\texttt{ATOM\_MAP} or \texttt{ATOM\_UNMAP}) or the \texttt{AST} (for
\texttt{ATOM\_ACTIVATE}). For higher-dimensional data mappings,
the \texttt{AMU} converts the mapping to a linear mapping at the \texttt{AAM}
granularity and broadcasts this mapping to all the hardware components that
require accurate information of higher-dimensional address mappings
(see \secref{sec:cache} for an example).

A hardware component determines the Atom ID of a specific physical
address (PA) by sending an \texttt{ATOM\_LOOKUP} request to the
\texttt{AMU}, which uses the PA as the index into the
\texttt{AAM}.
To avoid memory accesses for all the \texttt{ATOM\_LOOKUP} requests,
each \texttt{AMU} has an atom lookaside buffer
(ALB), which caches the results of recent \texttt{ATOM\_LOOKUP}
requests. The functionality of an ALB is similar to a TLB in an MMU,
so the \texttt{AMU} accesses the \texttt{AAM} \emph{only} on ALB
misses. The tags for the ALB are the physical page indexes, while the data are
the Atom IDs in the physical pages. In our evaluation, we find that a 256-entry
ALB can cover 98.9\%\footnote{\highlight{Does not include the Gramschmidt~\cite{polybench}
workload, which requires a more sophisticated caching policy than LRU to handle
large strides.}}
of the \texttt{ATOM\_LOOKUP} requests.

\subsubsection{\X in Virtualized Environments}
\label{sec:virtualized}
Virtualized environments employ virtual machines (VMs) or containers that
execute applications over layers of operating systems and hypervisors. The
existence of multiple address spaces that are seen by the guest and host
operating systems, along with more levels of abstraction between the application
and the underlying hardware resources, makes the design of hardware-software
mechanisms challenging. \X is, however, designed to seamlessly function in these
virtualized environments, as we describe next.

\textbf{\X Components.} The primary components of \X include the \texttt{AAM}, \texttt{AST},
the \texttt{PAT}s, and the \texttt{GAT}. Each of these
components
function with no changes in virtualized environments: 
\One \texttt{AAM}: The hardware-managed \texttt{AAM}, which maps physical addresses to atom IDs, is
indexed by the \emph{host} physical address. As a result, this table is \emph{globally
shared} across all processes running on the system irrespective of the presence of multiple
levels of virtualization. 
\Two \texttt{AST} and
\texttt{PAT}s: All atoms are tracked at the \emph{process level}
(irrespective of whether the processes
belong to the same or different VMs). The per-process hardware-managed tables
(\texttt{AST} and \texttt{PAT}s) are
reloaded during a context switch to contain the state and attributes of the
atoms that belong to the currently-executing process.
Hence, the functioning of these tables remains the same in the presence of
VMs or containers. 
\Three \texttt{GAT}: The \texttt{GAT} is software-managed and is
maintained by each \emph{guest} OS. During
context switches, a register is loaded with a host physical address that points
to the new process' \texttt{GAT} and \texttt{AST}. 


\sloppy{\textbf{\X Interfaces.} The three major interfaces (\texttt{CREATE},
\texttt{MAP/UNMAP}, and \texttt{ACTIVATE/DEACTIVATE}) require no changes for
operation in virtualized
environments. The \texttt{CREATE} operator is handled in software at compile time by
the guest OS and all created atoms are loaded into the \texttt{GAT} by the guest OS
at program load time. The \texttt{MAP/UNMAP} operator communicates directly with
the MMU to map the host physical address to the corresponding atom ID using the
\X ISA instructions. The \texttt{ACTIVATE/DEACTIVATE} operator
simply updates the \texttt{AST}, which contains the executing process' state. 

\textbf{Optimizations.} OS-based software optimizations (e.g., DRAM
placement in ~\secref{sec:dram}) require that the OS have visibility into the
available physical resources. The physical resources may however be abstracted
away from the guest OS in the presence of virtualization. In this case, the
resource allocation and its optimization needs to be handled by the hypervisor or host OS
for all the VMs that it is hosting. To enable the hypervisor/host OS to make
resource allocation decisions, the guest OS also communicates the attributes of
the application's atoms to the hypervisor.  For hardware optimizations (e.g.,
caching policies, data compression), the hardware components (e.g., caches,
prefetchers) retrieve the atom attributes for each process using the
\texttt{AAM} and \texttt{PATs}. This is the same mechanism irrespective of the
presence of virtualization. These components use application/VM IDs to
distinguish between addresses from different applica\-tions/VMs (similar to
prior work~\cite{pard-ma-asplos15} or modern commercial virtualization
schemes~\cite{intel-vt,amd-npt}).


}

\subsubsection{Overhead Analysis}
\label{sec:overheads} 
The overheads of \X fall into four categories: memory storage overhead, instruction
overhead, hardware area overhead, and context switch overhead, all of which are 
small:

\textbf{(1) Memory storage overhead.} The storage overhead comes from
the tables that maintain the attributes, status, and mappings of
atoms (\texttt{AAM}, \texttt{AST}, \texttt{GAT}, and \texttt{PAT}). As
\secref{sec:systemdesign} discusses, the \texttt{AST} is very small
(32B). The \texttt{GAT} and \texttt{PAT} are also small as the
attributes of each atom need 19B, so each \texttt{GAT} needs only
2.8KB assuming 256 atoms per application. \texttt{AAM} is the largest
table in \X, but it is still insignificant as it takes only 0.2\% of
the physical memory (e.g., 16MB on a 8GB system), and it can be made even
smaller by increasing the granularity of the address range unit
(\secref{sec:systemdesign}).

\textbf{(2) Instruction overhead.} There are instruction
overheads when applications invoke the \Xlib functions to create,
map/unmap, activate/deactivate atoms, which execute \X 
instructions. We find this overhead negligible because: \One \X
does \emph{not} use extra system calls to communicate with the OS, so these
operations are very lightweight; \Two the
program semantics or data mapping do \emph{not} change very frequently. Among the
workloads we evaluate, an additional 0.014\% instructions on average (at most,
0.2\%) are executed.

\textbf{(3) Hardware area overhead.} \X introduces two major hardware
components, \texttt{Attribute Translator} and \texttt{AMU}. We
evaluate the storage overhead of these two components (including the AMU-specific
registers) using
CACTI~6.5~\cite{cacti} at 14 $nm$ process technology, and find that their area is 0.144
$mm^2$, or 0.03\% of a modern Xeon E5-2698 CPU.

\textbf{(4) Context switch overhead.} \X introduces one
extra register for context switches{\textemdash}it stores the pointer to
\texttt{AST} and \texttt{GAT} (stored consecutively for each
application) in the \texttt{AMU}. \texttt{AAM} does
not need a context-based register because it is a global table. The OS 
does not save the AMU-specific registers for
\texttt{MAP/UNMAP} (Table~\ref{table:operators}) as the information is
saved in the \texttt{AAM}. One more register adds very small
overhead (two instructions, $\leq$ 1 ns) to the OS context switch (typically 3-5 $\mu$s).
\shepherd{Context switches also require flushing the ALBs and PATs. Because these structures are small, the overhead is also commensurately
small ({\textasciitilde}700 ns)}.

\subsection{Use Case 1: Cache Management}
Cache management is a well-known complex optimization
problem with substantial prior work in both software (e.g., code/compiler
optimizations, auto tuners, reuse hints) and
hardware (advanced replacement/insertion/partitioning policies).\remove{\footnote{We
discuss prior hint-based cache research in \secref{sec:relatedwork}.}} 
\X seeks
to \emph{supplement} both software and hardware approaches by providing key program semantics that are
challenging to infer at runtime.
As a concrete end-to-end example, we describe and evaluate
how \X enhances \emph{dynamic} policies
to improve the
portability and effectiveness of 
\emph{static software-based} cache optimizations under \emph{varying} cache space
availability (as a result of co-running applications or unknown cache size
in virtualized environments). 

Many software techniques \emph{statically} tune code by sizing the active working
set in the application to maximize cache locality and reuse{\textemdash}e.g., hash-join
partitioning~\cite{DatabaseMemoryAccess} in databases, cache
tiling~\cite{mehta-tile-taco13, automatic-yuki-cgo10,
  spl-xiong-pldi01, automated-whaley-pc01,polly-bondhugula-pldi08,
  tile-coleman-pldi95, stencil-henretty-sc13} in linear algebra and
stencils, cache-conscious data
layout~\cite{CacheSimilaritySearch} for similarity
search~\cite{SimilaritySearch}. 
\X improves the portability and
effectiveness of static optimizations when resource availability is unknown by
conveying the \emph{optimization intent} to
hardware{\textemdash}i.e., \X conveys which high-reuse working set (e.g., tile)
should be kept in the cache. It does \emph{not} dictate exactly what caching policy to
use to do this. The \emph{hardware cache}
leverages the conveyed information to keep the high-reuse working set of each
application in the cache by prioritizing
such data over other low-reuse data. 
In cases where the active working set does \emph{not} fit in 
the available cache space, the cache \emph{mitigates thrashing} by
\emph{pinning} part of the working set and then \emph{prefetches}
the rest based on the expressed access pattern. 
\subsubsection{Evaluation Methodology} 
We model and evaluate \X using
zsim~\cite{zsim} with a 
DRAMSim2~\cite{DRAMSim2} DRAM model.
We use the Polybench
suite~\cite{polybench}, a collection of linear algebra, stencil, and
data mining kernels. We use PLUTO~\cite{polly-bondhugula-pldi08}, a
polyhedral locality optimizer that uses \emph{cache tiling} to
statically optimize the kernels. We evaluate kernels that can be tiled
within three dimensions and a wide range of tile sizes (from 64B to 8MB), ensuring the total work is always the
same. 
\subsubsection{Evaluation Results}
\textbf{Overall performance.}
To understand the cache tiling challenge, in Figure~\ref{fig:cache_result_all},
we plot the execution time of 12
kernels, which are statically compiled with
different tile sizes. For each workload, we show the results of two
systems: \One \texttt{Baseline}, the baseline system with a high-performance cache replacement policy (DRRIP~\cite{rrip}) and a
multi-stride prefetcher~\cite{stride1} at L3; and \Two \texttt{\X},
the system with the aforementioned cache management and
prefetching mechanisms. 
\begin{figure*}[t] 
  \centering
  \includegraphics[width=0.99\textwidth]{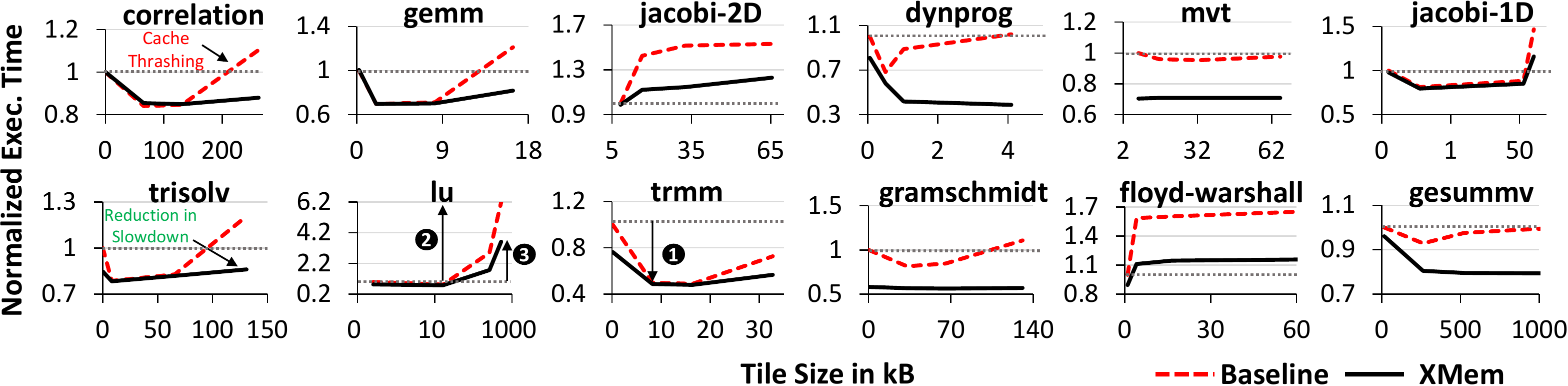}
  \caption{Execution time across different tile sizes (normalized to \texttt{Baseline} with the
  smallest tile size).}
  \label{fig:cache_result_all}
\end{figure*} 

\highlight{For the \texttt{Baseline} system, execution time varies significantly
with tile size, making tile size selection a challenging task. Small tiles significantly reduce the reuse in the
application and can be on average 28.7\% (up to 2$\times$ \ding{182}) \emph{slower} than
the \emph{best} tile size.
Many optimizations, hence, typically size the tile to be as big as what can fit in
the available cache space~\cite{defensive-bao-cgo13,tile-coleman-pldi95}. 
However, when an optimization makes incorrect assumptions regarding available
cache space (e.g., in virtualized
environments or due to co-running applications), the tile size
may \emph{exceed} the available cache space. We find that this can lead
to \emph{cache
thrashing} and severe slowdown (64.8\% on average, up to 7.6$\times$ \ding{183}), compared
to the performance with an optimized tile size. 
\texttt{\X}, however, significantly reduces this slowdown from cache thrashing
in the largest tile sizes to 26.9\% on average (up to 4.6$\times$~\ding{184}).}  
\texttt{\X}'s large improvement comes from accurate \emph{pinning} (that retains part of the
high-reuse working set in the cache) and more accurate prefetching (that fetches
the remaining working set).

\noindent \textbf{Performance portability.} To evaluate 
portability benefits from the reduced impact of cache thrashing in
large tile sizes, we run the following experiment. For each workload, we pick a
tile size optimized for a 2MB 
cache, and evaluate \emph{the same program binary} on a 2MB cache and 2 smaller caches (1MB and 512KB). Figure~\ref{fig:cache_result_portable} depicts the
\emph{maximum} execution time among these three cache sizes for both
\texttt{Baseline} and \texttt{\X}, normalized to \texttt{Baseline}
with a 2MB cache. \remove{We see that given the same program, \texttt{\X}
provides a much more consistent performance than \texttt{Baseline}
across different cache sizes. }When executing with less cache space,
we find that
\texttt{\X} increases the execution time by only 6\%, compared to the
\texttt{Baseline}'s 55\%. Hence, we conclude that by leveraging the program
semantics, \texttt{\X} greatly enhances the performance portability of
applications by reducing the impact of having less cache space than what the program is optimized for.
\begin{figure}[!t] 
  \centering
  \includegraphics[width=0.69\textwidth]{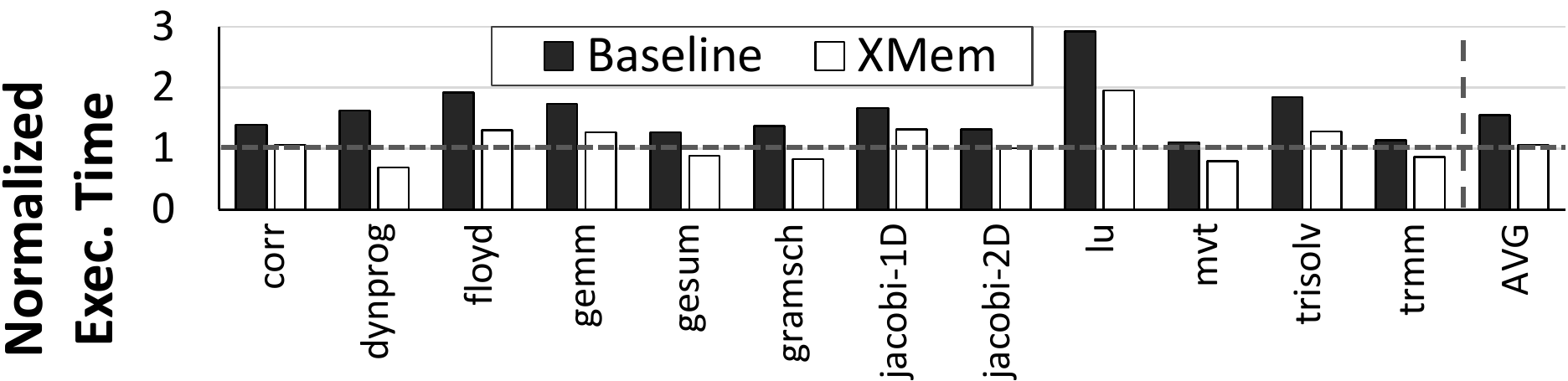}
  \caption{Maximum execution time with different cache
  sizes when code is optimized for a 2MB cache.}
  \label{fig:cache_result_portable}
\end{figure} 
\subsection{Use Case 2: Data Placement in DRAM}
\label{sec:dram}
Off-chip main memory (DRAM) latency is a major performance
bottleneck in modern CPUs~\cite{TailAtScale, ProfilingWarehouse,
  MemoryGap,memory-mutlu-imw13,runahead-mutlu-hpca03}. The performance impact of
  this latency is 
in large part determined by two factors: \One \emph{Row Buffer Locality (RBL)}~\cite{parbs-mutlu-isca08,tcm-kim-micro10}:
how
often requests access the \emph{same} DRAM row in a bank \emph{consecutively}. Consecutive accesses to
the same open row saves the long latency required to \emph{close} the already-open row and
\emph{open} the requested row. \Two \emph{Memory Level Parallelism
  (MLP)}~\cite{glew1998mlp,runahead-mutlu-hpca03}: the number of concurrent
  accesses to 
  different memory banks or channels. Serving requests in parallel \emph{overlaps} the latency of the 
  different requests.
  A key factor that
determines the DRAM access pattern{\textemdash}and thus RBL and MLP{\textemdash}is how the program's
data is mapped to the DRAM channels, banks, and
rows. This data placement is controlled by \One the OS (via the virtual-to-physical
address mapping)
and \Two the memory controller (via the mapping of physical addresses
to DRAM channels/banks/rows). 

To improve RBL and MLP, prior works use both the OS (e.g.,~\cite{micropages-sudan-asplos10,
  mcp-muralidhara-micro11, software-liu-pact12,
  irregularities-park-asplos13, bpm-liu-taco14, compiler-ding-micro14,
  improving-bheda-memsys16, improving-xie-hpca14, swhw-mi-npc10,
  balancing-jeong-hpca12, palloc-yun-rtas14, going-liu-isca14,
  locality-liu-aspdac17}) and the memory controller
(e.g.,~\cite{permutation-zhang-micro00,minimalist-kaseridis-micro11,
  xor-vandierendonck-taco05, breaking-zhang-jilp01,
  dream-ghasempour-memsys16, multiple-hillenbrand-apws17,
  impulse-carter-hpca99}) to introduce \emph{randomness} in how data is mapped
  to the DRAM channels/banks/rows 
or \emph{partition} banks/channels between different threads or
applications. While effective, these techniques are \emph{unaware} of the
different semantics of data structures in an application, and hence suffer from two
shortcomings. First, to determine the properties of data, an application
needs to be \emph{profiled} before execution or
pages need to be migrated \emph{reactively} based on runtime behavior. Second,
these techniques apply the \emph{same} mapping for \emph{all} data
structures within the application, even if RBL and
MLP vary significantly across different data structures.

\X enables distinguishing between data structures and provides key
access semantics to the OS. Together with the knowledge of the underlying
banks, channels, ranks, etc. and other co-running applications, the OS can create an intelligent mapping at the \emph{data
  structure} granularity. Based on the data structure access patterns, the OS
can \One improve RBL by \emph{isolating} data structures with high RBL from data structures that could cause interference if placed in the same bank and \Two improve MLP by \emph{spreading} out accesses to concurrently-accessed data structures across multiple banks
and channels.
\subsubsection{Evaluation Methodology}
\label{sec:dram_methodology}
We use zsim~\cite{zsim} and DRAMSim2~\cite{DRAMSim2} for evaluation. \remove{We implement the aforementioned mechanisms and
algorithms in the simulators. }We strengthen our baseline system in
three ways: \One We use the \emph{best}-performing physical DRAM
mapping, among all the seven mapping schemes in DRAMSim2 and the two proposed in~\cite{permutation-zhang-micro00,
  minimalist-kaseridis-micro11}, as our baseline; \Two We
\emph{randomize} virtual-to-physical address mapping, which is shown
to perform better than the Buddy
algorithm~\cite{irregularities-park-asplos13}; \Three For each
workload, we enable the L3 prefetcher \emph{only if} it
improves performance.
We evaluate a wide range of workloads
from SPEC CPU2006~\cite{SPEC}, Rodinia~\cite{rodinia}, and
Parboil~\cite{parboil} and show results for 27 memory intensive workloads (with L3
MPKI > 1)\remove{, as the performance of memory non-intensive workloads are
not affected by DRAM placement}. 
\subsubsection{Evaluation Results}
\label{sec:dram_results}

We evaluate three systems: \One \texttt{Baseline}, the strengthened
baseline system; \Two
\texttt{\X}, DRAM placement using \X; \Three
an
\emph{ideal} system that has \emph{perfect} RBL, which represents the
best performance possible by improving
RBL. Figure~\ref{fig:dram_result_speedup} shows the speedup of the
last two systems over \texttt{Baseline}. 
Figure~\ref{fig:dram_result_latency} shows the corresponding memory read latency,
normalized to \texttt{Baseline}. We make two observations.
\begin{figure}[h]
  \centering
 \begin{subfigure}[t]{0.49\linewidth}
\centering
  \includegraphics[width=1\textwidth]{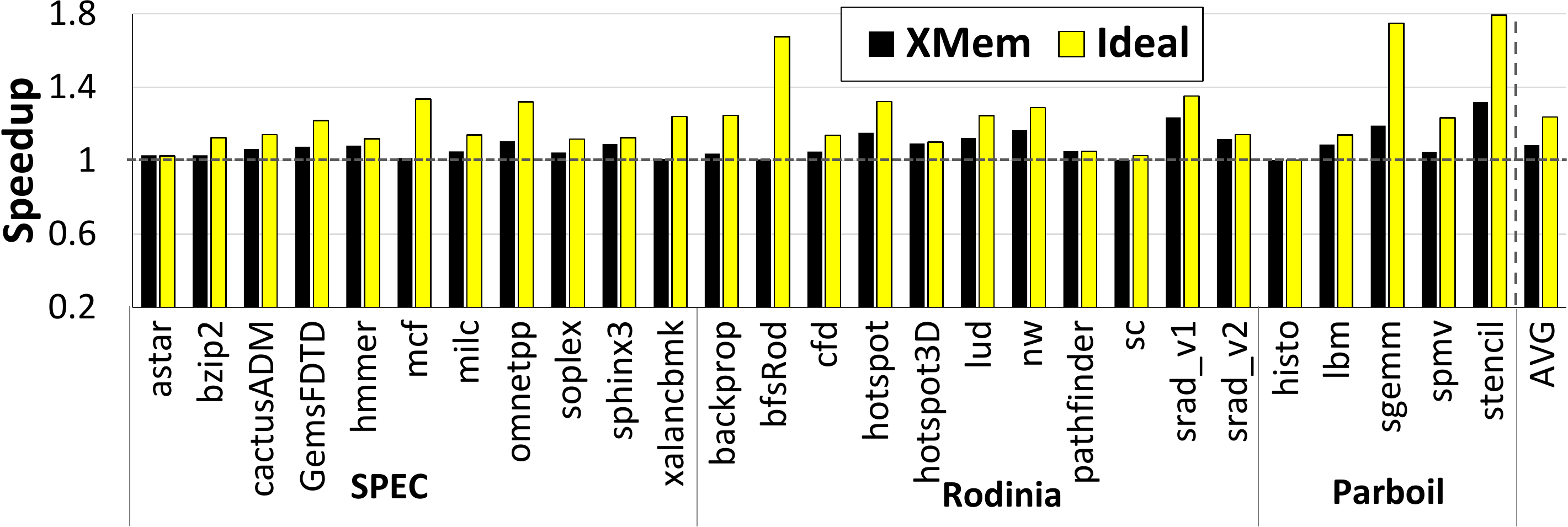}
  \caption{Speedup w/ \X-based DRAM placement.}
  \label{fig:dram_result_speedup}
\end{subfigure} 
 \begin{subfigure}[t]{0.49\linewidth}
  \centering
  \includegraphics[width=1\textwidth]{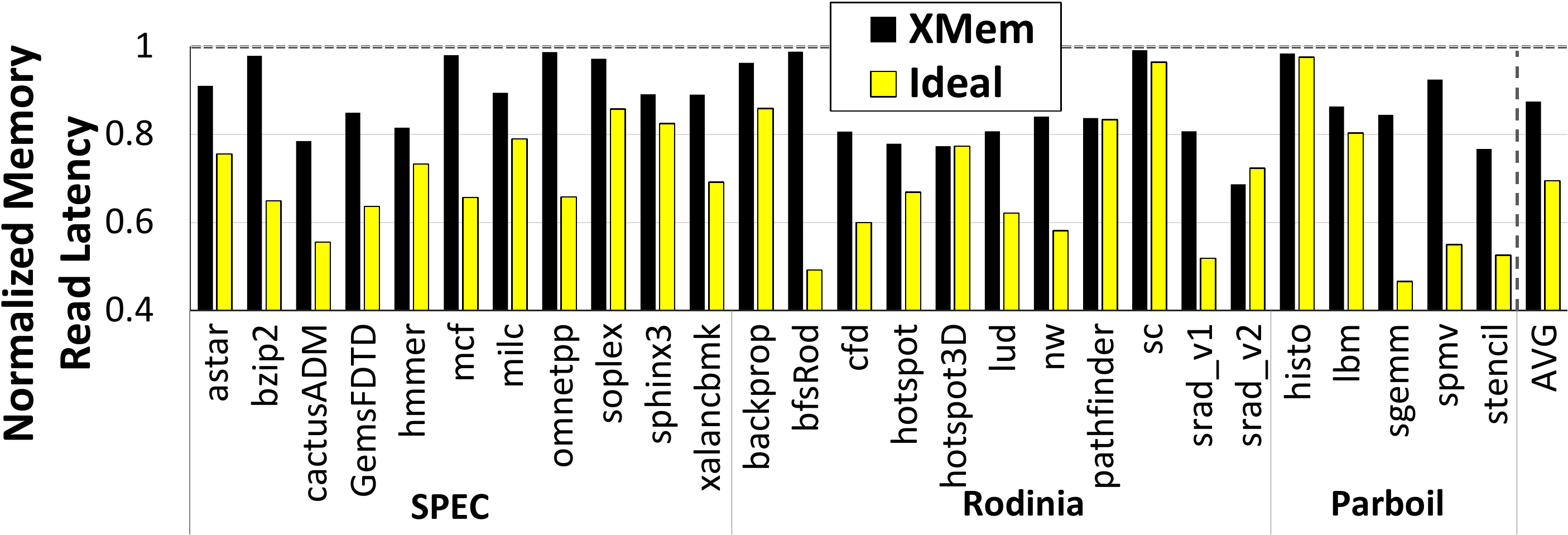}
  \caption{Normalized read latency with \X-based DRAM placement.}
  \label{fig:dram_result_latency}
\end{subfigure} 
 \caption{Leveraging \X for DRAM placement.}
\end{figure} 

First, \X-based DRAM placement improves performance across a 
range of workloads: by 8.5\% on average over \texttt{Baseline}, up to
31.9\%. It is a significant improvement as the absolute upper-bound
for \emph{any} DRAM row-buffer optimization is 24.4\%
(\texttt{Ideal}). Among the 27 workloads, only 5
workloads do not see much improvement{\textemdash}they either \One have less
than 3\% headroom to begin with (\texttt{sc} and \texttt{histo}) or
\Two are dominated by random accesses
(\texttt{mcf}, \texttt{xalancbmk}, and \texttt{bfsRod}). 
Second, the performance improvement of \X-based DRAM placement comes
from the significant reduction in average memory latency, especially 
read latency, which is usually on the critical path. On average,
\texttt{\X} reduces read latency by 12.6\%, up to
31.4\%. Write latency is reduced by 6.2\% (not
shown). 

We conclude that leveraging both the program semantics provided by \X
and knowledge of the underlying DRAM organization enables the OS to create intelligent DRAM mappings
at a fine (data
structure) granularity, thereby reducing memory latency and improving performance.

\subsection{Related Work}
To our knowledge, this is the first work to design a holistic and general
cross-layer interface to enable the entire system and
architecture to be aware of key higher-level program semantics that can be
leveraged in memory optimization. 
We now briefly discuss closely related prior
work specific to XMem. A more general comparison of approaches taken in XMem is
discussed in Section~\ref{sec:relatedwork}.

\textbf{Expressive programming models and runtime systems.} 
\highlight{Numerous software-only approaches tackle the 
disconnect between an application, the OS, and the underlying memory resources
via
programming models
and runtime systems
that allow explicit expression of data
locality and independence~\cite{x10-charles-oopsla05,chapel-chamberlain-ijhpca,sequoia-fatahalian-sc06,legiondependent-treichler-oopsla16,legionstructure-bauer-sc14,
realm-treichler-pact14,regent-slaughter-sc15,hpt-yan-lcpc09,hta-bikshandi-ppopp06,
legion-bauer-sc12, programming-guo-ppopp08, tida-unat-hpc16} in
the programming model.} This explicit expression enables the programmer and/or
runtime system to make effective 
memory placement decisions in a NUMA system or produce code that is
optimized to effectively leverage the cache hierarchy.
These approaches have several shortcomings. First, they are
entirely software-based and are hence limited to using the \emph{existing} interfaces
to the architectural resources. Second, unlike \X, which is general and only
hint-based, programming model-based approaches \emph{require} rewriting applications to suit the
model, while ensuring that program correctness is retained. Third, these
systems are specific to an application type (e.g., operations on tiles, arrays).
\X is a general interface that is \emph{not} limited to any programming
language, application, or architecture. These approaches are orthogonal to \X, and
\X can be built into them to enable a wider range of memory. 

\highlight{The Locality Descriptor~\cite{ldesc} is a cross-layer abstraction to express
data locality in GPUs. This abstraction is similar in spirit to \X in bridging
the semantic gap between hardware and software. However, the Locality Descriptor
is primarily designed to convey \emph{locality semantics} to leverage cache and
NUMA locality in GPUs. \X aims to convey \emph{general} program semantics to aid
memory optimization. This goal imposes 
different design challenges, requires
describing a different set of semantics, and requires optimizing a different set
of architectural techniques, leading to a very different
cross-layer design for the abstraction.} 

\textbf{Leveraging hints, annotations, and software management for memory
optimization.}
A large body of prior work aims to leverage the benefits of static program
information in the form of hints, annotations, or directives in 
memory optimization. These include \One hint-based
approaches, such as 
software prefetch instructions~\cite{intel-prefetching} and cache
bypass/insertion/eviction hints~\cite{swcache-jain-iccad01,compilerpartitioned-ravindran-lctes07,popt-gu-lcpc08,pacman-brock-ismm13,evictme-wang-pact02,generatinghints-beyls-jsysarch05,keepme-sartor-interact05,compilerassisted-yang-lcpc04,cooperativescrubbing-sartor-pact14,runtimellc-pan-sc15,prefetchtasklifetimes-papaefstathiou-ics13,radar-manivannan-hpca16,compile-beyls-epic2,modified-tyson-micro95}; \Two hardware-software
cooperative prefetch techniques~\cite{effective-chen-micro95,compilerassisted-vanderwiel-vlsi99,chiueh-sunder-sc94,integrated-gornish-icpp94,guided-wang-isca03,controlling-jain-csg01,hybrid-skeppstedt-icpp97,prefetchingtechnique-karlsson-hpca00,loopaware-fuchs-micro14,techniques-ebrahimi-hpca09}
that use compiler
analysis or annotations
to inform a hardware prefetch engine; and \Three program annotations to place
data in heterogeneous memories
(e.g.,~\cite{pageplacement-agarwal-asplos15,characterizing-luo-dsn14,flikker-liu-asplos11}).
\X differs from these works in several ways. First, many of these
approaches seek to inform hardware components with \emph{specific directives}
that override dynamic
policies by enforcing \emph{static} policies. This loss in dynamism introduces
challenges when the workload behavior changes, the underlying architecture
changes or is unknown (portability), or in the presence of co-running
applications~\cite{comparing-leverich-isca07,whirlpool-mukkara-asplos16,zorua}. \X
does \emph{not} \emph{direct} policy at any component but only provides
higher-level program semantics. The memory components can use this information
to \emph{supplement}
their dynamic management policies. 
Second, the approaches are \emph{specific} to an optimization (e.g., prefetching,
cache insertion/eviction). \X provides a \emph{general} holistic interface to communicate
program semantics that can be leveraged by a wide range of system/architectural
components for optimization. 

The closest work to ours is
Whirlpool~\cite{whirlpool-mukkara-asplos16}, which provides a memory allocator
to \emph{statically} classify data into pools. Each pool is then managed
differently at runtime to place
data efficiently in NUCA caches.
Whirlpool is similar to \X in the ability to classify data into similar types and in retaining the
benefits of dynamic management. However, \X is 
\One more versatile, as it
enables \emph{dynamically} classifying/reclassifying data and expressing more powerful program semantics
than just static data classification and 
\Two a general and holistic interface that
can be used for a wide range of use cases, including Whirlpool itself.
Several prior
works~\cite{swcache-chiou-dac00,jigsaw-beckmann-pact13,jenga-tsai-isca17,vls-cook-tr09,hybridcache-cong-ispled11,bic-fajardo-dac11,talus-beckmann-hpca15}
use runtime systems or the OS to aid in management of the cache. These
approaches are largely orthogonal to \X and can be used in conjunction with \X
to provide more
benefit.

\shepherd{\textbf{Tagged Architectures.} Prior work proposes to associate
software-defined metadata with each memory
location in the form of tagged/typed
memory~\cite{architectural-dhawan-asplos15,tagged-feustel-taco73,hardware-zeldovich-osdi08,mondrian-witchel-asplos02\remove{,raksha-dalton-isca07}}.
These proposals are typically used for fine-grained memory access protection,
debugging, etc., and usually incur \emph{non-trivial}
performance/storage overhead. In contrast, XMem aims to deliver \emph{general} program
semantics to many system/hardware components to aid in
performance optimization with \emph{low overhead}. 
To this end, \X is designed to enable a number of features and benefits that
cannot be obtained from tagged/typed architectures: \One a flexible and
extensible abstraction to dynamically describe program behavior with \Xlib; and \Two low-overhead interfaces to many hardware components to easily access the expressed semantics. 
PARD~\cite{pard-ma-asplos15} and Labeled RISC-V~\cite{labeled-yu} are tagged
architectures that enable labeling memory requests with tags to
applications, VMs, etc. These tags are used to convey an
application's QoS,
security requirements, etc., to hardware. \X is similar in that it provides an
interface to hardware to convey information from software. However, unlike these
works~\cite{pard-ma-asplos15,labeled-yu}, we design a new abstraction (the atom) to flexibly
express program semantics that can be seamlessly integrated into programming
languages, runtime systems, and modern ISAs. The atom lends itself to a
low-overhead implementation to convey software semantics to hardware
components \emph{dynamically} and at flexible granularities. \X can potentially
\emph{leverage} tagged
architectures to communicate atom IDs to
different hardware components. Hence, PARD and Labeled RISC-V are complementary
to XMem.}

%% file: tables/usecases.tex
\footnotesize
\begin{tabular}{@{}p{0.80in}>{\raggedright}p{1.8in}p{4.2in}@{}}
  \toprule
  \textbf{Memory \mbox{optimization}} &
  \textbf{Example semantics provided by XMem} &
  \textbf{Example Benefits of XMem} \\
  \toprule
  \multirow{1}{0.80in}{Cache management} &
  \One Distinguishing between data structures or pools of similar data; \Two Working
  set size; \Three Data reuse  &  Enables: \One applying different caching policies
  to different data structures or pools of data; \Two avoiding cache thrashing by
  \emph{knowing} the active working set size; \Three bypassing/prioritizing data
  that has
  no/high reuse.\\
  \midrule
  \multirow{1}{0.80in}{Page placement in
  DRAM e.g.,~\cite{irregularities-park-asplos13,mcp-muralidhara-micro11}} &
  \One Distinguishing between data structures; \Two Access pattern; \Three
  Access intensity  &  Enables page placement at the \emph{data structure} granularity
  to \One isolate data structures that have high row buffer locality and \Two
  spread out concurrently-accessed irregular data structures across banks and channels to
  improve parallelism.\\
  \midrule
  \multirow{1}{0.80in}{Cache/memory compression
e.g.,~\cite{lcp-micro, caba,
bdi, zca-dusser-sc09, MMCompression,
sc2-arelakis-isca14,exploiting-pekhimenko-hpca15,toggle-pekhimenko-hpca16}} &
  \One Data type: integer, float, char; \Two Data properties: sparse, pointer,\quad
  data index &Enables using a \emph{different compression algorithm} for each data
  structure based on data type and data properties, e.g., sparse data encodings, FP-specific
  compression, delta-based compression for
  pointers~\cite{bdi}.\\\\
  \midrule
  \multirow{1}{0.80in}{Data prefetching
  e.g.,~\cite{stride1,sms-somogyi-isca06,spatiotemporal-somogyi-isca09,bump-volos-micro14}} & \One Access pattern: strided, irregular,
  irregular but repeated (e.g., graphs), access stride; \Two Data type: index, pointer & 
  Enables \One \emph{highly accurate} software-driven prefetching while leveraging the benefits
  of hardware prefetching
  (e.g., by being memory bandwidth-aware, avoiding cache thrashing); 
  \Two using different prefetcher \emph{types} for different data structures:
  e.g., stride~\cite{stride1},
  tile-based~\cite{chiueh-sunder-sc94}, pattern-based~\cite{sms-somogyi-isca06,
spatiotemporal-somogyi-isca09,bump-volos-micro14,
semanticlocality-peled-isca15}, data-based for
indices/pointers~\cite{techniques-ebrahimi-hpca09,stateless-cooksey-asplos02}, etc. 
  \\ 
  \midrule
  \multirow{1}{0.80in}{DRAM cache management e.g.,~\cite{hma-meswani-hpca15,
chop-jiang-hpca10,unison-jevdjic-micro14, diestacked-jevdjic-isca13,
\remove{bimodal-gulur-micro14,}banshee-yu-micro17,rbl-yoon-iccd11,enabling-meza-cal12}} 
&
  \One Access intensity; \Two Data reuse; \Three
  Working set size  & \One Helps avoid cache thrashing by knowing working set
  size~\cite{banshee-yu-micro17}; \Two Better DRAM cache management via reuse behavior and
  access intensity
  information.\\\\\\ 
  \multicolumn{1}{c}{}\\
  \midrule
  \multirow{1}{0.80in}{Approximation in memory
  e.g.,\cite{load-miguel-micro14,doppelganger-miguel-micro15,
rollback-thwaites-pact14,
architecturalsupport-esmaeilzadeh-isca12,approximate-sampson-tocs14,
enerj-sampson-pldi11,rfvp-yazdanbakhsh-taco16}
  }& \One Distinguishing between
  pools of similar data; \Two Data properties: tolerance
  towards approximation  & Enables \One each memory component to track
  how approximable data is (at a fine granularity) to inform approximation techniques; \Two data placement in heterogeneous
  reliability memories~\cite{characterizing-luo-dsn14}. 
\\\\
  \midrule
  \multirow{1}{0.80in}{Data placement: NUMA systems
e.g.,\cite{unlocking-agarwal-hpca15,
trafficmanagement-dashti-asplos13}} & \One Data partitioning across threads
(i.e.,
relating data to threads that access it); \Two Read-Write properties & 
Reduces the need for profiling or data migration \One to co-locate data with
threads that access it and \Two to identify Read-Only data, thereby enabling techniques
such as replication. \\
  \midrule
  \multirow{1}{0.80in}{Data placement: hybrid
  memories
  e.g.,~\cite{exploringhybrid-wang-pact13,datatiering-dulloor-eurosys16,utility-li-cluster17}} &
  \One Read-Write properties (Read-Only/Read-Write); \Two Access intensity;
  \Three Data structure size; \Four Access pattern & Avoids the need for
  profiling/migration 
  of data in hybrid memories to \One effectively manage
  the asymmetric read-write properties in NVM (e.g., placing Read-Only data in the
  NVM)~\cite{exploringhybrid-wang-pact13,datatiering-dulloor-eurosys16}; \Two
  make tradeoffs between data structure "hotness" and size to allocate
  fast/high bandwidth memory~\cite{pageplacement-agarwal-asplos15}; and
  \Three leverage row-buffer locality in placement based on access
  pattern~\cite{rbl-yoon-iccd11}.\\ \\ 
  \midrule
  \multirow{1}{0.80in}{Managing NUCA 
  systems e.g.,~\cite{whirlpool-mukkara-asplos16,rnuca-hardavellas-isca09} 
  } &\One Distinguishing 
  pools of similar data; \Two Access intensity; \Three Read-Write or Private-Shared properties& 
  \One Enables using different cache policies for different data pools (similar
  to~\cite{whirlpool-mukkara-asplos16}); \Two Reduces the need for reactive
  mechanisms that detect
  sharing and read-write characteristics to inform cache policies. 
\\
  \bottomrule
\end{tabular}

%% file: tables/operators.tex
\footnotesize
\begin{tabular}{@{}p{0.6in}>{\raggedright}p{3.95in}p{2.2in}@{}}
  \toprule
  \textbf{\X   \mbox{Op}} &
  \textbf{\Xlib Functions (Application Interface)} &
  \textbf{\X ISA Insts (Architecture Interface)} \\
  \toprule
  \multirow{1}{*}{\texttt{CREATE}} &
  \texttt{AtomID CreateAtom(data\_prop, access\_pattern, reuse, rw\_characteristics)}  &  No ISA instruction required\\
  \midrule
  \multirow{4}{*}{\texttt{MAP/UNMAP} } &
 \texttt{AtomMap(atom\_id, start\_addr, size, map\_or\_unmap)} 
 
 \texttt{Atom2DMap(atom\_id, start\_addr, lenX, sizeX, sizeY, map\_or\_unmap)}
 \texttt{Atom3DMap(atom\_id, start\_addr, lenX, lenY, sizeX, sizeY, sizeZ, map\_or\_unmap)}
 &
  \texttt{ATOM\_MAP AtomID, Dimensionality}
  
  \texttt{ATOM\_UNMAP AtomID, Dimensionality}
  
  \emph{Address ranges specified in
  AMU-specific registers} \\
  \midrule
  \multirow{2}{0.6in}{\texttt{ACTIVATE/  
  DEACTIVATE}} &
  \texttt{AtomActivate(atom\_id)} 

\texttt{AtomDeactivate(atom\_id)} 
  &  \texttt{ATOM\_ACTIVATE  AtomID} 

   \texttt{ATOM\_DEACTIVATE  AtomID} 
  \\
  \bottomrule
\end{tabular}

%% file: sections/xmem_impl.tex
\subsection{Summary}
This work makes the case for richer cross-layer interfaces to bridge the semantic gap
between the application and the underlying system and architecture. To this end,
we introduce \Xfull (\X), a holistic cross-layer
interface that communicates higher-level program semantics from the application to
different system-level and architectural components (such as caches,
prefetchers, and memory controllers) to aid in memory optimization. \X 
improves the performance and portability of a wide range of software and hardware memory
optimization techniques by enabling them to leverage key
semantic information that is otherwise unavailable. 
We evaluate and
demonstrate \X's benefits for two use cases: \One static software cache
optimization, by
leveraging data locality semantics, and \Two OS-based page placement in DRAM, by leveraging the
ability to distinguish between data structures and their access patterns.
We conclude that \X provides a versatile, rich, and low overhead interface to 
bridge the semantic gap in order to enhance memory system optimization. We hope
\X encourages future work to 
explore re-architecting the traditional interfaces to enable many other benefits that are not possible
today.

%% file: sections/ldesc.tex
\section{The Locality Descriptor}
\label{sec:ldesc}

This chapter proposes a rich cross-layer abstraction to address a key challenge
in modern GPUs: leveraging data locality. We demonstrate how existing
abstractions in the GPU programming model are insufficient to expose the
significant data locality exhibited by GPU programs. We then propose a flexible
abstraction that enables the application to express data locality concisely and,
thus, enables the driver and hardware to leverage any data locality to improve
performance.

\subsection{Overview}

\highlight{Graphics Processing Units (GPUs)} have evolved into powerful programmable machines
that deliver high performance and energy efficiency to many
important classes of applications today. Efficient use of memory system resources
is critical to fully harnessing the massive computational
power offered by a GPU. A key contributor to this efficiency is
\emph{data locality}{\textemdash}both \One \emph{reuse} of data within the
application in the cache hierarchy (\emph{reuse-based locality}) and \Two
\highlight{placement of
data \emph{close} to the computation that uses it in a non-uniform memory
access
(NUMA) system} (\emph{NUMA locality})~\cite{mcmgpu-arunkumar-isca17, numagpu-milic-micro17, tom-hsieh-isca16, toward-kim-sc17}.\remove{ By exploiting different forms of data locality,
we can make more efficient use of both on-chip cache real estate, and on- and off-chip
bandwidth in memory systems that are expected to have increasingly NUMA characteristics}

Contemporary GPU programming models (e.g., CUDA~\cite{nvidia-cuda},
OpenCL~\cite{amd-opencl}) are designed to harness
the massive computational power of a GPU by enabling explicit expression of
\emph{parallelism} and control of \emph{software-managed memories}
(scratchpad memory and register file). However, there is no clear explicit way
to express and exploit \emph{data locality}{\textemdash}i.e., \emph{data reuse}, to
better utilize the hardware-managed cache hierarchy, or \emph{NUMA locality},
to\remove{more} 
efficiently use a NUMA
memory system.

\textbf{Challenges with Existing Interfaces.} Since there is no explicit interface in
the programming model  
to express and exploit data locality, expert programmers use
various techniques such as software
scheduling~\cite{locality-li-asplos17} and prefetch/bypass
hints~\cite{efficient-xie-iccad13,cuda-cache-op} to carefully manage locality 
to obtain high performance.
However, all such software approaches are significantly limited for
three reasons.  
First, exploiting data locality is a challenging task, requiring a range of
hardware mechanisms such as thread scheduling~\cite{largewarp,improving-lee-hpca14, locality-li-asplos17, nmnl-pact13,
owl-jog-asplos13, laperm-wang-isca16, twinkernels-gong-cgo17,
improving-chen-cal17, cachehierarchy-lai-tc15, enabling-wu-ics15}, cache
bypassing/prioritization~\cite{adaptive-tian-gpgpu15, locality-li-ics15,
adaptive-li-sc15, efficient-liang-tcadics,survey-mittal-jlpea16,
ctrlc-lee-iccd16, selectively-zhao-icpads16, coordinated-xie-hpca15,
prioritybased-li-hpca15, efficient-xie-iccad13,
medic}, and prefetching~\cite{manythread-lee-micro10, apogee-sethia-pact13,
cta-jeon-ceng14, threadaware-liu-taco16,
spareregister-lakshminarayana-hpca14,orchestrated-jog-isca13},
\highlight{to which software has \emph{no easy access}}. 
Second, GPU programs exhibit many different types of data locality, e.g.,
inter-CTA (reuse of data across Cooperative Thread Arrays or thread blocks),
inter-warp and intra-warp locality. \highlight{Often, \emph{multiple} different techniques are required to exploit
each type of locality, as a 
single technique in isolation is
insufficient~\cite{accesspattern-koo-isca17,orchestrated-jog-isca13,apres-oh-isca16,locality-li-asplos17,medic}.} 
Hence, 
software-only approaches quickly become \emph{tedious} and difficult programming tasks.
 Third, any software optimization employing fine-grained ISA instructions to manage
caches or manipulating thread indexing to alter CTA
scheduling is \emph{not portable} to a different architecture with a
different CTA scheduler, \remove{different number of SMs, }different cache sizes,
etc~\cite{zorua}. 

At the same time, software-transparent architectural techniques miss critical
program semantics regarding locality inherent in the algorithm. 
For example, CTA scheduling is used to improve
data locality by scheduling CTAs that share data at the same core. This requires
knowledge of \emph{which} CTAs share data{\textemdash}\highlight{knowledge that 
\emph{cannot} easily be inferred by the architecture~\cite{locality-li-asplos17,
improving-chen-cal17}.} Similarly, NUMA locality is created by placing data close
to the threads that use it. This requires a priori knowledge of \emph{which} threads
access \emph{what} data to avoid expensive reactive page
migration~\cite{automatic-cabezas-ics15, automatic-cabezas-pact14}. 
Furthermore, many architectural techniques, such as prefetching or cache
bypassing/prioritization, would benefit from
knowledge of the application's access semantics. 

\begin{figure}[t]
  \centering
  \includegraphics[width=0.69\textwidth]{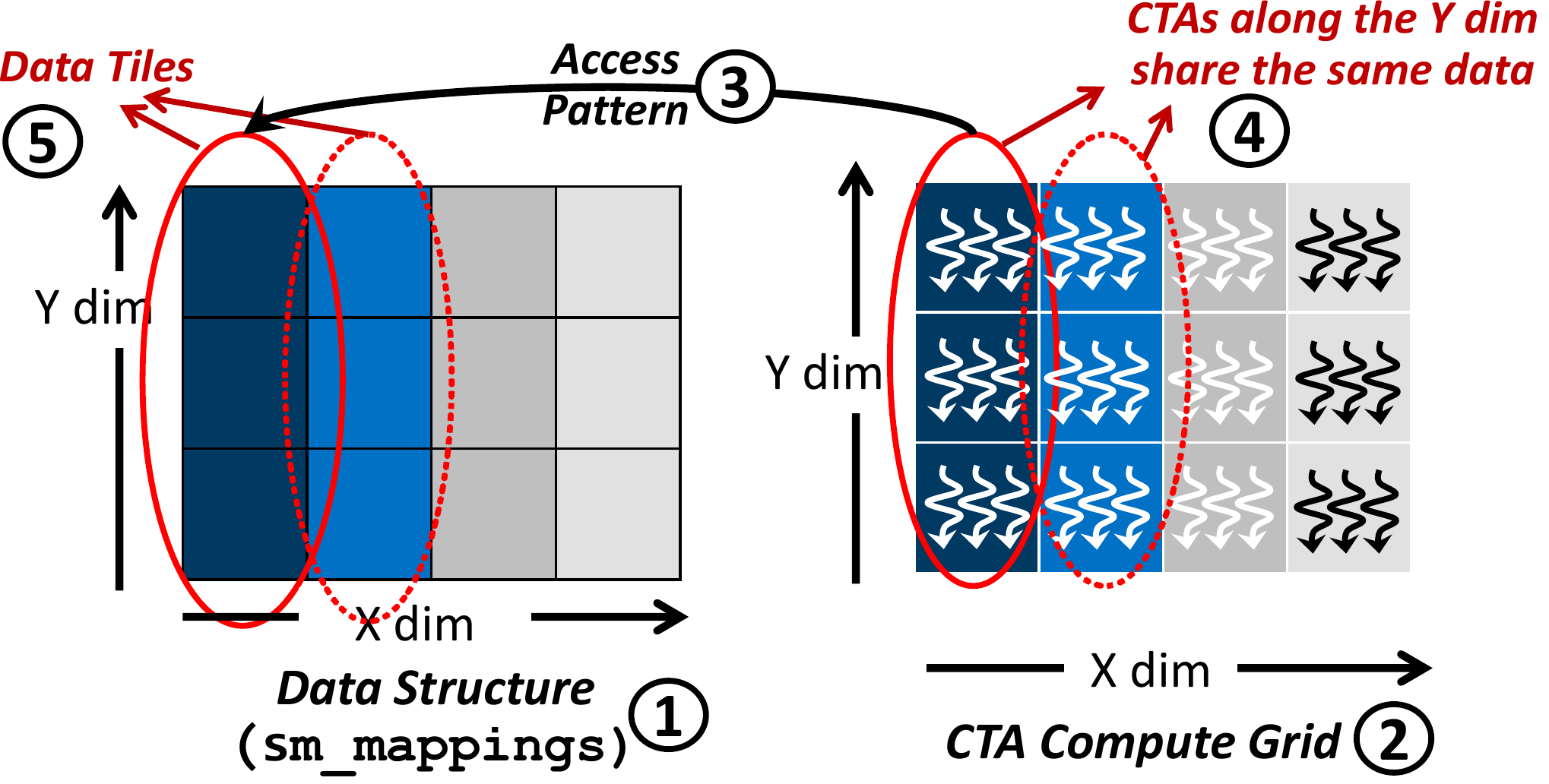}
  \caption{Inter-CTA data locality in \histo (Parboil).}
  \label{fig:introfig1}
\end{figure}

\textbf{A Case Study.} As a motivating example, we examine a common locality pattern of
CTAs sharing data (\emph{inter-CTA locality}), seen in the \histo benchmark
(Parboil~\cite{parboil}). \histo has a predominantly accessed data structure
(\texttt{sm\_mappings}).
Figure~\ref{fig:introfig1} depicts
how this data structure \ding{172} is accessed by the CTA grid \ding{173}. All threads in GPU programs
are partitioned into a multidimensional grid of CTAs. CTAs with the same color
access the same data range (also colored the same) \ding{174}. As depicted, there is
plentiful reuse of data between CTAs \ding{175} and the workload has a very deterministic
access pattern. 

Today, however, exploiting
reuse-based
locality or NUMA locality for this workload, \emph{at any level of the compute
stack}, is a challenging task.
The hardware architecture, on the one hand, misses key program information: knowledge of
\emph{which CTAs} access the same data \ding{175}, so they can be scheduled at
the same SM \highlight{(Streaming Multiprocessor)}; and
knowledge of \emph{which data} is
accessed by those CTAs \ding{176}, so that data can be placed at the same NUMA zone. 
The programmer/compiler, on the other hand, has \emph{no easy access} to hardware techniques such as
CTA scheduling or data placement. Furthermore, optimizing for locality is a
tedious task as a \emph{single technique} alone is insufficient to exploit
locality (\xref{sec:motivation_ldesc}). For example, to exploit NUMA locality, we need to
coordinate \emph{data placement} with CTA scheduling to place data close to the
CTAs that access it.
\emph{Hence, neither the programmer, the compiler, nor hardware
techniques can easily exploit the plentiful data locality in this workload.}  

\begin{figure}[h]
  \centering
  \includegraphics[width=0.68\textwidth]{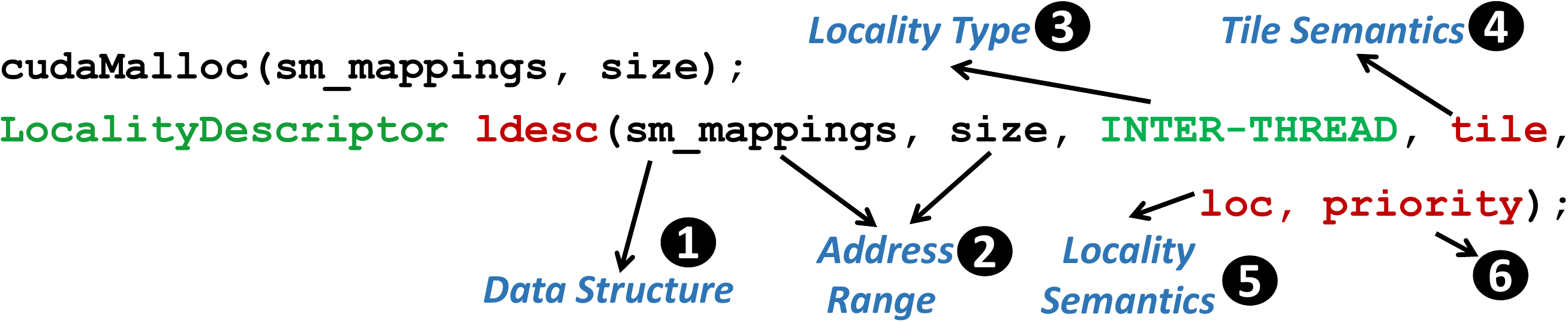}
  \caption{The Locality Descriptor specification for \histo.}
  \label{fig:introfig2}
\end{figure}

\textbf{Our Approach.} To address these
challenges, we introduce the Locality Descriptor: a \emph{cross-layer} 
abstraction to \emph{express} and \emph{exploit} different forms of data locality that
all levels of the compute stack{\textemdash}from application to architecture{\textemdash}recognize. 
The Locality Descriptor \One introduces a flexible and portable interface that enables the programmer/software to \emph{explicitly}
express and optimize for data locality and \Two enables the hardware to
\emph{transparently} 
coordinate a range of architectural techniques (\highlight{such as CTA scheduling, cache
management, and data
placement}), guided by the knowledge of \emph{key
program semantics}.   
Figure~\ref{fig:introfig2} shows how the programmer or compiler can use the
Locality Descriptor to leverage both reuse-based locality and NUMA locality.
We briefly summarize how the Locality Descriptor works here, and provide an end-to-end
description in the rest of the chapter. 

First, each instance of a Locality Descriptor describes a single \emph{data structure's} locality
characteristics (in Figure~\ref{fig:introfig2}, \texttt{sm\_mappings} \ding{182}) and conveys
the corresponding address range \ding{183}. 
Second, we define several fundamental \emph{locality types} as a contract 
between the software and the architecture. The locality type, which can be 
\texttt{INTER-THREAD}, \texttt{INTRA-THREAD}, or \texttt{NO-REUSE}, drives
the underlying optimizations used to exploit it. In \texttt{histo},
\interthread~\ding{184} 
describes inter-CTA locality. \highlight{The locality type~\ding{184} and the
\emph{locality
semantics}~\ding{186}, such as access pattern, inform the architecture to use CTA scheduling 
and other techniques that exploit the corresponding locality type 
(described in \xref{sec:abstraction_overview}).} 
Third, we partition the data structure into \emph{data tiles} that are used to
relate data to the
threads that access it. In Figure~\ref{fig:introfig1}, each data range that has the same color (and
is hence accessed by the same set of CTAs) forms a data tile. Data tiles and the
threads they access are described
by the \emph{tile semantics} \ding{185}
(\xref{sec:abstraction_overview}), which informs the architecture \emph{which CTAs}
to schedule together and \emph{which
data} to place at the same NUMA zone. 
Fourth, we use a software-provided \emph{priority} \ding{187} to reconcile
optimizations between {Locality Descriptor}s for different data structures in the same
program if they 
require \emph{conflicting} optimizations (e.g., different CTA scheduling
strategies). 

\highlight{We evaluate the benefits of using {Locality Descriptor}s to exploit different forms of both
\emph{reuse-based} locality and \emph{NUMA} locality.} 
We demonstrate that {Locality Descriptor}s effectively leverage program semantics to improve performance by 26.6\% on average (up to 46.6\%) when exploiting reuse-based
locality in the cache hierarchy, and by 53.7\% (up to 2.8X) when exploiting NUMA
locality. 

The major \textbf{contributions} of this work are: 
\begin{itemize}
\item This is the first work to propose a holistic cross-layer approach to
explicitly \emph{express} and \emph{exploit} data locality in GPUs as a first class entity in both the programming model and
the hardware architecture. 
\item We design the Locality Descriptor, which enables \One the 
software/programmer to describe data locality in an architecture-agnostic
manner and \Two the architecture to leverage key program semantics and
coordinate many architectural techniques transparently to the software. We
architect an end-to-end extensible design to connect five
architectural techniques (CTA scheduling, cache bypassing, cache prioritization,
data placement, prefetching) to the Locality Descriptor programming abstraction.  
\item We comprehensively evaluate the efficacy and versatility of the {Locality Descriptor} in leveraging different
types of reuse-based and NUMA locality, and demonstrate significant performance improvements over state-of-the-art approaches.  
\end{itemize}

%
\subsection{Motivation}
\label{sec:motivation_ldesc} 
We use two case studies to motivate our work: \One Inter-CTA locality, where
different CTAs access the same data and \Two NUMA locality in a GPU
with a NUMA memory system. 

\subsubsection{\highlight{Case Study 1: Inter-CTA Locality}} 
A GPU kernel is formed by a \emph{compute grid}, which is
a 3D grid of Cooperative Thread Arrays (CTAs). Each
CTA, in turn, comprises a 3D array of threads. Threads are scheduled for execution 
at each Streaming Multiprocessor (SM) at a CTA granularity. 
Inter-CTA locality~\cite{improving-lee-hpca14, locality-li-asplos17,
laperm-wang-isca16, twinkernels-gong-cgo17, improving-chen-cal17,
cachehierarchy-lai-tc15, enabling-wu-ics15} 
is data reuse that exists when multiple CTAs access the same data.
CTA scheduling~\cite{improving-lee-hpca14, locality-li-asplos17,
laperm-wang-isca16, twinkernels-gong-cgo17, improving-chen-cal17,
cachehierarchy-lai-tc15, enabling-wu-ics15} 
is a technique that is used to schedule CTAs that share data
at the same SM to exploit inter-CTA locality at the per-SM local L1 caches. 

\begin{figure}[h]
  \centering
  \includegraphics[width=0.68\textwidth]{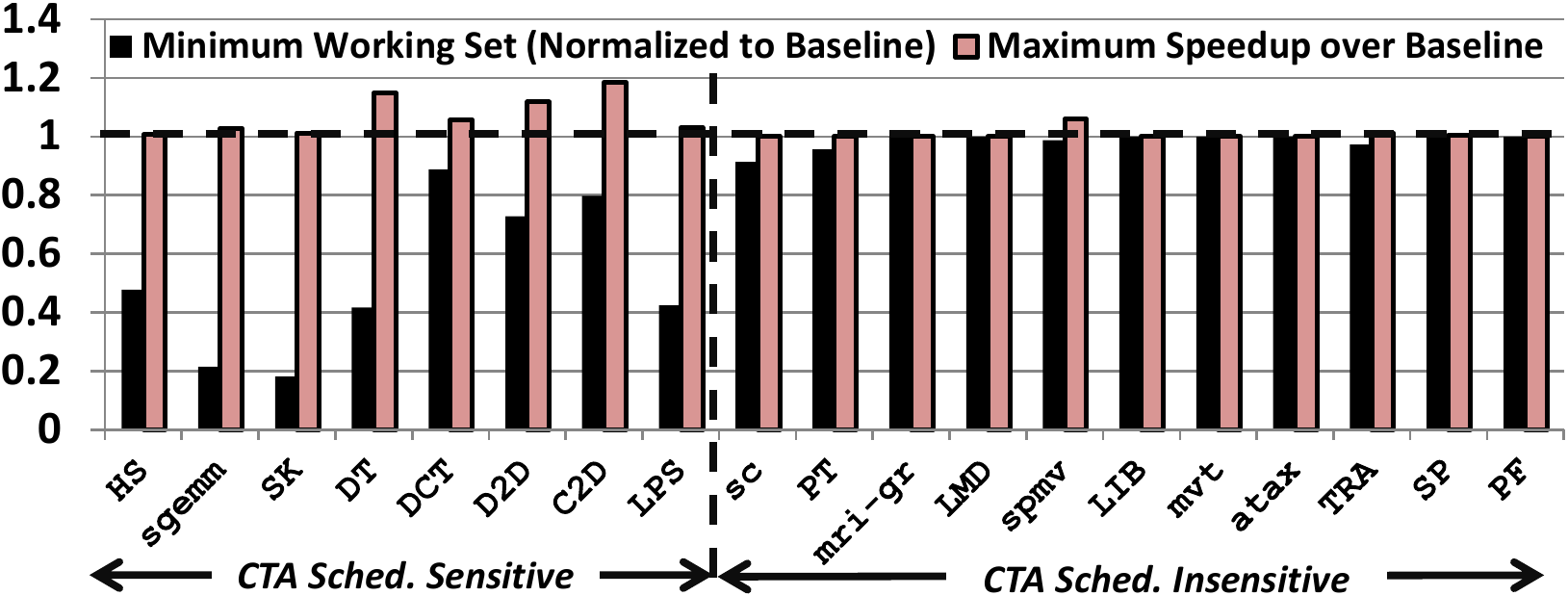}
  \caption{CTA scheduling: performance and working set.}
  \label{fig:motiv-sched}
\end{figure}

To study the impact of CTA scheduling, we evaluate 48 scheduling strategies, each
of which groups (i.e., clusters)
CTAs differently: either along the grid's X, Y, or Z dimensions, or in different
combinations of the three. 
The goal of CTA scheduling for locality is to maximize sharing between CTAs at
each SM and effectively \emph{reduce} the amount of data accessed by each SM. Hence, as
a measure of how well CTA scheduling improves locality for each workload, in
Figure~\ref{fig:motiv-sched} we plot the \emph{minimum} working set at each SM (normalized to baseline)
across all 48 scheduling strategies.
We define \emph{working set} as the average number of uniquely
accessed cache lines at each SM. A smaller working set implies fewer capacity
misses, more sharing, and better locality. 
Figure~\ref{fig:motiv-sched} also shows the \emph{maximum} performance
improvement among all evaluated scheduling strategies for each benchmark. 

\begin{figure}[h]
  \centering

  \includegraphics[width=0.68\textwidth]{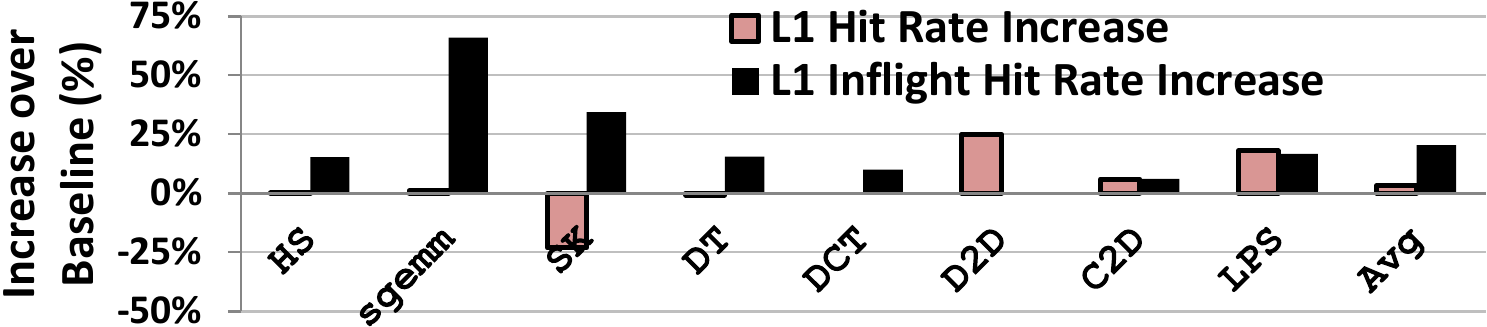}
  \caption{CTA scheduling: L1 hit rate and L1 inflight hit rate.}
  \label{fig:motiv-cache}
\end{figure}

Figure~\ref{fig:motiv-sched} shows that even though CTA scheduling
significantly reduces the working set of CTA-scheduling-sensitive applications
(on the left) by 54.5\%, it has almost no impact on performance (only 3.3\% on
average across all applications).
To understand this minimal impact on performance, in Figure~\ref{fig:motiv-cache} we plot the 
corresponding increase in L1 hit rate for the specific scheduling strategy that
produced the smallest working set (only for the scheduling-sensitive workloads). We also plot the increase in \emph{inflight hit rate}, which we
measure as the number of MSHR hits, i.e., another thread already accessed the
same cache line, but \highlight{the line} has not yet been retrieved from memory and hits
at the MSHRs.

Figure~\ref{fig:motiv-cache} shows that CTA scheduling has little
impact in improving the L1 hit rate (by 3\% on average) with the
exception of \texttt{D2D} and \texttt{C2D}. This explains the minimal performance
impact. CTA scheduling, however, a substantially
increases L1 inflight hit rate (by 20\% on average). 
\highlight{This indicates that even though there is higher data locality due to
more threads
sharing the same data, these threads wait for the same
data at the \emph{same time}. 
As a result, the increased
locality simply causes more threads to \emph{stall}, rather than improving
hit
rate.}
Hence, while CTA scheduling is very effective in \emph{exposing} data locality, we
still need to address other challenges (e.g., threads stalling together) to
obtain
performance gains from improved data locality. 
\highlight{Furthermore, determining \emph{which} scheduling strategy to use is another
challenge, as each application
requires a \emph{different} strategy to maximize locality based on the program's
sharing pattern.} 
\remove{In such
cases, a potential technique that can be used to reduce stalls and improve
performance is \emph{prefetching}. The increased criticality of the data (as a
result of more exposed locality and sharing) and the increased stall time makes
prefetching a good candidate to improving the benefits of CTA scheduling. \thoughts{EIMAN-11-2-17:
I'm not sure here that we want to start talking about prefetching, it makes it sound
like we're just motivating the use of prefetching. Whereas what you're doing here should
just be saying that trying to do scheduling alone is not enough, i don't think you need
to start talking about how you're going to solve it and make the solution sound
obvious in any way here. You can probably do without the last 2-3 sentences here and
just have something that leads you into the next paragraph. Nandita: Fixed.}}

In summary, to exploit inter-CTA locality \One
the hardware-controlled CTA scheduler needs to know \emph{which CTAs access the same
data}, to choose an appropriate scheduling strategy (this requires knowledge of
program semantics) and \Two \highlight{a scheduling strategy that \emph{exposes} locality
in the cache
is \emph{not} necessarily sufficient
for translating locality into performance (we need to coordinate other
techniques).}  

\subsubsection{\highlight{Case Study 2: NUMA Locality}} 
\label{sec:motiv-numalocality} 

\begin{figure}[h]
  \centering
  \includegraphics[width=0.69\textwidth]{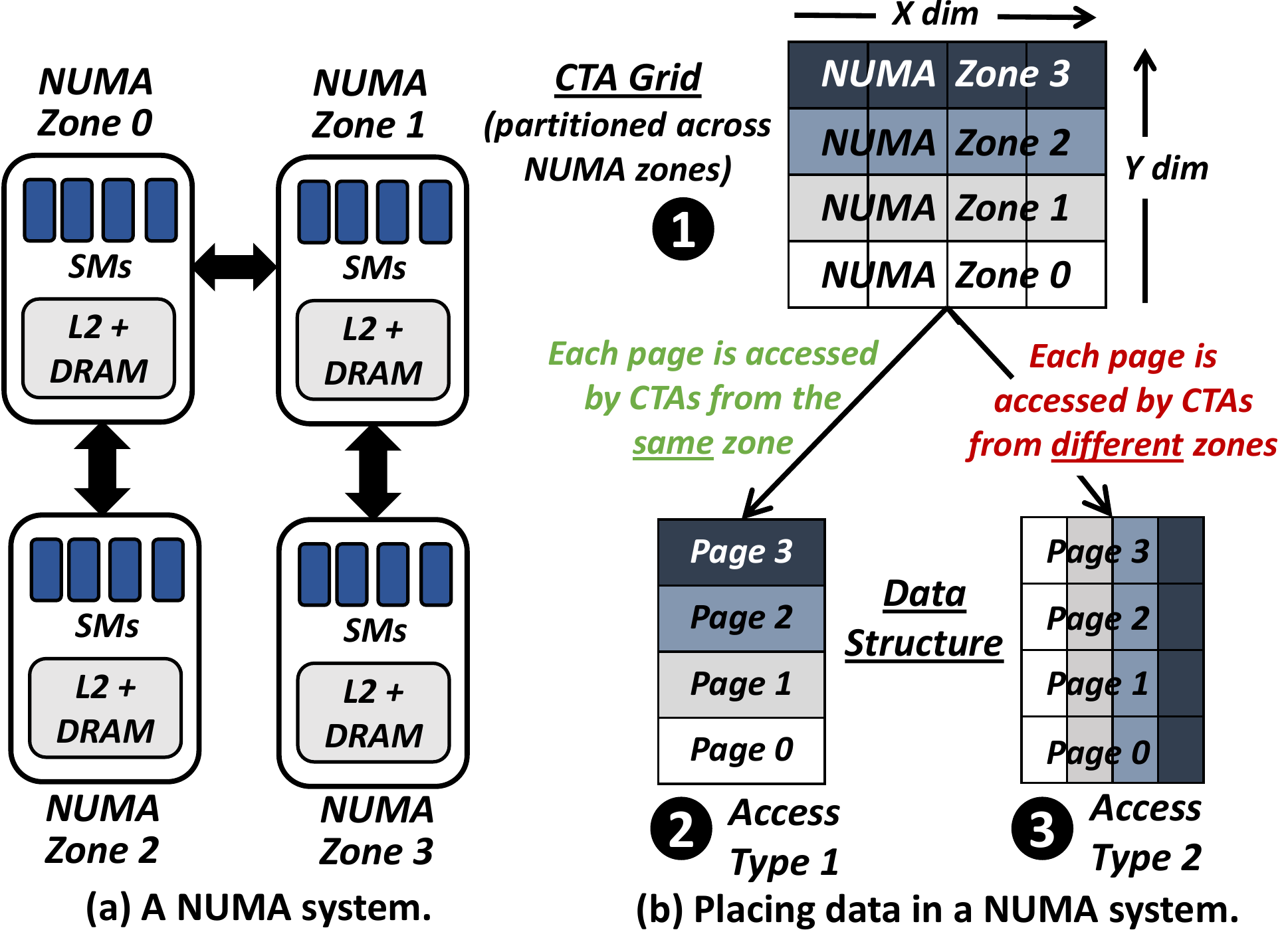}
  \caption{NUMA locality.}
  \label{fig:motiv-numa}
\end{figure}

For continued scaling, future GPUs are expected to \highlight{employ 
non-uniform memory access (NUMA) memory systems. This can be in the form of multiple memory
stacks~\cite{tom-hsieh-isca16,toward-kim-sc17}, unified virtual address spaces in
multi-GPU/heterogeneous} 
systems~\cite{unlocking-agarwal-hpca15,
automatic-cabezas-ics15, towards-sakai-candar16,
achieving-kim-ppopp11, transparent-lee-pact13,
automatic-cabezas-pact14,umh-ziabari-taco16} or multi-chip
modules, where SMs and memory modules are
partitioned into \emph{NUMA zones} or \emph{multiple GPU modules}~\cite{mcmgpu-arunkumar-isca17,
numagpu-milic-micro17}. Figure~\ref{fig:motiv-numa} depicts the system evaluated
in~\cite{mcmgpu-arunkumar-isca17} with four NUMA zones. \highlight{A 
request to a remote NUMA zone goes over the 
lower bandwidth inter-module interconnect, has higher
latency, and incurs more traffic compared to \emph{local
requests}~\cite{mcmgpu-arunkumar-isca17}}. To maximize performance and efficiency, we need
to control \One how data is placed across NUMA zones and \Two how CTAs
are scheduled to maximize local accesses. 

To understand why this is
a challenging task, let us consider the heuristic-based hardware mechanism proposed
in~\cite{mcmgpu-arunkumar-isca17}, 
where the CTA
grid is partitioned across the 4 NUMA zones such that contiguous CTAs
are scheduled at the same SM. Data is
placed at the
page granularity (64KB)
at the NUMA zone where it is \emph{first accessed},
based on the heuristic that consecutive CTAs are likely to share the same page(s). 
Figure~\ref{fig:motiv-numa} 
depicts a CTA grid (\ding{182}), which is partitioned between NUMA zones in this
manner{\textemdash}consecutive CTAs along the X dimension are scheduled at the
same NUMA zone. 
\highlight{This heuristic-based mechanism works well for Access Type 1
(\ding{183}),} where CTAs that
are scheduled at the same NUMA zone access the same page(s) (the color
scheme depicts which CTAs access what data). However, for Access Type 2
(\ding{183}), this policy fails as a single page is shared by CTAs that are scheduled
at \emph{different zones}.
\highlight{Two challenges cause this policy to fail.
First, suboptimal scheduling: the simple
scheduling policy~\cite{mcmgpu-arunkumar-isca17} does not \emph{always} co-schedule CTAs that share the same pages at the
same zone. This happens when scheduling is \emph{not} coordinated with the
application's access pattern. Second, large and fixed page granularity: more CTAs than what can be scheduled at a single zone
may access the \emph{same} page. This happens when there are fine-grained accesses
by \emph{many} CTAs to each page and when different data structures are accessed
by the CTAs in \emph{different} ways.\remove{ The same scheduling strategy may not co-schedule
CTAs that share pages for \emph{all} data structures.}}  
For these reasons (as we evaluate in
\xref{sec:eval_numa}),
a heuristic-based approach is often ineffective at exploiting NUMA locality. 
%
\subsubsection{Other Typical Locality Types}

\highlight{We describe other locality types, caused by different access
patterns, and  
require other optimizations for locality next.}

\textbf{Inter-warp Locality.} Inter-warp locality is data reuse between
warps that belong to the same/different CTAs. This type of locality occurs in 
stencil programs (workloads such as \texttt{hotspot}~\cite{rodinia} and 
\texttt{stencil}~\cite{parboil}),
where each thread accesses a set of neighboring data
elements, leading to data reuse between neighboring warps. Inter-warp locality is also
a result of misaligned
accesses to cache lines by threads in a warp~\cite{accesspattern-koo-isca17,manythread-lee-micro10,
revealing-koo-iiswc15}, since data is always fetched at the 
cache line granularity (e.g., \texttt{streamcluster}~\cite{rodinia} and 
\texttt{backprop}~\cite{rodinia}). 
Inter-warp locality has short reuse
distances~\cite{accesspattern-koo-isca17} as nearby warps are typically 
scheduled together and caching policies such as LRU can
exploit a significant portion of this locality. However, potential for 
improvement exists using techniques such as inter-warp
prefetching~\cite{apres-oh-isca16,manythread-lee-micro10,locality-li-asplos17} or CTA scheduling to
co-schedule CTAs that share data~\cite{locality-li-asplos17}. 

\textbf{Intra-thread Locality.\@} This is reuse of data by the \emph{same thread}
(seen in \texttt{LIBOR}~\cite{sdk} and 
\texttt{lavaMD}~\cite{rodinia}), where
each thread operates on its own working set. Local memory
usage in the program is also an example of this type of locality.
The key challenge
here is \emph{cache thrashing} because \One the overall working set of
workloads with this locality type \highlight{is large} due to lack of sharing among threads and \Two the reuse
distance per thread \highlight{is large} as hundreds of threads are swapped in and
out by the GPU's multithreading before the data is reused by the same thread. Techniques that have
been proposed to address cache thrashing include cache bypassing or
prioritization (e.g. pinning) of different
forms~\cite{adaptive-tian-gpgpu15, locality-li-ics15,
adaptive-li-sc15, efficient-liang-tcadics,survey-mittal-jlpea16,
ctrlc-lee-iccd16, selectively-zhao-icpads16, coordinated-xie-hpca15,
prioritybased-li-hpca15,
efficient-xie-iccad13,medic,mask} and/or warp/CTA
throttling~\cite{tor-micro12,nmnl-pact13,locality-li-asplos17,Kayiran-micro2014}.

\subsubsection{Key Takeaways \& Our Goal} 
In summary, locality in GPUs can be of different forms depending on
the GPU program. Each locality type presents \emph{different} challenges that
need to be addressed. \highlight{Tackling each challenge often
requires coordination of 
\emph{multiple} techniques (such as CTA scheduling and cache bypassing), many of which software has
no easy access to.} Furthermore, to be effective, some of these
techniques (e.g., CTA scheduling, memory placement) require knowledge of \emph{program
semantics}, which is prohibitively difficult to infer at run time. 

Our goal is to design a holistic \emph{cross-layer} abstraction{\textemdash}that all levels
of the compute stack recognize{\textemdash}to express and exploit the different
forms of data
locality. Such an abstraction should enable connecting a range of 
architectural techniques with the locality properties exhibited by the program. In doing so, the abstraction should 
\highlight{\One provide the programmer/software a simple, yet powerful interface to express
data locality and
\Two enable architectural techniques to leverage
key program semantics to optimize for locality.} 
\subsection{Locality Descriptor: Abstraction}
\label{sec:approach}

\begin{figure}[h]
  \centering
  \includegraphics[width=0.69\textwidth]{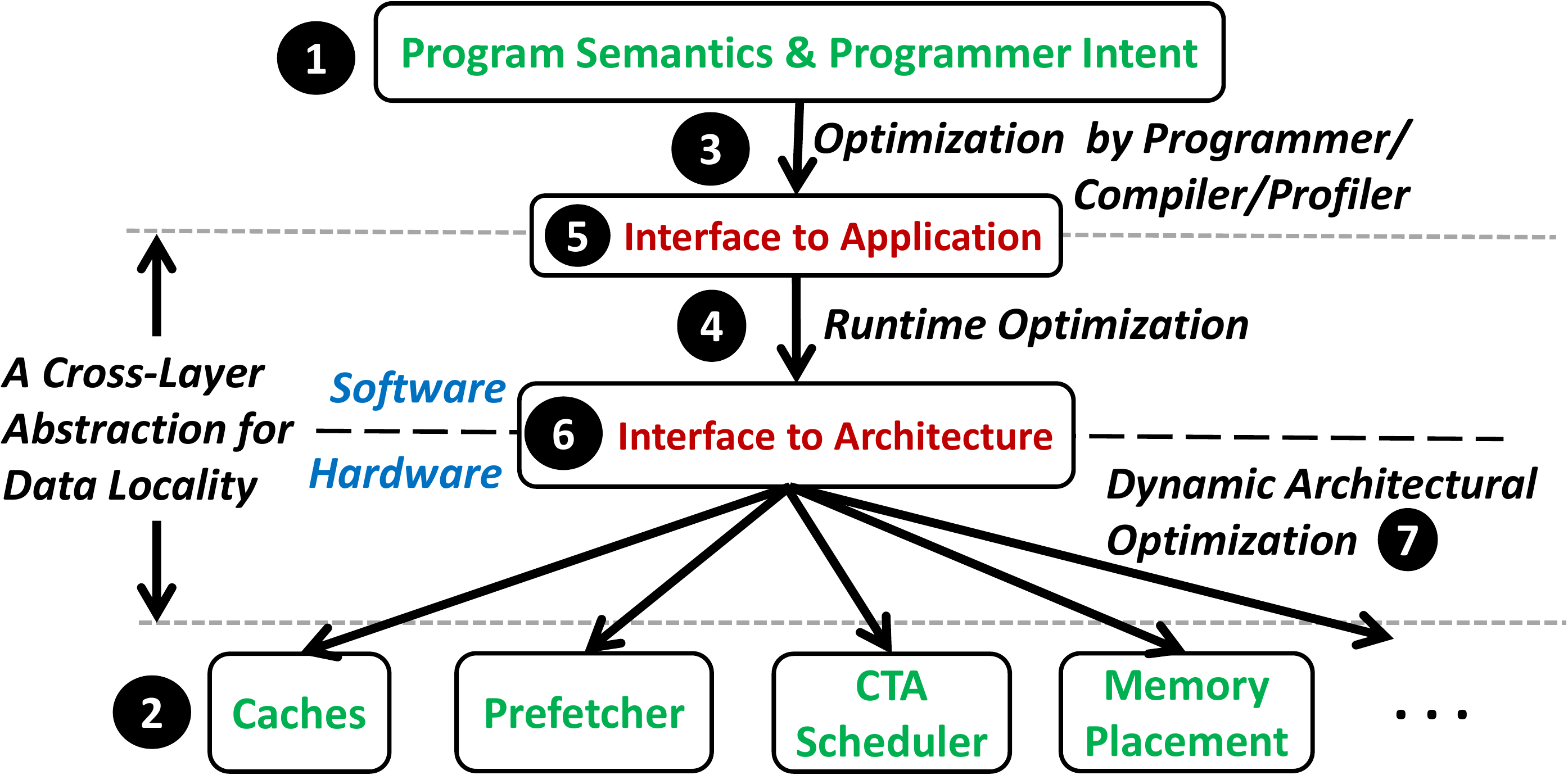}
  \caption{Overview of the proposed abstraction.}
  \label{fig:overview}
\end{figure}

Figure~\ref{fig:overview} depicts an overview of our proposed abstraction.  
The goal is to connect program semantics and programmer intent (\ding{182})  with the
underlying architectural mechanisms (\ding{183}). 
By doing so, we enable optimization at different levels of the stack: 
\One as an additional knob for
static code tuning by the programmer, compiler, or autotuner (\ding{184}), \Two runtime
software optimization (\ding{185}), and \Three
dynamic architectural optimization (\ding{188}) using a combination of
architectural techniques.
This abstraction interfaces with a parallel GPU programming model like CUDA
(\ding{186}) and
conveys key program semantics to the architecture through low overhead
interfaces (\ding{187}). 


\subsubsection{Design Goals}
\label{sec:design_goals}
We set three goals that drive the design of our proposed
abstraction:
\One \emph{Supplemental and hint-based
only}: The abstraction should be an optional add-on to optimize for
\emph{performance}, requiring no change to the rest of the program, nor should
it impact the program's 
\emph{correctness}. \Two \emph{Architecture-agnosticism}: The abstraction 
should abstract away any low-level details of the architecture (e.g., cache
size, number of SMs, caching policy). Raising the abstraction level improves
portability, reduces programming effort, and enables architects to flexibly
design and improve techniques across GPU generations, transparently to the
software. \Three \emph{Generality and flexibility}: The abstraction should flexibly
describe a wide range of locality types typically seen in GPU programs. It should be
easy to extend what the abstraction can express and the underlying architectural
techniques that can benefit from it.

\subsubsection{Example Program}
\label{sec:example}

\begin{figure}[h]
  \centering
  \includegraphics[width=0.58\textwidth]{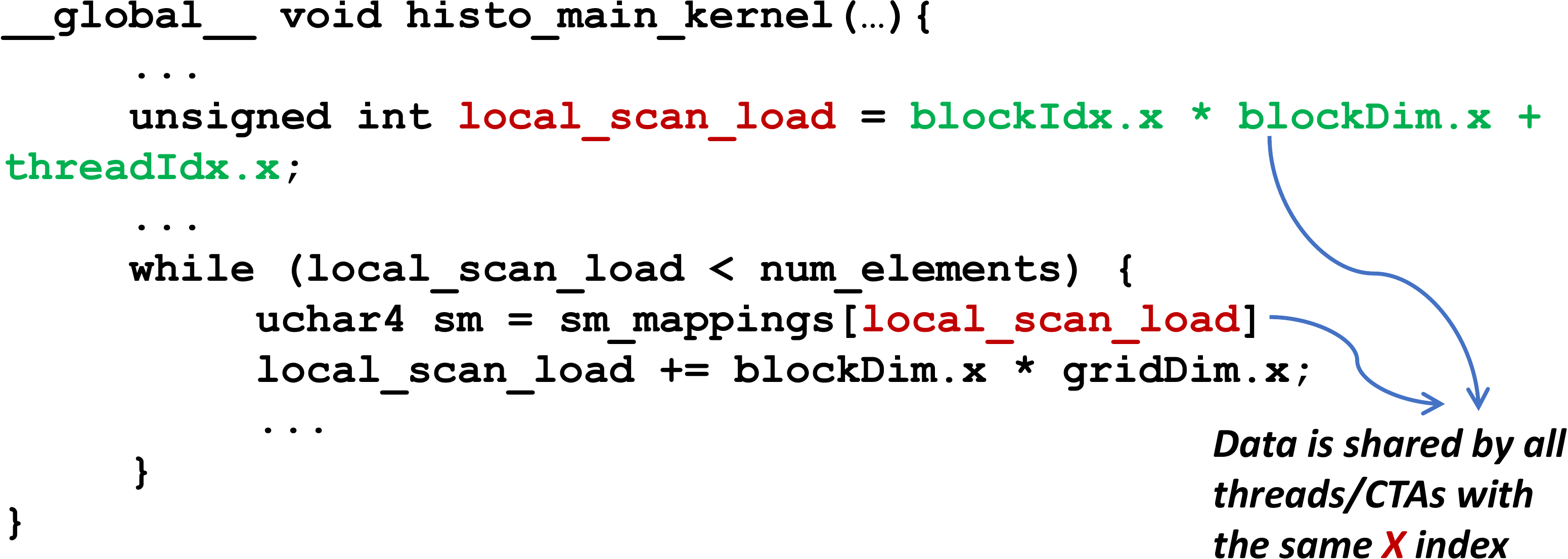}
  \caption{Code example from \histo (Parboil).}
  \label{fig:histocode}
\end{figure}

We describe the Locality Descriptor with the help of the \histo (Parboil~\cite{parboil})
workload example described
in \xref{sec:intro}. 
We begin with an overview of how a Locality Descriptor is specified for \texttt{histo} and
then describe the key ideas behind the Locality Descriptor's components. 
Figure~\ref{fig:histocode} depicts a code
example from this application. The primary data structure 
is \texttt{sm\_mappings}, which is indexed by a function of the thread
and block index only along the $X$ dimension. Hence, the threads that have
the same index along the $X$ dimension access the same part of
this data structure.

\begin{figure}[h]
  \centering
  \includegraphics[width=0.68\textwidth]{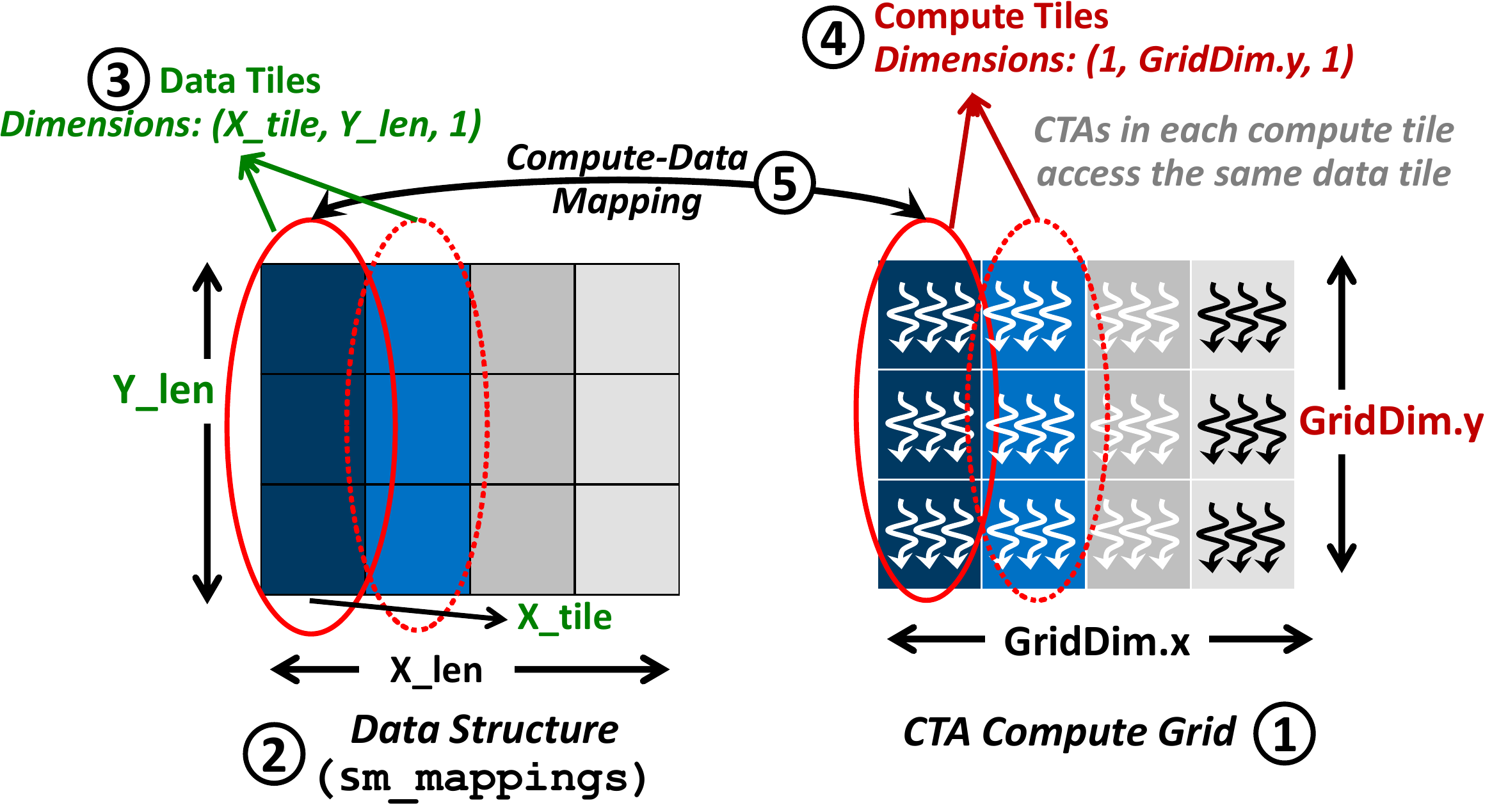}
  \caption{Data locality and compute/data tiles in \histo.}
  \label{fig:histo_depict_tile}
\end{figure}

\begin{figure}[h]
  \centering
  \includegraphics[width=0.58\textwidth]{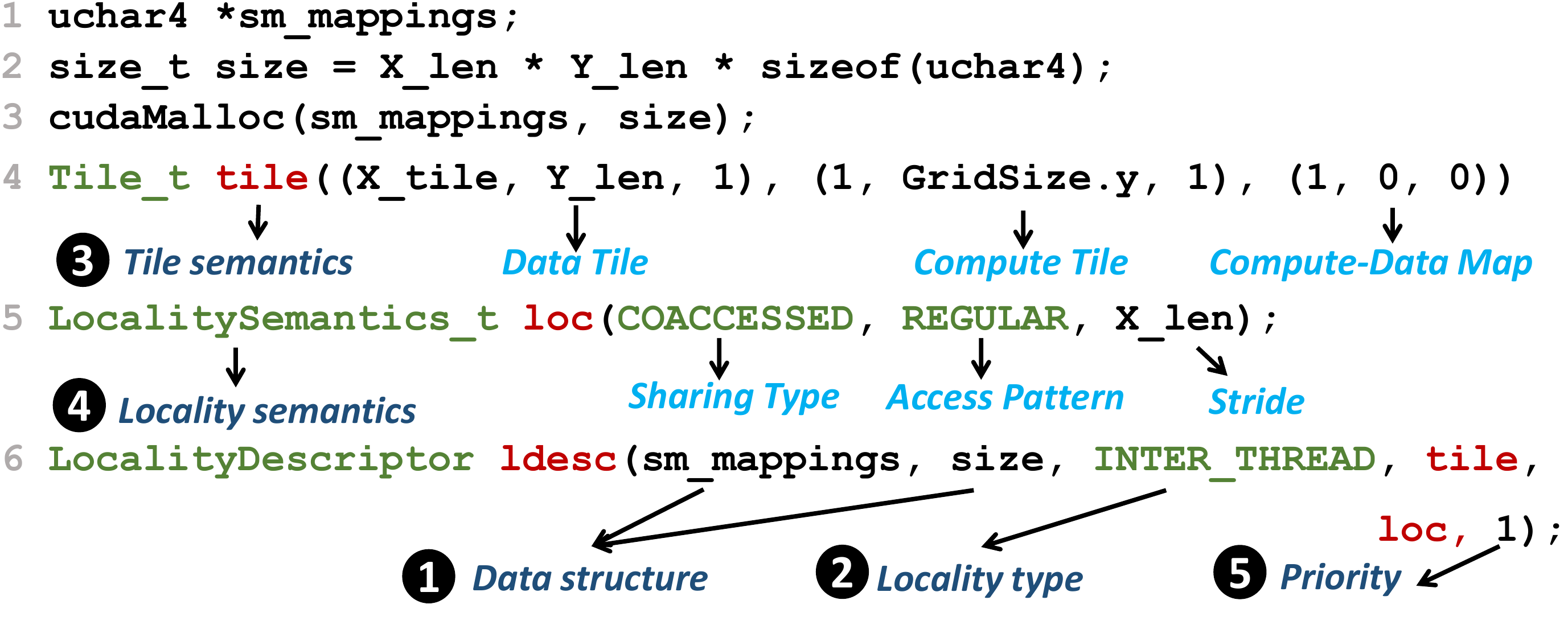}
  \caption{Locality Descriptor example for \histo.}
  \label{fig:histo_depict}
\end{figure}

Figure~\ref{fig:histo_depict_tile} depicts the data locality in
this application in more detail. \ding{172} is the CTA grid and \ding{173} is the
\texttt{sm\_mappings} data structure. The CTAs that
are colored the same access the same data range (also colored the
same). As \xref{sec:intro} discusses, in order to describe locality with the 
Locality Descriptor abstraction, we partition each data structure in
\emph{data tiles} \ding{174} that group data shared by the same CTAs. In addition,
we partition the 
CTA grid along the X dimension into \emph{compute tiles}~\ding{175} to group together CTAs
that access the same data tile. \highlight{We then relate the compute and data tiles with a
\emph{compute-data} mapping \ding{176} to describe which compute tile
accesses which data tile.} Figure~\ref{fig:histo_depict} depicts the
code example to express
the locality in this example with a Locality Descriptor. As \xref{sec:intro}
describes, the key components of a Locality Descriptor are: the associated data
structure (\ding{182}),
its locality type (\ding{183}), tile semantics (\ding{184}), locality
semantics (\ding{185}),
and its priority (\ding{186}). We now describe each component in
detail. 

\subsubsection{An Overview of Key Ideas and Components}
\label{sec:abstraction_overview}

\begin{figure}[h]
  \centering
  \includegraphics[width=0.48\textwidth]{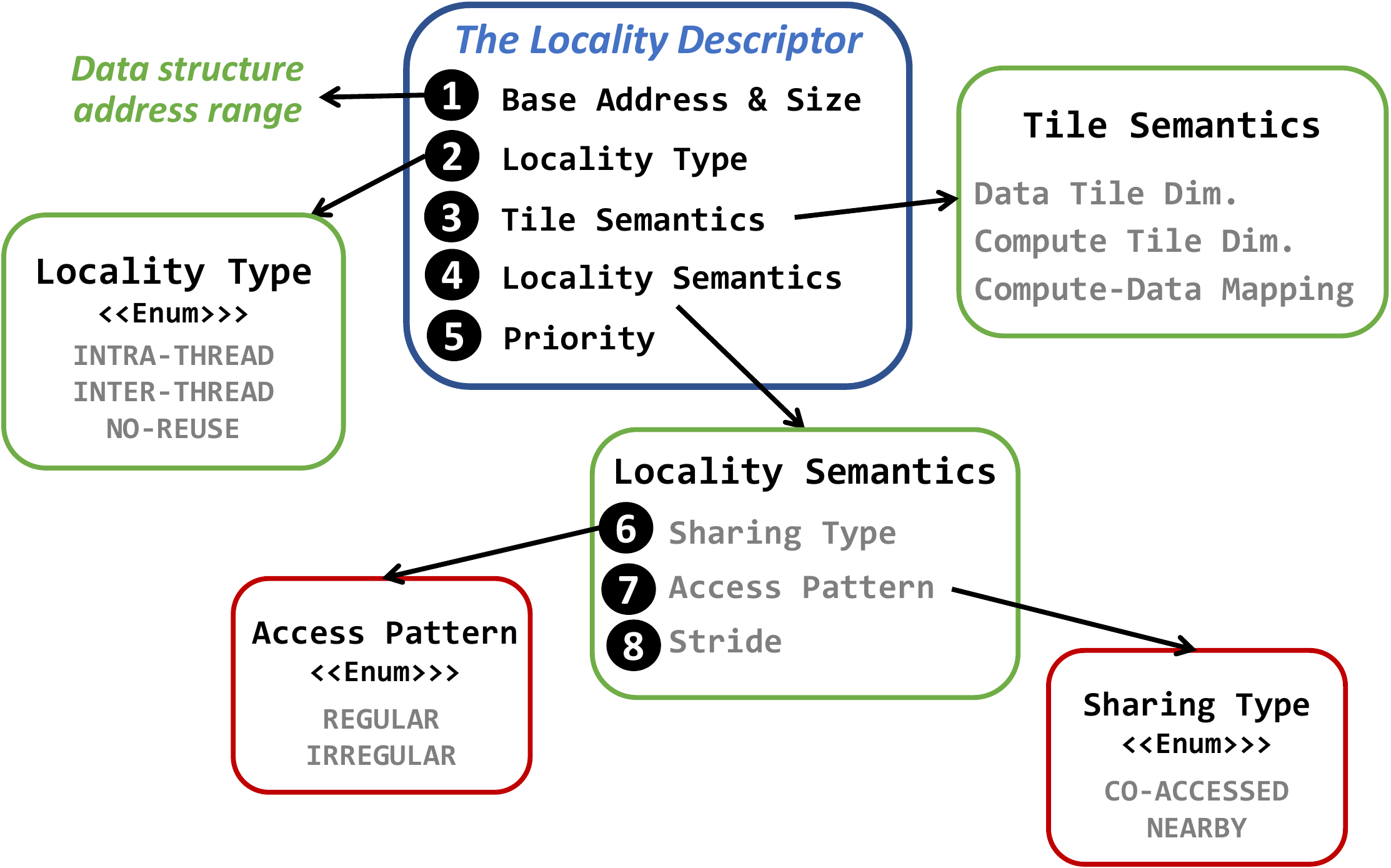}
  \caption{Overview of the Locality Descriptor.}
  \label{fig:abstraction_overview}
\end{figure}

Figure~\ref{fig:abstraction_overview} shows an overview of the components of the Locality Descriptor. 
We now describe the five key components and the key insights behind their design.

\textbf{Data Structure (\ding{182}).}~We build the abstraction around the program's \emph{data structures}
(each specified with its base address and size). Each instance of the Locality Descriptor describes the locality
characteristics of a single data structure.
Designing the Locality Descriptor around the program's data structures is advantageous for
two reasons. First, it ensures \emph{architecture-agnosticism} as a
data structure is a software-level concept, easy for the programmer to reason
about. 
Second, it is natural to tie locality properties to data
structures because in GPU programming models, \emph{all} threads typically
access \highlight{a given} 
data structure in the same way. 
For example, some data structures are simply
streamed through by all threads with no reuse. Others are heavily reused
by groups of threads.

\textbf{Locality Type (\ding{183}).}
~Each instance of the Locality Descriptor has an explicit \emph{locality type}, which forms
a contract or \emph{basis of understanding} between software and hardware. 
This design
choice leverages the known
observation that locality type often determines the underlying
optimization mechanisms (\xref{sec:motivation_ldesc}). Hence,
software need only specify locality type and the system/architecture
transparently employs a different set of architectural techniques
based on the specified type. 
We provide three
fundamental types: \One \intrathread: when the reuse of data is by the
same thread itself, \Two \interthread: when the reuse of data is due
to sharing of data between different threads (e.g., inter-warp or
inter-CTA locality), and \Three \noreuse: when there is no reuse of
data (NUMA locality can still be exploited, as described below). 
If a data structure has multiple locality
types (e.g., if a data structure has both \emph{intra-thread} and
\emph{inter-thread} reuse), multiple {Locality Descriptor}s with different types can be specified for that data
structure. We discuss how these cases are handled in \xref{sec:detailed_design}. 


\textbf{Tile Semantics (\ding{184}).}
~As data locality is essentially the outcome of how computation
accesses data, we need to express the \emph{relation between compute
  and data}. To do this, we first need a \emph{unit} of computation
and data as a basis. To this end, we \remove{create a flexible unit
of compute and data by \emph{partitioning}}partition the data structure into a
number of \emph{data tiles (\dtiles)} and the compute grid into a number of
\emph{compute tiles (\ctiles)}. Specifically, a \dtile is a 3D range of
data elements in a data structure and a \ctile is a 3D range of threads or CTAs
in the 3D compute grid 
(e.g., \ding{174} and \ding{175} in
Figure~\ref{fig:histo_depict_tile}).

This design provides two major benefits. First, it provides a
\emph{flexible and architecture-agnostic} scheme to express
locality types. For example, to express \interthread locality, a \mbox{\dtile} is
the range of data shared by a set of CTAs; and each such set of CTAs forms a \ctile. 
To express \intrathread locality, a \ctile is just a single thread and the \dtile is the
range of data that is reused by that single thread. Second, such
decomposition is \emph{intrinsic} and conceptually similar to the
\highlight{existing} 
hierarchical tile-based GPU programming model. Tile partitioning can hence be
done easily by the programmer or the compiler using techniques such
as~\cite{automatic-cabezas-ics15,automatic-cabezas-pact14}. For
irregular data structures (e.g, graphs), which cannot be easily
partitioned, the Locality Descriptor can be used to describe the entire data
structure. This imposes little limitation as such data structures
exhibit \highlight{an irregular type of} locality that cannot be easily described by
software. \remove{Hence, we trade off a little generality (i.e., fined-grain
optimization of irregular data structures) to significantly reduce
design complexity.}

We further reduce complexity in expression by stipulating only an
\emph{imprecise} description of locality. There are two primary
instances of this. First, we use a simple 1:1
mapping between \ctile and \dtile. This is a
non-limiting simplification because data locality is fundamentally
about \emph{grouping} threads and data based on sharing. If multiple
\ctiles access the same \dtile, a bigger \ctile should simply be specified. In
an extreme case, where the entire data structure is
shared by all threads, we should only have one \ctile and one \dtile. In
another case, where there is only intra-thread locality (no sharing
among threads), there is a natural 1:1 mapping between each thread and its
working set. This simplification
would be an approximation in cases with irregular sharing, which is
out of the Locality Descriptor's scope because of the high complexity and low
potential\remove{ in these cases}. Second,
\ctile and \dtile partitioning implies that the grouping of threads and data
needs to be \emph{contiguous}{\textemdash}a \ctile cannot access a set of 
data elements that is interleaved with data accessed by a different
\ctile. This is, again, a non-limiting
requirement as contiguous data elements are typically accessed
by neighboring threads to maximize spatial locality and reduce memory
traffic. If there is an interleaved mapping between \ctiles
and the \dtiles they access, the
\ctiles and \dtiles can be approximated by 
merging them into bigger tiles \highlight{until they are contiguous}. This design drastically
reduces the expression complexity and covers typical GPU applications.

Specifically, the tile semantics are expressed in three parts: \One
\emph{\dtile dimensions}: The number of data elements (in each dimension) that
form a \dtile. Depending on the data structure, the unit could be any data
type. In the \histo example (\ding{174} in
Figure~\ref{fig:histo_depict_tile}), the \dtile dimensions are
\texttt{(X\_tile, Y\_len, 1)}, where \texttt{X\_tile} is
the range accessed by a single \ctile along the $X$ dimension,
\texttt{Y\_len} is the full length \highlight{of the data structure} (in data elements) along the
$Y$ dimension. \Two \emph{\ctile dimensions}: The number of CTAs in each
dimension that form a \ctile. The compute tile dimensions in the \histo
example (\ding{175} in Figure~\ref{fig:histo_depict_tile}) are
\texttt{(1, GridDim.y, 1)}: 1 CTA along the $X$ dimension,
\texttt{GridDim.y} is the length of the whole grid along the $Y$
dimension, and since this is a 2D grid, the $Z$
dimension is one. \Three \emph{Compute-data map}: We use a simple function to
rank which order to traverse \ctiles first in the 3D compute grid as we
traverse the \dtiles in a data structure in
$X{\rightarrow}Y{\rightarrow}Z$ order. For example, the mapping
function \texttt{(3,1,2)} implies that when \dtiles are traversed in the
$X{\rightarrow}Y{\rightarrow}Z$ order, and the \ctiles are traversed in the
$Y{\rightarrow}Z{\rightarrow}X$ order. In our \histo example, this
mapping (\ding{176} in Figure~\ref{fig:histo_depict_tile}) is simply
\texttt{(1,0,0)} as the \ctiles need only be traversed along the $X$
dimension. This simple function saves runtime overhead, but more complex
functions can also be used. 

\textbf{Locality Semantics (\ding{185}).}
~This component describes
the type of reuse in the data structure as well as the access pattern.
\xref{sec:detailed_design} describes how this information is
used for optimization. This component has two parts: Sharing Type (\ding{187})
and Access Pattern (\ding{188}). There are two options for Sharing Type
(\ding{186}) to reflect the
  typical sharing patterns. \texttt{COACCESSED}
  indicates that the \emph{entire} \dtile is shared by all the
  threads in the corresponding \ctile. \texttt{NEARBY} indicates
  that the sharing is more irregular, with nearby threads in the
  \ctile accessing nearby data elements (the form of sharing
  seen due to misaligned accesses to cache lines or stencil-like
  access patterns~\cite{accesspattern-koo-isca17,
    locality-li-asplos17}). Sharing type can be extended to include
  other sharing types between threads (e.g., partial sharing). \thoughts{EIMAN-10-30-17: what
    other types would that be? Would be good to give a sentence saying
    what other types, and another couple of sentences say how this is
    easily extensible to include those, if that's
    possible.}\thoughts{KEVIN-11-12-17: Add other partial sharing
    types}
Access Pattern~(\ding{187}) is primarily used
  to inform the prefetcher and includes whether the access pattern is
  \texttt{REGULAR} or \texttt{IRREGULAR}, along with a stride (\ding{189}) within
  the \dtile for a \texttt{REGULAR} access pattern.

  \textbf{Priority (\ding{186}).}
  Multiple {Locality Descriptor}s may require \emph{conflicting} optimizations (e.g.,
  different CTA scheduling strategies). We ensure that these
  conflicts are rare by using a conflict resolution mechanism
  described in \xref{sec:detailed_design}. When a conflict
  \emph{cannot} be resolved, we use a software-provided \emph{priority} to
  give precedence to certain {Locality Descriptor}s. This design gives the software
  more control in optimization, and ensures the key data structure(s)
  are prioritized. \tech{This priority is also used to give precedence to a certain
  \emph{locality type}, when there are multiple {Locality Descriptor}s with different types
  for the same data structure. } 
\subsection{Locality Descriptor: Detailed Design}
\label{sec:detailed_design}

We detail the design of the programmer/compiler interface
(\ding{186} in Figure~\ref{fig:overview}), runtime optimizations
(\ding{185}), and the architectural interface and mechanisms (\ding{183}
and \ding{188}){\textemdash}CTA scheduling, memory placement, cache
management, and prefetching.

\subsubsection{The Programmer/Compiler Interface}
The Locality Descriptor can be specified in the code after the data structure is
initialized and copied to global memory. Figure~\ref{fig:histo_depict} is an example. 
If the semantics of a data structure
change between kernel calls, its Locality Descriptor can be re-specified between
kernel invocations. 


The information to specify the Locality Descriptor can be extracted in three
ways. First, the compiler can use static analysis to determine forms
of data locality, \emph{without} programmer intervention, using
techniques \highlight{like}~\cite{improving-chen-cal17,locality-li-asplos17} for inter-CTA
locality. Second, the programmer can annotate the program (as was done in this
work), which is particularly useful 
when the programmer
wishes to hand-tune code for performance and 
to specify the \emph{priority} \highlight{ordering} of data
structures when resolving potential optimization conflicts 
(\xref{sec:abstraction_overview}).
Third, software tools such as auto-tuners or
profilers~\cite{toward-davidson-iwapc10, script-khan-taco13,automatic-sato-sat11} can determine data locality and access
patterns via dynamic analysis. 

During compilation, the compiler extracts the variables that determine the address range of
each Locality Descriptor, so the system can resolve the virtual addresses at run time. The
compiler
then summarizes the Locality Descriptor semantics corresponding to these address ranges and
places this information in the object file. 

\subsubsection{Runtime Optimization}
\label{sec:overall_algorithm}

At run time, the GPU driver and runtime system determine how to
exploit the locality characteristics expressed in the {Locality Descriptor}s based on the
specifics of the underlying architectural components (e.g., number of SMs, NUMA zones). 
Doing so includes determining the: \One CTA scheduling strategy, \Two 
caching policy (prioritization and bypassing), \Three data placement
strategy across NUMA zones, and \Four prefetching strategy. In this work, we provide an algorithm
to coordinate these techniques.
Both, the set of techniques used and the algorithm to coordinate them, are
extensible. As such, more architectural techniques can be added and the
algorithm can be enhanced.
We first
describe the algorithm
that determines \emph{which} architectural techniques are employed for different
locality types, and then detail each architectural
technique in the following subsections. 

\begin{figure}[h]
  \centering
  \includegraphics[width=0.58\textwidth]{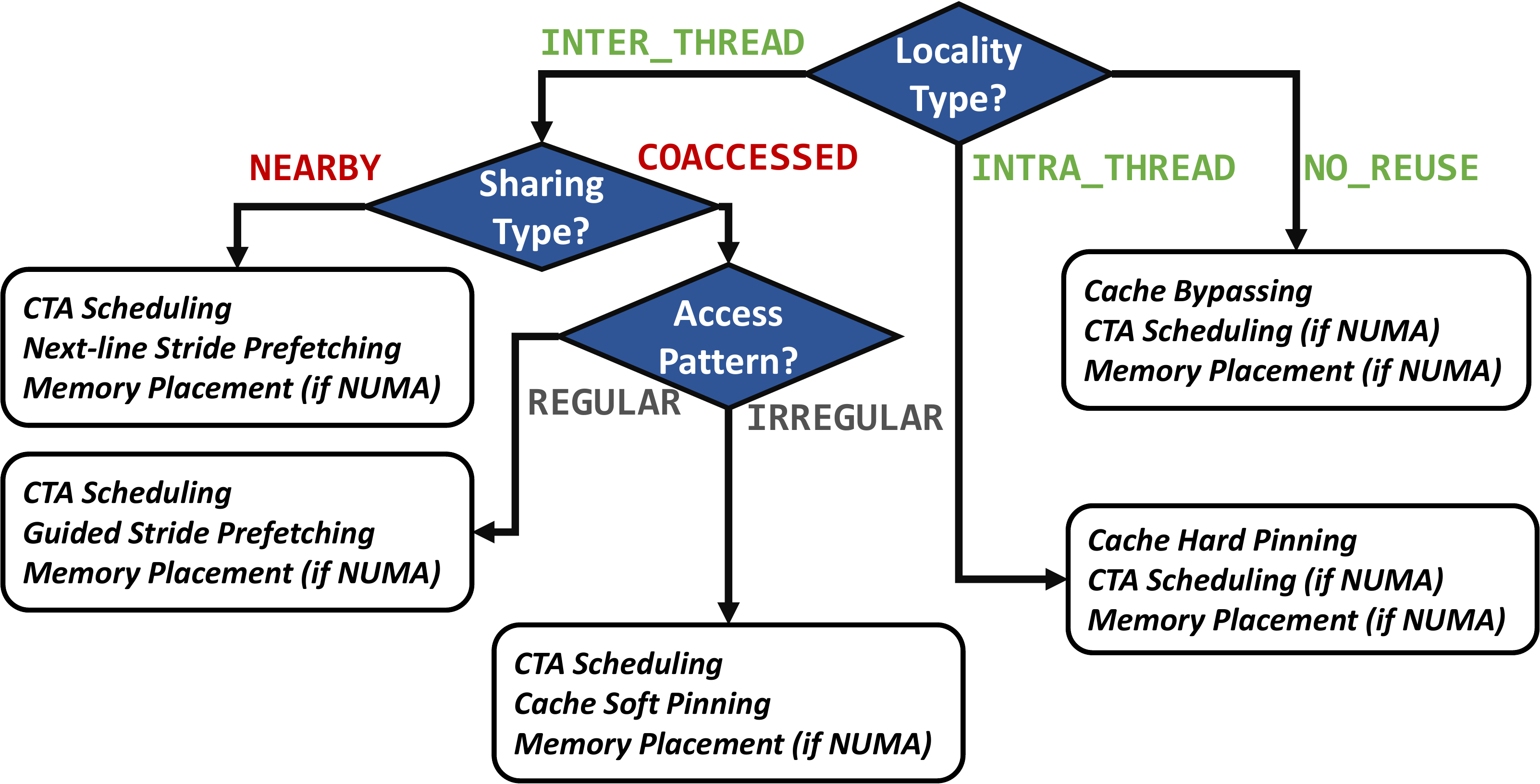}
  \caption{\highlight{Flowchart of architectural optimizations leveraging
  {Locality Descriptor}s.}}
  \label{fig:overall_algorithm}
\end{figure}

Figure~\ref{fig:overall_algorithm} depicts the \highlight{flowchart} that
determines which optimizations are employed.
The algorithm depicted works based on the three \emph{locality types}. First,
for \interthread {Locality Descriptor}s, we employ CTA scheduling (\xref{sec:cta_scheduling}) to expose locality. We 
also use other techniques based on the \emph{access pattern} and
the \emph{sharing type}: \One For \highlight{\texttt{COACCESSED} sharing with a
\texttt{REGULAR} access pattern}, we use guided stride
prefetching (\xref{sec:prefetcher}) to overlap the long latencies when many threads are stalled
together waiting on
the same data; \Two For \highlight{\texttt{COACCESSED} sharing with a
\texttt{IRREGULAR} access pattern}, we
employ cache prioritization using \emph{soft pinning} (\xref{sec:cache}) to keep
data in the cache long enough to exploit locality; \Three For 
\texttt{NEARBY} sharing, we use simple nextline prefetching tailored
to the frequently-occurring access pattern. Second, for an \intrathread
Locality Descriptor, we employ a thrash-resistant caching policy, \emph{hard pinning}
(\xref{sec:cache}),
to keep a part of the working set in the cache. Third, for a \noreuse
Locality Descriptor, we use cache bypassing as the data is not reused. In a NUMA system,
\emph{irrespective of the locality type}, we employ CTA scheduling and memory placement to
minimize accesses to remote NUMA zones. If there are conflicts between
different data structures, they are resolved using the priority order,
\highlight{as
described in \xref{sec:cta_scheduling} and \xref{sec:placement}.}  

\subsubsection{CTA Scheduling}
\label{sec:cta_scheduling}

\begin{figure}[h]
  \centering
  \includegraphics[width=0.68\textwidth]{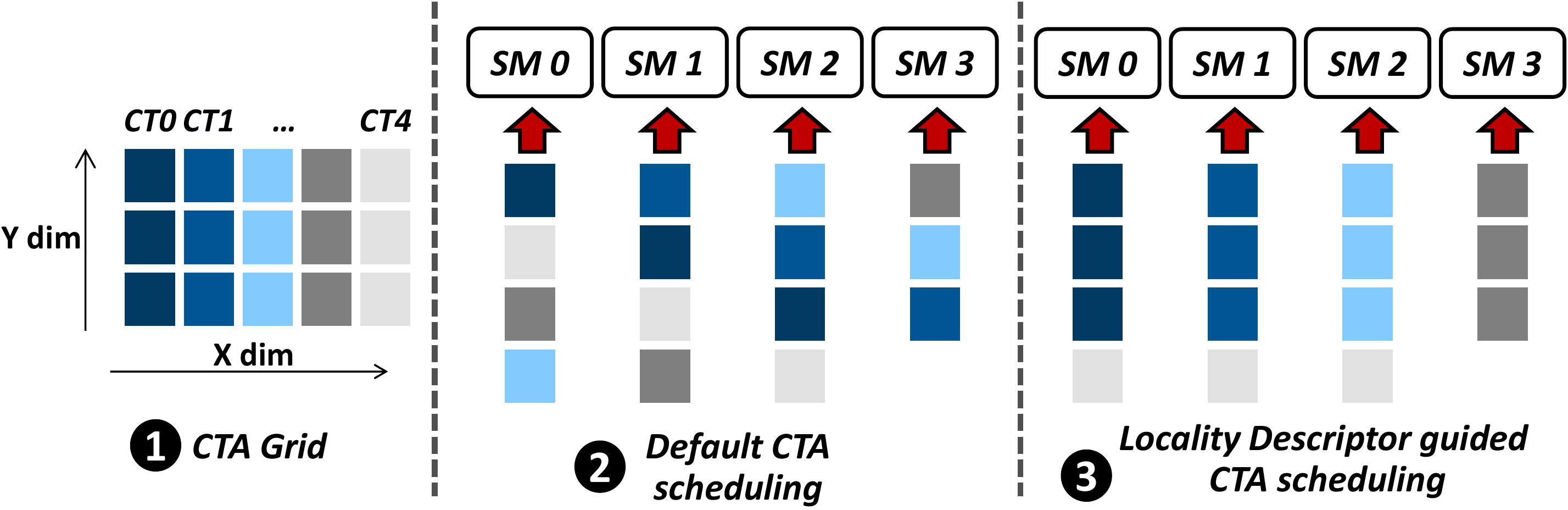}
  \caption{CTA scheduling example.}
  \label{fig:scheduler}
\end{figure}

Figure~\ref{fig:scheduler} depicts an example of CTA
scheduling for the CTA grid (\ding{182}) from our example
(\texttt{histo}, \xref{sec:example}). The default CTA scheduler
(\ding{183}) traverses one dimension at a time ($X \rightarrow Y
\rightarrow Z$), and schedules CTAs at each SM in a round robin
manner, ensuring load balancing across SMs. Since this approach does \emph{not}
consider locality, the default scheduler schedules CTAs that access the same data
at \emph{different} SMs (\ding{183}).

\label{sec:scheduler} 

The Locality Descriptor guided CTA scheduling (\ding{184}) shows how we expose locality by
grouping CTAs in each \ctile into a cluster. Each cluster is then scheduled at
the same SM. In this example, we spread the last \ctile (CT4) across
three SMs to trade off locality for parallelism. To enable such
application-specific scheduling, we need an algorithm to use the
{Locality Descriptor}s to drive a CTA scheduling policy that \One schedules CTAs from the
same \ctile~\highlight{(that share data)} together to expose locality, \Two ensures all SMs are fully
occupied, and \Three resolves conflicts between multiple {Locality Descriptor}s. We
use Algorithm~\ref{alg:cta_scheduling} to form \emph{CTA clusters},
and schedule each formed cluster at the same SM in a non-NUMA
system.\footnote{This algorithm optimizes only for the L1 cache, but it
  can be extended to optimize for the L2 cache as well.} In a NUMA system,
we first \emph{partition} the CTAs across the different NUMA zones
(see \xref{sec:placement}), and then use
Algorithm~\ref{alg:cta_scheduling} \emph{within} each NUMA zone.

\begin{algorithm}
\caption{\small Forming CTA clusters using {Locality Descriptor}s}
\label{alg:cta_scheduling}
\begin{scriptsize}
\begin{algorithmic}[1]
\State \textbf{Input:} $LDesc_{1...N}$: all N {Locality Descriptor}s, sorted by
priority (highest first)
\State \textbf{Output:} $CLS$ = ($CLS_X, CLS_Y, CLS_Z$):
\highlight{the final cluster dimensions}
\For{$i = 1$ to $N$} \Comment{Step 1: Split \ctiles into 2 to ensure each $LDesc$
has enough \ctiles for
  all SMs (to load balance)}
\While{\var{CT\_NUM}($LDesc_i$) < $SM_{NUM}$ and
  \var{CT\_DIM}($LDesc_i$) != (1, 1, 1,)}
\State Divide the \ctile of $LDesc_i$ into 2 along the largest dimension
\EndWhile
\EndFor
\State $CLS \gets$ \var{CT\_DIM}($LDesc_1$) \Comment{Each cluster is now formed
by each of the highest
  priority $LDesc$'s \ctiles after splitting}
\For{$i = 2$ to $N$} \Comment{Step 2: Merge the \ctiles of lower priority
  $LDesc$s to form larger clusters to also leverage locality from lower
  priority $LDesc$s} 
\For{$d$ in $(X, Y, Z)$}
\State $MCLS_d \gets CLS_d \times$
\var{MAX} (\var{FLOOR}(\var{CT\_DIM}($LDesc_i$) / 
$CLS_d$) , 1) \Comment{\texttt{Merge \ctiles along each dimension}}
\EndFor
\If{\var{CT\_NUM}($MCLS$) $\geq$ $SM_{NUM}$} \Comment{Ensure there are enough
\ctiles for all SMs} 
\State $CLS \gets MCLS$
\EndIf
\EndFor
\end{algorithmic}
\end{scriptsize}
\end{algorithm}

The algorithm first ensures that each Locality Descriptor has enough \ctiles for all SMs. If
that is not the case, it \texttt{split}s \ctiles (lines 3--7), to ensure we have enough clusters to
occupy all SMs. Second, the algorithm uses the \ctiles of the highest priority
Locality Descriptor as the initial CTA clusters (line 8), and then attempts to
\texttt{merge} the lower-priority {Locality Descriptor}s (lines 9--16).\footnote{Although the
algorithm starts by optimizing the highest priority \X, it is designed to
find a scheduling strategy that is optimized for \emph{all} {\X}s. Only when no such
strategy can be found (i.e., when there are conflicts), is the highest priority
\X prioritized over others.} \emph{Merging} 
tries to find a cluster that also groups CTAs with shared data in other lower-priority 
{Locality Descriptor}s while keeping the clusters larger than the number of SMs (first step). By
scheduling the merged cluster at each SM, the system can expose
locality for multiple data structures. \highlight{The GPU driver runs 
Algorithm~\ref{alg:cta_scheduling} before launching the kernel to determine the
CTA scheduling policy.} 

\subsubsection{Cache Management}
\label{sec:cache}

The Locality Descriptor enables the cache to distinguish reuse patterns of
different data structures and apply policies
accordingly. We use two caching mechanisms that can be
further extended. First, \emph{cache bypassing} (e.g.,~\cite{adaptive-tian-gpgpu15, locality-li-ics15,
adaptive-li-sc15, efficient-liang-tcadics,survey-mittal-jlpea16,
ctrlc-lee-iccd16, selectively-zhao-icpads16, coordinated-xie-hpca15,
prioritybased-li-hpca15, efficient-xie-iccad13}), which does not insert data that has no
reuse (\noreuse locality type) into the cache. Second, \emph{cache
  prioritization}, which inserts some data structures into the cache
with higher priority than the others. We implement this in two ways:
\One hard pinning and \Two soft pinning. \emph{Hard pinning} is a mechanism
to prevent cache thrashing due to large working sets by ensuring that part of
the working set stays in the cache. We implement hard pinning by
inserting all hard-pinned data with the highest priority
and evicting a specific cache way (e.g., the 0th way) when
\emph{all} cache lines in the same set have the highest priority. Doing so
protects the cache lines in other cache ways from being repeatedly evicted. We use a
timer to automatically reset all priorities to \emph{unpin} these pinned lines
periodically. \emph{Soft
pinning},
on the other hand, simply prioritizes one data structure over others
without any policy to control thrashing. As \xref{sec:overall_algorithm}
discusses, we use hard pinning for data with \intrathread locality type,
which usually has a large working set as there is very limited sharing
among threads. We use soft pinning for data with \interthread locality type
to ensure that this data is retained in the cache until other threads that share
the data access
it.

\subsubsection{Prefetching}
\label{sec:prefetcher}

As \xref{sec:motivation_ldesc} discusses, using CTA scheduling \highlight{alone} to
expose locality hardly improves performance, as the
increased locality causes more threads to stall, waiting for the
\emph{same critical data} at the \emph{same time} (see the L1 inflight hit
rate, Figure~\ref{fig:motiv-sched}). \highlight{As a result, the memory latency
to this critical data becomes the performance bottleneck, since there are
too few threads left to hide the memory latency.} We address this
problem by employing a hardware prefetcher guided by the Locality Descriptor to prefetch the
\emph{critical data} ahead of time. We employ a prefetcher only for
{\interthread}~{Locality Descriptor}s \highlight{because the
data structures they describe are shared} by multiple threads, and hence, are more critical to
avoid stalls. The prefetcher is triggered when an access misses the
cache on these data structures. The prefetcher is instructed based on
the access pattern and the sharing type. As Figure~\ref{fig:overall_algorithm} shows, there are two cases. First, for \texttt{NEARBY}
sharing, the prefetcher is directed to simply prefetch the next cache line. 
Second, for \texttt{COACCESSED} sharing with a \texttt{REGULAR} access pattern, the prefetched address is a function of
\One the access stride, \Two the number of bytes that are accessed at
the same time (i.e., the width of the data tile), and, \Three the size of the
cache, as prefetching too far ahead means more data needs to be
retained in the cache. The address to prefetch is calculated as:
\texttt{current address +
  (L1\_size/(number\_of\_active\_tiles*data\_tile\_width) *
  stride)}. 
The \texttt{number\_of\_active\_tiles} is the number of \dtiles that the
prefetcher is actively prefetching. 
The equation decreases the prefetch distance when there are more active
{\dtiles}~\highlight{to reduce thrashing}. This form of controlled
prefetching avoids excessive use of memory bandwidth by only prefetching 
data that is shared by many threads, and has high accuracy as it is informed by
the Locality Descriptor.

\subsubsection{Memory Placement}
\label{sec:placement}
As \xref{sec:motiv-numalocality} discusses, exploiting locality
on a NUMA system requires coordination between CTA scheduling and
memory placement such that CTAs access local data within
each NUMA zone. There are two major challenges (depicted in
Figure~\ref{fig:motiv-numa} in \xref{sec:motiv-numalocality}): \One how to
partition data among NUMA zones at a fine
granularity. A paging-based mechanism
(e.g.,~\cite{mcmgpu-arunkumar-isca17}) does not solve this problem as a 
large fixed page size is typically ineffective
(\xref{sec:motiv-numalocality}), while small page sizes
are prohibitively expensive to
manage~\cite{mosaic}, and \Two
how to partition CTAs among NUMA zones to exploit locality among \emph{multiple} data
structures \highlight{that} may be accessed differently by the CTAs in the program. To
address these two challenges, we use a \emph{flexible} data
mapping scheme, \highlight{which} we describe below, and a CTA partitioning algorithm that leverages this scheme.

\begin{figure}[h]
  \centering
  \vspace{-2mm}
  \includegraphics[width=0.68\textwidth]{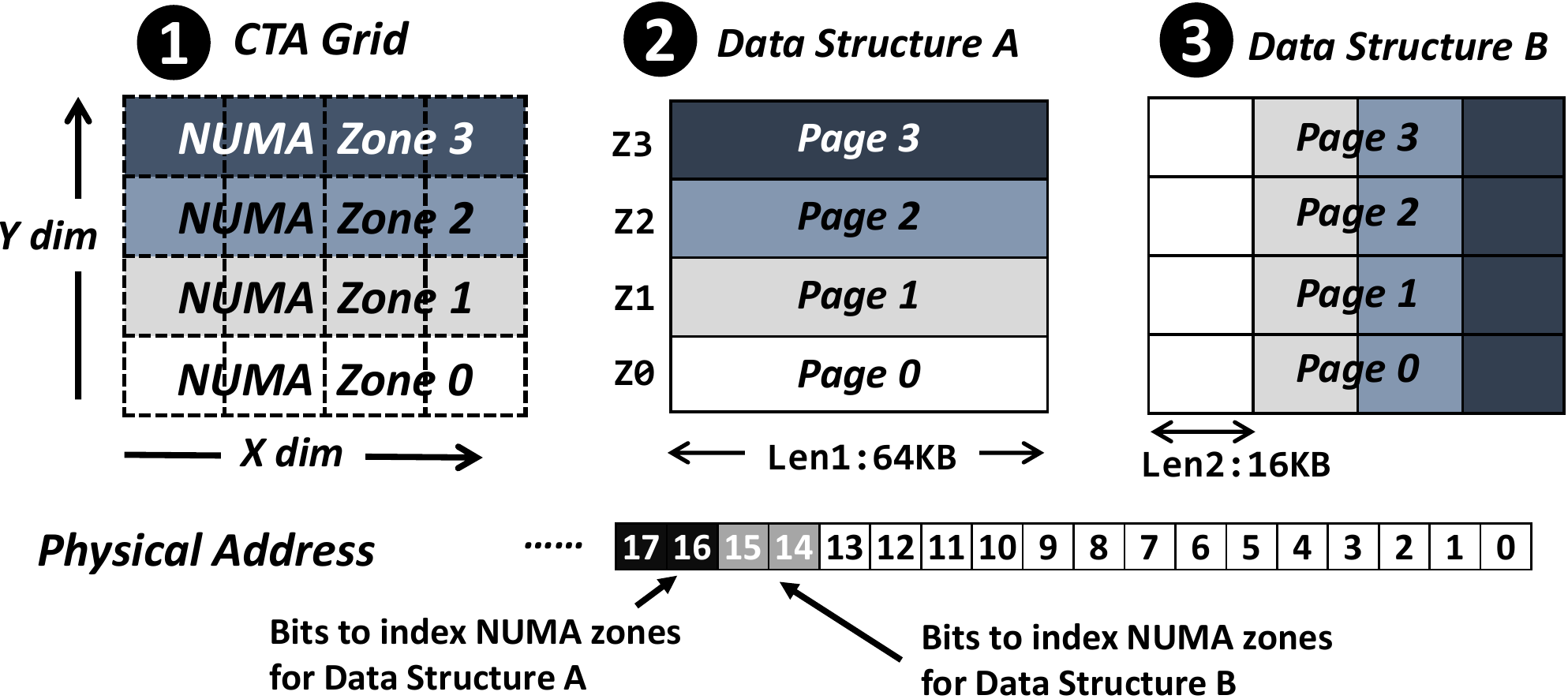}
  \caption{Memory placement with {Locality Descriptor}s.}
  \label{fig:placement}
  \vspace{-2mm}
\end{figure}

\textbf{Flexible Fine-granularity Data Mapping.} We enhance the mapping between physical addresses and NUMA zones to
enable data partitioning at a flexible granularity, smaller than a page
(typically 64KB). Specifically, we use consecutive bits within the
physical address itself to index the NUMA zone (similar
to~\cite{tom-hsieh-isca16} in a different context). We allow using a
\emph{different} set of bits for different 
data structures. Thus, each data structure can be partitioned across NUMA
zones at a \emph{different} granularity.\footnote{We
  limit the bits that can be chosen to always preserve the minimum DRAM burst size (128B) 
  by always specifying a size between 128B-64KB (bits
  7-16). We always use
  bit 16/17 for granularities larger than 64KB as we can flexibly map
  virtual pages to the desired NUMA zone using the page table. We
  enable flexible bit mapping by modifying the hardware address
  decoder in the memory controller.} Figure~\ref{fig:placement} shows
how this is done for the example in \xref{sec:motiv-numalocality}. As the figure
shows, CTAs in each NUMA zone \ding{182} access
the same page (64KB) for data structure A \ding{183}, but they
only access the same \emph{quarter-page (16KB)} for data structure B
\ding{184}. If we partition data across NUMA zones \highlight{only at the page
granularity~\cite{mcmgpu-arunkumar-isca17}}, most accesses to
data structure B would access remote NUMA zones. With our
mechanism, we can choose bits 16-17 (which interleaves data between NUMA zones at a 64KB granularity) and
bits 14-15 (which interleaves data at a 16KB granularity) in the physical
address to index the NUMA zone for data structures A
and B respectively. \highlight{Doing so} results in all accesses to be in the \emph{local} NUMA
zone for \emph{both} data structures.

This design has two constraints. First, we can partition
data only at a power-of-two granularity. Our findings, however, show that this is
not a limiting constraint because \One regular GPU kernels typically exhibit power-of-two
strides across CTAs (consistent with~\cite{tom-hsieh-isca16}); and
\Two even with non-power-of-two strides, this approach is still reasonably
effective compared to page-granularity placement (as shown quantitatively in \xref{sec:eval_numa}). Second, to avoid cases where a data structure
is \emph{not} aligned with the interleaving granularity, we require that the
GPU runtime align data structures at the page granularity.

\textbf{CTA Scheduling Coordinated with Data Placement.} 
To coordinate memory placement with CTA scheduling, we use a simple
greedy search algorithm (Algorithm~\ref{alg:numa_cta_memory}) that partitions the CTAs
across the NUMA zones and selects the most effective address mapping bits for each data
structure. We provide a brief overview here. 

The algorithm evaluates the efficacy of all possible
address mappings for the data structure described by the highest-priority Locality Descriptor
(line 4). This is done by determining which N consecutive bits between
bit 7-16 in the physical address are the most effective bits to index NUMA zones for that
data structure (where N is the base-2-log of the number of NUMA zones). 
To determine which mapping is the most effective, the algorithm first determines the
corresponding CTA partitioning scheme for that address mapping using the
\texttt{NUMA\_PART} function (line 5). The \texttt{NUMA\_PART} function simply
schedules each \ctile at the NUMA zone where the \dtile it accesses
is placed (based on the address mapping \highlight{that is} being tested). The 1:1 \ctile/\dtile compute
mapping in the Locality Descriptor gives us the information to easily do this. To evaluate
the effectiveness or \emph{utility} of each address mapping and the corresponding CTA partitioning scheme, we use
the \texttt{COMP\_UTIL} function (line 7). This function calculates the ratio of
local/remote accesses for each mapping. 
\begin{algorithm}
\caption{\small \highlight{CTA partitioning and memory placement for NUMA}}
\label{alg:numa_cta_memory}
\begin{scriptsize}
\begin{algorithmic}[1]
\State \textbf{Input:} $LDesc_{1...N}$: all N {Locality Descriptor}s, sorted by
priority (highest first)
\State \textbf{Output 1:} $CTA\_NPART$:
the final CTA partitioning for NUMA zones
\State \textbf{Output 2:} $MAP_{1...N}$:
the address mapping bits for each $LDesc$
\For{$b\_hi = $ 7 to 16} \Comment{Test all possible mappings for the highest-priority $LDesc$} 
\State $CTA\_PART_{b\_hi} \gets$ \texttt{NUMA\_PART}($LDesc_1$, $b\_hi$)
\Comment{Partition the CTAs based on the address mapping being
evaluated}
\State $best\_util\_all \gets 0$ \Comment{$best\_util\_all$:
  the current best utility}
\State $util_{b\_hi} \gets$ \texttt{COMP\_UTIL}($N$, $LDesc_1$,
$CTA\_PART_{b\_hi}$, $b\_hi$) \Comment{Calculate the utility of the CTA
partitioning scheme $+$ address mapping}
\For{$i$ = 2 to $N$} \Comment{Test other $LDesc$s} 
\State $TMAP_i \gets$ 7 \Comment{$TMAP$:
  temporary mapping}
\State $best\_util \gets$ 0 \Comment{$best\_util$:
  the utility with the best mapping}
\For{$b\_lo$ = 7 to 16} \Comment{Test all possible address mappings}
\State $util \gets $ \texttt{COMP\_UTIL}($N - i + 1$, $LDesc_i$,
$CTA\_PART_{b\_hi}$, $b\_lo$) \Comment{Calculate overall best mapping} 
\If {$util > best\_util$} 
\State $TMAP_i \gets b\_lo$; $best\_util \gets util$ \Comment{update
  the best mapping}
\EndIf
\EndFor
\State $util_{b\_hi} \gets util_{b\_hi} + best\_util$ \Comment{\highlight{update the new
best utility}}  
\EndFor
\If {$util_{b\_hi} > best\_util\_all$} 
\State $MAP \gets TMAP$; $MAP_1 \gets b\_hi$;
\State $best\_util\_all \gets util_{b\_hi}$; $CTA\_NPART \gets CTA\_PART_{b\_hi}$
\EndIf
\EndFor
\end{algorithmic}
\end{scriptsize}
\end{algorithm}

Since we want a CTA partitioning scheme that is effective for \emph{multiple} data structures, we
also evaluate how \emph{other data structures} can be mapped, based on each CTA
partitioning scheme tested for the high-priority data structure (line 8). 
Based on which of the tested mappings has the highest overall utility, we
finally pick the CTA partitioning scheme and an address mapping scheme for each
data structure (line 12).

The GPU driver runs Algorithm~\ref{alg:numa_cta_memory} 
when all the dynamic information is available at run time (i.e.,
number of NUMA zones, CTA size, data structure size, etc.). The
overhead is negligible because: \One most GPU kernels have only
several data structures (i.e., small $N$), and \Two the two core
functions (\texttt{NUMA\_PART} and \texttt{COMP\_UTIL}) are very
simple due to the 1:1 \ctile/\dtile mapping.

The Locality Descriptor method is more flexible and versatile than a first-touch page
migration scheme~\cite{mcmgpu-arunkumar-isca17}, which \One requires demand
paging to be enabled, \Two is limited to a \emph{fixed} page size,
\Three always schedules CTA in a fixed manner. 
With the 
knowledge of how CTAs access data (i.e., the \dtile-\ctile compute mapping) and the ability to control and coordinate both the CTA scheduler and flexibly place
data, our approach \highlight{provides} a powerful substrate to leverage NUMA locality. 
\subsection{Methodology}
\label{sec:methodology}


We model the entire Locality Descriptor framework in GPGPU-Sim
3.2.2~\cite{GPGPUSim}.  To isolate the effects of the cache locality
versus NUMA locality, we evaluate them separately: we evaluate
reuse-based locality using an existing single-chip non-NUMA system
configuration (based on Fermi GTX 480); and we use a
futuristic NUMA system (based on~\cite{mcmgpu-arunkumar-isca17}) to
evaluate NUMA-based locality. We use the system parameters
in~\cite{mcmgpu-arunkumar-isca17}, but with all compute and bandwidth
parameters (number of SMs, memory bandwidth, inter-chip
interconnect bandwidth, L2 cache) scaled by 4 to ensure that the evaluated
workloads have sufficient parallelism to saturate the
compute units.  Table~\ref{tab:param} summarizes the major system parameters.
We use GPUWattch~\cite{gpuwattch} to model GPU power
consumption. 

\begin{table}[h] 
      \caption{Major parameters of the simulated systems.}
	\centering
	\begin{tabular}{@{}|p{1.75in}|p{4in}|@{}} \hline
Shader Core &  1.4 GHz; GTO
scheduler~\cite{tor-micro12}; 2 schedulers per SM \\ & Round-robin CTA
scheduler \\\hline
SM Resources    &  Registers: 32768; Scratchpad: 48KB, L1: 32KB, 4 ways \\ \hline
Memory Model  & FR-FCFS scheduling~\cite{frfcfs,zuravleff1997controller}, 16
banks/channel \\ \hline
Single Chip System            &  15 SMs; 6 memory channels; L2: 768KB, 16
ways   \\ \hline
Multi-Chip System            & \specialcell{4 \highlight{GPMs (GPU
Modules) or NUMA
zones}; \\64~SMs 
(16 per module); 32 memory channels; \\L2:~4MB, 16 ways; Inter-GPM
Interconnect: 192 GB/s; \\DRAM Bandwidth: 768 GB/s (192 GB/s per module)}
\\ \hline
 \end{tabular}%
\label{tab:param}%
\end{table}%


We evaluate workloads from the CUDA SDK~\cite{sdk},
Rodinia~\cite{rodinia}, Parboil~\cite{parboil} and PolybenchGPU~\cite{polybench}
benchmark suites.
We run each kernel either to completion or up to 1B instructions.\remove{ For each application, we
make sure the amount of work is the same across all evaluated
configurations.}
Our major performance metric is \highlight{instruction throughput (IPC).} 
From the workloads in Table~\ref{table:applications}, we use
cache-sensitive workloads (i.e., workloads where increasing the L1 by
4$\times$ improves performance more than 10\%), to evaluate reuse-based
locality. We use memory bandwidth-sensitive workloads (workloads that
improve performance by more than 40\% with 2$\times$ memory
bandwidth), to evaluate NUMA locality.  \thoughts{EIMAN-11-3-17: Where
  does the 40\% number come from? Is it just arbitrary?}

\bgroup
\setlength{\tabcolsep}{0.2em}
\begin{table}[h]
\fontfamily{ptm}\selectfont
\centering
  \caption{Summary of Applications}
\begin{tabular}{@{}|l|l|@{}} \hline
\textbf{Name (Abbr.)} & \textbf{{Locality Descriptor} types
(\xref{sec:abstraction_overview})} \\ \hhline{|=|=|}
Syrk (SK)~\cite{polybench}  & \interthread \texttt{(COACCESSED, REGULAR)}, \noreuse \\ \hline 
Doitgen (DT)~\cite{polybench} & \interthread \texttt{(COACCESSED, REGULAR)}, \noreuse
\\ \hline 
dwt2d (D2D)~\cite{rodinia}  & \interthread \texttt{(NEARBY, REGULAR)} \\ \hline  
Convolution-2D (C2D)~\cite{polybench}  & \interthread \texttt{(NEARBY)} \\ \hline 
\specialcell{Sparse Matrix Vector \\Multiply (SPMV)~\cite{parboil}}  &
\specialcell{\intrathread,\\ \interthread
\texttt{(COACCESSED, IRREGULAR)}} \\ \hline  
LIBOR (LIB)~\cite{sdk}  & \intrathread \\ \hline  
LavaMD (LMD) \cite{rodinia}  & \specialcell{\intrathread,\\ \interthread \texttt{(COACCESSED,
REGULAR)}}\\ \hline  
histogram (HS)~\cite{parboil}  & \interthread \texttt{(COACCESSED, REGULAR)}\\ \hline  
atax (ATX)~\cite{polybench}  & \noreuse,
\interthread \texttt{(COACCESSED, REGULAR)}\\ \hline  
mvt (MVT)~\cite{polybench}  & \noreuse, \interthread \texttt{(COACCESSED,
REGULAR)}\\ \hline  
particlefilter (PF)~\cite{rodinia}  & \noreuse\\ \hline  
streamcluster (SC)~\cite{rodinia}  & \noreuse,
\interthread \texttt{(NEARBY)}\\ \hline  
transpose (TRA)~\cite{sdk}  & \noreuse\\ \hline  
Scalar Product (SP)~\cite{sdk}  & \noreuse\\ \hline  
Laplace Solver (LPS)~\cite{sdk}  & \noreuse, \intrathread\\ \hline  
pathfinder (PT)~\cite{rodinia}  & \noreuse \\ \hline  
\end{tabular}
\label{table:applications}
\end{table}
\egroup

\subsection{Evaluation}
\subsubsection{Reuse-Based (Cache) Locality}

We evaluate six configurations:
\One \texttt{Baseline}: our baseline system with the default CTA
scheduler.
\Two \texttt{BCS}: a heuristic-based
CTA scheduler based on BCS~\cite{improving-lee-hpca14}, which schedules two 
consecutive CTAs at the same SM. \Three \texttt{LDesc-Sched}: the \LDescriptor-guided CTA scheduler,
which uses the \LDescriptor semantics and algorithm. 
\highlight{Compiler techniques such as ~\cite{improving-chen-cal17,
locality-li-asplos17} can produce the same benefits.} 
\Four \texttt{LDesc-Pref}: the \LDescriptor-guided prefetcher. Sophisticated classification-based prefetchers such
as~\cite{apres-oh-isca16}, can potentially obtain similar benefits.
\emph{(v)} \texttt{LDesc-Cache}: the \LDescriptor-guided cache prioritization and
bypassing scheme. \emph{(vi)} \texttt{LDesc}: \highlight{our
proposed scheme, which uses the \LDescriptor to 
distinguish between the different locality types and selectively employs
different (scheduling, prefetching, caching, and data placement) optimizations.}

\begin{figure}[h]
  \centering
 \begin{subfigure}[t]{0.49\linewidth}
  \centering
  \includegraphics[width=1\textwidth]{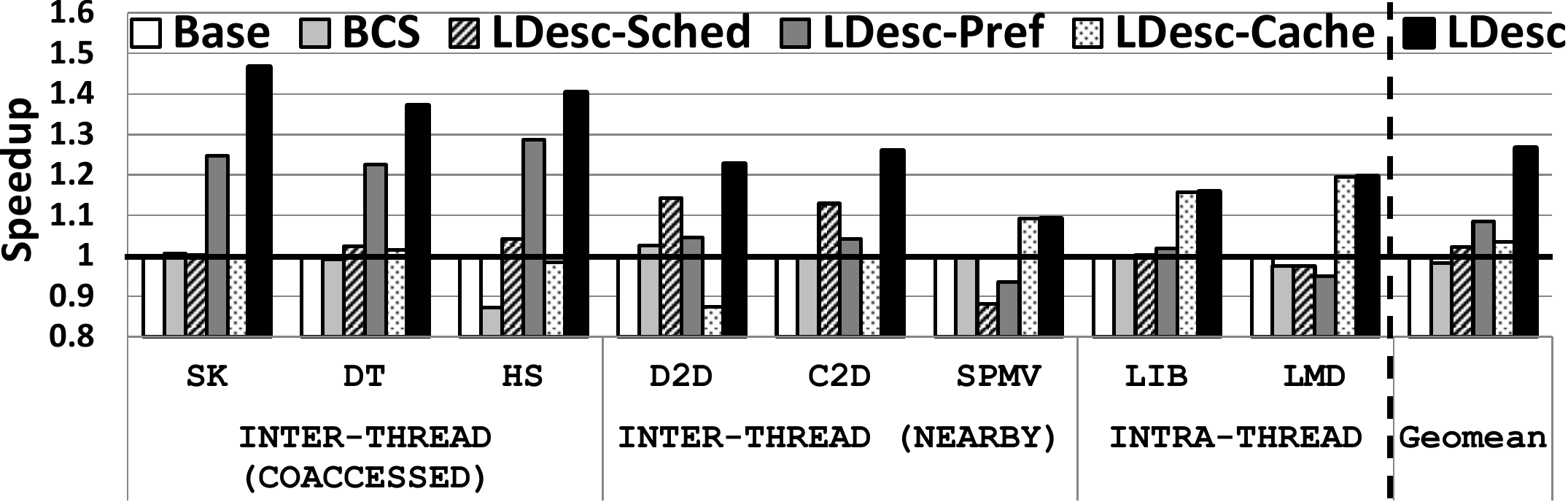}
  \caption{Normalized performance with {\LDescriptor}s.}
  \label{fig:performance-reuse}
\end{subfigure} 
 \begin{subfigure}[t]{0.49\linewidth}
  \centering
  \includegraphics[width=1\textwidth]{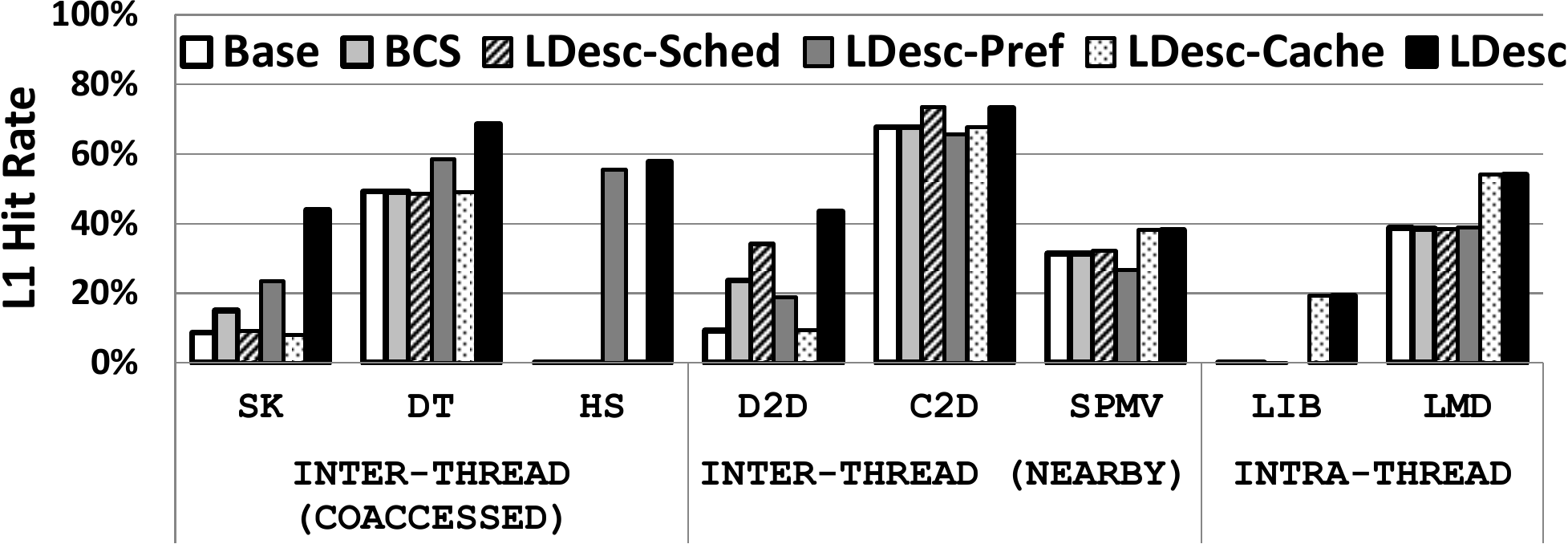}
  \caption{L1 hit rate with {\LDescriptor}s.}
  \label{fig:l1-hitrate}
\end{subfigure} 
  \caption{Leveraging reuse-based (cache) locality.}
\end{figure}

Figure~\ref{fig:performance-reuse} depicts the speedup over
\texttt{Baseline} across all configurations. 
\texttt{LDesc} improves performance by 26.6\% on average (up to 46.6\%) over
\texttt{Baseline}. \texttt{LDesc} always performs either as well as or better than any of the techniques
in isolation.
Figure~\ref{fig:l1-hitrate} shows the L1 hit rate for different
configurations. \highlight{\texttt{LDesc}'s performance improvement comes from a 41.1\% 
improvement in average hit rate (up to 57.7\%) over
\texttt{Baseline}.  
We make three observations that provide insight into \texttt{LDesc}'s
effectiveness.} 

First, different applications benefit from \emph{different} optimizations. Applications
with \interthread type of locality (\texttt{SK, DT, HS, D2D, C2D}) benefit from CTA
scheduling and/or prefetching. 
However, \texttt{LIB, LMD, SPMV} do not benefit from CTA scheduling as
there is little inter-CTA reuse to be exploited. 
Similarly, prefetching 
significantly hurts performance in these workloads (\texttt{LIB, LMD, SPMV}) as
the generated prefetch requests
exacerbate the memory
bandwidth bottleneck. As a result, significant performance degradation occurs
when the working set is too large (e.g., when
there is no sharing and only \intrathread reuse, as in \texttt{LMD}
and \texttt{LIB}) or where the access patterns in
the major data structures are not sufficiently regular (\texttt{SPMV}).
Cache prioritization and bypassing is very effective in workloads with
\intrathread reuse (\texttt{LIB, LMD}), but is largely
ineffective and can even hurt performance in workloads such as \texttt{D2D} and 
\texttt{HS} when a non-critical data structure or too many data structures are
prioritized in the cache. 
Since \texttt{LDesc} is able to distinguish between locality types, it is able to
select the best combination of optimizations for each application.   

Second, a single optimization is very often \emph{insufficient} to exploit locality.
For the \interthread applications (\texttt{SK, DT, HS, D2D, C2D}),
\texttt{LDesc}-guided CTA
scheduling significantly reduces the L1 working set (by 67.8\% on
average, not graphed). However, this
does \emph{not} translate into significant performance improvement when
scheduling is applied by itself (only 2.1\% on average). 
This is because of an 17\% average increase in L1 inflight hit rate as
a result of more threads accessing the same data. These threads wait on the same
shared data at the \emph{same time}, and hence cause increased stalls at the
core. The benefit of increased locality is thus lost. 
Prefetching (\texttt{LDesc-Pref}) is an effective technique to alleviate
effect. However, prefetching by itself significantly increases the memory
traffic and this hinders its ability to improve performance when applied alone.
When combined with scheduling, however, prefetching effectively reduces
the long memory latency stalls. Synergistically, CTA scheduling reduces the overall memory traffic
by minimizing the working set. For the \interthread \texttt{NEARBY}
workloads (\texttt{C2D, D2D}), CTA scheduling co-schedules CTAs with overlapping
working sets. This allows more effective prefetching between the CTAs for the
critical high-reuse data. In the cases described above, prefetching and CTA scheduling work
better synergistically than in isolation, and \texttt{LDesc} is able to effectively
combine and make use of multiple techniques depending on the locality type. 

Third, \texttt{LDesc}-guided CTA scheduling is significantly more
effective than the heuristic-based approach, \texttt{BCS}. This is because \texttt{LDesc}
tailors the CTA scheduling policy for each application by clustering
CTAs based on the locality characteristics of each data structure.
\highlight{Similarly, the \texttt{LDesc} prefetcher and replacement policies
are highly effective, because they leverage program semantics from the \LDescriptor.} 

\textbf{Conclusions.} We make the following conclusions: \One The \LDescriptor is an
effective and versatile mechanism to
leverage reuse-based locality to improve GPU performance and energy efficiency; 
\Two Different locality types require different
optimizations{\textemdash}a single mechanism or set of mechanisms do not work
for all locality types. We demonstrate that the {\LDescriptor} can effectively connect
different locality types with the underlying architectural optimizations. 
\Three The \LDescriptor enables the hardware architecture to leverage  
the program's locality semantics to provide significant performance benefits
over heuristic-based approaches such as the \texttt{BCS} scheduler.

\subsubsection{NUMA Locality}
\label{sec:eval_numa}
\begin{figure}[h]
  \centering
  \includegraphics[width=0.65\textwidth]{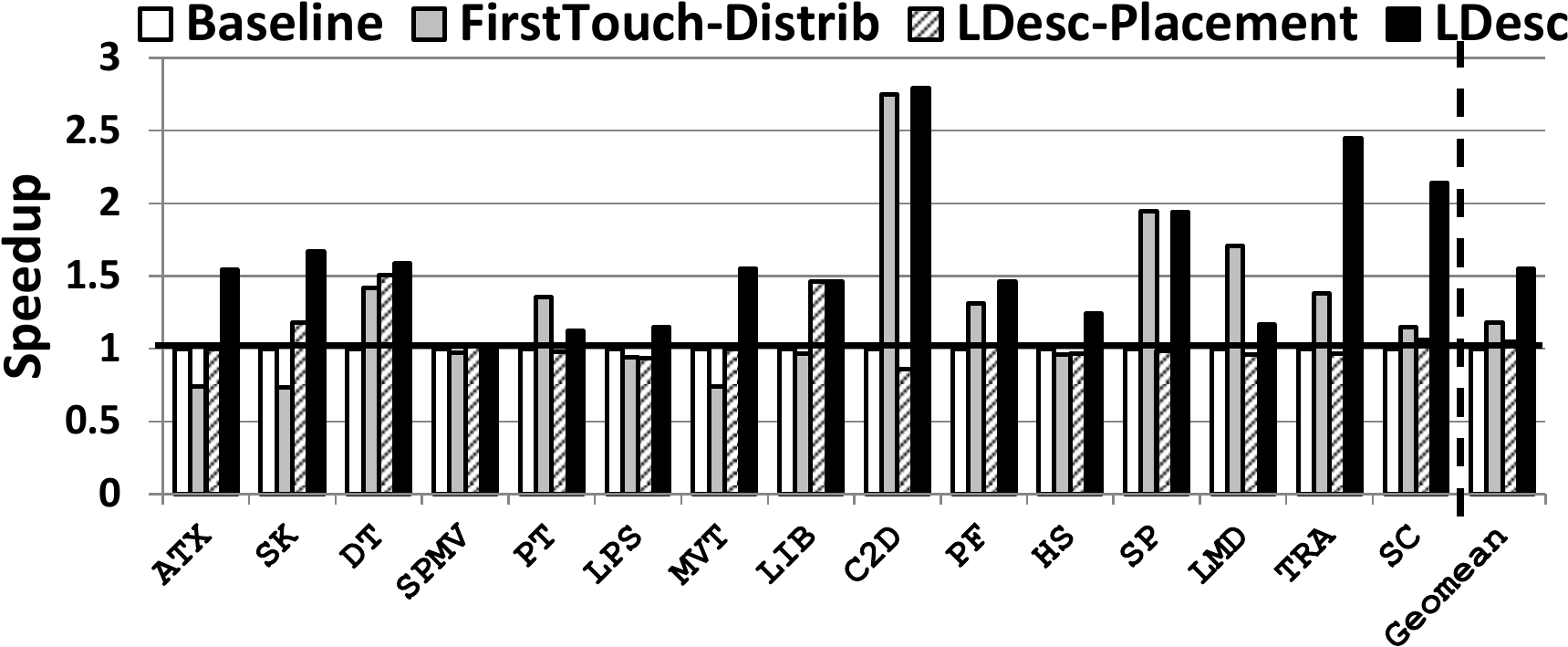}
  \caption{NUMA Locality: Normalized performance.}
  \label{fig:performance-numa}
\end{figure}

To evaluate the benefits of the \LDescriptor in exploiting NUMA locality,
Figure~\ref{fig:performance-numa} compares four
different mechanisms: \One \texttt{Baseline}: The baseline system which uses a static XOR-based
address hashing mechanism~\cite{permutation-zhang-micro00} to randomize data placement
across NUMA zones. \Two \texttt{FirstTouch-Distrib}: \highlight{The state-of-the-art mechanism
proposed in~\cite{mcmgpu-arunkumar-isca17}, where each page (64KB) is placed at the NUMA 
zone where it is first accessed.} This scheme also employs a heuristic-based
distributed scheduling strategy where the compute grid is partitioned equally
across the NUMA zones such that \emph{contiguous} CTAs are placed in the same NUMA
zone. 
\Three \texttt{LDesc-Placement}: The memory placement mechanism
based on the semantics of the
{\LDescriptor}s, but \emph{without} the accompanying CTA scheduling strategy.  
\Four \texttt{LDesc}: The {\LDescriptor}-guided memory placement mechanism with
the coordinated CTA scheduling
strategy. 
We draw two conclusions from the figure. 

\begin{figure}[h]
  \centering
  \includegraphics[width=0.65\textwidth]{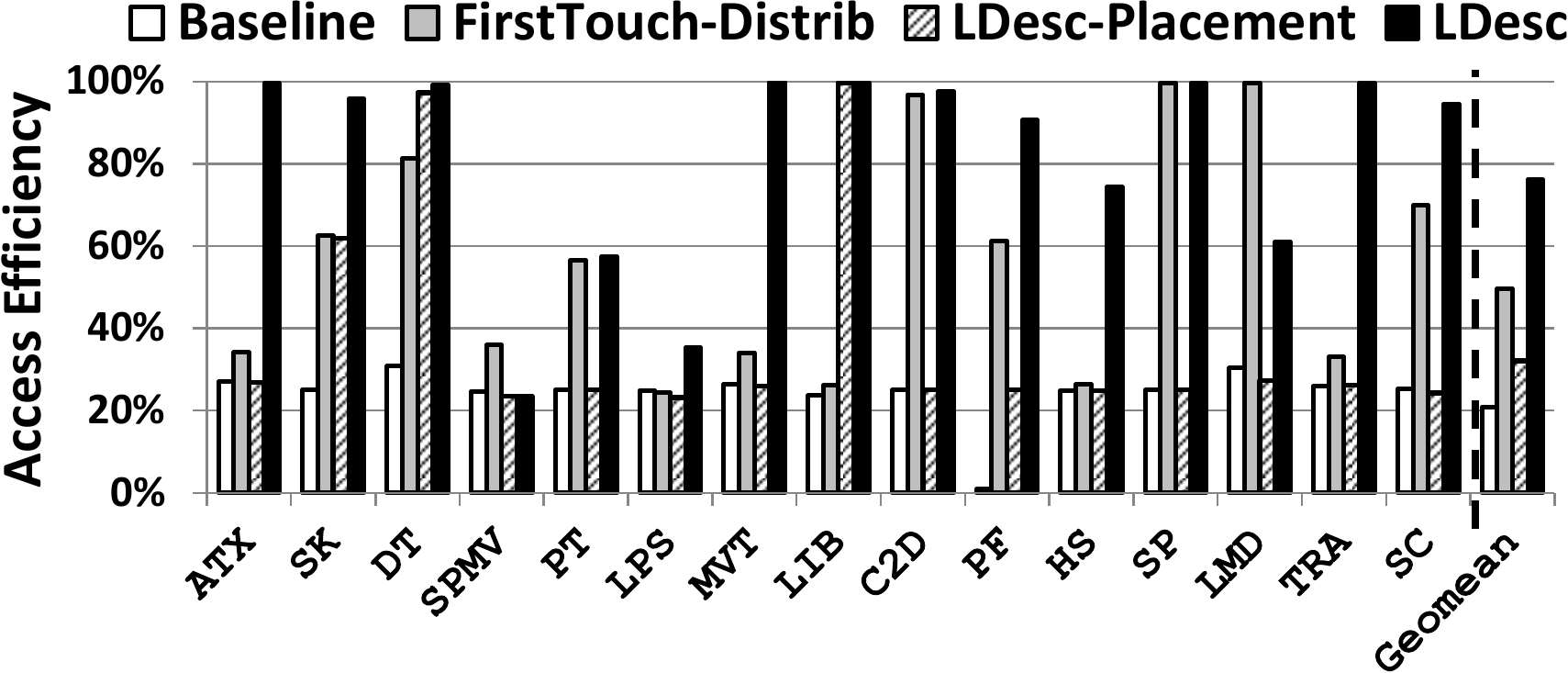}
  \caption{\highlight{NUMA access efficiency.}}
  \label{fig:access-efficiency}
\end{figure}

\begin{figure}[h]
  \centering
  \includegraphics[width=0.65\textwidth]{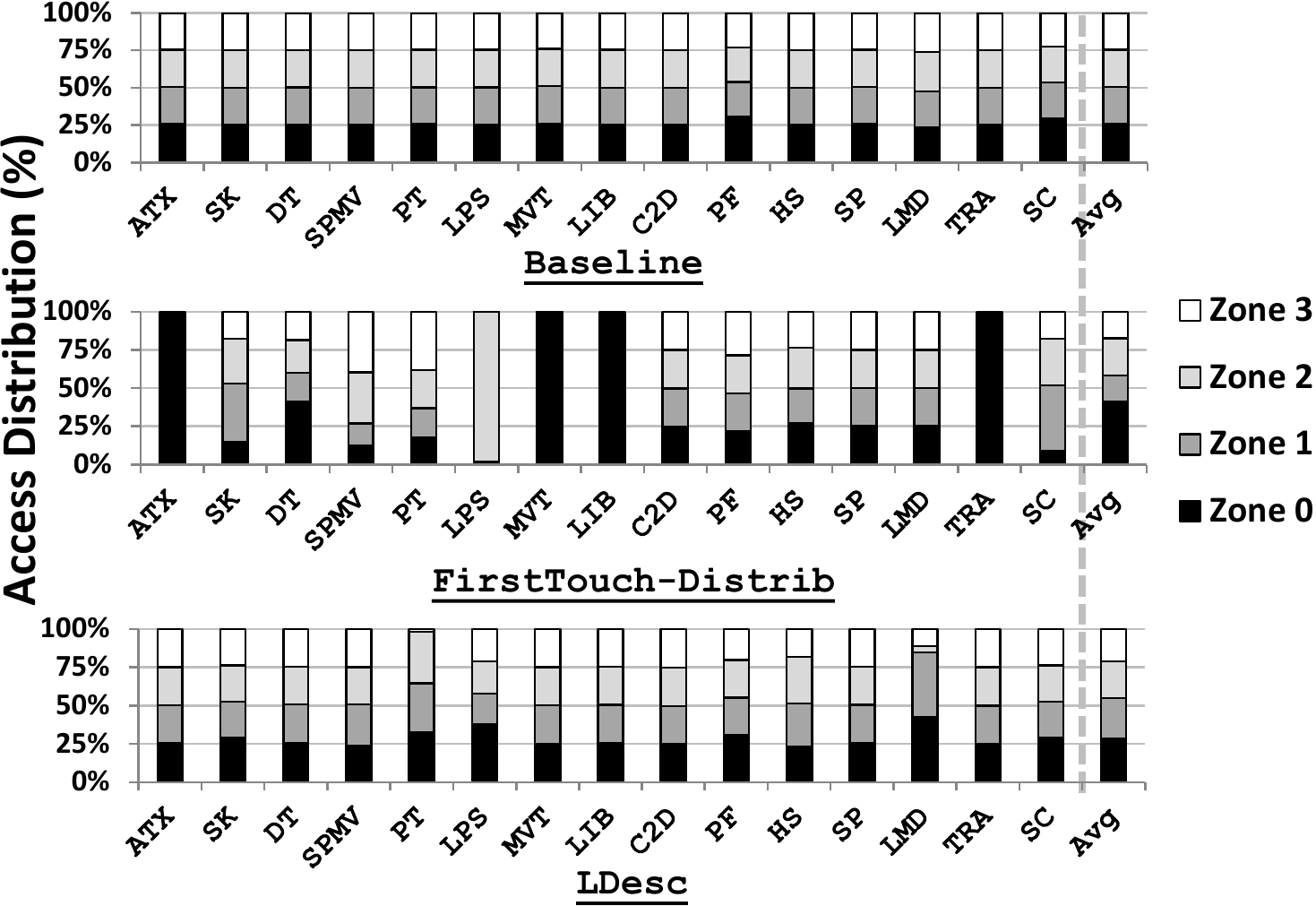}
  \caption{\highlight{NUMA zone access distribution.}} 
  \label{fig:access-skew}
\end{figure}

\textbf{Conclusion 1.\@} 
\texttt{LDesc} is an effective mechanism in NUMA
data placement, outperforming \texttt{Baseline} by
53.7\% on average (up to 2.8$\times$) and \texttt{FirstTouch\_Distrib} by 31.2\%
on average (up to 2.3$\times$). The performance impact of NUMA placement is primarily
determined by two factors: \One \emph{Access efficiency} (plotted in
Figure~\ref{fig:access-efficiency}), which is defined as the fraction of total memory accesses
that are to the local NUMA zone (higher is better).
Access efficiency determines the amount of
traffic across the  interconnect between NUMA zones as well as the latency of
memory accesses. \Two \emph{Access distribution} (plotted in
Figure~\ref{fig:access-skew}) across NUMA zones. Access
distribution determines the effective memory bandwidth being utilized by the
system{\textemdash}a non-uniform distribution of accesses across
NUMA zones may lead to underutilized bandwidth in one or more zones, which can create a new performance bottleneck and degrade performance. 

The static randomized mapping in \texttt{Baseline} aims to
balance access distribution across NUMA zones (with an average distribution
of {\textasciitilde}25\% at each zone), but is \emph{not} optimized to maximize access
efficiency (\highlight{only 22\% on average}).
\texttt{FirstTouch-Distrib} on the other hand, has higher access efficiency
in some workloads (e.g., \texttt{SP, PT\remove{, C2D, LMD}}) 
by ensuring that a page is placed where the CTA that accesses it first is scheduled (49.7\%
on average).
However, \texttt{FirstTouch-Distrib} is still ineffective for many workloads for three
reasons: \One Large page granularity (64KB) often leads to high skews in access
distribution when pages are shared between many CTAs, e.g., \texttt{ATX, MVT,
LIB\remove{, LPS, TRA}} (Figure~\ref{fig:access-skew}). This is because a majority of
pages are placed in the NUMA zone where the CTA that is
furthest ahead in execution is scheduled. \Two
\texttt{FirstTouch-Distrib} has low access
efficiency when the heuristic-based scheduler does \emph{not} schedule CTAs that access
the same pages at the same NUMA zone (e.g., \texttt{DT, HS, SK}). \Three
\texttt{FirstTouch-Distrib} 
has low
access
efficiency when each CTA irregularly accesses a large number of pages because
data \emph{cannot} be partitioned between the NUMA zones at a fine granularity
(e.g.,
\texttt{SPMV}). 

\texttt{LDesc} interleaves data at a fine granularity
depending on how each data structure is partitioned between CTAs and schedules
those CTAs accordingly.\remove{. \texttt{LDesc}
then schedules CTAs such that it predominantly accesses data in the same NUMA
zone.} If a data
structure is shared among more CTAs than what can be scheduled at a single zone,
the data structure is partitioned
across NUMA zones, as \texttt{LDesc} favors parallelism over locality. 
Hence, \texttt{LDesc} tries to improve access efficiency
while reducing skew in access distribution in the presence of a large amount of
data sharing. As a result,
\texttt{LDesc} has an average access efficiency of 76\% and access distribution
close to 25\% across the NUMA zones (Figure~\ref{fig:access-skew}). \texttt{LDesc} is
less effective in access efficiency in cases where the data structures are irregularly
accessed (\texttt{SPMV}) or when non-power-of-two data tile sizes lead to
imperfect data partitioning (\texttt{LPS, PT, LMD, HS}). 

\textbf{Conclusion 2.\@} From Figure~\ref{fig:performance-numa}, we see that 
\texttt{LDesc} is \emph{largely ineffective} without
coordinated CTA scheduling. 
\texttt{LDesc-Placement} retains the \texttt{LDesc} benefit in reducing the
skew in access distribution (not graphed). However, \emph{without}
coordinated CTA scheduling\remove{ each CTA at the appropriate NUMA zone where the data it accesses is
placed, the} access efficiency is very low (32\% on average).

We conclude that the \LDescriptor approach is an effective strategy for data
placement in a NUMA environment by \One leveraging locality
semantics in intelligent data placement \emph{and} CTA scheduling and \Two
orchestrating the two techniques using a single interface. 

\subsection{Related Work}
To our knowledge, this is the first work to propose a cross-layer
abstraction that enables the software/programmer
to flexibly express and exploit different forms of data locality
in GPUs. This enables leveraging program semantics to transparently
coordinate architectural techniques that are critical to improving performance and
 energy efficiency.  
We now briefly discuss closely related prior
work specific to the Locality Descriptor. A more general comparison of the approach taken in this
work is
discussed in Section~\ref{sec:relatedwork}.

\textbf{Improving Cache Locality in GPUs.} There is a large body of research
that aims to improve cache locality in GPUs using a range of hardware/software
techniques such as CTA scheduling~\cite{improving-lee-hpca14, locality-li-asplos17, nmnl-pact13,
owl-jog-asplos13, laperm-wang-isca16, twinkernels-gong-cgo17,
improving-chen-cal17, cachehierarchy-lai-tc15, enabling-wu-ics15},
prefetching~\cite{manythread-lee-micro10, apogee-sethia-pact13, apres-oh-isca16,
cta-jeon-ceng14, threadaware-liu-taco16,
spareregister-lakshminarayana-hpca14,caba},
warp scheduling~\cite{cawa-lee-isca15, locality-zhang-fgcs17,
tor-micro12}, cache bypassing~\cite{adaptive-tian-gpgpu15, locality-li-ics15,
adaptive-li-sc15, efficient-liang-tcadics,survey-mittal-jlpea16,
ctrlc-lee-iccd16, selectively-zhao-icpads16, coordinated-xie-hpca15,
prioritybased-li-hpca15, efficient-xie-iccad13}, and other cache management
schemes~\cite{accesspattern-koo-isca17,
reducing-choi-gpgpu12, adaptive-chen-micro14, medic,
efficient-khairy-gpgpu15, dacache-wang-ics15, warppool-kloosterman-micro15,
apres-oh-isca16, sacat-khairy-tpds17, iacm-kim-iccd16, locality-zhang-spa16,
characterizing-jia-ics12, revealing-koo-iiswc15, owl-jog-asplos13,
orchestrated-jog-isca13, orchestrating-mu-vlsi14, adaptive-zheng-cal15,
efficient-gaur-micro13}. Some of these works orchestrate multiple techniques
~\cite{orchestrated-jog-isca13, owl-jog-asplos13,
accesspattern-koo-isca17, apres-oh-isca16, iacm-kim-iccd16,
locality-li-asplos17, cta-jeon-ceng14, adaptive-zheng-cal15} to leverage synergy between optimizations. However, these prior approaches are
either hardware-only, software-only, or focus on optimizing a single technique.
\highlight{Hence, they are 
limited \One by what is possible with the information that can be solely \emph{inferred} in 
hardware, \Two by existing software interfaces that limit what optimizations are
possible, or \Three in terms of the range of 
optimizations that can be used.} 
In contrast, the Locality Descriptor provides a new, portable and flexible interface to the
software/programmer. This interface allows
easy access to hardware techniques in order to leverage data locality.
Furthermore, 
\highlight{all the above prior approaches} are largely orthogonal to the Locality Descriptor as they can use the Locality Descriptor to enhance 
their efficacy with the knowledge of program semantics. 

\highlight{The closest work to ours is ACPM~\cite{accesspattern-koo-isca17}, an
architectural cache management technique that identifies
intra-warp/inter-warp/streaming locality and selectively applies cache pinning 
or bypassing based on the detected locality type. This work is limited to the 
locality types that can be inferred by hardware, and it does \emph{not} tackle inter-CTA locality or NUMA 
locality, both of which require a priori knowledge of program semantics 
and hardware-software codesign.} 

\textbf{Improving Locality in NUMA GPU Systems.} 
A range of hardware/software techniques to enhance NUMA locality have been proposed in different contexts in GPUs:
multiple GPU modules~\cite{mcmgpu-arunkumar-isca17}, 
multiple memory stacks~\cite{tom-hsieh-isca16}, 
and multi-GPU systems with unified
virtual addressing
~\cite{toward-kim-sc17, 
automatic-cabezas-ics15, towards-sakai-candar16,
achieving-kim-ppopp11, transparent-lee-pact13, automatic-cabezas-pact14}.
We already qualitatively and quantitatively compared against
\texttt{FirstTouch-Distrib}~\cite{mcmgpu-arunkumar-isca17} in our evaluation. 
Our memory placement technique
is similar to the approach taken\remove{ to data locality with
multiple memory stacks} in TOM~\cite{tom-hsieh-isca16}. In TOM, frequent
power-of-two strides seen in GPU kernels are leveraged to use consecutive bits
in the address to index a memory stack. TOM, however,
\One \highlight{is the state-of-the-art technique} targeted at near-data processing and does \emph{not} require coordination with
CTA scheduling, \Two relies on a profiling run to identify the index bits, and
\Three does \emph{not} allow using different index bits for different data structures.  
Techniques to improve locality in multi-GPU systems~\cite{toward-kim-sc17, 
automatic-cabezas-ics15, towards-sakai-candar16,
achieving-kim-ppopp11, transparent-lee-pact13, automatic-cabezas-pact14} use
profiling and compiler analysis to partition the compute grid and 
data across multiple GPUs. These works are similar to the Locality Descriptor in terms of the partitioning used for
forming data and compute tiles and, hence, can easily leverage {Locality Descriptor}s to
further
exploit reuse-based locality and NUMA locality in a single GPU.

\textbf{\highlight{Expressive Programming Models/Runtime Systems/Interfaces.}\@} 
In the context of multi-core CPUs and distributed/heterogeneous systems, there have been numerous software-only approaches\remove{ in the form of programming models
and runtime systems}
that allow explicit expression of data
locality~\cite{x10-charles-oopsla05,chapel-chamberlain-ijhpca,sequoia-fatahalian-sc06,\remove{legiondependent-treichler-oopsla16,legionhierarchical-treichler-oopsla13,}legionstructure-bauer-sc14,
realm-treichler-pact14,regent-slaughter-sc15,hpt-yan-lcpc09,hta-bikshandi-ppopp06,
legion-bauer-sc12}, data
independence~\cite{\remove{legionhierarchical-treichler-oopsla13,legiondependent-treichler-oopsla16,}legionstructure-bauer-sc14,realm-treichler-pact14,regent-slaughter-sc15,legion-bauer-sc12}
or even tiles~\cite{programming-guo-ppopp08,tida-unat-hpc16}, to
enable the runtime to \highlight{perform} NUMA-aware placement or produce code that is
optimized to better exploit the cache hierarchy. These approaches \One are
software-only; hence, they do not have access to many architectural techniques
that are key to exploiting locality and \Two do not tackle the GPU-specific
challenges in exploiting data locality.
These works are largely orthogonal to ours and can use 
{Locality Descriptor}s to leverage hardware techniques to exploit reuse-based locality and NUMA locality in GPUs.

\subsection{Summary}

This chapter \highlight{demonstrates the benefits of}
an \emph{explicit abstraction} for data locality in GPUs that is recognized by all layers
of the compute stack, from the programming model to the hardware architecture. 
We introduce the Locality Descriptor, a rich cross-layer abstraction to explicitly express
and \highlight{effectively} leverage data locality in GPUs. 
The Locality Descriptor \One provides the 
software/programmer a flexible and portable interface to optimize for data 
locality without any knowledge of the underlying architecture and \Two enables the architecture to leverage 
program semantics to \highlight{optimize and coordinate multiple hardware
techniques in a manner that is transparent to the programmer.}  
The key idea is to design the abstraction around the program's data structures
and specify locality semantics based on how the program accesses each data
structure. 
We evaluate and demonstrate 
the performance benefits of {Locality Descriptor}s from effectively leveraging different types of reuse-based locality in 
the cache hierarchy and NUMA locality in a NUMA memory system. 
We conclude that by providing a flexible and powerful cross-cutting interface,
the Locality Descriptor enables leveraging a critical yet \highlight{challenging} factor in harnessing a GPU's
computational power, data locality.

%% file: sections/zorua.tex
\section{Zorua}
\label{sec:zorua}

Modern GPU programming models directly manage several on-chip hardware resources in GPU
that are critical to performance, e.g., registers, scratchpad memory, and thread
slots. This chapter describes how this \emph{tight coupling} between the
programming model and hardware resources causes significant challenges in
programmability and performance portability, and heavily constrains the
hardware's ability to aid in resource management. We propose a new framework,
Zorua, that \emph{decouples} the programming model from the management of
hardware resources by effectively virtualizing these resources. We demonstrate how
this virtualization significantly addresses the programmability, portability, and
efficiency challenges in managing these on-chip resources by enabling the
hardware to assist in their allocation and resource management. 

\subsection{Overview}

Modern Graphics Processing Units (GPUs) have evolved into powerful
programmable machines over the last decade, offering high performance
and energy efficiency for many classes of applications
by concurrently executing thousands of threads. In order to execute,
each thread requires several major on-chip resources: \One registers,
\two scratchpad memory (if used in the program), and
\three a thread slot in the thread scheduler that keeps all the
bookkeeping information required for execution.

Today, these hardware resources are {\em statically} allocated to threads based on several parameters{\textemdash}the number of threads
per thread block, register usage per thread, and scratchpad usage per
block. We refer to these static application parameters as the
\emph{resource specification} of the application. This resource specification
forms a critical component of modern GPU programming models (e.g.,
CUDA~\cite{cuda},
OpenCL~\cite{opencl}). The static
allocation over a fixed set of hardware resources based on the
software-specified resource specification creates a
\emph{tight coupling} between the program (and the programming model) 
and the physical hardware resources. As a result of this tight coupling, for
each application, there are only a few optimized resource specifications that
maximize resource utilization. Picking a suboptimal specification
leads to underutilization of resources and hence, very often,
performance degradation. This leads to three key
difficulties related to obtaining good performance on modern GPUs: programming
ease, portability, and resource inefficiency (performance). 

\textbf{Programming Ease.} First, the burden falls upon the
programmer to optimize the resource specification.
For a naive programmer, this is a very challenging
task~\cite{OptGPU1,OptGPU2,OptGPU3,toward-davidson-iwapc10,OptGPU4,jpeg-occ, asplos-sree}.
This is because, in addition to selecting a specification suited to an
algorithm, the programmer needs to be aware of the details of the GPU architecture
to 
fit the specification to the underlying hardware resources. This
\emph{tuning} is easy to get wrong because there are \emph{many}
highly suboptimal performance points in the specification space, and
even a minor deviation from an optimized specification can lead to a
drastic drop in performance due to lost parallelism. We refer to such
drops as \emph{performance cliffs}. We analyze the effect of
suboptimal specifications on real systems for 20 workloads 
(Section~\ref{sec:perf-cliffs}), and experimentally demonstrate that
changing resource specifications can produce as much as a 5$\times$
difference in performance due to the change in parallelism. Even a
minimal change in the specification (and hence, the resulting allocation) of one resource can result in a significant
performance cliff, degrading performance by as much as 50\%
(Section~\ref{sec:perf-cliffs}).

\textbf{Portability.} Second, different GPUs have varying quantities
of each of the resources. Hence, an optimized specification on one GPU
may be highly suboptimal on another. In order to determine the extent
of this portability problem, we run 20 applications on three
generations of NVIDIA GPUs: Fermi, Kepler, and Maxwell
(Section~\ref{sec:motivation:port}). An example result demonstrates
that highly-tuned code for Maxwell or Kepler loses as much as 69\%
of its performance on Fermi. This lack of \emph{portability} necessitates
that the programmer \emph{re-tune} the resource specification of the application for
\emph{every} new GPU generation. This
problem is especially significant in virtualized environments, such as cloud or
cluster computing, where the same program may run on a wide range of GPU
architectures, depending on data center composition and hardware availability.

\textbf{Performance.} Third, for the programmer who chooses to employ software
optimization tools (e.g., auto-tuners) or manually tailor the program to fit the
hardware,
performance is still constrained by the \emph{fixed, static} resource
specification. It is well
known~\cite{virtual-register,gebhart-hierarchical,compiler-register,shmem-multiplexing,caba,
caba-bc, virtual-thread}
that the on-chip resource requirements of a GPU application vary throughout
execution. Since the program (even after auto-tuning) has to {\em statically}
specify its {\em worst-case} resource requirements, severe
\emph{dynamic underutilization} of several GPU
resources~\cite{virtual-register,compiler-register,gebhart-hierarchical,kayiran-pact16,caba,caba-bc} ensues,
leading to suboptimal performance (Section~\ref{sec:underutilized_resources}).

\textbf{Our Goal.} To address these three challenges at the same time, we propose to
\emph{decouple} an application's resource specification from the available hardware
resources by \emph{virtualizing} all three major resources in a holistic manner. This
virtualization provides the illusion of \emph{more} resources to the GPU programmer and
software than physically available, and enables the runtime system and the
hardware to {\em dynamically} manage multiple physical resources in a manner that is transparent
to the programmer, thereby alleviating dynamic underutilization.

Virtualization is a concept that has been applied to the management of hardware
resources in many contexts (e.g.,~\cite{virtual-memory1,
virtual-memory2,
virtualization-1,virtualization-2,ibm-360,pdp-10,how-to-fake,vmware-osdi02}),
providing various benefits. We believe that applying the general principle of
virtualization to the management of \emph{multiple} on-chip resources in GPUs offers
the opportunity to alleviate several important challenges in modern GPU
programming, which are described above. However, at the same time, effectively adding a new level
of indirection to the management of multiple latency-critical GPU resources
introduces several new challenges
(see Section~\ref{sec:virt-challenges}). This necessitates the design of a new
mechanism to effectively address the new challenges and enable the benefits of
virtualization. In this work, we introduce a new framework,
\emph{Zorua},\footnote{Named after a Pok\'{e}mon~\cite{pokemon} with the power of illusion, able to
take different shapes to adapt to different circumstances (not unlike our proposed
framework).} to
decouple the programmer-specified resource specification of an application from
its physical on-chip hardware resource allocation by effectively virtualizing
the multiple
on-chip resources in GPUs. 

\textbf{Key Concepts.} The virtualization strategy used by Zorua is
built upon two key concepts. First, to mitigate performance cliffs when we do
not have enough physical resources, we \emph{oversubscribe} resources by a small
amount at runtime, by leveraging their dynamic underutilization and maintaining a
swap space (in main memory) for the extra resources required. Second, Zorua
improves utilization by determining the runtime resource requirements of an
application. It then allocates and deallocates resources dynamically, managing
them \One \emph{independently} of each other to maximize their utilization; and \two in a \emph{coordinated} manner, to enable efficient execution of each
thread with all its required resources available.

\textbf{Challenges in Virtualization.} Unfortunately, oversubscription means
that latency-critical resources, such as registers and scratchpad, may be
swapped to memory at the time of access, resulting in high overheads in performance
and energy. This leads to two critical challenges in designing a framework to
enable virtualization. The first challenge is to effectively determine the
\emph{extent} of virtualization, i.e., by how much each resource appears to be
larger than its physical amount, such that we can
minimize oversubscription while still reaping its benefits. This is difficult as
the resource requirements continually vary during runtime. The second challenge
is to minimize accesses to the swap space. This requires \emph{coordination} in
the virtualized management of \emph{multiple resources}, so that enough of each
resource is available
on-chip when needed.

\textbf{Zorua}. In order to address these challenges, Zorua employs a
hardware-software codesign that comprises three components: \One
\textbf{\emph{the compiler}} annotates the program to specify the resource needs
of {\em each phase} of the application; \two \textbf{\emph{a runtime system}},
which we refer to as the \textbf\emph{coordinator}, 
uses the compiler annotations to dynamically manage the virtualization of the
different on-chip resources; and \three \textbf{\emph{the hardware}}
employs mapping tables to locate a virtual resource in the physically available
resources or in the swap space in main memory. The coordinator plays the key role
of scheduling threads {\em only when} the expected gain in thread-level
parallelism outweighs the cost of transferring oversubscribed resources from the
swap space in memory, and coordinates the oversubscription and allocation of
multiple on-chip resources.

\textbf{Key Results.} We evaluate Zorua with many resource specifications
for eight applications across three GPU architectures (Section~\ref{sec:eval}).
Our experimental results show that Zorua \One reduces the range in performance for
different resource specifications by 50\% on average (up to 69\%), by alleviating
performance cliffs, and hence eases the burden on the programmer to provide
optimized resource specifications, \two improves performance for code with optimized
specification by 13\%  on average (up to 28\%), and \three enhances portability by reducing
the maximum porting performance loss by 55\% on average (up to 73\%) for three different
GPU architectures. We conclude that decoupling the resource specification and
resource management via virtualization significantly eases programmer burden,
by alleviating the need to provide optimized specifications and enhancing
portability, while still improving or retaining performance for programs that already have
optimized specifications.
 
\textbf{Other Uses.} We believe that Zorua offers the opportunity to address
several other key challenges in GPUs today, for example: (i) By providing an new level of
indirection, Zorua provides a natural way to enable dynamic and fine-grained
control over resource partitioning among {\em multiple GPU kernels and
applications}. (ii) Zorua can be utilized for {\em low-latency preemption} of GPU
applications, by leveraging the ability to swap in/out resources from/to
memory in a transparent manner. (iv) Zorua provides a simple mechanism to provide
dynamic resources to support other programming paradigms such as nested
parallelism, helper threads, etc. and even system-level tasks. (v) The dynamic
resource management scheme in Zorua improves the energy efficiency and scalability
of expensive on-chip resources (Section~\ref{sec:applications}).

The main \textbf{contributions} of this work are: \begin{itemize}

\item{This is the first work that takes a holistic approach to
decoupling a GPU application's resource specification from its physical on-chip
resource allocation via the use of virtualization. We develop a comprehensive
virtualization framework that provides \emph{controlled} and \emph{coordinated}
virtualization of \emph{multiple} on-chip GPU resources to maximize the efficacy
of virtualization.}

\item{We show how to enable efficient oversubscription of multiple GPU resources with dynamic
fine-grained allocation of resources and swapping mechanisms into/out of main
memory. We provide a hardware-software cooperative framework that \One controls the extent of
oversubscription to make an effective tradeoff between higher thread-level
parallelism due to virtualization versus the latency and capacity overheads of
swap space usage, and \two coordinates the virtualization for multiple on-chip
resources, transparently to the programmer.}

\item{We demonstrate that by providing the illusion of having more resources
than physically available, Zorua \One reduces programmer burden, providing
competitive performance for even suboptimal resource specifications, by reducing
performance variation across different specifications and by alleviating
performance cliffs; \two reduces performance loss when the program with its resource
specification tuned for one GPU platform is ported to a different platform; and
\three retains or enhances performance for highly-tuned code by improving
resource utilization, via dynamic management of resources.}

\end{itemize}

\subsection{Motivation: Managing On-Chip Resources and Parallelism in GPUs}
\label{sec:motivation} 
The amount of parallelism that the GPU can provide for any application depends on the
utilization of on-chip resources by threads within the application. As a
result, suboptimal usage of these resources may lead to loss in the parallelism
that can be achieved during program execution. This loss in parallelism 
often leads to significant degradation in performance, as GPUs primarily use
fine-grained multi-threading~\cite{burtonsmith,cdc6600} to hide the long latencies during execution. 

The granularity of synchronization -- i.e., the number of threads in a thread
block -- and the amount of scratchpad memory used per thread block is determined by
the programmer while adapting any algorithm or application for execution on a
GPU. 
\highlight{This choice involves a complex tradeoff between
minimizing data movement, by using \emph{larger} scratchpad memory sizes, and
reducing the
inefficiency of synchronizing a large number of threads, by using
\emph{smaller} scratchpad memory and thread block sizes. 
A similar tradeoff exists when determining the number of registers
used by the application.
Using \emph{fewer} registers
minimizes hardware register usage and enables higher parallelism during
execution, whereas using \emph{more} registers avoids expensive accesses to
memory.} The resulting application parameters -- the number of
registers, the amount of scratchpad memory, and the number of threads per thread
block -- dictate the on-chip resource requirement and hence, determine the
parallelism that can be obtained for that application on
any GPU. 

In this section, we study the performance implications of different choices of
resource specifications for GPU applications to demonstrate the key issues we aim to alleviate.
\subsubsection{Performance Variation and Cliffs} \label{sec:perf-cliffs}
To understand the impact of resource specifications and the resulting
utilization of physical resources on GPU performance, we conduct an experiment
on a Maxwell GPU system (GTX 745) with 20 GPGPU workloads from the CUDA
SDK~\cite{sdk}, Rodinia~\cite{rodinia}, GPGPU-Sim benchmarks~\cite{GPGPUSim},
Lonestar~\cite{lonestar}, Parboil~\cite{parboil}, and US DoE application
suites~\cite{DBLP:conf/sc/VillaJOBNLSWMSKD14}. We use the NVIDIA profiling tool
(NVProf)~\cite{sdk} to determine the execution time of each application
kernel\ignore{(detailed methodology is in Section~\ref{sec:methodology})}.
We sweep the three parameters of the specification{\textemdash}number of threads in a
thread block, register usage per thread, and scratchpad memory usage per thread
block{\textemdash}for each workload, and measure their impact on execution time. \ignore{We
ensure that the application performs the same amount of work for each
specification.}

\begin{figure}[h] \centering
\includegraphics[width=0.48\textwidth]{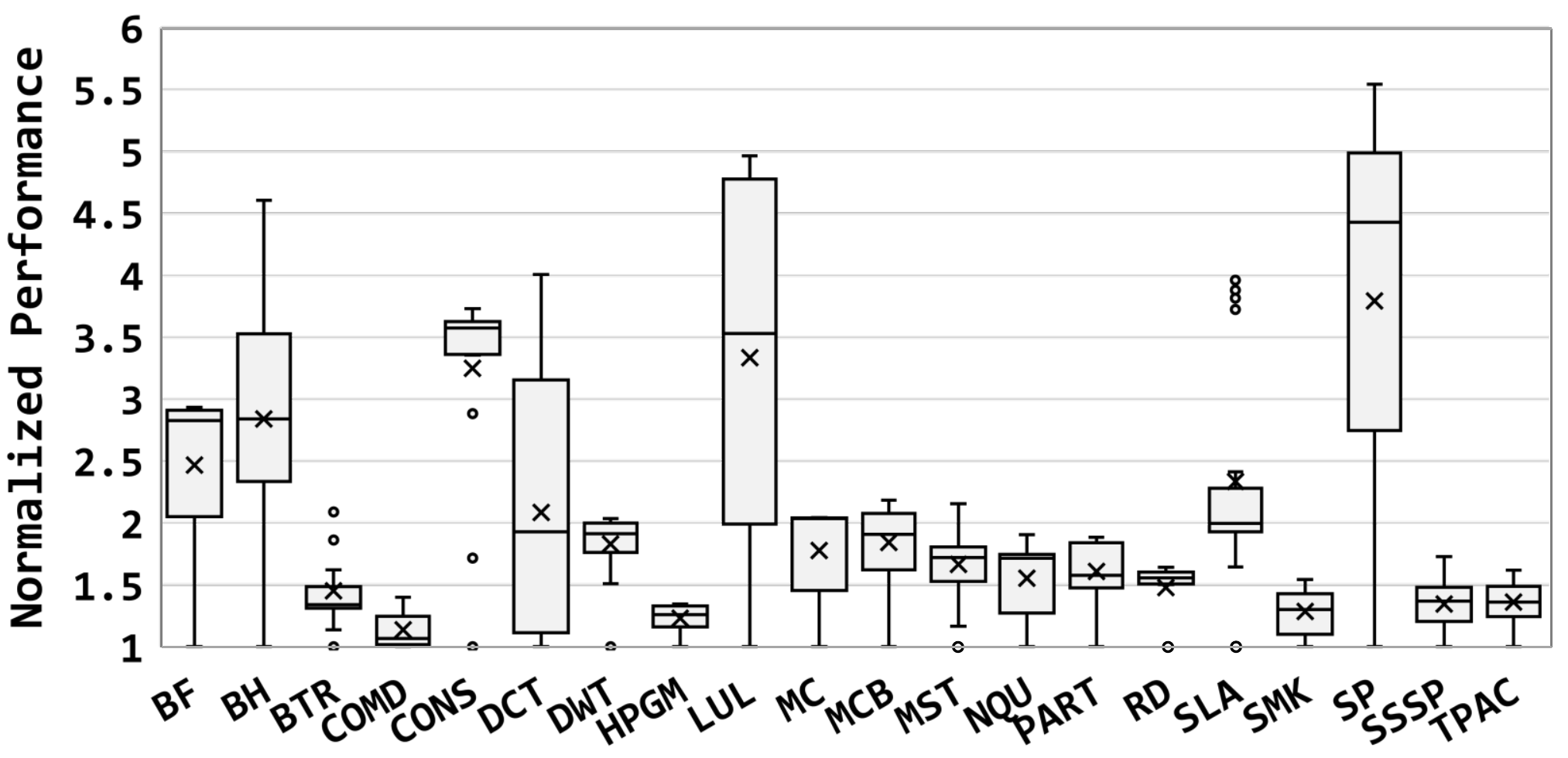}
\caption{\small{Performance variation across specifications.}}
\label{fig:summary-variation} \end{figure}

Figure~\ref{fig:summary-variation} shows a summary of variation in performance
(higher is better), normalized to the slowest specification for each application, across all
evaluated specification points for each application\ignore{Our
technical report~\cite{zorua-tr} contains more detail on the evaluated ranges.} in a Tukey box
plot~\cite{mcgill1978variations}. The boxes in the box plot represent the range
between the first quartile (25\%) and the third quartile (75\%). 
The whiskers extending from the
boxes represent the maximum and minimum points of the distribution, or
1.5$\times$ the
length of the box, whichever is smaller. Any points that lie more than
1.5$\times$
the box length beyond the box are considered to be
outliers~\cite{mcgill1978variations}, and are plotted as
individual points. The line in the middle of the box represents the median,
while the ``X'' represents the average. 

\label{sec:motivation:cliffs} 

We can see that there is
significant variation in performance across different specification points (as
much as 5.51$\times$ in \emph{SP}), proving the importance of optimized resource
specifications. In some applications (e.g., \emph{BTR, SLA}), few points
perform well, and these points are significantly better than others, suggesting that it would be
challenging for a programmer to locate these high performing specifications and obtain
the best performance. Many workloads (e.g., \emph{BH, DCT, MST}) also have
higher concentrations of specifications with suboptimal performance in
comparison to the best performing point, implying that, without effort, it is
likely that the programmer will end up with a resource specification that leads
to low performance. 

There are several sources for this performance variation. One important source is the
loss in thread-level parallelism as a result of a suboptimal resource
specification. Suboptimal specifications that are \emph{not} tailored to fit the
available physical resources lead to the underutilization of resources.
This causes a drop in the number of threads that can be executed concurrently,
as there are insufficient resources to support their execution. Hence, better
and more balanced utilization of resources enables higher thread-level
parallelism. Often, this loss in
parallelism from resource underutilization
manifests itself in what we refer to as a \emph{performance cliff}, where a
small deviation from an optimized specification can lead to
significantly worse performance, i.e., there is very high variation in
performance between two specification points that are nearby. To demonstrate the
existence and analyze the behavior of performance cliffs, we examine two representative workloads more closely. 

\begin{figure}[h]
\centering \begin{subfigure}[h]{0.99\linewidth} 
\centering
\includegraphics[width=0.49\textwidth]{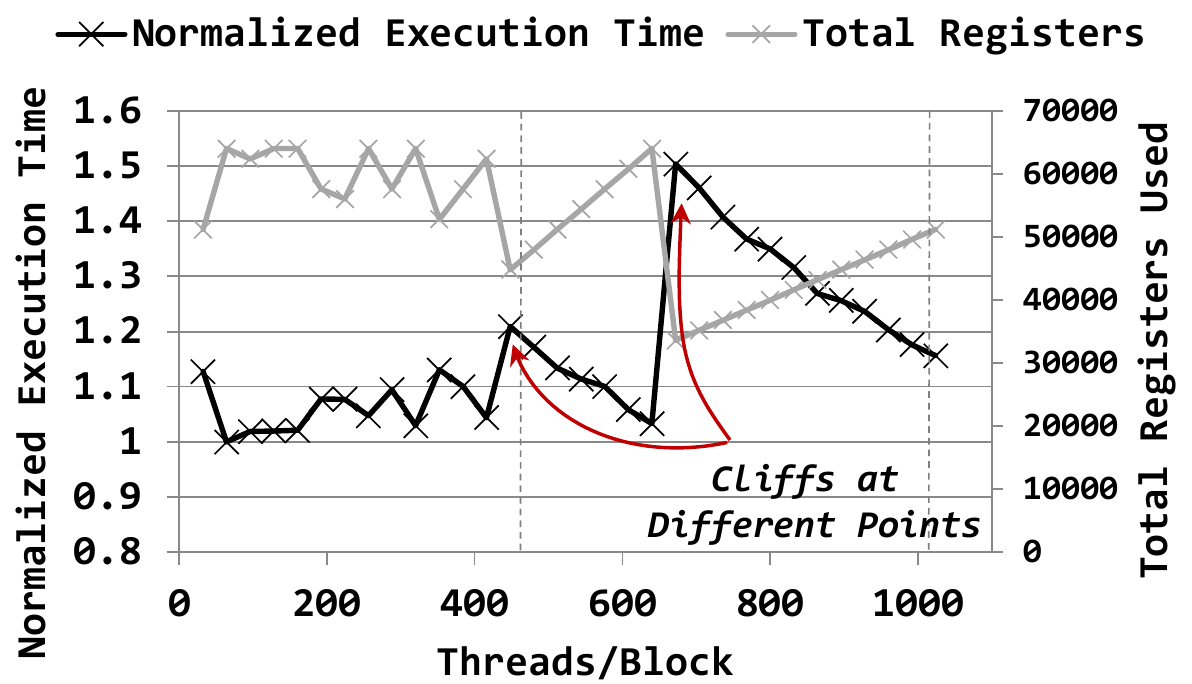}
\caption{Threads/block sweep}
\label{fig:mst-time}
\end{subfigure}
\begin{subfigure}[h]{0.99\linewidth}
\centering
\includegraphics[width=0.49\textwidth]{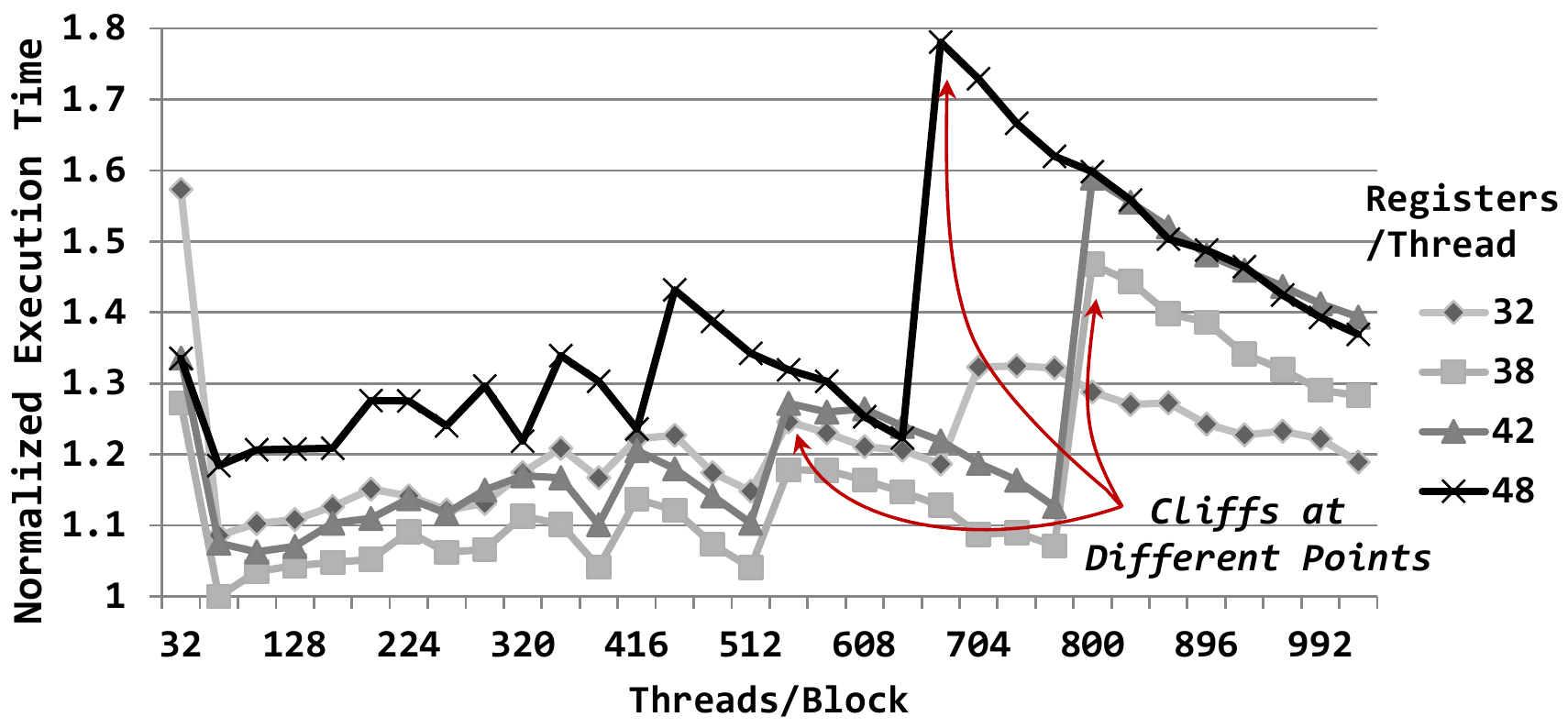} 
\caption{Threads/block \& Registers/thread sweep} 
\label{fig:mst-registers} 
\end{subfigure}
\caption{Performance cliffs in \emph{Minimum Spanning Tree} (\emph{MST}).} 
\label{fig:mst-cliff}
\end{figure}

Figure~\ref{fig:mst-time} shows \One how the application execution time changes;
and \two how the corresponding number of registers, statically used, changes when the number of
threads per thread block increases from 32 to 1024 threads, for \emph{Minimum
Spanning Tree (MST)}~\cite{lonestar}. We make two observations.
 
First, let us focus on the execution time between 480 and 1024 threads per
block. As we go from 480 to 640 threads per block, execution time gradually
decreases. Within this window, the GPU can support two thread blocks running
concurrently for \emph{MST}. The execution time falls because the increase in the number of threads per block
improves the overall throughput (the number of thread blocks running
concurrently remains constant at two, but each thread block does more work in
parallel by having more threads per block). However, the corresponding total number of registers used by the blocks
also increases. At 640 threads per block, we reach the point where the total
number of available registers is not large enough to support two blocks. As a
result, the number of blocks executing in parallel drops from two to one,
resulting in a significant increase (50\%) in execution time, i.e., the
\emph{performance cliff}.\footnote{Prior
work~\cite{warp-level-divergence} has studied performing resource allocation at the finer warp
granularity, as opposed to the coarser granularity of a
thread block. As we discuss in
Section~\ref{sec:relatedwork} and demonstrate in Section~\ref{sec:eval}, this does
\emph{not} solve the problem of performance cliffs.}
We see many of these cliffs earlier in the graph as
well, albeit not as drastic as the one at 640 threads per block.

Second, Figure~\ref{fig:mst-time} shows the existence of performance cliffs when
we vary \emph{just one} system parameter{\textemdash}the number of threads per block. To make
things more difficult for the programmer, other parameters (i.e., registers per thread or scratchpad memory
per thread block) also need to be decided at the same time\ignore{, which makes the
programmer's task much harder in avoiding such cliffs}. Figure~\ref{fig:mst-registers}
demonstrates \highlight{that performance cliffs also exist} when the \emph{number of registers per
thread} is varied from 32 to 48.\footnote{We note that the
register usage reported by the compiler may vary from the actual runtime
register usage~\cite{sdk}, hence slightly altering the points at which cliffs
occur.} As this figure shows, performance cliffs now occur at \emph{different points} for \emph{different
registers/thread curves}, which makes optimizing resource specification, so as to avoid
these cliffs, much harder for the programmer.
 
\begin{figure}[h] \centering
\includegraphics[width=0.49\linewidth]{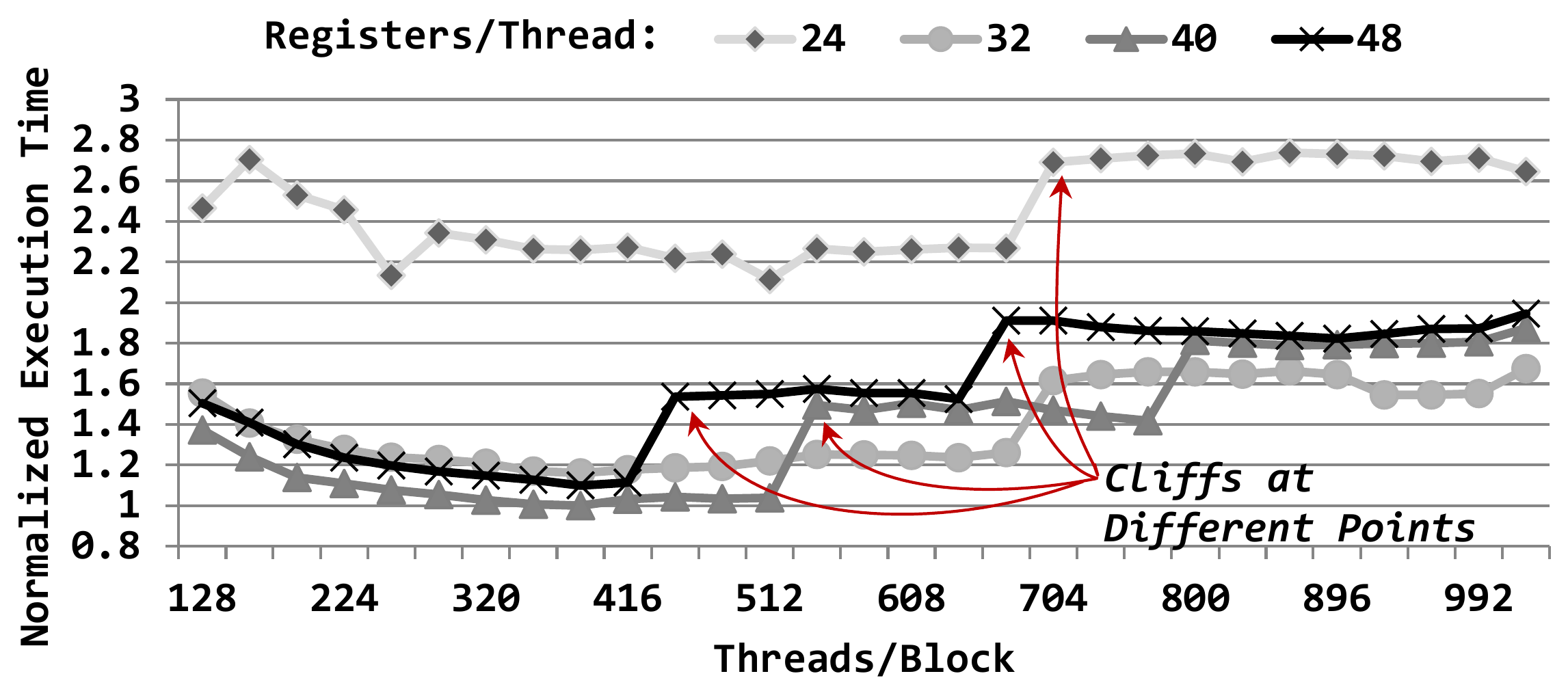}
\caption{Performance cliffs in \emph{Barnes-Hut}
(\emph{BH}).}
\label{fig:bh-motiv} \end{figure}

\highlight{\emph{Barnes-Hut (BH)} is another application that
exhibits very significant performance cliffs depending on the number of threads
per block and registers per thread.  Figure~\ref{fig:bh-motiv}
plots the variation in performance with the number of threads per block
when \emph{BH} is compiled for a range of register sizes (between 24 and 48 registers
per thread). We make two observations from the figure.  First, 
similar to \emph{MST}, we observe a significant variation in
performance that manifests itself in the form of performance cliffs.
Second, we observe that the points at which the performance cliffs occur
change \newII{greatly} depending on the number of registers assigned to each thread
during compilation.}

\highlight{We conclude that performance cliffs are pervasive across GPU programs,
and occur due to fundamental limitations of existing GPU hardware resource
managers, where resource management is static, coarse-grained, and tightly coupled to the
application resource specification. Avoiding performance cliffs by
determining more optimal resource specifications is a challenging
task, because the occurrence of these cliffs depends on several factors,
including the application characteristics, input data, and the underlying
hardware resources.}

\ignore{This phenomenon is not an artifact
of an application, but rather
a fundamental limitation of the GPU hardware that
distributes existing limited resources in a static and coarse-grain fashion. 
}

\subsubsection{Portability} 
\label{sec:motivation:port}

As we show in Section~\ref{sec:perf-cliffs}, tuning GPU applications to achieve good
performance on a given GPU is already a challenging task. To make things worse,
even after this tuning is done by
the programmer for one particular GPU architecture, it has to be \emph{redone} for
every new GPU generation (due to changes in the available physical resources
across generations) to ensure that good performance is retained.
We demonstrate this \emph{portability problem} by running sweeps of the three
parameters of the resource specification on various workloads, on three real GPU
generations: Fermi (GTX 480),
Kepler (GTX 760), and Maxwell (GTX 745).
\begin{figure}[h] 
\centering 
\begin{subfigure}[h]{0.49\linewidth} 
\centering
\includegraphics[width=0.90\textwidth]{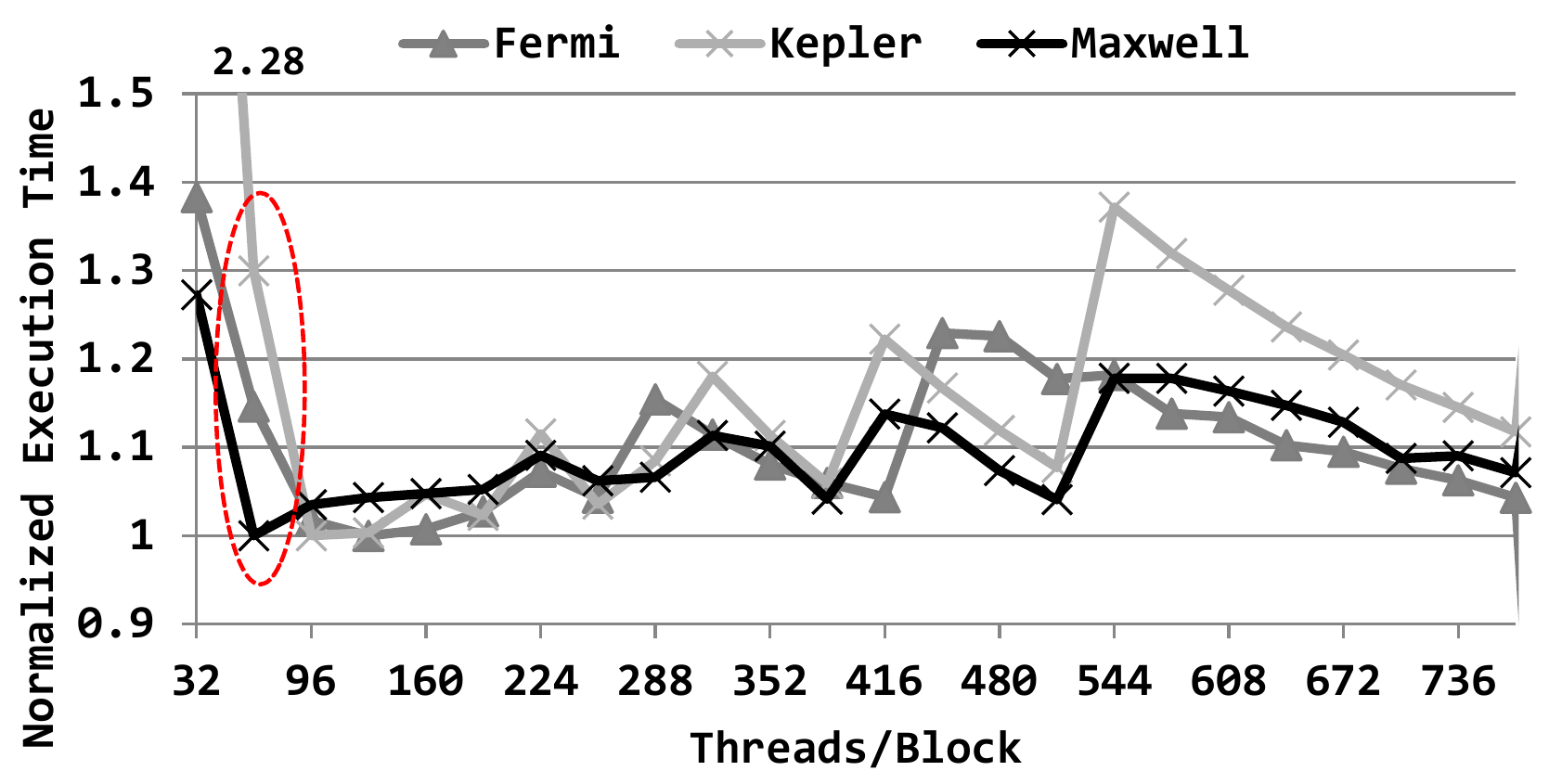} 
\caption{\emph{MST}}
\label{fig:mst-port} 
\end{subfigure} 
\begin{subfigure}[h]{0.49\linewidth}
\centering 
\includegraphics[width=0.90\textwidth]{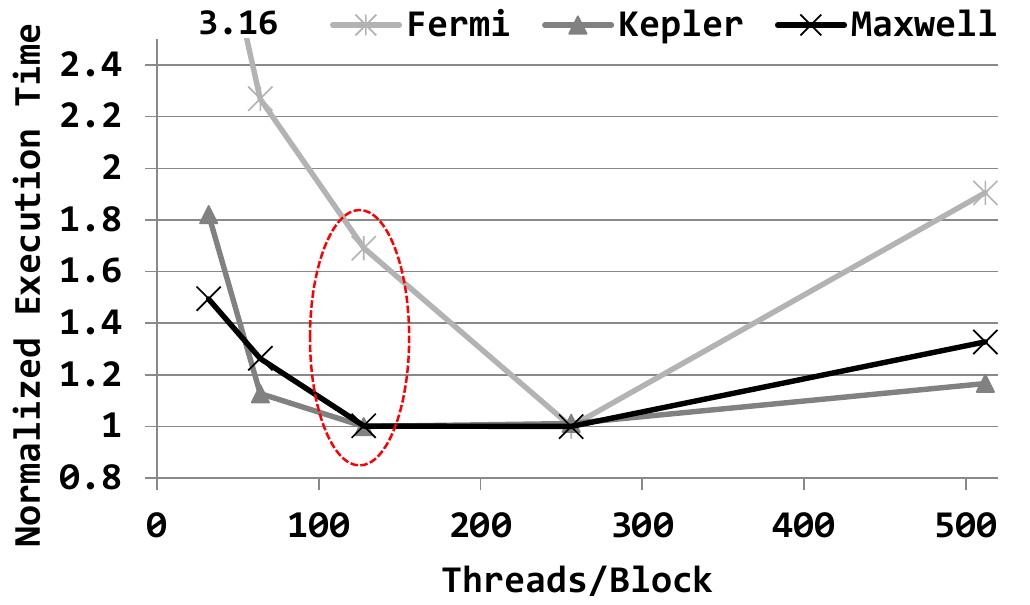}
\caption{\emph{DCT}} 
\label{fig:dct-port} 
\end{subfigure} 
\begin{subfigure}[h]{0.49\linewidth}
\centering 
\includegraphics[width=0.90\textwidth]{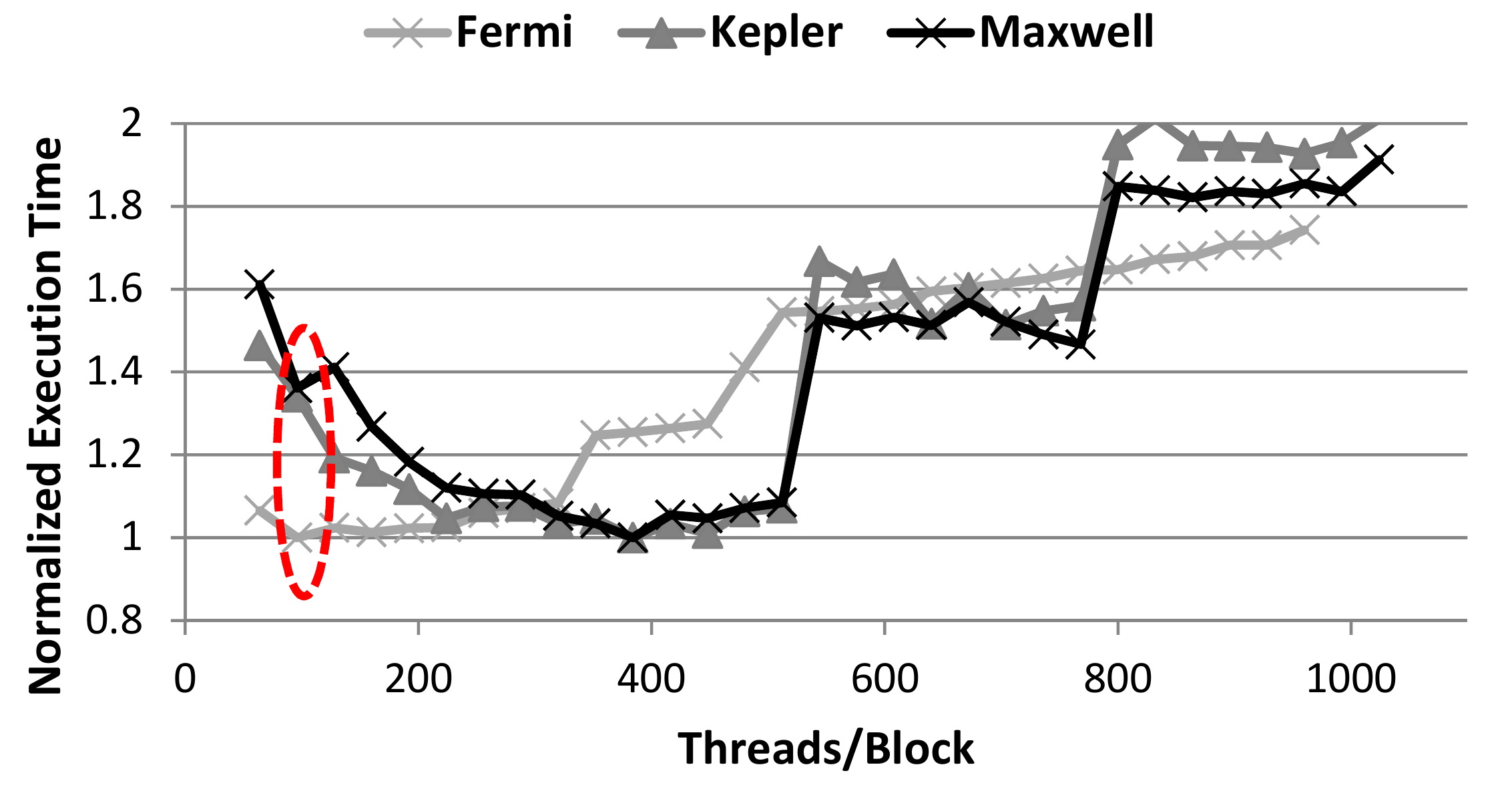}
\caption{\emph{BH}} 
\label{fig:bh-port} 
\end{subfigure} 
\caption{Performance variation across different GPU generations (Fermi, Kepler, and Maxwell) for
\emph{MST}, \emph{DCT}, and \emph{BH}.}
\label{fig:port} 
\end{figure}

Figure~\ref{fig:port} shows how the optimized performance points change between
different GPU generations for two representative applications (\emph{MST} and
\emph{DCT}). For every generation, results are normalized to the lowest
execution time for that particular generation. As we can see in
Figure~\ref{fig:mst-port}, the best performing points for different generations occur at
\emph{different} specifications because the application behavior changes with the variation in
hardware resources. 
For \emph{MST}, the \emph{Maxwell} architecture performs best at 64
threads per block. However, the same specification point is not efficient
for either of the other generations (\emph{Fermi} and \emph{Kepler}), producing 
15\% and 30\% lower performance, respectively, compared to the best specification
for each generation.
For \emph{DCT} (shown in Figure~\ref{fig:dct-port}),
both \emph{Kepler} and \emph{Maxwell} perform best at 128 threads per
block, but using the same specification for \emph{Fermi} would lead to a 
69\% performance loss. Similarly, for \emph{BH} (Figure~\ref{fig:bh-port}), the
optimal point for 
\emph{Fermi} architecture is at 96 threads per block. \highlight{However, using the
same configuration for the two later GPU architectures -- \emph{Kepler} and
\emph{Maxwell} could lead to very suboptimal performance results.
Using the same configuration results in as much as a 34\% performance loss
on \emph{Kepler}, and a 36\% performance loss on \emph{Maxwell}.

We conclude that the tight coupling between the programming model and the
underlying resource management in hardware imposes a significant challenge in
performance portability. To avoid suboptimal performance, an application has to
be \emph{retuned} by the programmer to find an optimized resource specification
for \emph{each}
GPU generation.
}

\subsubsection{Dynamic Resource Underutilization}
\label{sec:underutilized_resources}

Even when a GPU application is \emph{perfectly} tuned for a particular GPU
architecture, the on-chip resources are
typically not fully
utilized~\cite{virtual-register,gebhart-hierarchical,compiler-register,shmem-multiplexing,unified-register,
caba,asplos-sree,spareregister-lakshminarayana-hpca14, mask, mosaic, ltrf-sadrosadati-asplos18}.
For example, it is well known that while the compiler conservatively allocates registers to
hold the \emph{maximum number} of live values throughout the execution, the number of
live values at any given time is well below the maximum for large portions of application execution time.
To determine the magnitude of this \emph{dynamic
underutilization},\footnote{\revision{Underutilization of registers occurs in two
major forms{\textemdash}\emph{static}, where registers are unallocated
throughout
execution~\cite{caba,warp-level-divergence,unified-register,spareregister-lakshminarayana-hpca14,energy-register,mosaic,mask,ltrf-sadrosadati-asplos18},
and \emph{dynamic}, where utilization of the registers drops during runtime as a
result of early completion of warps~\cite{warp-level-divergence}, short register
lifetimes~\cite{virtual-register,gebhart-hierarchical,compiler-register} and
long-latency operations~\cite{gebhart-hierarchical,compiler-register}. We do not
tackle underutilization from long-latency operations (such as memory
accesses) in this work, and leave the exploration of alleviating this type of
underutilization to future work.}} we conduct an experiment where we measure
the dynamic usage (per epoch) of both scratchpad memory and registers for
different applications with \emph{optimized} specifications in our workload pool. 
\begin{figure}[h]
  \centering
  \begin{subfigure}[h]{0.99\linewidth}
  \centering
  \includegraphics[width=0.49\textwidth]{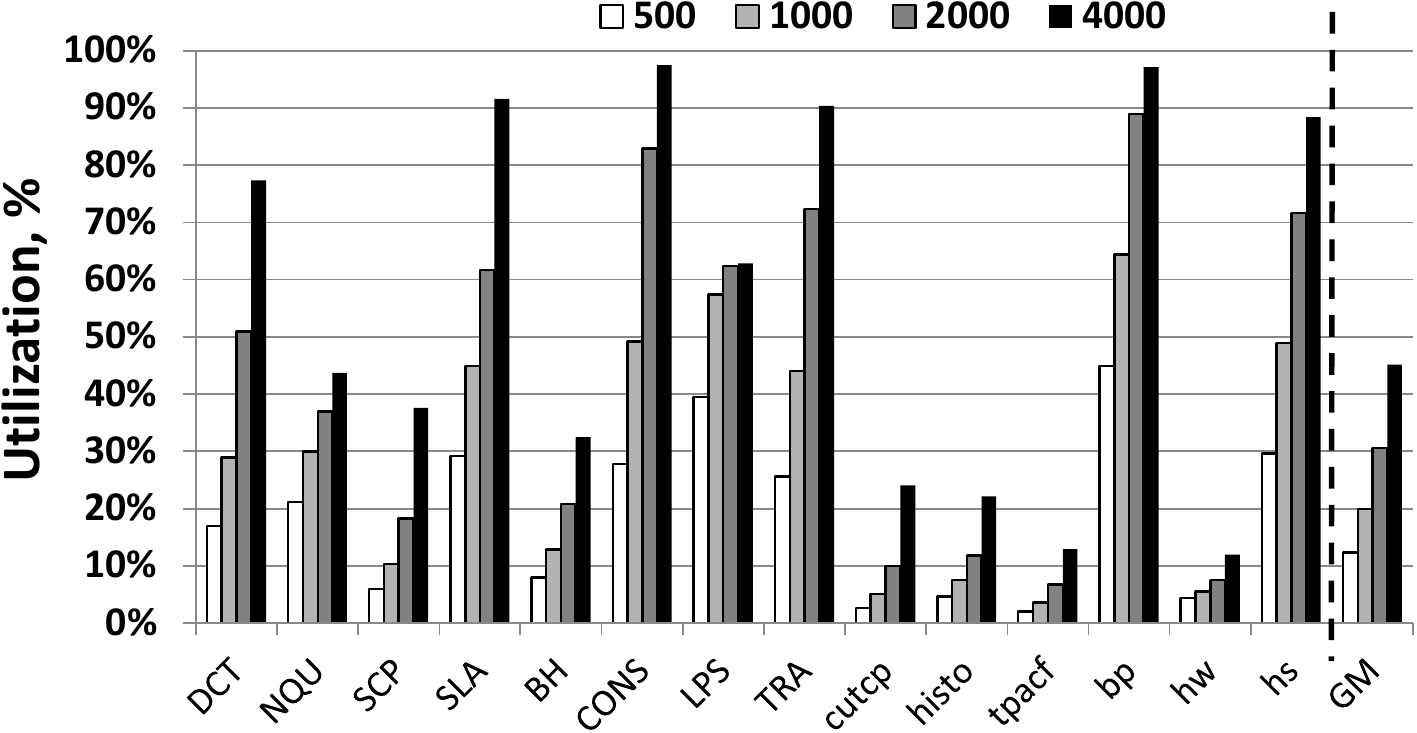}
  \caption{Scratchpad memory}
  \label{fig:scratchpad}
  \end{subfigure}
  \begin{subfigure}[h]{0.99\linewidth}
  \centering
  \includegraphics[width=0.49\textwidth]{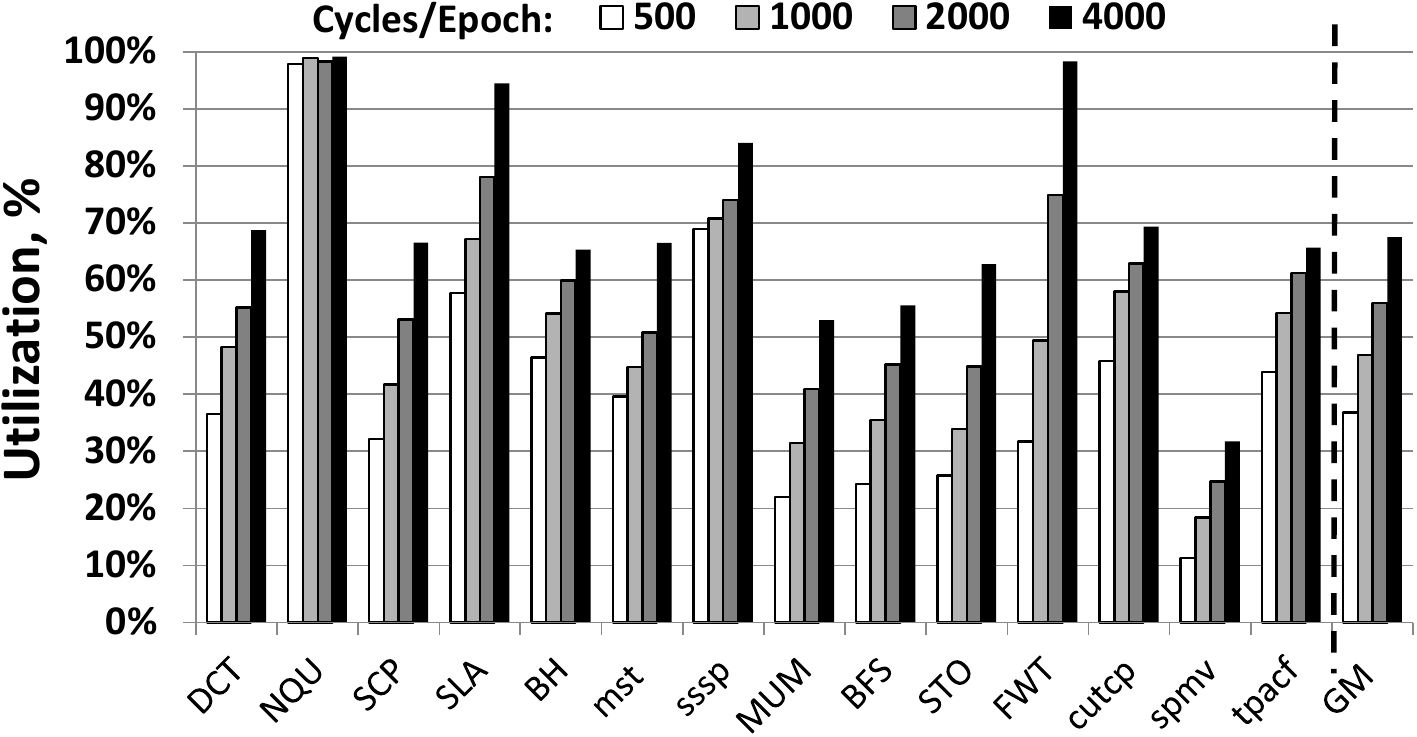}
  \caption{Registers}
  \label{fig:registers}
  \end{subfigure}
  \caption{Dynamic resource utilization for different length epochs.}
  \label{fig:utilization}
\end{figure} 

We vary the length of epochs from
500 to 4000 cycles. Figure~\ref{fig:utilization} shows the results of this
experiment for \emph{(i)}~scratchpad memory (Figure~\ref{fig:scratchpad}) and
\emph{(ii)}~on-chip registers (Figure~\ref{fig:registers}).
We make two major observations from these figures.

\highlight{First, for relatively small epochs (e.g., 500 cycles), the average utilization
of resources is very low (12\% for scratchpad memory and 37\% for registers).
Even for the largest epoch size that we analyze (4000 cycles), the
utilization of scratchpad memory is still less than 50\%, and the utilization of
registers is less than 70\%.} This observation clearly suggests that there is an opportunity
for a better dynamic allocation of these resources that could allow higher
effective GPU parallelism.

Second, there are several noticeable applications, e.g., \emph{cutcp}, \emph{hw},
\emph{tpacf}, where utilization of the scratchpad memory is always lower than
15\%. This dramatic underutilization due to static resource allocation can lead
to significant loss in potential performance benefits for these applications.

In summary, we conclude that existing static on-chip resource allocation in GPUs
can lead to significant resource underutilization that can lead to suboptimal performance
and energy waste.

\subsubsection{Our Goal}
As we see above, the tight coupling between the resource specification and
hardware resource allocation, and the resulting heavy dependence of performance on
the resource specification, creates a number of challenges.
In this work, our goal is to alleviate these challenges by providing a mechanism
that can \One 
 ease the burden on the programmer by ensuring reasonable
performance, \emph{regardless of the resource specification}, by successfully
avoiding performance cliffs, while retaining performance for code with
optimized specification; \two enhance portability by minimizing the variation
in performance for optimized specifications across different GPU generations; and
\three maximize dynamic resource utilization even in highly optimized code to further
improve performance. We make two key observations from our studies above to help
us achieve this goal.

\textbf{Observation 1: }\emph{Bottleneck Resources.} We find that performance cliffs occur when
the amount of any resource required by an application exceeds the
physically available amount of that resource. This resource becomes a
\emph{bottleneck}, and limits the amount of parallelism that the GPU can
support. If it were possible to provide the application with a \emph{small
additional amount} of the bottleneck resource, the application can see a
significant increase in parallelism and thus avoid the performance cliff. \ignore{For
example, an application may require 18KB of scratchpad memory per thread block.
When run on a system with a 32KB scratchpad memory, the application can only run
one thread block at a time. If we could find a way to effectively provide 4KB
more of scratchpad, the application could run two blocks concurrently.}


\textbf{Observation 2: }\emph{Underutilized Resources.} As discussed in
Section~\ref{sec:underutilized_resources}, there is significant underutilization
of resources at runtime. These underutilized resources could be employed to
support more parallelism at runtime, and thereby alleviate the aforementioned
challenges.

We use these two observations to drive our resource virtualization solution,
which we describe 
next.

\subsection{Our Approach: Decoupling the Programming Model from Resource Management}

In this work, we design Zorua, a framework that provides the illusion of more GPU
resources than physically available by decoupling the resource specification
from its allocation in the hardware resources. We introduce a new level of
indirection by virtualizing the on-chip resources to allow the hardware to
manage resources transparently to the programmer.

The virtualization provided by Zorua builds upon two \emph{key concepts} to
leverage the aforementioned observations. First, when there are insufficient
physical resources, we aim to provide the illusion of the required amount by
\emph{oversubscribing} the required resource. We perform this oversubscription
by leveraging the dynamic underutilization as much as possible, or by spilling to a
swap space in memory. This oversubscription essentially enables the illusion of
more resources than what is available (physically and statically), and supports
the concurrent execution of more threads. Performance cliffs are mitigated by
providing enough additional resources to avoid drastic drops in parallelism.
Second, to enable efficient oversubscription by leveraging underutilization, we
dynamically allocate and deallocate physical resources depending
on the requirements of the application during execution. We manage the
virtualization of each resource \emph{independently} of other resources to maximize
its runtime utilization.

\begin{figure}[h] \centering
\includegraphics[width=0.49\textwidth]{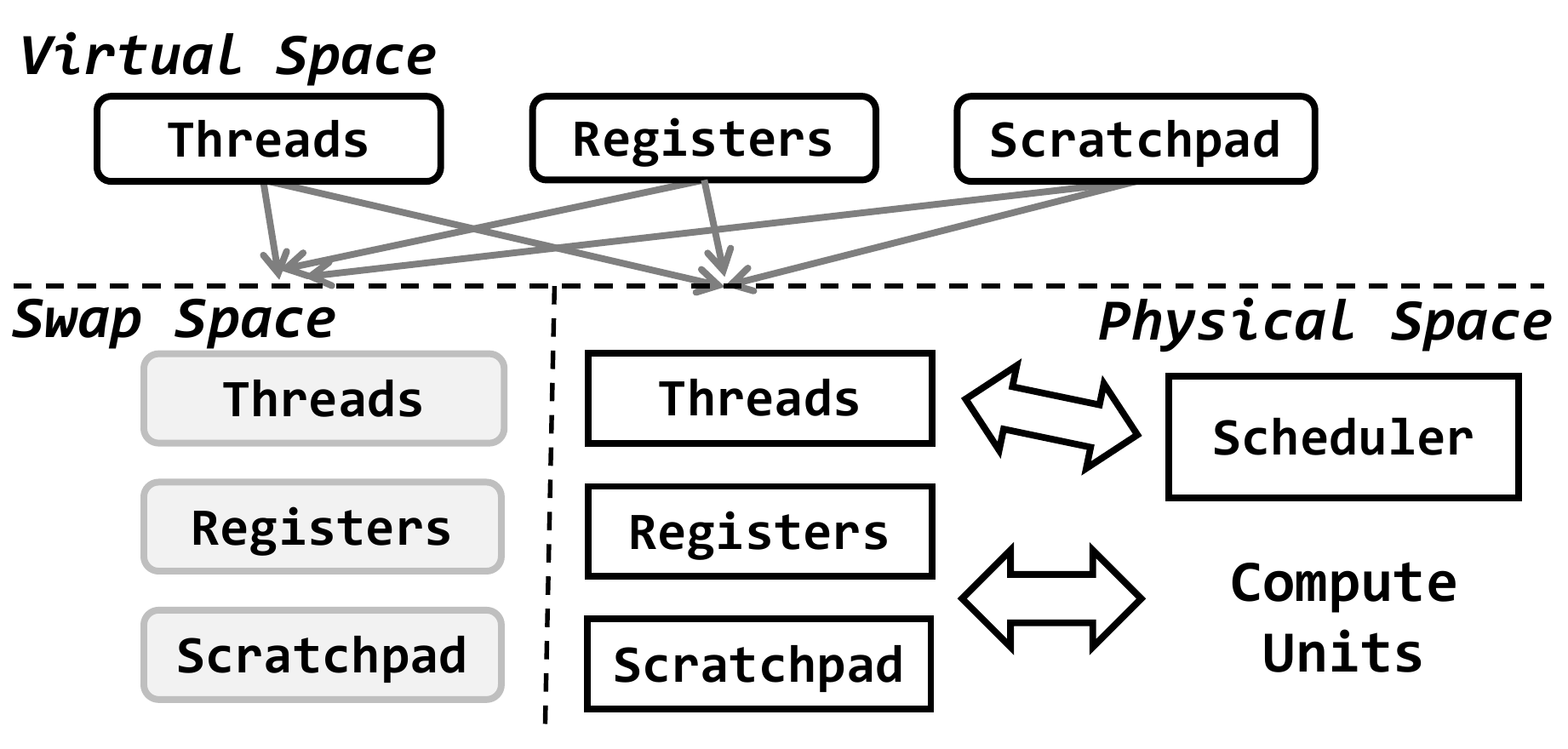}
\caption{High-level overview of Zorua.} \label{fig:overview_zorua} \end{figure}

Figure~\ref{fig:overview_zorua} depicts the high-level overview of the virtualization
provided by Zorua. 
The \emph{virtual space} refers to the \emph{illusion} of the quantity of available
resources. The \emph{physical space} refers to the \emph{actual} hardware resources
(specific to the GPU architecture), and the \emph{swap space} refers to the resources
that do not fit in the physical space and hence are \emph{spilled} to other
physical locations.
For the register file and scratchpad memory, the swap space is mapped to global
memory space in the memory hierarchy. For threads, only
those that are mapped to the physical space are available for scheduling
and execution at any given time. If a thread is mapped to the swap space, its
state (i.e., the PC and the SIMT stack) is saved in memory. Resources in the
virtual space can be freely re-mapped between the physical and swap
spaces to maintain the illusion of the virtual space resources.
 
In the baseline architecture, the thread-level parallelism that can be
supported,
and hence the throughput obtained from the GPU, depends on the quantity of
\emph{physical resources}. With the virtualization enabled by Zorua, the
parallelism that can be supported now depends on the quantity of \emph{virtual
resources} (and how their mapping into the physical and swap spaces is managed). Hence, the size of the virtual space for each resource plays the key
role of determining the parallelism that can be exploited. Increasing
the virtual space size enables higher parallelism, but leads to higher swap
space usage. It is critical to minimize accesses to the swap space to avoid the
latency overhead and capacity/bandwidth contention associated with accessing the
memory hierarchy. 

In light of this, there are two key challenges that need to be addressed to
effectively virtualize on-chip resources in GPUs. We now discuss these
challenges and provide an overview of how we address them.

\subsubsection{Challenges in Virtualization} \label{sec:virt-challenges}
\textbf{Challenge 1:}\emph{ Controlling the Extent of Oversubscription.} A key
challenge is to determine the \emph{extent} of oversubscription, or the size of
the virtual space for each resource. As discussed above, increasing the size of
the virtual space enables more parallelism. Unfortunately, it could
also result in more spilling of resources to the swap space. Finding the tradeoff
between more parallelism and less overhead is challenging, because the dynamic
resource requirements of each thread tend to significantly fluctuate throughout
execution. As a result, the size of the virtual space for each resource
needs to be \emph{continuously} tuned to allow the virtualization to adapt to
the runtime requirements of the program. 

\textbf{Challenge 2:} \emph{Control and Coordination of Multiple Resources.}
\ignore{Another critical challenge is to efficiently map the continuously varying
virtual resource space to the physical and swap spaces. 
This is critical as it is very important to minimize accesses to
the swap space. Accessing the swap space for the register file or scratchpad
involves expensive accesses to global memory, due to the added
latency and contention. Also, only those threads that are mapped to the
physical space are available to the warp scheduler for selection. Furthermore, each
thread requires \emph{multiple} resources for execution, each of which may be mapped to
the physical or swap space. It is critical to \emph{coordinate} the mapping of
these different virtual resources to ensure that a thread has all the
resources required at any given time mapped to the physical space, to enable
execution with minimal overhead. Thus, an effective virtualization framework
must coordinate the allocation of \emph{multiple} physical resources.}
Another critical challenge is to efficiently map the continuously varying
virtual resource space to the physical and swap spaces. 
This is important for two reasons. First, it is critical to minimize accesses to
the swap space. Accessing the swap space for the register file or scratchpad
involves expensive accesses to global memory, due to the added
latency and contention. Also, only those threads that are mapped to the
physical space are available to the warp scheduler for selection. Second, each
thread requires multiple resources for execution. It is critical to
\emph{coordinate} the allocation and mapping of
these different resources to ensure that an executing thread has \emph{all} the
required resources allocated to it, while minimizing accesses
to the swap space. Thus, an effective virtualization framework
must coordinate the allocation of \emph{multiple} on-chip resources.

\subsubsection{Key Ideas of Our Design}

To solve these challenges, Zorua employs two key ideas. First, we leverage the software
(the compiler) to provide annotations with information regarding the resource
requirements of each \emph{phase} of the application. This information enables the
framework to make intelligent dynamic decisions, with respect to both the size of
the virtual space and the allocation/deallocation of resources (Section~\ref{sec:key_idea_phases}).

Second, we use an adaptive runtime system to control the allocation of resources
in the virtual space and their mapping to the physical/swap spaces. This allows us to \One
dynamically alter the size of the virtual space to change the extent of
oversubscription; and \two continuously coordinate the allocation of multiple
on-chip resources and the mapping between their virtual and physical/swap
spaces, depending
on the varying runtime requirements of each thread
(Section~\ref{sec:key_idea_coordinator}).

\textbf{Leveraging Software Annotations of Phase Characteristics.}
\label{sec:key_idea_phases} 
We observe that the runtime variation in resource requirements
(Section~\ref{sec:underutilized_resources}) typically occurs at 
the granularity of \emph{phases} of a few tens of instructions. This variation
occurs because different parts of kernels perform different operations that
require different resources. For example, loops that primarily load/store data
from/to 
scratchpad memory tend to be less register heavy. Sections of
code that perform specific computations (e.g., matrix transformation, graph
manipulation), can either be register heavy or primarily operate out of
scratchpad. Often, scratchpad memory is used for only short
intervals~\cite{shmem-multiplexing}, e.g., when data exchange between threads is
required, such as for a reduction operation. 

\begin{figure}[h] \centering
\includegraphics[width=0.69\textwidth]{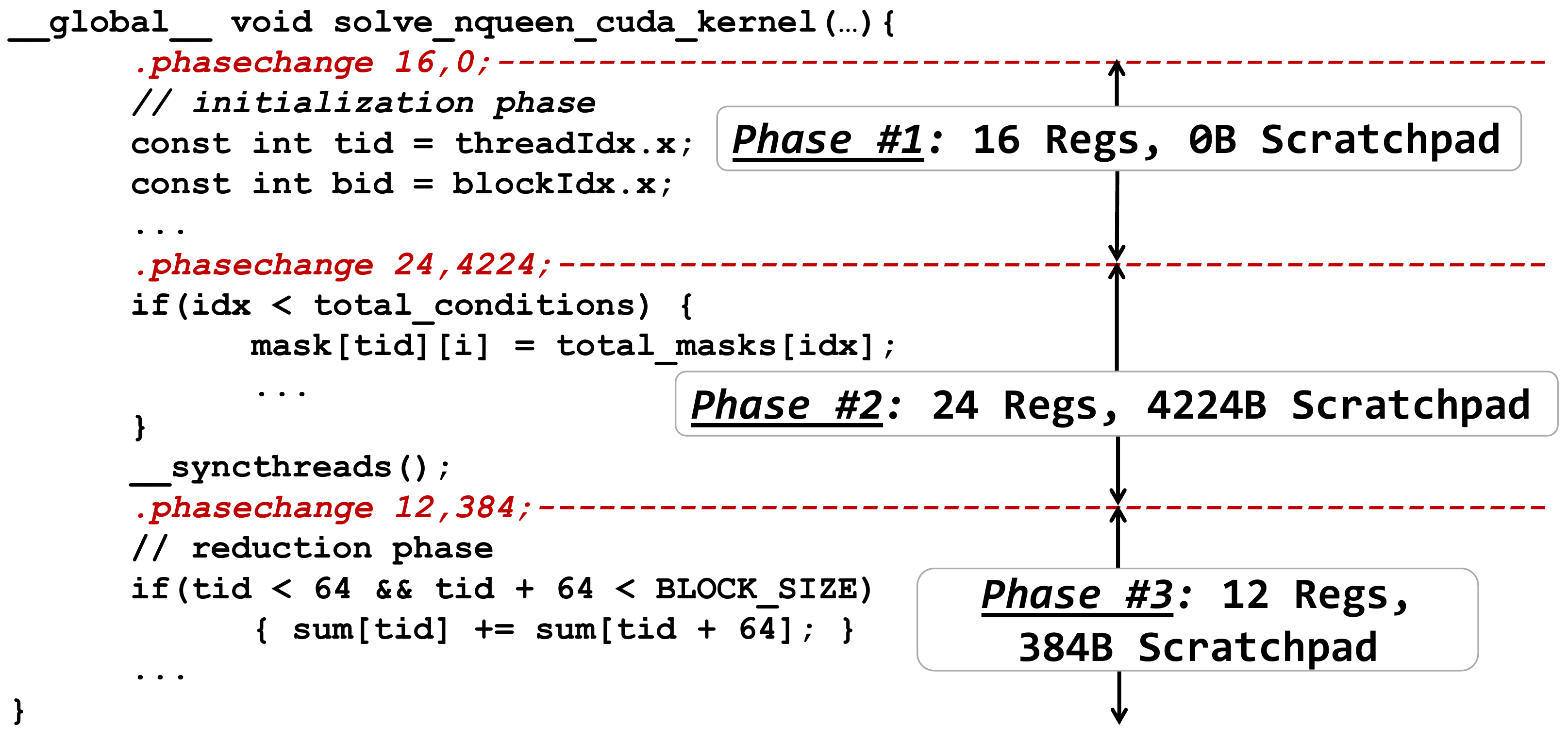}
\caption{Example phases from \emph{NQU}.} \label{fig:phases} \end{figure}

Figure~\ref{fig:phases} depicts a few example phases from the \emph{NQU}
(\emph{N-Queens
Solver})~\cite{NQU} kernel. \emph{NQU} is a scratchpad-heavy application, but it does
not use the
scratchpad at all during the initial computation phase. During its second phase, it performs
its primary computation out of the scratchpad, using as much as 4224B. During its
last phase, the scratchpad is used only for reducing results, which requires
only 384B. There is also significant variation in the maximum
number of live registers in the different phases.

\begin{figure}[h] \centering
\includegraphics[width=0.69\textwidth]{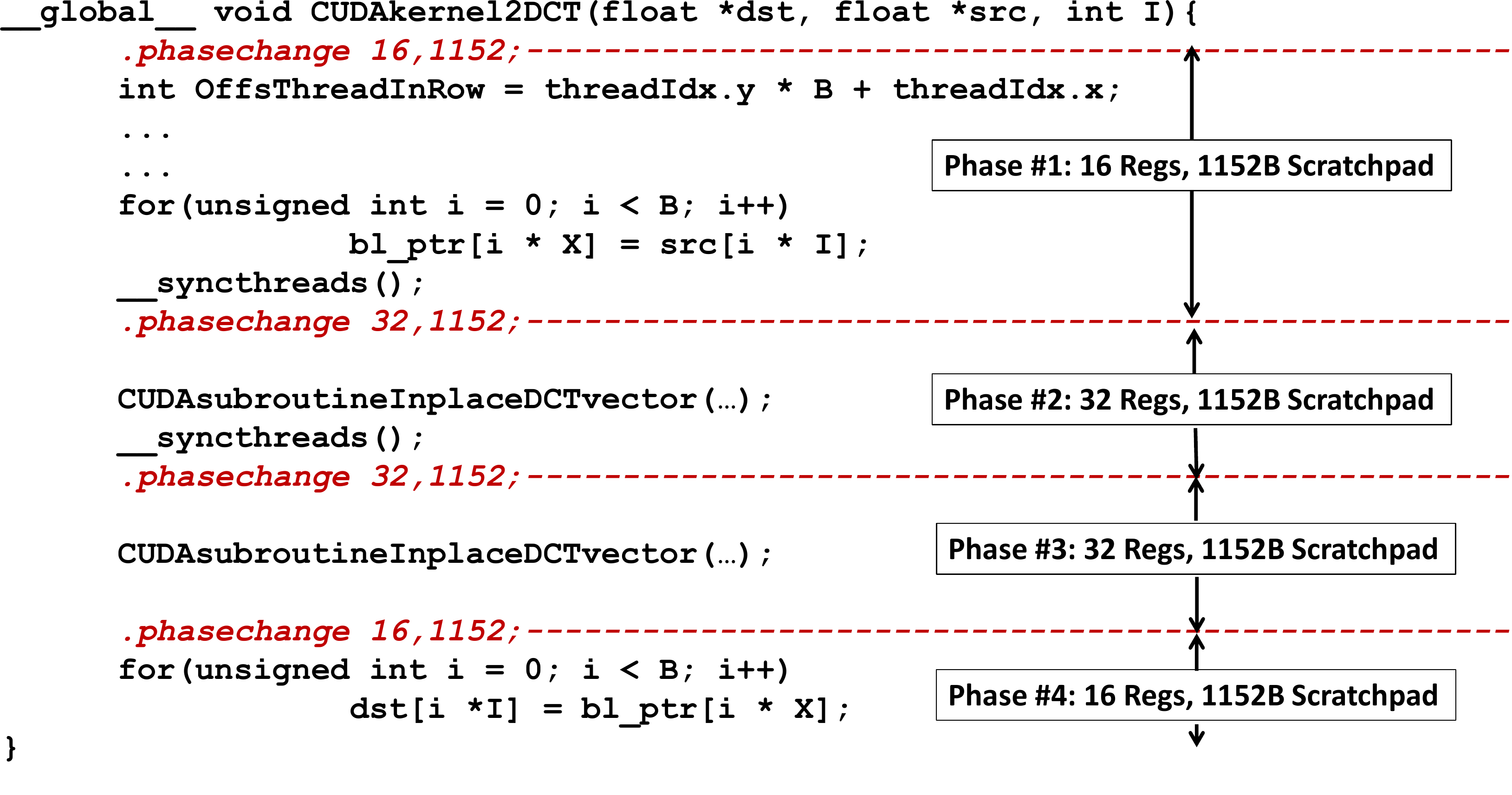}
\caption{Example phases from \emph{DCT}} \label{fig:phases2} \end{figure} 

Another example of phase variation from the \emph{DCT} (\emph{Discrete Fourier
Transform}) kernel is depicted in Figure~\ref{fig:phases2}. \emph{DCT} is both
register and scratchpad-intensive. The scratchpad memory usage \highlight{does not vary} in
this kernel. However, the register usage significantly
varies -- the register usage increases by 2X in the second and third phase in
comparison with the first and fourth phase.

In order to capture both the resource requirements as well as their variation
over time, we 
partition the program into a number of \emph{phases}. A phase is a
sequence of instructions with sufficiently different resource requirements than
adjacent phases. Barrier or fence operations also indicate a change in
requirements for a different reason{\textemdash}threads that are waiting at a barrier do
not immediately require the thread slot that they are holding. 
We interpret barriers and fences as phase boundaries since they potentially alter
the utilization of their thread slots. The compiler inserts special instructions
called \emph{phase specifiers} to mark the start of a new phase. Each phase
specifier contains information regarding the resource requirements of the next
phase. Section~\ref{sec:phase_specifiers} provides more detail on the semantics
of phases and phase specifiers. 
 
A phase forms the basic unit for resource allocation and
de-allocation, as well as for making oversubscription decisions. It offers
a finer granularity than an \emph{entire thread} to make such decisions.
The phase specifiers provide information on the \emph{future resource usage} of
the thread at a phase boundary. This enables \One preemptively controlling the
extent of oversubscription at runtime, and \two dynamically allocating and
deallocating resources at phase boundaries to maximize utilization of the
physical resources.

\begin{figure}[h] \centering
\includegraphics[width=0.59\textwidth]{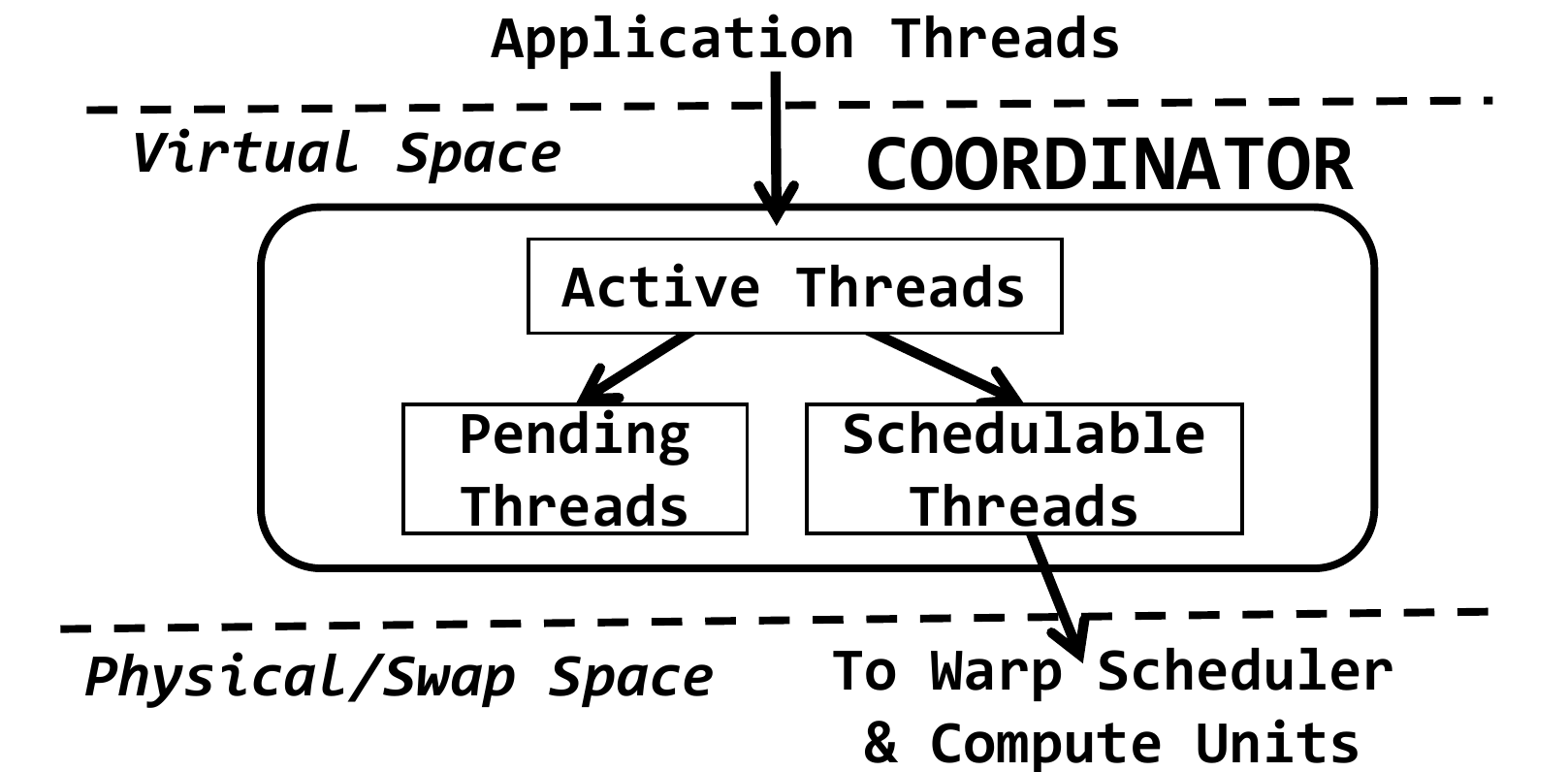}
\caption{Overview of the coordinator.} \label{fig:coordinator} \end{figure}

\textbf{Control with an Adaptive Runtime System.}
\label{sec:key_idea_coordinator}
Phase specifiers provide information to make oversubscription and allocation/deallocation
decisions. However, we still need a way to make decisions on the
extent of oversubscription and appropriately allocate resources at runtime. To this
end, we use an adaptive runtime system, which we refer to as the
\emph{coordinator}. Figure~\ref{fig:coordinator} presents an overview of the
coordinator.

The virtual space enables the illusion of a larger amount of each of the
resources than what is physically available, to adapt to different application
requirements. This illusion enables higher thread-level parallelism than what
can be achieved with solely the fixed, physically available resources, by
allowing more threads to execute concurrently. The size of the virtual space at a given
time determines this parallelism, and those threads that are effectively
executed
in parallel are referred to as \emph{active threads}. All active threads have
thread slots allocated to them in the virtual space (and hence can be executed),
but some of
them may not be mapped to
the physical space at a given time. As discussed previously,
the resource requirements of each application continuously change during
execution. To adapt to these runtime changes, the coordinator leverages
information from the phase specifiers to make decisions on oversubscription. The
coordinator makes these decisions at
every phase boundary and thereby controls the size of the virtual space for each resource
(see
Section~\ref{sec:walkthrough}).

To enforce the determined extent of oversubscription, the coordinator allocates
all the required resources (in the virtual space) for only a \emph{subset} of
threads from the active threads. Only these dynamically selected threads, referred to
as \emph{schedulable threads}, are available to the warp scheduler and compute
units for execution. The coordinator, hence, dynamically partitions the active threads into
\emph{schedulable threads} and the \emph{pending threads}. 
Each thread is swapped
between \emph{schedulable} and \emph{pending} states, depending on the
availability of resources in the virtual space.
Selecting only a subset of threads to execute at any time ensures that
the determined size of
the virtual space is not exceeded for any resource, and helps coordinate the
allocation and mapping of multiple on-chip resources to minimize expensive data
transfers between the physical and swap spaces (discussed in
Section~\ref{sec:mechanism}). 
%

\subsubsection{Overview of Zorua} In summary, to effectively address the challenges
in virtualization by leveraging the above ideas in design, Zorua employs a
software-hardware codesign that comprises three components: \One
\textbf{\emph{The compiler}} annotates the program by adding special
instructions (\emph{phase specifiers}) to partition it into \emph{phases} and to
specify the resource needs of each phase of the application. \two
\textbf{\emph{The coordinator}}, a hardware-based adaptive runtime system, 
uses the compiler annotations to dynamically allocate/deallocate resources for
each thread at phase boundaries. The coordinator plays the key role of
continuously controlling the extent of the oversubscription (and hence the size
of the virtual space) at each phase boundary.\ignore{by altering the number of threads
that are executing at any given time.} \three \textbf{\emph{Hardware
virtualization support}} includes a mapping table for each resource to
locate each virtual resource in either the physical space or the swap
space in main memory, and the machinery to swap resources between the physical
and swap spaces.

\subsection{Zorua: Detailed Mechanism}
\label{sec:mechanism}
We now detail the operation and implementation of the various components of the Zorua
framework. 
\subsubsection{Key Components in Hardware}
Zorua has two key hardware components: \One the \emph{coordinator}
that contains queues to buffer the \emph{pending threads} and
control logic to make oversubscription and resource management
decisions, and \two \emph{resource mapping tables} to map each of the
resources to their corresponding physical or swap spaces.

\begin{figure}[t]
 \centering
 \includegraphics[width=0.69\textwidth]{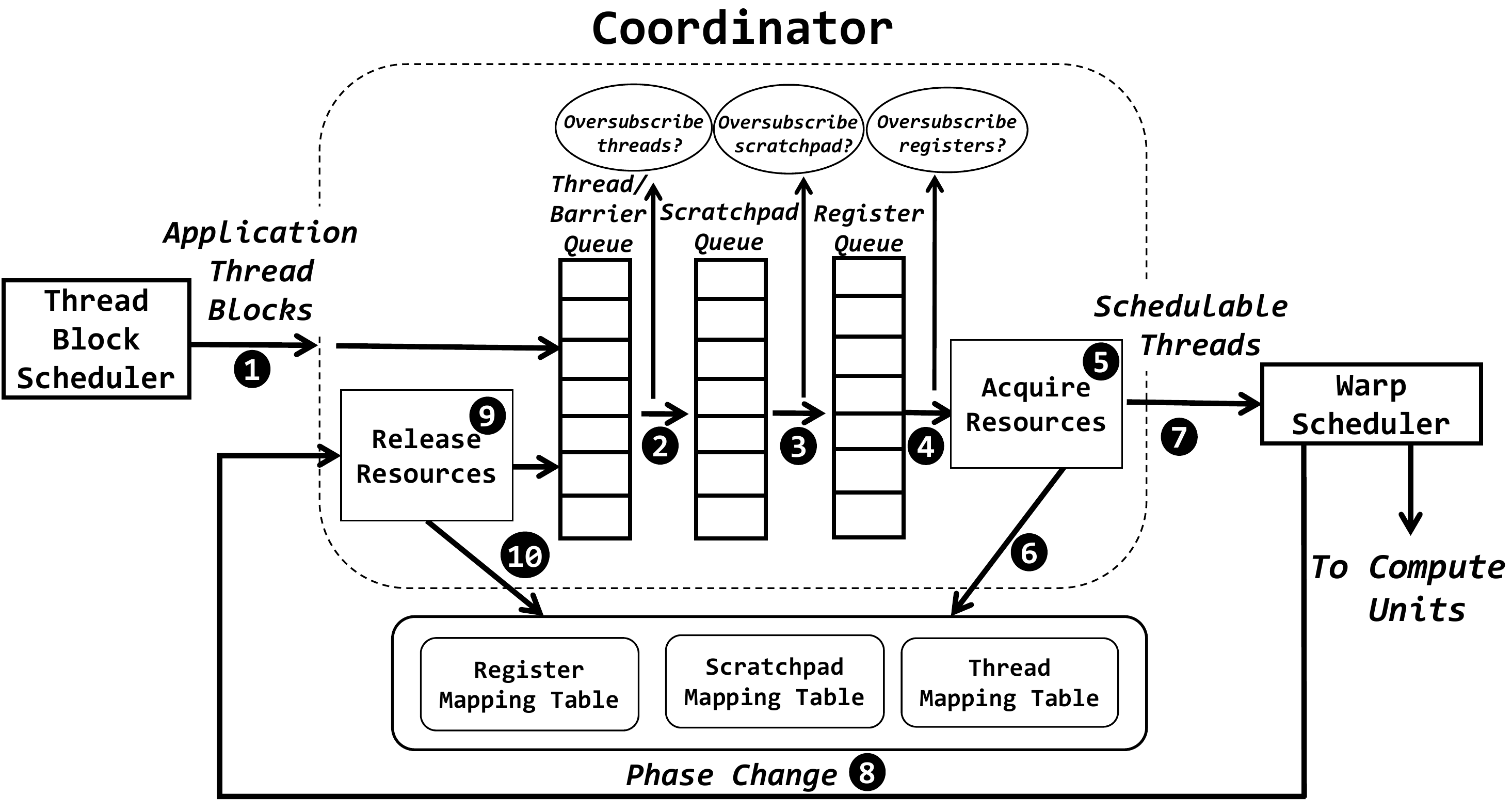}
 \caption{Overview of Zorua in hardware.} 
 \label{fig:full_overview}
\end{figure}

Figure~\ref{fig:full_overview} presents an overview of the hardware components
that are added to each SM. The coordinator interfaces with the thread block
scheduler (\ding{182}) to schedule new blocks onto an SM. It also
interfaces with the warp schedulers by providing a list of
\emph{schedulable warps} (\ding{188}).\footnote{We use an additional
 bit in each warp slots to indicate to the
 scheduler whether the warp is schedulable.} The resource mapping
tables are accessible by the coordinator and the compute
units. We present a detailed walkthrough of the operation of Zorua
and then discuss its individual components in more detail. \ignore{In
 addition, the coordinator interfaces with the on-chip resources
 using per-resource \emph{mapping tables} which are used to control
 the mapping of resources and their oversubscription at runtime
 (described in Section~\ref{sec:virtualizing_resources}). }

\subsubsection{Detailed Walkthrough}
The coordinator is called into action by three events: \One
a new thread block is scheduled at the SM for execution, \two a
warp undergoes a phase change, or \three a warp or a thread block reaches the end of execution. 
Between these
events, the coordinator performs no action and execution proceeds as
usual. We now walk through the sequence of actions performed by the
coordinator for each type of event.
\label{sec:walkthrough}


\textbf{Thread Block: Execution Start.} When a thread block is
scheduled onto an SM for execution (\ding{182}), the coordinator first
buffers it. 
The primary decision that the coordinator makes is to determine
whether or not to make each thread available to the scheduler for
execution. The granularity at which the coordinator makes decisions is
that of a warp, as threads are scheduled for execution at the
granularity of a warp (hence we use \emph{thread
  slot} and \emph{warp slot} interchangeably). Each warp requires
three resources: a thread slot, registers, and potentially
scratchpad. The amount of resources required is determined by the phase
specifier (Section~\ref{sec:phase_specifiers}) at the start of
execution, which is placed by the compiler into the code. The coordinator must
supply each warp with \emph{all} its required
resources in either the physical or swap space before presenting it to the warp scheduler for execution.

To ensure that each warp is furnished with its resources and
to coordinate potential oversubscription for each resource, the
coordinator has three queues{\textemdash}\emph{thread/barrier, scratchpad,
 and register queues}. The three queues together essentially house
the \emph{pending threads}. Each warp must traverse each queue
(\ding{183}~\ding{184}~\ding{185}), as described next, before becoming eligible to be
scheduled for execution. The coordinator allows a warp to traverse
a queue when \emph{(a)} it has enough of the corresponding resource
available in the physical space, or \emph{(b)} it has an insufficient
resources in the physical space, but has decided to oversubscribe and allocate
the resource in the swap space. The total size of the resource allocated in the physical
and swap spaces cannot exceed the determined virtual space size. The coordinator determines the availability of
resources in the physical space using the mapping tables
(see Section~\ref{sec:virtualizing_resources}). If there is an insufficient
amount of a resource in the physical space, the coordinator needs to decide
whether or not to increase the virtual space size for that particular resource by
oversubscribing and using swap space. We describe the
decision algorithm in Section~\ref{sec:oversubscription_decision}. If
the warp cannot traverse \emph{all} queues, it is left waiting in the first
(\emph{thread/barrier}) queue
until the next coordinator event. Once a warp has traversed \emph{all} the
queues, the coordinator acquires all the resources required for the warp's
execution (\ding{186}). The corresponding mapping tables for each
resource is updated (\ding{187}) to assign resources to the warp, as
described in Section~\ref{sec:virtualizing_resources}.

\textbf{Warp: Phase Change.} At each phase change (\ding{189}), the warp is removed
from the list of schedulable warps and is returned to the coordinator
to acquire/release its resources. Based on the information in its
phase specifier, the coordinator releases the
resources that are no longer live and hence are no longer required (\ding{190}).
The coordinator updates the mapping tables to free these resources
(\ding{191}). The warp is then placed into a specific queue, depending on which
live resources it retained from the previous phase and which new resources it
requires. The warp then
attempts to traverse the remaining queues
(\ding{183}~\ding{184}~\ding{185}), as described above. A warp that
undergoes a phase change as a result of a barrier instruction is
queued in the \emph{thread/barrier queue} (\ding{183}) until all warps
in the same thread block reach the barrier.
  
\textbf{Thread Block/Warp: Execution End.} When a warp completes
execution, it is returned to the coordinator to release any
resources it is holding. Scratchpad is released only when
the entire thread block completes execution. When the coordinator has
free warp slots for a new thread block, it requests the thread block
scheduler (\ding{182}) for a new block.

 
\textbf{Every Coordinator Event.} At any event, the coordinator
attempts to find resources for warps waiting at the queues, to enable them to
execute. Each warp in each queue (starting from the \emph{register queue}) is
checked for the availability of the required resources. If the coordinator is
able to allocate resources in the physical or swap space without exceeding the
determined size of virtual space, the warp is allowed to traverse the queue.
 
\subsubsection{Benefits of Our Design}
\textbf{Decoupling the Warp Scheduler and Mapping Tables from the Coordinator.} Decoupling the
warp scheduler from the coordinator enables Zorua to use any scheduling
algorithm over the schedulable warps to enhance performance. One case
when this is useful is when increasing parallelism degrades
performance by increasing cache miss
rate or causing memory
contention~\cite{tor-micro12,nmnl-pact13,Kayiran-micro2014}. Our decoupled
design allows this challenge to be addressed
independently from the coordinator using more intelligent scheduling
algorithms~\cite{tor-micro12,Kayiran-micro2014,largewarp, medic} and cache management
schemes~\cite{locality-li-ics15,coordinated-xie-hpca15,prioritybased-li-hpca15,medic}. Furthermore,
decoupling the mapping tables from the coordinator allows
easy integration of any implementation of the mapping tables that may
improve efficiency for each resource.

\ignore{
\textbf{Throttling Warps.} 
Even in cases where the overhead of oversubscription
is outweighed by the increase in parallelism that it offers, increasing
parallelism in some cases can lead to performance degradation by increasing
the cache miss rate in cache-sensitive workloads
~\cite{tor-micro12,nmnl-pact13}. 
The decoupling between the coordinator and the thread scheduler
offers a opportunity to address this challenge independent of the Zorua framework.
More intelligent scheduling algorithms~\cite{tor-micro12} and cache management schemes~\cite{x,y,z} can be used
along with Zorua to maximize the caching performance. 
}

\textbf{Coordinating Oversubscription for Multiple Resources.}
The queues help ensure that a warp is allocated
\emph{all} resources in the virtual space before execution. 
They \One ensure an ordering in resource allocation to avoid deadlocks, and
\two enforce priorities between resources. In our evaluated approach, we use
the following order of priorities: threads, scratchpad, and registers. We prioritize scratchpad over
registers, as scratchpad is shared by all warps in a block and hence has a higher
value by enabling more warps to execute. We prioritize threads over scratchpad,
as it is wasteful to allow warps stalled at a barrier to acquire other
resources{\textemdash}other
warps that are still progressing towards the barrier may be starved of the resource they
need. Furthermore, managing each resource independently allows
different oversubscription policies for each resource and enables fine-grained
control over the size of the virtual space for that resource.\ignore{Our
technical report~\cite{zorua-tr} has a more detailed discussion of these
benefits.} 

\textbf{Flexible Oversubscription.} Zorua's design can flexibly
enable/disable swap space usage, as the dynamic fine-grained
management of resources is independent of the swap space. Hence, in cases where the
application is well-tuned to utilize the available resources, 
swap space usage can be disabled or minimized, and Zorua can still improve
performance by reducing dynamic underutilization of resources. Furthermore, 
different oversubscription algorithms can be flexibly employed to manage the
size of the virtual space for each resource (independently or cooperatively).
These algorithms can be  
designed for different purposes, e.g., minimizing swap space usage, improving
fairness in a multikernel setting, reducing energy, etc. In
Section~\ref{sec:oversubscription_decision},
we describe an example algorithm to improve performance by making a
good tradeoff between improving parallelism and reducing swap space usage.

\textbf{Avoiding Deadlocks.} 
\highlight{A resource allocation deadlock could happen if resources are
distributed among too many threads, such that \emph{no} single thread is able to
obtain enough necessary resources for execution.}
Allocating resources using \emph{multiple} ordered queues
helps avoid deadlocks in resource allocation \highlight{in
three ways}. \highlight{First, new resources are allocated to a warp only once the warp has
traversed \emph{all} of the queues.} This ensures that resources are not wastefully
allocated to warps that will be stalled anyway. Second, \highlight{a warp is} allocated
resources based on how many resources it already has, i.e. how many queues it
has already traversed. Warps that \highlight{already hold} multiple live
resources are prioritized in allocating new resources over warps that \highlight{do
\emph{not}
hold} any resources. Finally, if there are insufficient resources to maintain a
minimal level of parallelism (e.g., 20\% of SM occupancy in our evaluation), the coordinator
handles this rare case by simply oversubscribing resources to ensure that there
is no deadlock in allocation. 

\textbf{Managing More Resources.} \highlight{Our} design also allows flexibly adding more
resources to be managed by the virtualization framework, for example, thread
block slots. Virtualizing a new resource with Zorua simply requires adding a new
queue to the coordinator and a new mapping table to manage the virtual to
physical mapping.

\subsubsection{Oversubscription Decisions}
\label{sec:oversubscription_decision}
\textbf{Leveraging Phase Specifiers.} Zorua leverages the information
provided by phase specifiers (Section~\ref{sec:phase_specifiers})
to make oversubscription decisions for each
phase. For each resource, the coordinator checks whether allocating
the requested quantity according to the phase specifier would cause
the total swap space to exceed an \emph{oversubscription
 threshold}, or \emph{o\_thresh}. This threshold essentially dynamically sets the size of
 the virtual space for each resource. The coordinator allows
oversubscription for each resource only within its threshold. \emph{o\_thresh}
is dynamically determined to
adapt to the characteristics of the workload, and tp ensure good performance by achieving a good tradeoff between the overhead of
oversubscription and the benefits gained from parallelism.

\textbf{Determining the Oversubscription Threshold.} In order to make
the above tradeoff, we use two architectural statistics: \One idle time at the
cores,
\emph{c\_idle}, as an indicator for potential performance
benefits from parallelism; and \two memory idle time (the idle cycles when
all threads are stalled waiting for data from memory or the memory
pipeline), \emph{c\_mem}, as an indicator of a saturated memory
subsystem that is unlikely to benefit from more parallelism.\footnote{This
is similar to the approach taken by prior work~\cite{nmnl-pact13} to
estimate the performance benefits of increasing parallelism.} We use
Algorithm~\ref{alg:determining_threshold} to determine
\emph{o\_thresh} at runtime. Every \emph{epoch}, the
change in \emph{c\_mem} is compared with the change in
\emph{c\_idle}. If the increase in \emph{c\_mem} is greater, this
indicates an increase in pressure on the memory subsystem, suggesting
both lesser benefit from parallelism and higher overhead from
oversubscription. In this case, we reduce \emph{o\_thresh}. On the
other hand, if the increase in \emph{c\_idle} is higher, this is
indicative of more idleness in the pipelines, and higher potential
performance from parallelism and oversubscription. We increase
\emph{o\_thresh} in this case, to allow more oversubscription and
enable more parallelism. Table~\ref{table:thresholds} describes the variables
used in Algorithm~\ref{alg:determining_threshold}.\ignore{We include more
detail on these variables in our technical report~\cite{zorua-tr}.}

\begin{algorithm} 
{
\begin{algorithmic}[1]
\State \emph{o\_thresh} $=$ \emph{o\_default} \Comment{Initialize threshold} 
\For{\emph{each epoch}}
\State \emph{c\_idle\_delta $=$ (c\_idle $-$ c\_idle\_prev)} \Comment{Determine the
change in c\_idle and c\_mem from the previous epoch}
\State \emph{c\_mem\_delta $=$ (c\_mem $-$ c\_mem\_prev)}
\If{\emph{(c\_idle\_delta $-$ c\_mem\_delta) > c\_delta\_thresh}}
\Comment{Indicates more idleness and potential for benefits from parallelism}
\State \emph{o\_thresh $+=$ o\_thresh\_step}
\EndIf
\If{\emph{(c\_mem\_delta $-$ c\_idle\_delta) > c\_delta\_thresh}}
\Comment{Traffic in memory is likely to outweigh any parallelism benefit}
\State \emph{o\_thresh $-=$ o\_thresh\_step}
\EndIf
\EndFor
\caption{\small{Determining the oversubscription threshold}}
\label{alg:determining_threshold}
\end{algorithmic} 
} 
\end{algorithm}

\begin{table}[h]
\centering 
\begin{tabular}{ll} \toprule
\textbf{Variable} & \textbf{Description} \\ \midrule 
\emph{o\_thresh} & oversubscription threshold (dynamically determined)
\\ \cmidrule(rl){1-2} 
\emph{o\_default} & initial value for \emph{o\_thresh}
, (experimentally determined \\ & to be 10\% of
total physical resource) 
\\ \cmidrule(rl){1-2} 
\emph{c\_idle} & core cycles when no threads are issued to the core \\
& (but the pipeline is not stalled)~\cite{nmnl-pact13} \\ \cmidrule(rl){1-2}
\emph{c\_mem} & core cycles when all warps are waiting for data \\ & from memory or
stalled at the memory pipeline \\ \cmidrule(rl){1-2}
\emph{*\_prev} & the above statistics for the previous epoch \\ \cmidrule(rl){1-2}
\emph{c\_delta\_thresh} & threshold to produce change in \emph{o\_thresh}
\\ & (experimentally determined to be 16) 
\\ \cmidrule(rl){1-2}
\emph{o\_thresh\_step} & increment/decrement to \emph{o\_thresh}
, experimentally
\\ & determined to be 4\% of the total physical resource
\\ \cmidrule(rl){1-2}
\emph{epoch} & interval in core cycles to change \emph{o\_thresh}\\
 & (experimentally determined to be 2048)\\ 
\bottomrule
\end{tabular} 
\caption{Variables for oversubscription}
\label{table:thresholds}	
\end{table}
 
\subsubsection{Virtualizing On-chip Resources}
\label{sec:virtualizing_resources}
A resource can be in either the physical space, in
which case it is mapped to the physical on-chip resource, or the swap
space, in which case it can be found in the
memory hierarchy. Thus, a resource is effectively virtualized, and we need
to track the mapping between the virtual and physical/swap spaces. We
use a \emph{mapping table} for each resource to determine \One whether the resource
is in the physical or swap space, and \two the location of the
resource within the physical on-chip hardware. The compute units
access these mapping tables before accessing the real
resources. An access to a resource that is mapped to the swap space is converted
to a global
memory access that is addressed by the logical resource ID and
warp/block ID (and a base register for the swap space of the resource). In addition to the mapping tables, we use two registers
per resource to track the amount of the resource that is \One
free to be used in physical space,\ignore{(indicating availability of
 resources on-chip)} and \two mapped in swap space.\ignore{(indicating the extent of
 oversubscription)} These two counters enable the coordinator to
make oversubscription decisions
(Section~\ref{sec:oversubscription_decision}). We now go into more
detail on virtualized resources in Zorua.\footnote{Our implementation of
a virtualized resource aims to minimize complexity. This implementation
is largely orthogonal to the
 framework itself, and one can envision other 
 implementations (e.g.,~\cite{virtual-register,shmem-multiplexing,virtual-thread}) for
 different resources.}

\begin{figure}[h] \centering
 	\begin{subfigure}[h]{0.49\linewidth} 
		\centering
 		\includegraphics[width=0.9\textwidth]{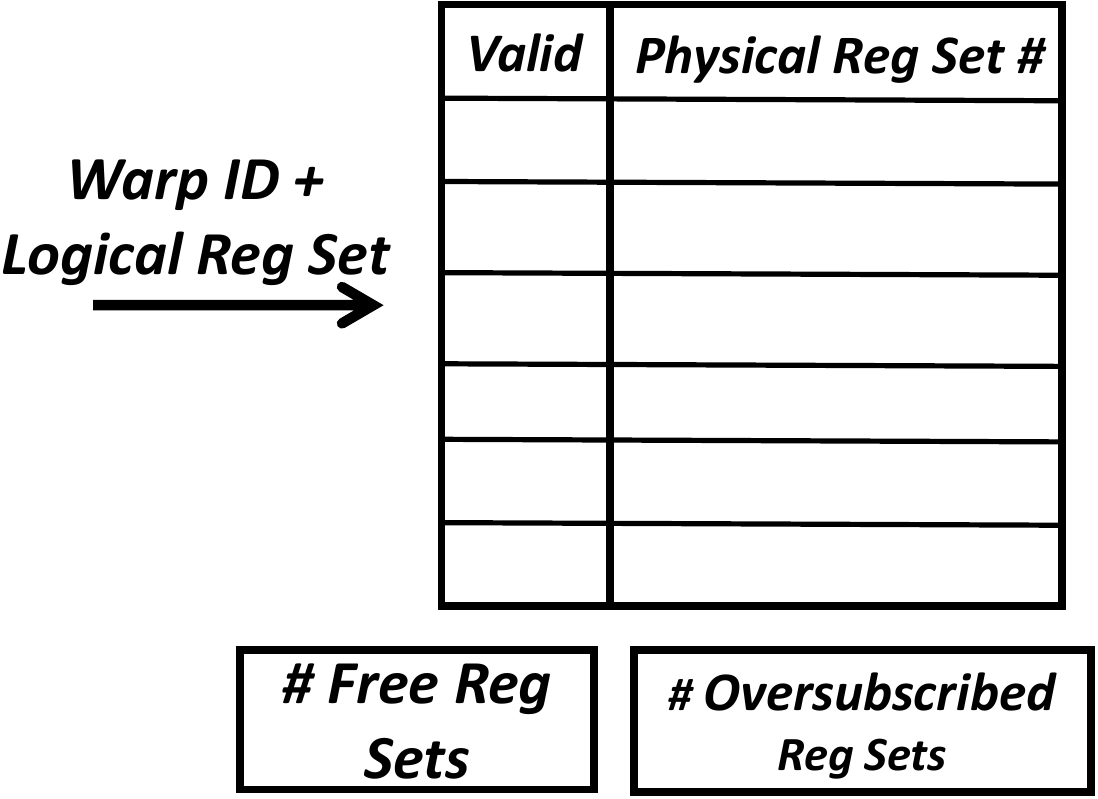}
 		\caption{Register Mapping Table}
	\end{subfigure}%
	\hfill
 	\begin{subfigure}[h]{0.49\linewidth} 
		\centering
 		\includegraphics[width=1\textwidth]{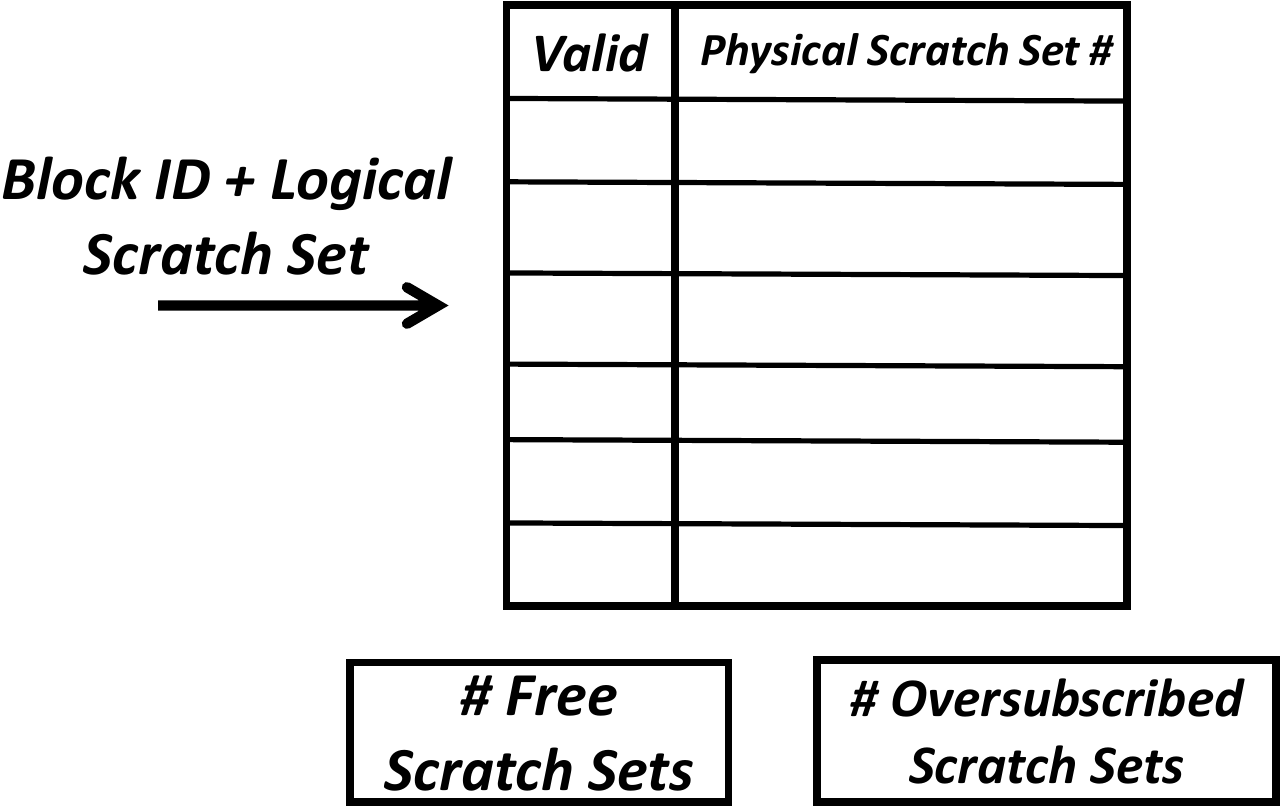}
 		\caption{Scratchpad Mapping Table}
	\end{subfigure}
	\caption{Mapping Tables}
 	\label{fig:mapping_table}
\end{figure}

\textbf{Virtualizing Registers and Scratchpad Memory.}
In order to minimize the overhead of large mapping tables, we map
registers and scratchpad at the granularity of a \emph{set}. The size
of a set is configurable by the architect{\textemdash}we use 4*\emph{warp\_size}\footnote{We
 track registers at the granularity of a warp.} for the register mapping table, and 1KB for
scratchpad\ignore{ in our experiments}. \ignore{The size of a set is a tradeoff
 between the size of the table and the granularity of
 allocation/deallocation and hence, maximizing
 utilization. } Figure~\ref{fig:mapping_table} depicts the
 tables for the registers and scratchpad. The register mapping table
is indexed by the warp ID and the logical register set number
(\emph{logical\_register\_number / register\_set\_size}). The
scratchpad mapping table is indexed by the block ID and the logical
scratchpad set number (\emph{logical\_scratchpad\_address /
 scratchpad\_set\_size}). Each entry in the mapping table contains
the physical address of the register/scratchpad content in the
physical register file or scratchpad. The valid bit indicates whether
the logical entry is mapped to the physical space or the swap
space.\ignore{(in which case it is converted into a global memory
 access).} With 64 logical warps and 16 logical thread blocks, the register mapping table
takes 1.125 KB ($64 \times 16 \times 9$ bits, or 0.87\% of the register
file) and the scratchpad mapping table takes 672 B ($16 \times 48 \times 7$
bits, or 1.3\% of the scratchpad).

\ignore{The
physical register file is 128KB in size which implies 256 registers
sets. The physical shared memory size is 48KB which implies 48 scratch
sets. For the register file, the total size of the table is \emph{64
 entries X (log2(256 register sets) + 1 valid bit) X 16 maximum
 register sets per thread = 1.12KB} ~= 0.87\% of the register
file. For the scratchpad table, the total size of the table is
\emph{16 entries X (log2(48 scratch sets) + 1 valid bit) X 48 sets per
 block = 672B} ~= 1.3\% of the scratchpad memory size. The mapping
table is used to determine the location of the register during the
operand collection stage of execution. \todo{Nandita}{Discuss latency
 overhead.}}

\textbf{Virtualizing Thread Slots.}
Each SM is provisioned with a fixed number of \emph{warp slots}, which
determine the number of warps that are considered for execution every
cycle by the warp scheduler. In order to oversubscribe warp slots, we
need to save the state of each warp in memory before remapping the
physical slot to another warp. This state includes the bookkeeping
required for execution, i.e., the warp's PC (program counter) and
the SIMT stack, which holds divergence information for each executing
warp.\ignore{The SIMT stack includes the next PC, active mask, and
 reconvergent PC which is a total of 384B with a maximum stack depth
 of 32~\cite{dynamic-warp,rhu2013dpe}. } The thread slot mapping
table records whether each warp is mapped to a physical slot or swap space.
The 
table is indexed by the logical warp ID, and stores the address of the
physical warp slot that contains the warp. \ignore{A valid bit
 indicates the whether the slot can be found in the physical space or
 in swap space. }In our baseline design with 64 logical warps, this
mapping table takes 56 B ($64 \times 7$ bits).

\ignore{In our
baseline design, there are 48 physical warp slots per SM, and we
support upto 64 warps in the virtual space. The total size of the
thread slot mapping table is \emph{64 entries X (log2(48) + 1 valid
 bit) = 56 bytes.}}

\subsubsection{Handling Resource Spills} If the coordinator has oversubscribed any
 resource, it is possible that the resource can be found either 
\highlight{\emph{(i)}~on-chip (in the
 physical space) or \emph{(ii)}~in the swap space in the memory hierarchy}. As described
 above, the location of any virtual resource is determined by the mapping
 table for each resource. \highlight{If the resource is found on-chip, the mapping
 table provides the physical location in the register file and scratchpad
 memory.} If the resource is in the swap space, the access to
 that resource is converted to a global memory load that is addressed \highlight{either} by the
 \highlight{\emph{(i)}~thread block ID and logical register/scratchpad set, in the case of  
registers or scratchpad memory; or \emph{(ii)} logical warp ID, in the case of warp slots}.
The oversubscribed resource is typically found in the L1/L2 cache but in the
worst case, could be in memory. 
When the coordinator chooses to oversubscribe any resource beyond what is
available on-chip, the least frequently accessed resource set is spilled to the
memory hierarchy \highlight{using a simple store operation}. 

\subsubsection{Supporting Phases and Phase Specifiers}
\label{sec:phase_specifiers}
\textbf{Identifying phases.} The compiler partitions each application
into phases based on the liveness of registers and scratchpad
memory. To avoid changing phases too often, the compiler uses
thresholds to determine phase boundaries. In our evaluation, we define
a new phase boundary when there is \One a 25\% change in the number of live
registers or live scratchpad content, and \two a minimum of 10
instructions since the last phase boundary. To simplify hardware
design, the compiler draws phase boundaries only where there is no control
divergence.\footnote{The phase boundaries for the applications in our pool
easily fit this restriction,
but the framework can be extended to support control divergence if needed.}


Once the compiler partitions the application into phases, it inserts
instructions{\textemdash}\emph{phase specifiers}{\textemdash}to specify the
beginning of each new phase and convey information to the
framework on the number of registers and scratchpad memory
required for each phase. As described in Section~\ref{sec:key_idea_phases}, a barrier or
a fence instruction also implies a phase change, but the compiler does
not insert a phase specifier for it as the resource requirement does
not change.

\begin{figure}[h]
 \centering
 \includegraphics[width=0.49\textwidth]{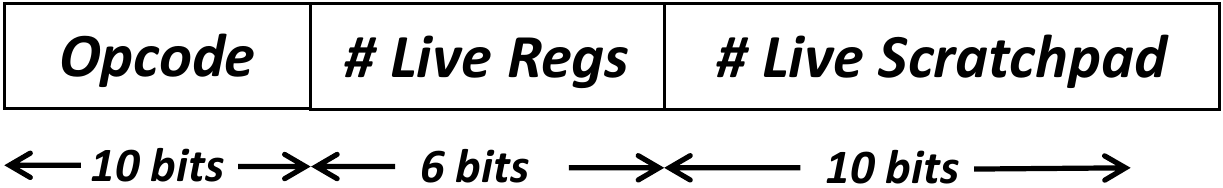}
 \caption{Phase Specifier}
 \label{fig:phase_specifier}
 \end{figure}

\textbf{Phase Specifiers.} The phase specifier instruction contains
fields to specify \One the number of live registers and
\two the amount of scratchpad memory in
bytes, both for the next phase. Figure~\ref{fig:phase_specifier} describes the
fields in the phase specifier instruction. The instruction decoder sends this information to the coordinator along with the phase
change event. The coordinator keeps this information in the
corresponding warp slot.

\subsubsection{Role of the Compiler and Programmer} 
 The compiler plays an important role, annotating the code with
 phase specifiers to convey information to the coordinator regarding
 the resource requirements of each phase. The compiler, however, does
 {\em not} alter the size of each thread block or the scratchpad memory
 usage of the program. The resource specification provided by the
 programmer (either manually or via auto-tuners) is retained to guarantee correctness. For registers, the
 compiler follows the default policy or uses directives as specified by the user. One could
 envision more powerful, efficient resource allocation with a
 programming model that does {\em not} require \emph{any} resource
 specification and/or compiler policies/auto-tuners that are \emph{cognizant} of the
 virtualized resources. 
 \subsubsection{Implications to the Programming Model and Software
 Optimization} Zorua offers several new opportunities and implications
 in enhancing the programming model and software optimizations (via libraries,
 autotuners, optimizing compilers, etc.) which we briefly
 describe below. \highlight{We leave these ideas for exploration in future work.}

\textbf{Flexible programming models for GPUs and heterogeneous
systems.}
State-of-the-art high-level programming languages and models still
assume a fixed amount of on-chip resources and hence, with the help of
the compiler or the runtime system, are required to find \emph{static} resource
specifications to fit the application to the desired GPU. Zorua, by
itself, also still requires the programmer to specify resource specifications to
ensure correctness{\textemdash}albeit they are not required to be highly optimized for a given
architecture. However, 
by providing a flexible but dynamically-controlled view of the on-chip hardware resources, Zorua changes the abstraction of the on-chip resources
that is offered to the programmer and software. This offers the opportunity to
rethink resource management in GPUs from the ground up. One could envision more powerful resource
allocation and better programmability with programming models that do
\emph{not} require static
resource specification, leaving the compiler/runtime system and the underlying
virtualized framework to completely handle \emph{all} forms of on-chip resource
allocation, unconstrained by the fixed physical resources in a specific
GPU, entirely at runtime. This is especially
significant in future systems that are likely to support a wide range of
compute engines and accelerators, making it important to be able to write
high-level code that can be partitioned easily, efficiently, and at a
fine granularity across any accelerator, \emph{without} statically tuning any code segment to run efficiently on the GPU.

\textbf{Virtualization-aware compilation and autotuning.} Zorua changes the
contract between the hardware and software to provide a more powerful
resource abstraction (in the software) that is
\emph{flexible and dynamic}, by pushing some more functionality into the hardware,
which can more easily react to the runtime resource requirements of the
\highlight{running} program. We can re-imagine compilers
 and autotuners to be more intelligent, leveraging this new abstraction and,
 hence the virtualization, to
 deliver more efficient
and high-performing code optimizations 
that are \emph{not} possible with the fixed and static
abstractions of today. They could, for example, \emph{leverage} the
oversubscription and dynamic management that Zorua provides to tune the code to
more aggressively use resources that are underutilized at runtime. As we
demonstrate in this work, static optimizations are limited by the fixed view of
the resources that is available to the program today. Compilation frameworks
that are cognizant of the \emph{dynamic} allocation/deallocation of resources
provided by Zorua could make more efficient use of the available resources. 

\textbf{Reduced optimization space.} \highlight{Programs written for applications
in machine learning, computer graphics, computer vision, etc., 
typically follow the \emph{stream} programming paradigm, where the code is
decomposed into
many \emph{stages} in an \emph{execution pipeline.} Each stage processes only a part
of the input data in a pipelined fashion to make better use of the caches. 
A key challenge in writing complex pipelined code is finding \emph{execution schedules}
(i.e., how the work should be partitioned across stages)
and optimizations that perform best for \emph{each} pipeline stage from a
prohibitively large space of potential solutions.} This requires complex tuning
algorithms or profiling runs that are both computationally intensive and
time-consuming. The search for optimized specifications has to be done when
there is a change in input data or in the underlying architecture. By pushing
some of the resource management
functionality to the hardware, Zorua reduces this search space for optimized
specifications by
making it less sensitive to the wide space of resource specifications.

\subsection{Evaluation}
\label{sec:eval}

We evaluate the effectiveness of Zorua by studying three different mechanisms: (i)~\emph{Baseline}, the
baseline GPU
that schedules kernels and manages resources at the thread block level;
(ii)~\emph{WLM} (Warp Level Management), a state-of-the-art mechanism for GPUs
to schedule kernels and manage registers at the warp level~\cite{warp-level-divergence}; and (iii)~\emph{Zorua}.
For our evaluations, we run each application on 8{\textendash}65 (36 on average) 
different resource specifications\ignore{Our technical
report~\cite{zorua-tr} provides the specifications.}
(the ranges are in
Table~\ref{table:applications}). 

\subsubsection{Effect on Performance Variation and Cliffs}
\label{sec:eval:var}

\begin{figure}[h]
 \centering 
 \includegraphics[width=0.69\textwidth]{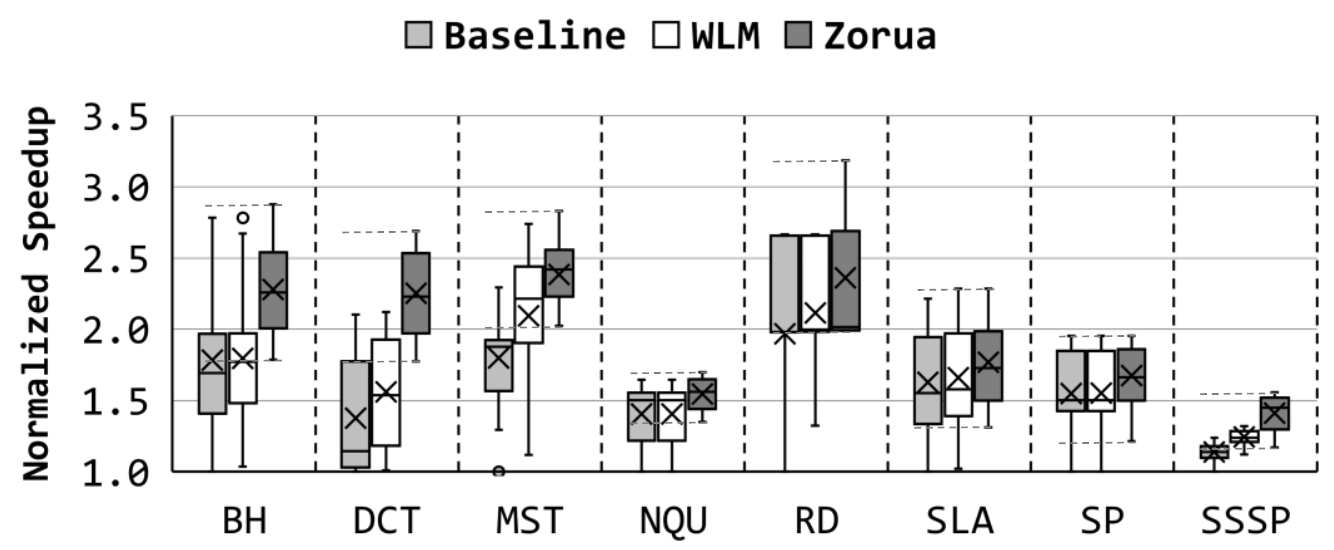}
 \caption{Normalized performance distribution.}
 \label{fig:performance_range}
\end{figure}

We first examine how Zorua alleviates the high variation in performance by reducing
the impact of resource specifications on resource utilization.
Figure~\ref{fig:performance_range} presents a Tukey 
box plot~\cite{mcgill1978variations} 
(see Section~\ref{sec:motivation} for a 
description of the presented box plot),
illustrating the performance distribution (higher is better) for each
application (for all different application resource specifications we
evaluated), normalized to the slowest Baseline operating point \emph{for that
application}. 
We make two major
observations.

\ignore{The boxes in the box plot represent the range between the first
quartile (25\%) and the third quartile (75\%). The whiskers extending from the
boxes represent the maximum and minimum points of the distribution, or 1.5X the
length of the box, whichever is smaller. Any points that lie more than 1.5X 
the box length beyond the box are considered to be
outliers~\cite{mcgill1978variations}, and are plotted as
individual points. The line in the middle of the box represents the median,
while the ``X'' represents the average.}

First, we find that Zorua significantly reduces the \emph{performance range} across
all evaluated resource specifications. Averaged across all of our applications, the 
worst resource specification for Baseline achieves 96.6\% lower performance 
than the best performing resource specification. For WLM~\cite{warp-level-divergence},
this performance range reduces only slightly, to
88.3\%. With Zorua, the performance range drops significantly, to 48.2\%.
We see
drops in the performance range for \emph{all} applications except \emph{SSSP}. With \emph{SSSP}, the
range is already small to begin with (23.8\% in Baseline), and Zorua
exploits the dynamic underutilization, which improves performance but also adds a small amount of variation.

Second, while Zorua reduces the performance range, it also preserves or improves performance
of the best performing points. As we examine in more detail in
Section~\ref{sec:eval:perf}, the reduction in performance range occurs as
a result of improved performance mainly at the lower end of the distribution.

\begin{figure}[h]
 \centering
 \begin{subfigure}[t]{0.335\linewidth}
 \centering
 \includegraphics[width=1.0\textwidth]{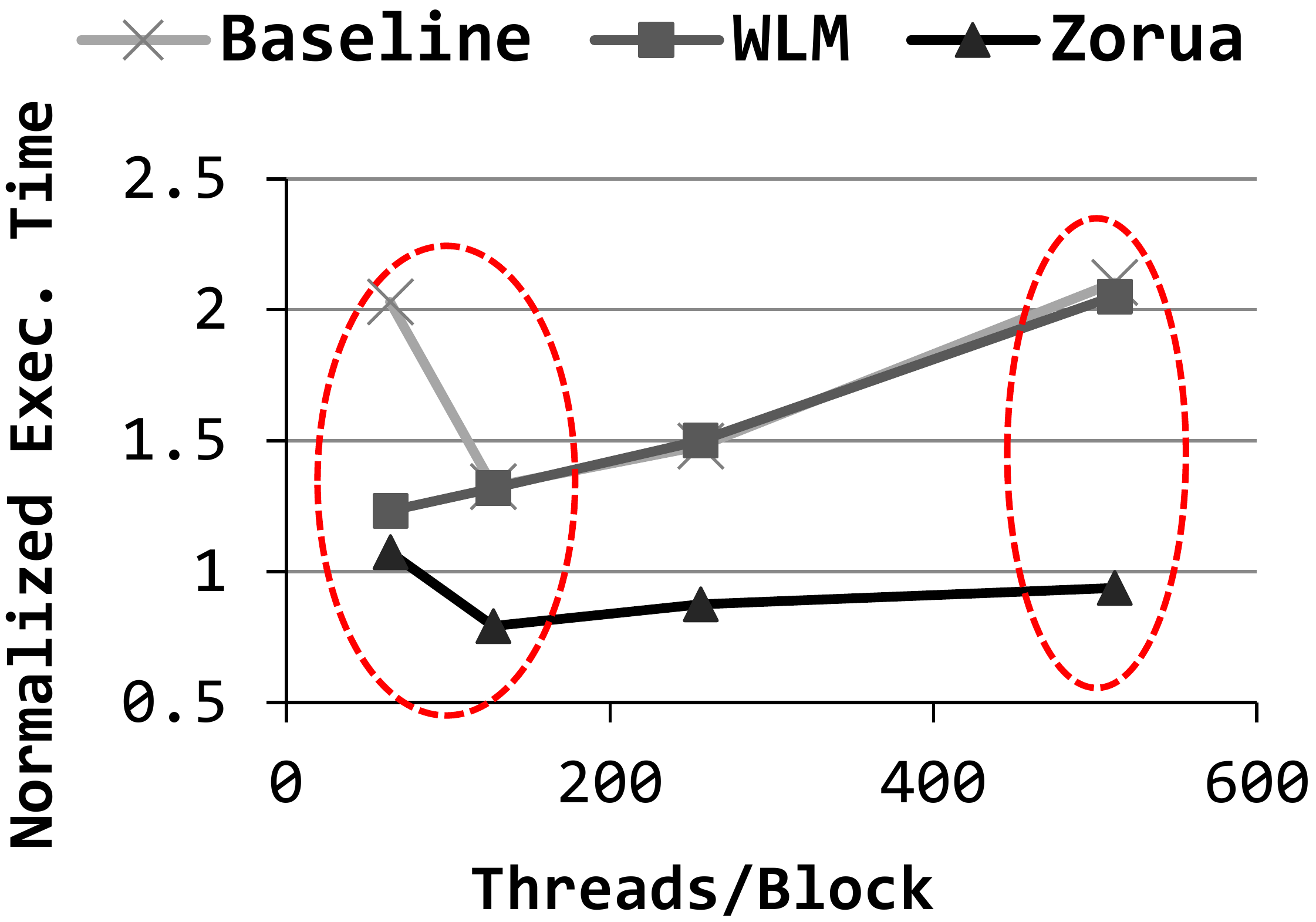}
 \caption{\emph{DCT}} 
 \label{fig:performance_cliff_result_dct}
 \end{subfigure}
 \begin{subfigure}[t]{0.318\linewidth}
 \centering
 \includegraphics[width=1.0\textwidth]{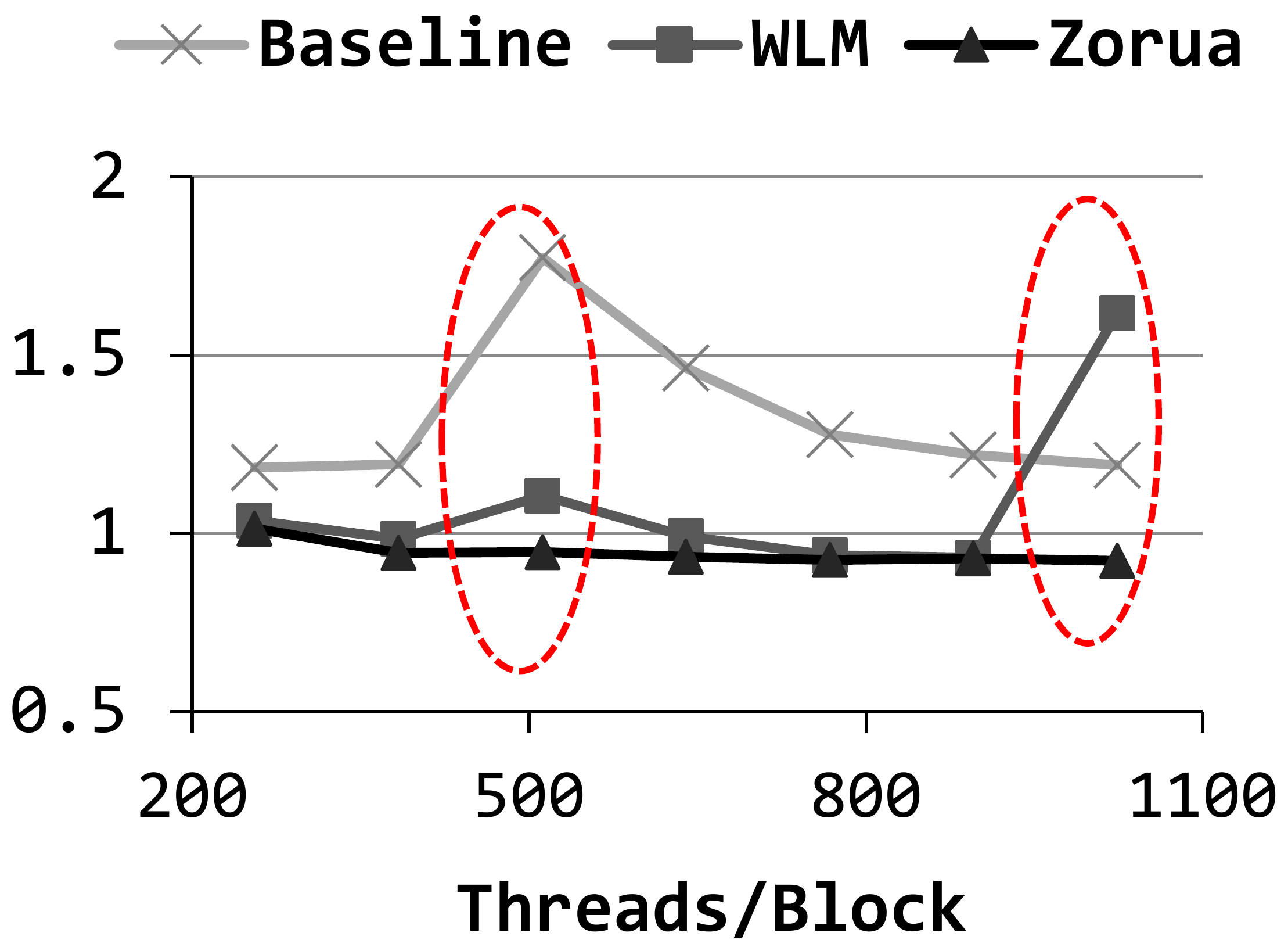} 
 \caption{\emph{MST}}
 \label{fig:performance_cliff_result_mst}
 \end{subfigure}
 \begin{subfigure}[t]{0.316\linewidth}
 \centering
 \includegraphics[width=1.0\textwidth]{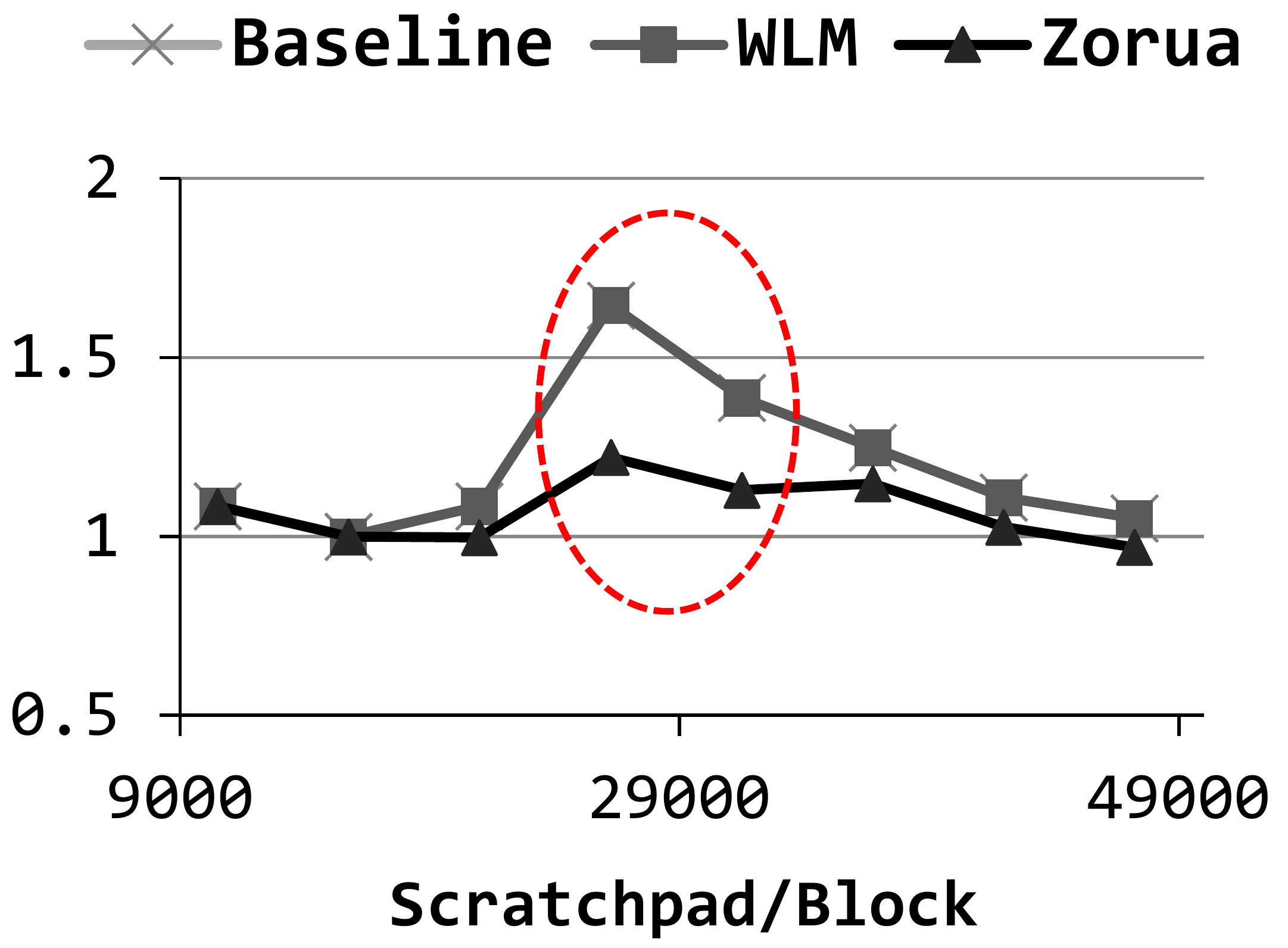}
 \caption{\emph{NQU}}
 \label{fig:performance_cliff_result_nqu}
 \end{subfigure}
 \caption{Effect on performance cliffs.}
 \label{fig:performance_cliff_result}
\end{figure}

To gain insight into how Zorua reduces the performance range and improves
performance for the worst performing points, we analyze how it reduces
performance cliffs. With Zorua, we ideally want to \emph{eliminate} the cliffs we 
observed in Section~\ref{sec:motivation:cliffs}. We study the tradeoff 
between resource specification and execution time for three representative 
applications: 
\emph{DCT} (Figure~\ref{fig:performance_cliff_result_dct}), \emph{MST}
(Figure~\ref{fig:performance_cliff_result_mst}), and
\emph{NQU} (Figure~\ref{fig:performance_cliff_result_nqu}).
For all three figures, we normalize execution time to the \emph{best} execution time
under Baseline. Two observations are in order.


First, Zorua successfully mitigates the performance cliffs that occur
in Baseline. For example, \emph{DCT} and \emph{MST} are both sensitive to the thread block size, as shown in 
Figures~\ref{fig:performance_cliff_result_dct}
and~\ref{fig:performance_cliff_result_mst}, respectively. We have circled the
locations at which cliffs exist in Baseline. Unlike Baseline, Zorua
maintains more steady execution times across the  
number of threads per block, 
employing 
oversubscription to overcome the loss in parallelism due to insufficient on-chip
resources. We see
similar results across all of our applications.

Second, we observe that while WLM~\cite{warp-level-divergence} can reduce some of the cliffs by mitigating
the impact of large block sizes, many cliffs still
exist under WLM (e.g., \emph{NQU} in Figure~\ref{fig:performance_cliff_result_nqu}).
This cliff in \emph{NQU} occurs as a result of insufficient scratchpad memory, which
cannot be handled by warp-level management. Similarly, the cliffs for \emph{MST}
(Figure~\ref{fig:performance_cliff_result_mst}) also persist with WLM because \emph{MST}
has a lot of barrier operations, and the additional warps scheduled by WLM
ultimately stall, waiting for other warps within the same block
to acquire resources. We find that, with oversubscription, Zorua is able to 
smooth out those cliffs that WLM
is unable to eliminate.

Overall, we conclude that Zorua \One reduces the performance variation across
resource specification points, so that performance depends less on the
specification provided by the programmer; and \two can alleviate the
performance cliffs experienced by GPU applications.

\subsubsection{Effect on Performance}
\label{sec:eval:perf}

As Figure~\ref{fig:performance_range} shows,
Zorua either retains or improves the best performing point for each application,
compared to the Baseline. Zorua improves the best performing point for each
application by 12.8\% on average, and by as much as 27.8\%
(for \emph{DCT}). This improvement comes from the improved parallelism obtained by exploiting the
dynamic underutilization of resources, which exists \emph{even for optimized
specifications}. Applications such as \emph{SP} and \emph{SLA} have little dynamic
underutilization, and hence do not show any performance improvement.
\emph{NQU} \emph{does}
have significant dynamic underutilization, but Zorua does not improve the best
performing point as the overhead of oversubscription outweighs the benefit,
and Zorua dynamically chooses not to oversubscribe. We conclude that even for
many specifications that are optimized to fit the hardware resources, 
Zorua is able to further improve performance. 

We also note that, in addition to reducing performance variation and
improving performance for optimized points, Zorua improves performance
by 25.2\% on average for all resource specifications across all evaluated
applications.

\ignore{We present a deeper look into
these results in Section~\ref{sec:deeper_look}.}

\subsubsection{Effect on Portability}
\label{sec:eval:port}

As we describe in Section~\ref{sec:motivation:port},
performance cliffs often behave differently across different GPU architectures, and can
significantly shift the best performing resource specification point.
We study how Zorua can ease the burden of performance tuning if an application 
has been already tuned for one GPU model, and is later ported to another GPU. 
To understand this, we define a new
metric, \emph{porting performance loss}, that quantifies the performance impact
of porting an application without re-tuning it. To calculate this, we first
normalize the execution time of each specification point to the execution time
of the best performing specification point. We then pick a
source GPU architecture (i.e., the architecture that the GPU was tuned for) and
a target GPU architecture (i.e., the architecture that the code will run on),
and find the point-to-point drop in performance for all points whose
performance on the source GPU comes within 5\% of the performance at the
best performing specification point.\footnote{We include any point within 5\%
of the best performance as there are often multiple points close to the
best point, and the programmer may choose any of them.}

\begin{figure}[h]
 \centering 
 \includegraphics[width=0.88\textwidth]{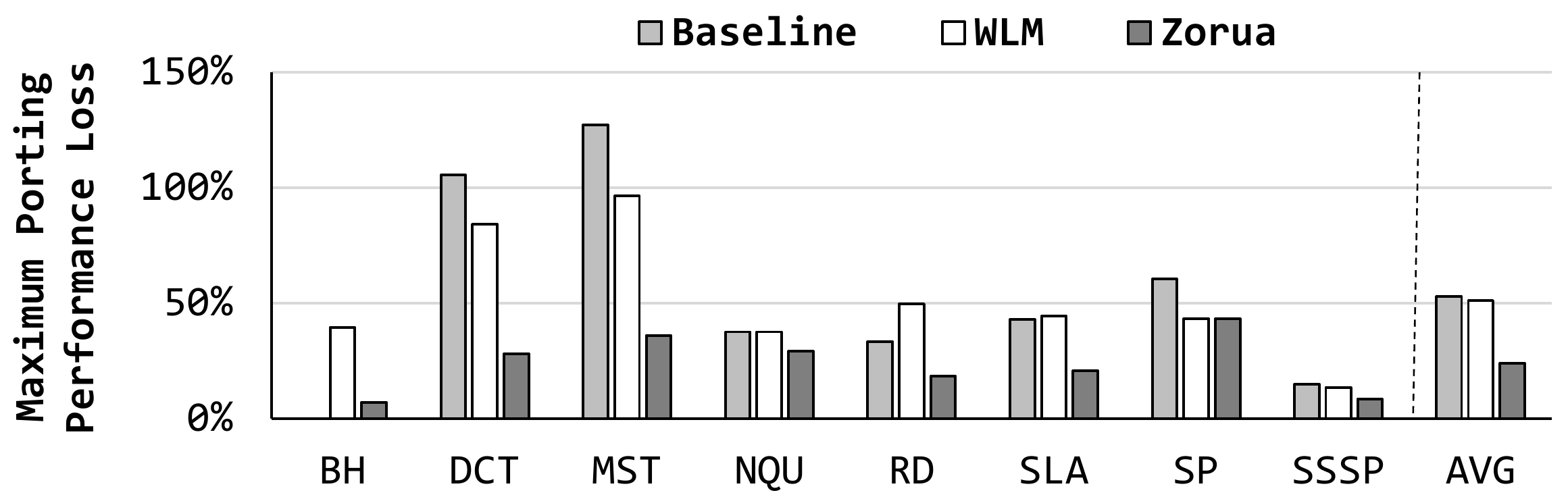}
 \caption{Maximum porting performance loss.}
 \label{fig:portability_result_overall}
\end{figure}

Figure~\ref{fig:portability_result_overall} shows the \emph{maximum} porting
performance loss for each application, across any two pairings of our three
simulated GPU architectures (Fermi, Kepler, and Maxwell). We find that Zorua greatly 
reduces the maximum porting performance loss that occurs under both Baseline
and WLM for all but one of our applications. On average, the maximum porting performance 
loss is 52.7\% for Baseline, 51.0\% for WLM, and only 23.9\% for Zorua.

Notably, Zorua delivers significant improvements in portability for applications 
that previously suffered greatly when ported to another GPU, such as \emph{DCT} and 
\emph{MST}. For both of these applications, the performance variation differs so much
between GPU architectures that, despite tuning the application on the source
GPU to be within 5\% of the best achievable performance, their performance on the target
GPU is often more than twice as slow as the best achievable performance on the target 
platform. Zorua significantly lowers this porting performance loss down to 28.1\% for \emph{DCT} and 36.1\% for \emph{MST}. We also observe
that for \emph{BH}, Zorua
actually increases the porting performance loss slightly with respect to the
Baseline. This is because for Baseline, there are only two points that perform within the 
5\% margin for our metric, whereas with Zorua, we have five points that fall in
that range. Despite this, the increase in porting performance loss for \emph{BH} is low, deviating
only 7.0\% from the best performance.

\begin{figure}[h]
 \centering 
 \includegraphics[width=0.88\textwidth]{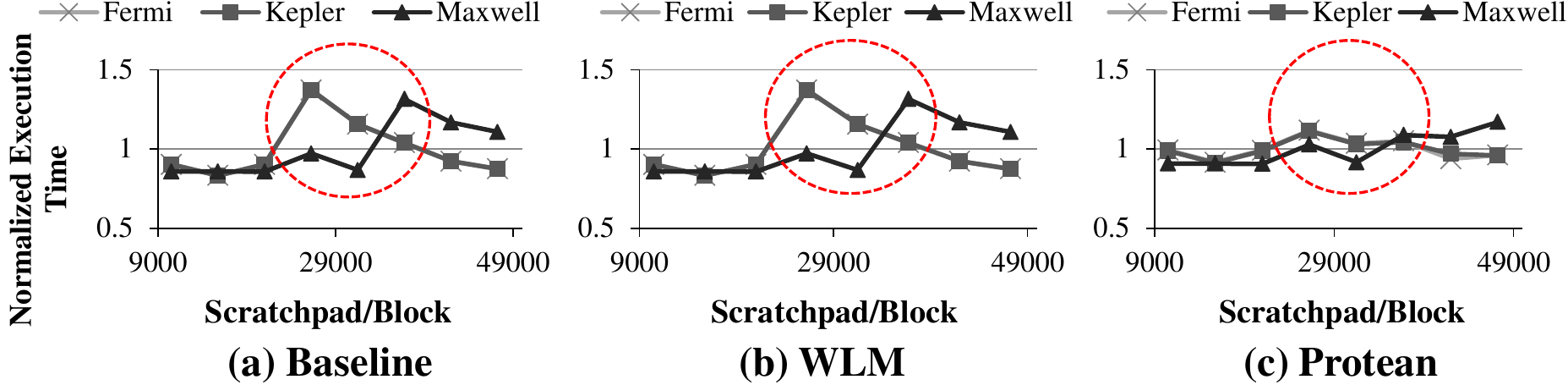}
 \caption{Impact on portability (NQU).}
 \label{fig:portability_nqu}
\end{figure}

\begin{figure}[h]
 \centering 
 \includegraphics[width=0.88\textwidth]{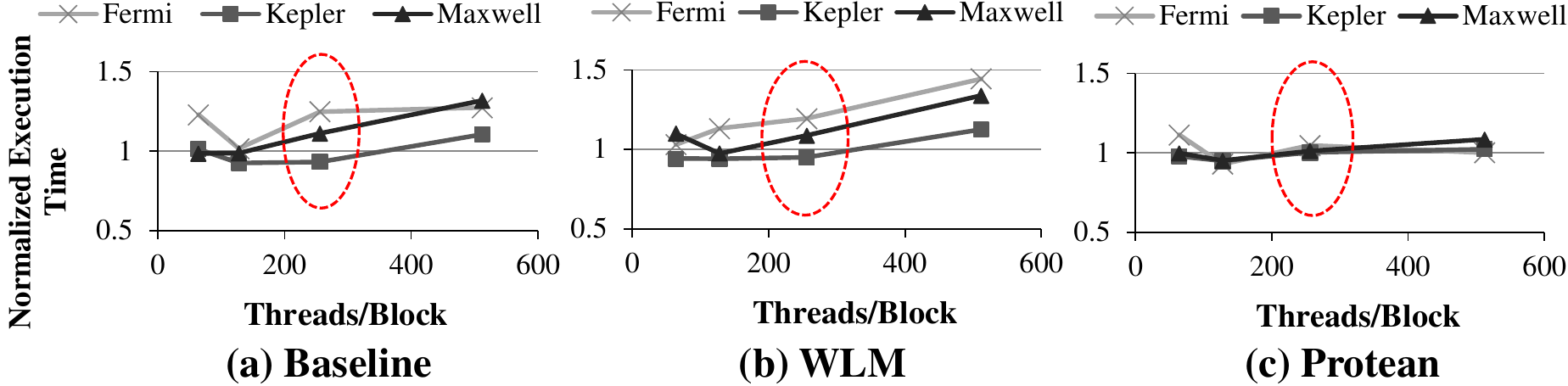}
 \caption{Impact on portability (DCT)}
 \label{fig:portability_dct}
\end{figure}

To take a closer look into the portability benefits of Zorua, we run experiments
to obtain the performance sensitivity curves for each application using
different GPU architectures. Figures~\ref{fig:portability_nqu} and \ref{fig:portability_dct} depict the
execution time curves while sweeping a single resource specification for \emph{NQU} and
\emph{DCT} for the three evaluated GPU architectures -- Fermi, Kepler, and
Maxwell. We make two major observations from the figures. 

First, Zorua significantly alleviates the presence of performance cliffs and
reduces the
performance variation across \emph{all} three evaluated architectures, thereby
reducing the impact of both resource specification and underlying architecture
on the resulting performance curve. In comparison, WLM is unable to make a
significant impact on the performance variations and the cliffs remain for all
the evaluated architectures.  

Second, by reducing the performance variation across all three GPU generations,
Zorua significantly reduces the \highlight{\emph{porting performance loss}}, i.e., the loss in
performance when code optimized for one GPU generation is run on another (as
highlighted within the figures). 

We conclude that Zorua enhances portability of applications by reducing the
impact of a change in the hardware resources for a given resource
specification. For applications that have 
already been tuned on one platform, Zorua significantly lowers the penalty of not 
re-tuning for another platform, allowing programmers to save development time.

\subsubsection{A Deeper Look: Benefits \& Overheads}
\label{sec:deeper_look}
\begin{figure}[h]
 \centering 
 \includegraphics[width=0.67\textwidth]{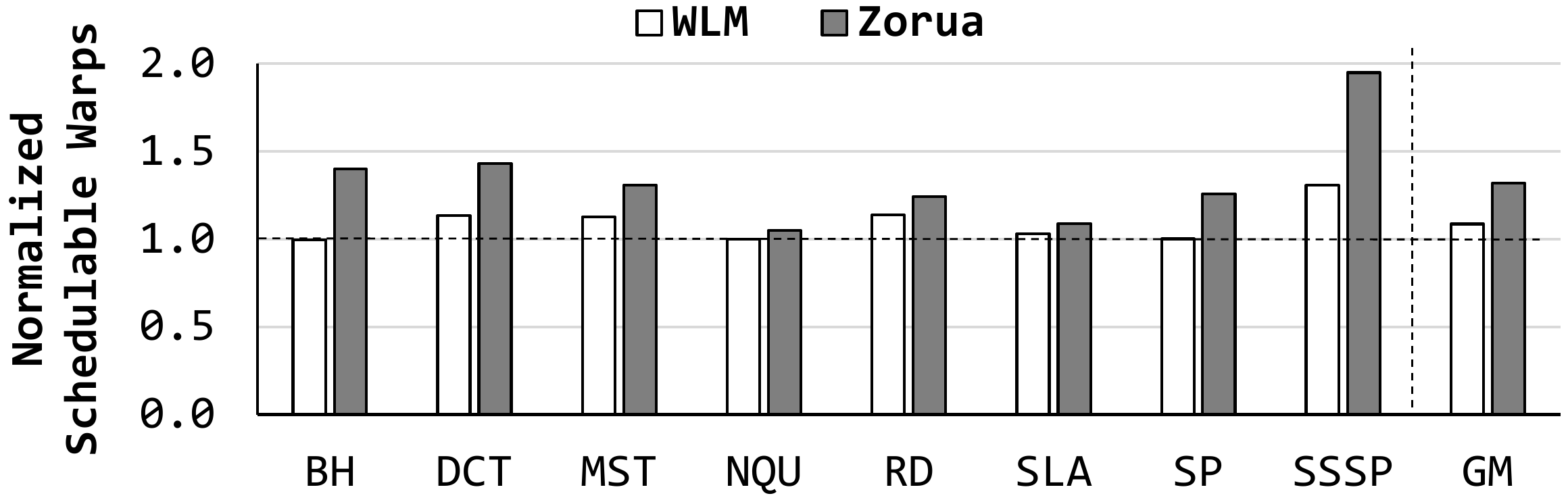}
 \caption{Effect on schedulable warps.}
 \label{fig:active_warps}
\end{figure}

To take a deeper look into how Zorua is able to provide the above benefits, 
in Figure~\ref{fig:active_warps}, we show the 
number of \emph{schedulable warps} (i.e., warps that are available to be
scheduled by the warp scheduler at any given time excluding warps
waiting at a barrier), averaged across all of specification points. 
On average, Zorua increases the number of
schedulable 
warps by 32.8\%, significantly more than WLM (8.1\%), which is constrained by the
fixed amount of available resources. We conclude that by oversubscribing and
dynamically managing resources, Zorua is able to improve thread-level parallelism, and hence performance. 

\begin{figure}[h]
 \centering 
 \includegraphics[width=0.67\textwidth]{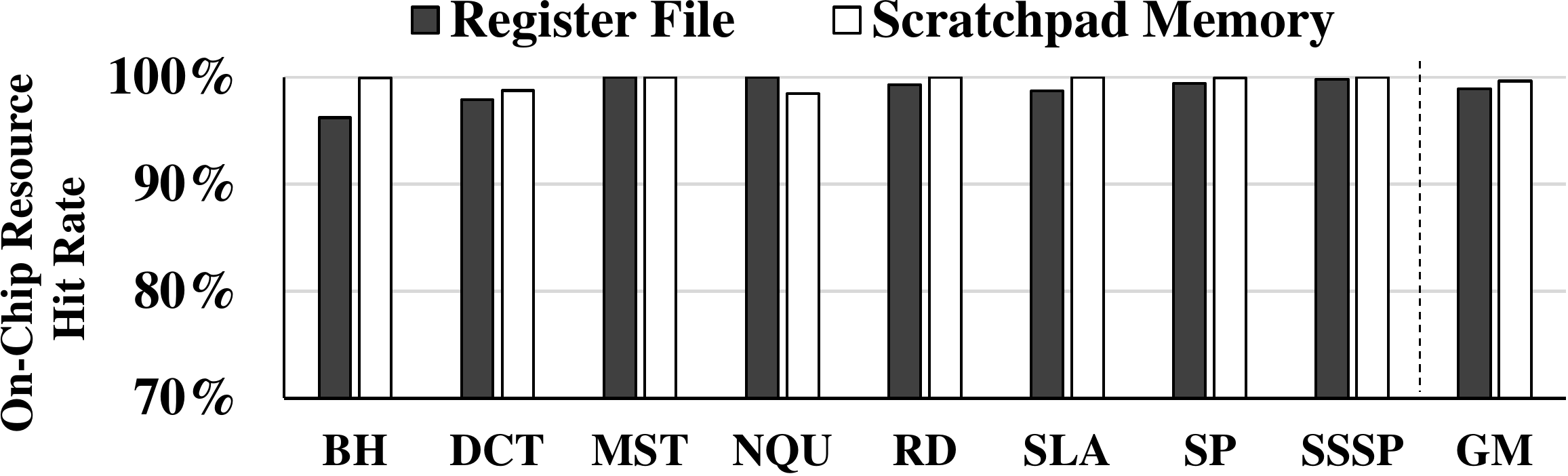}
 \caption{Virtual resource hit rate in Zorua}
 \label{fig:resource_hit_rate}
\end{figure}

We also find that the overheads due to resource swapping and
contention do not significantly impact the performance of Zorua.
Figure~\ref{fig:resource_hit_rate} depicts resource hit rates for each
application, i.e., the fraction of all resource accesses that were found on-chip
as opposed to making a potentially expensive off-chip access. The
oversubscription mechanism (directed by the coordinator) is able to keep resource hit rates
very high, with an average hit rate of 98.9\% for the
register file and 99.6\% for scratchpad memory.

\begin{figure}[h]
 \centering 
 \includegraphics[width=0.67\textwidth]{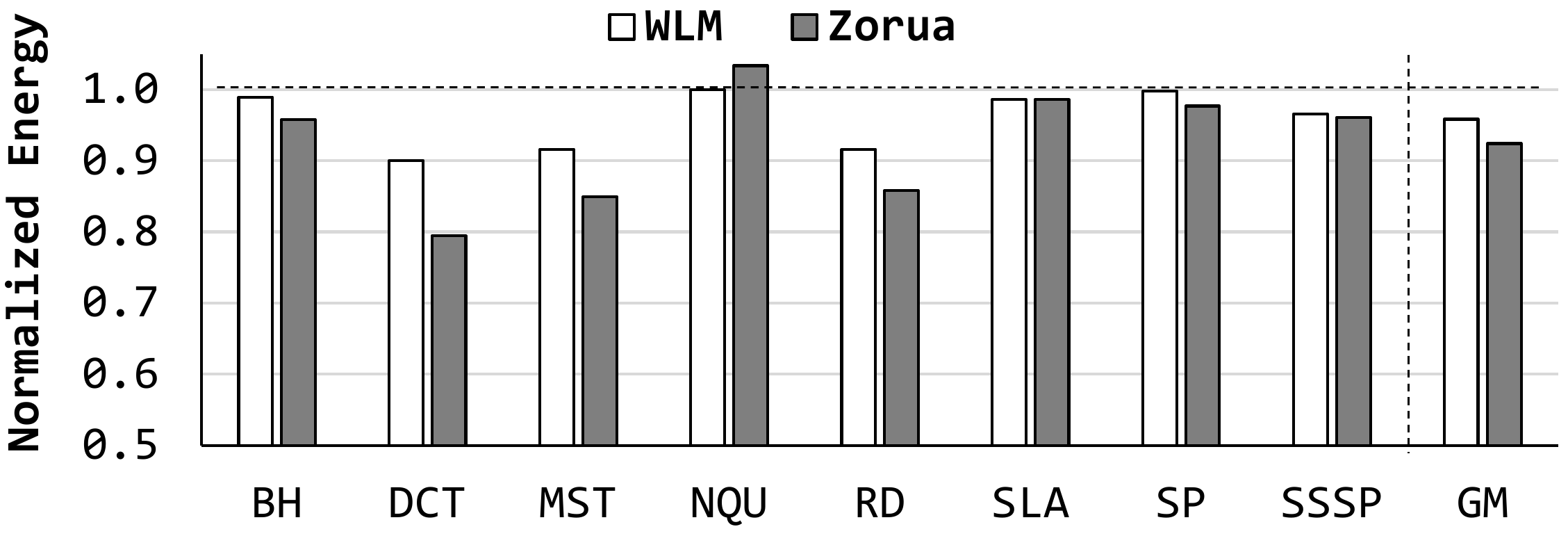}
 \caption{Effect on energy consumption.}
 \label{fig:energy}
\end{figure}

Figure~\ref{fig:energy} shows the average reduction in total system energy consumption
of WLM and Zorua over Baseline for each application (averaged across the 
individual energy consumption over Baseline for each evaluated specification
point). We observe that Zorua
reduces the total energy consumption across all of our applications, except for 
\emph{NQU} (which has a small increase of 3\%). Overall, Zorua provides a 
mean energy reduction of 7.6\%, up to 20.5\% for \emph{DCT}.\footnote{We note that the energy consumption can be
reduced further by appropriately optimizing the oversubscription algorithm. We
leave this exploration to future work.} We conclude that Zorua is an energy-efficient virtualization framework for GPUs.


We estimate the die area overhead of Zorua with CACTI
6.5~\cite{wilton1996cacti}, using the same
40nm process node as the GTX 480 \ignore{(Fermi)}, which our system closely models. We
include all the overheads from the coordinator and the resource
mapping tables (Section \ref{sec:mechanism}). The total
area overhead is 0.735 $mm^2$ for all 15 SMs, which is only
0.134\% of the die area of the GTX 480.

\subsection{Other Applications}
\label{sec:applications}
By providing the illusion of more resources than physically available, Zorua provides the
opportunity to help address other important challenges in GPU computing today.
We discuss several such opportunities in this section.
\subsubsection{Resource Sharing in Multi-Kernel or
Multi-Programmed Environments}
Executing multiple kernels or applications within the same SM can improve
resource utilization and
efficiency~\cite{asplos-sree,simultaneous-sharing,fine-grain-hotpar,kernelet,app-aware-GPGPU-2014,mosaic,mask,rachata-isca}. 
Hence, providing support to enable fine-grained sharing and partitioning
of resources is critical for future GPU systems. 
This is especially true in environments where multiple different applications
may be consolidated on the same GPU, e.g. in clouds or clusters. 
By providing a flexible view of each of the resources, Zorua provides a natural way
to enable dynamic and fine-grained control over resource partitioning and
allocation among multiple kernels. Specifically, 
Zorua provides several key benefits
for enabling better performance and efficiency in multi-kernel/multi-program
environments. First, selecting the optimal resource specification for an
application is challenging in virtualized environments (e.g., clouds), as it is unclear which
other applications may be running alongside it. Zorua can improve efficiency in resource
utilization \emph{irrespective} of the application specifications and of other kernels
that may be executing on the same SM. Second, Zorua manages the different
resources independently and at a fine granularity, using a dynamic runtime system
(the coordinator). This enables the maximization of
resource utilization, while providing the ability to control the partitioning of
resources at runtime to provide QoS, fairness, etc., by leveraging the coordinator. 
Third, Zorua enables oversubscription of the different resources. This obviates the
need to alter the application specifications~\cite{asplos-sree,kernelet} in
order to ensure there are sufficient resources to co-schedule kernels on the
same SM, and hence enables concurrent kernel execution
transparently to the programmer. 

\subsubsection{Preemptive Multitasking}
A key challenge in enabling true multiprogramming in GPUs is
enabling rapid preemption of
kernels~\cite{isca-2014-preemptive,simultaneous-sharing,chimera}. Context
switching on GPUs incurs a very high latency and overhead, as a result of the large amount of
register file and scratchpad state that needs to be saved 
before a new kernel can be executed. Saving state at a very coarse
granularity (e.g., the entire SM state) leads to very high preemption
latencies. Prior work proposes context minimization~\cite{chimera,igpu}
or context switching at the granularity of a thread
block~\cite{simultaneous-sharing} to improve response time during preemption. Zorua
enables fine-grained management and oversubscription of on-chip resources. It can be naturally extended to enable quick
preemption of a task via intelligent management of the swap space and the
mapping tables (complementary to approaches taken by prior
work~\cite{chimera,igpu}).
\subsubsection{Support for Other Parallel Programming Paradigms}
The fixed static resource allocation for each thread in
modern GPU architectures requires statically dictating the
resource usage for the program throughout its execution. Other forms
of parallel execution that are \emph{dynamic} (e.g., Cilk~\cite{cilk},
staged execution~\cite{marshaling,bis, joao.isca13}) require more flexible allocation
of resources at runtime, and are hence more challenging to enable.
Examples of this include \emph{nested parallelism}~\cite{nested}, where a kernel can dynamically spawn new kernels or
thread blocks, and \emph{helper threads}~\cite{caba} to utilize idle resource at runtime to
perform different optimizations or background tasks in parallel. Zorua makes it
easy to enable these paradigms by providing on-demand dynamic allocation of
resources. Irrespective of whether threads in the programming model are created
statically or dynamically, Zorua allows allocation of the required resources on
the fly to support the execution of these threads. The resources are simply
deallocated when they are no longer required. Zorua also enables
\emph{heterogeneous}
allocation of resources -- i.e., allocating different amounts of resources to
different threads. The current resource allocation model, in line with a GPU's
SIMT architecture, treats all threads the
same and allocates the same amount of resources. Zorua makes it easier to
support execution paradigms where each concurrently-running thread
executes different code at the same time, hence requiring different
resources. This includes helper
threads, multiprogrammed execution, nested parallelism, etc. Hence, with Zorua, applications are no
longer limited by a GPU's fixed SIMT model which only supports a fixed,
statically-determined number of homogeneous threads as a result of the resource
management mechanisms that exist today. 

\subsubsection{Energy Efficiency and Scalability}
To support massive parallelism, on-chip resources are a precious
and critical resource. However, these resources \emph{cannot} grow arbitrarily large as
GPUs continue to be area-limited and on-chip memory tends to be extremely power
hungry and area
intensive~\cite{energy-register,virtual-register,compiler-register,warped-register,virtual-thread,
ltrf-sadrosadati-asplos18}.
Furthermore, complex thread schedulers that can select a thread for execution
from an increasingly large thread pool are required in order to support an
arbitrarily large number of warp slots. Zorua enables using smaller
register files, scratchpad memory and less complex or fewer thread schedulers to
save power and area while still retaining or improving parallelism.

\subsubsection{Error Tolerance and Reliability}
The indirection offered by Zorua, along with the dynamic
management of resources, could also enable better reliability and simpler
solutions towards error tolerance in the on-chip resources. The
virtualization framework trivially allows remapping resources with hard or soft
faults such that no virtual resource is mapped to a faulty physical resource.
Unlike in the baseline case, faulty resources would not impact the
number of the resources
seen by the thread scheduler while scheduling threads for execution. A few
unavailable faulty registers,
warp slots, etc., could significantly reduce the number of the threads that
are scheduled concurrently (i.e., the runtime parallelism).

\subsubsection{Support for System-Level Tasks on GPUs}
As GPUs become increasingly general purpose, a key requirement is better
integration with the CPU operating system, and with complex distributed software
systems such as those employed for large-scale distributed machine
learning~\cite{tensorflow, gaia} or
graph processing~\cite{graphlab, tesseract}. If GPUs are architected to be
first-class compute engines,
rather than the slave devices they are today, they can be programmed and utilized in the
same manner as a modern CPU. 
This integration requires the GPU execution model to support system-level
tasks like interrupts, exceptions, etc. and more
generally provide support for access to distributed file systems,
disk I/O, or network communication. Support for these tasks and
execution models require dynamic provisioning of resources for execution of
system-level code. Zorua provides a building block to enable this.

\subsubsection{Applicability to General Resource Management in Accelerators} 
Zorua uses a program \emph{phase} as the granularity for managing resources.
This allows handling resources across phases
\emph{dynamically}, while leveraging \emph{static} information regarding resource requirements
from the software by inserting annotations at phase boundaries. Future work could potentially investigate the applicability
of the same approach to manage
resources and parallelism in \emph{other} accelerators (e.g.,
processing-in-memory accelerators~\cite{pim-enabled, tesseract, tom-hsieh-isca16, impica,
ambit, googlepim-asplos18, shaw1981non, boroumand2016pim, kim.bmc18,
guo-wondp14, stone1970logic, zhang-2014, kogge.iccp94, patterson.ieeemicro97,
pattnaik.pact16, ghose.pim.bookchapter18, seshadri.bookchapter17, akin.isca15,
seshadri.cal15,concurrent-datastructures}
or direct-memory access engines~\cite{rowclone,decoupled-dma, chang.hpca16}) that require efficient dynamic management of large
amounts of particular critical resources. 

\subsection{Related Work}
To our knowledge, this is the first work to propose a holistic framework to
decouple a GPU application's resource specification from its physical on-chip
resource allocation by virtualizing multiple on-chip resources. This
enables the illusion of more resources than what physically exists to the
programmer, while the hardware resources are managed at runtime by employing a
swap space (in main memory), transparently to the programmer.

We briefly discuss prior work related to aspects specific to Zorua (a more general discussion is in Section~\ref{sec:relatedwork}): 
\One virtualization of resources,
\two more efficient management of
on-chip resources. 
 
\textbf{Virtualization of Resources.}
 \emph{Virtualization}
 ~\cite{virtual-memory1,virtual-memory2,virtualization-1,virtualization-2}
 is a concept designed to provide the illusion, to the software and
 programmer, of more resources than what truly exists in physical
 hardware. It has been applied to the management of hardware
 resources in many different contexts
 ~\cite{virtual-memory1,virtual-memory2,virtualization-1,virtualization-2,vmware-osdi02,how-to-fake,pdp-10,ibm-360},
 with virtual memory~\cite{virtual-memory1, virtual-memory2}
 being one of the oldest forms of virtualization that is commonly
 used in high-performance processors today. Abstraction of hardware
 resources and use of a level of indirection in their management
 leads to many benefits, including improved utilization,
 programmability, portability, isolation, protection, sharing, and 
 oversubscription.

In this work, we apply the general principle of virtualization to the
management of multiple on-chip resources in modern
GPUs. Virtualization of on-chip resources offers the opportunity to
alleviate many different challenges in modern GPUs. However, in this
context, effectively adding a level of indirection introduces new
challenges, necessitating the design of a new virtualization
strategy. There are two key challenges. First, we need to dynamically determine
the \emph{extent} of the virtualization to reach an effective tradeoff
between improved parallelism due to oversubscription and
the latency/capacity overheads of swap space usage. Second,
we need to coordinate the virtualization of \emph{multiple} latency-critical
on-chip resources. To our knowledge, this is the first work to propose
a holistic software-hardware cooperative approach to virtualizing
multiple on-chip resources in a controlled and coordinated manner that
addresses these challenges, enabling the different benefits provided
by virtualization in modern GPUs.

Prior works propose to virtualize a specific on-chip resource for
specific benefits, mostly in the CPU context. For example, in CPUs,
the concept of virtualized registers was first used in the IBM
360~\cite{ibm-360} and DEC PDP-10~\cite{pdp-10} architectures to allow
logical registers to be mapped to either fast yet expensive physical
registers, or slow and cheap memory. More recent works~\cite{how-to-fake,cpu-virt-regs-1,cpu-virt-regs-2}, propose to virtualize registers to increase the effective
register file size to much larger register counts. This increases the
number of thread contexts that can be supported in a multi-threaded
processor~\cite{how-to-fake}, or reduces register spills and
fills~\cite{cpu-virt-regs-1,cpu-virt-regs-2}.
Other works propose to virtualize on-chip
resources in CPUs
(e.g.,~\cite{vls-cook-tr09,spills-fills-kills,hierarchical-scheduling-windows,twolevel-hierarchical-registerfile,virtual-physical-registers-hpca98}). In
GPUs, Jeon et al.~\cite{virtual-register} propose to virtualize the
register file by dynamically allocating and deallocating physical
registers to enable more parallelism with smaller, more
power-efficient physical register files. 
Concurrent to this work, Yoon et al.~\cite{virtual-thread} propose an approach to virtualize thread slots
to increase thread-level parallelism. 
These works propose
specific virtualization mechanisms for a single resource for specific
benefits. None of these works provide a cohesive virtualization
mechanism for \emph{multiple} on-chip GPU resources in a
controlled and coordinated manner, which forms a key contribution of
this work. 

\textbf{Efficient Resource Management.}
Prior works aim to improve parallelism by increasing resource utilization
using
hardware-based~\cite{warp-level-divergence,shmem-multiplexing,unified-register,virtual-register,improving-lee-hpca14,owl-jog-asplos13,orchestrated-jog-isca13,medic-ausavarungnirun-pact15,rachata-isca}
and software-based~\cite{shmem-multiplexing,stash-komuravelli-isca15,
asplos-sree,onchip-allocation,fine-grain-hotpar}
approaches.
%
Among these works, the closest to ours
are~\cite{virtual-register,virtual-thread}
(discussed earlier),~\cite{shmem-multiplexing} and~\cite{warp-level-divergence}. 
These approaches propose efficient techniques to
dynamically manage a single resource, and can be used along with Zorua to improve
resource efficiency further. 
Yang et
al.~\cite{shmem-multiplexing} aim to maximize utilization of the scratchpad with software
techniques, and by dynamically allocating/deallocating scratchpad. 
Xiang et al.~\cite{warp-level-divergence} propose
to improve resource utilization by scheduling threads at the finer granularity
of a warp rather than a thread block. This approach can help alleviate
performance cliffs, but not 
in the presence of synchronization or scratchpad
memory, nor does it address the dynamic underutilization within a thread during
runtime. We quantitatively compare to this approach in the evaluation and demonstrate
Zorua's benefits over it. 

Other works leverage resource
underutilization to improve energy
efficiency~\cite{warped-register,energy-register,compiler-register,virtual-register,grape}
or perform other useful
work~\cite{caba,spareregister-lakshminarayana-hpca14}. These works are
complementary to Zorua.

\subsection{Summary}

We propose Zorua, a new framework that decouples the application resource
specification from the allocation in the physical hardware resources (i.e.,
registers, scratchpad memory, and thread slots) in GPUs. Zorua
encompasses a holistic virtualization strategy to effectively
virtualize multiple latency-critical on-chip resources in a
controlled and coordinated manner. We demonstrate that by providing
the illusion of more resources than physically available, via dynamic
management of resources and the judicious use of a swap space in
main memory, Zorua enhances \One \emph{programming ease} (by reducing
the performance penalty of suboptimal resource specification), \two
\emph{portability} (by reducing the impact of different hardware
configurations), and \three \emph{performance} for code with an
optimized resource specification (by leveraging dynamic
underutilization of resources). We conclude that Zorua is an
effective, holistic virtualization framework for GPUs.

We believe
that the indirection provided by {Zorua}'s virtualization mechanism
makes it a generic framework that can address other challenges in
modern GPUs. For example, Zorua can enable fine-grained resource
sharing and partitioning among multiple kernels/applications, as well as 
low-latency preemption of GPU programs.
Section~\ref{sec:applications} details many other applications of the Zorua
framework. We hope that future work
explores these promising directions, building on the insights and
the framework developed in this work.

%% file: sections/caba.tex
\section{Assist Warps}
\label{sec:caba}
In this chapter, we propose a helper thread abstraction in GPUs to automatically
leverage idle compute and memory bandwidth. We demonstrate significant idleness
in GPU resources, even when code is highly optimized for any given architecture.
We then demonstrate how a rich hardware-software abstraction can enable
programmers to leverage idle compute and memory bandwidth to perform
light-weight tasks, such as prefetching, data compression, etc. 
\subsection{Overview}
GPUs employ
fine-grained multi-threading to hide the high memory access latencies
with thousands of concurrently running threads~\cite{keckler}.  GPUs
are well provisioned with different resources (e.g., SIMD-like
computational units, large register files) to support the execution of
a large number of these hardware contexts. \green{Ideally, if the demand
for all types of resources is properly balanced, all these
resources should be fully utilized by the application}\comm{Ideally, all these
resources should be fully utilized by the application if the demand
for all types of resources is properly balanced}. Unfortunately, this
balance is very difficult to achieve in practice.

As a result, bottlenecks in program execution, e.g., limitations in
memory or computational bandwidth, lead to long stalls and idle
periods in the shader pipelines of modern
GPUs~\cite{largewarp,orchestrated-jog-isca13,owl-jog-asplos13,equalizer}.  \green{Alleviating
these bottlenecks with optimizations implemented in
dedicated hardware requires significant engineering cost and effort.
Fortunately, the resulting under-utilization of on-chip computational
and memory resources from these imbalances in application
requirements, offers some new opportunities. For
example, we can use these resources for efficient integration of
\emph{hardware-generated threads} that perform useful work to
accelerate the execution of the primary threads.} Similar \emph{helper
  threading} ideas have been proposed in the context of
general-purpose
processors~\cite{ssmt,ssmt2,assisted-execution,ht4,ddmt,ht20,assisted-execution-04}
to either extend the pipeline with more contexts or use spare hardware
contexts to pre-compute useful information that aids main code
execution (e.g., to aid branch prediction, prefetching, etc.).

We believe that the general idea of helper threading can lead to even
more powerful optimizations and new opportunities in the context of
modern GPUs than in CPUs because (1) the abundance of on-chip
resources in a GPU obviates the need for idle hardware
contexts~\cite{ht5,ht4} or the addition of more storage (registers,
rename tables, etc.) and compute units~\cite{slice,ssmt} required to
handle more contexts and (2) the relative simplicity of the GPU
pipeline avoids the complexities of handling register renaming,
speculative execution, precise interrupts, etc.~\cite{ssmt2}. However,
\green{GPUs} that execute and manage thousands of thread contexts at
the same time pose new challenges for employing helper threading,
which must be addressed carefully\comm{that prevents from direct usage
  of related prior work}. First, the numerous regular program threads
executing in parallel could require an equal or larger number of
helper threads that need to be managed at low cost. Second, the compute and
memory resources are dynamically partitioned between threads in GPUs,
and resource allocation for helper threads should be cognizant of
resource interference and overheads. Third, lock-step execution and
complex scheduling{\textemdash}which are characteristic of GPU
architectures{\textemdash}exacerbate the complexity of fine-grained
management of helper threads.

In this work, we describe a new, flexible framework for bottleneck
acceleration in GPUs via helper threading (called \emph{Core-Assisted
  Bottleneck Acceleration} or \SADA), which exploits the
aforementioned new opportunities while effectively handling the new
challenges.  \SADA performs acceleration by generating special
warps{\textemdash}\emph{\helperwarps}{\textemdash}that can execute
code to speed up application execution and system tasks. \comm{For example, \SADA can
  be used to perform compression/decompression of the data that is
  transferred between main memory and cores, thereby alleviating the
  off-chip memory bandwidth bottleneck.}To simplify the support of the
numerous assist threads with CABA, we manage their execution at the
granularity of a \emph{warp} and use a centralized mechanism to track
the progress of each \emph{assist warp} throughout its execution. To
reduce the overhead of providing and managing new contexts for each
generated thread, as well as to simplify scheduling and data
communication, an assist warp \emph{shares the same context} as the
regular warp it assists.  Hence, the regular warps are overprovisioned
with \emph{available registers} to enable each of them to host its own
assist warp.

\textbf{Use of CABA for compression.} We illustrate an important use case for
the CABA framework: alleviating the memory bandwidth bottleneck by enabling
\emph{flexible data compression} in the memory hierarchy.
The basic idea is to have assist warps that (1) compress cache blocks before
they are written to memory, and (2) decompress cache blocks before they are
placed into the cache.  

\green{\SADA-based compression/decompression provides several benefits
  over a purely hardware-based implementation of data compression for
  memory.} First, CABA primarily employs hardware that is already
available on-chip but is otherwise underutilized. In contrast,
hardware-only compression implementations require \emph{dedicated logic} for
specific algorithms.  Each new algorithm (or a modification of an
existing one) requires engineering effort and incurs hardware
cost. Second, different applications tend to have distinct data
patterns~\cite{bdi} that are more efficiently compressed with
different compression algorithms.\comm{Hence, application awareness in
  terms of choice of algorithm enables better exploitation of data
  compression.} CABA offers versatility in algorithm choice as we find
that many existing hardware-based compression algorithms (e.g.,
Base-Delta-Immediate (BDI) compression~\cite{bdi}, Frequent Pattern
Compression (FPC)~\cite{fpc}, and C-Pack~\cite{c-pack}) can be
implemented using different assist warps with the CABA
framework. Third, not all applications benefit from data
compression. Some applications are constrained by other bottlenecks
(e.g., oversubscription of computational resources), or may operate on
data that is not easily compressible.  As a result, the benefits of
compression may not outweigh the cost in terms of additional latency
and energy spent on compressing and decompressing data. In these
cases, compression can be easily disabled by CABA, and the CABA
framework can be used in other ways to alleviate the current
bottleneck.




\textbf{Other uses of CABA.} The generality of \green{CABA} enables
its use in alleviating other bottlenecks with different optimizations.
We discuss two examples: (1) \green{using assist warps to perform
  \emph{memoization}} to eliminate redundant computations that have
the same or similar inputs~\cite{dynreuse,Arnau-memo,danconnors}, by
storing the results of frequently-performed computations in the main
memory hierarchy (i.e., by converting the computational problem into a
storage problem) and, (2) \green{using the idle memory pipeline to
  perform opportunistic \emph{prefetching}} to better overlap
computation with memory access. Assist warps offer a hardware/software
interface to implement hybrid prefetching
algorithms~\cite{techniques-ebrahimi-hpca09} with varying degrees of complexity. We also
briefly discuss other uses of CABA for (1) redundant multithreading, (2)
speculative precomputation, (3) handling interrupts, and (4) profiling and
instrumentation.


\textbf{Contributions.} In this work, we make the following contributions:


\begin{itemize}

\item We introduce the \emph{\SADAfull (\SADA) Framework}, which can
  mitigate different bottlenecks in modern GPUs by using
  underutilized system resources for \emph{assist warp} execution.

\item We provide a detailed description of how our framework can be
  used to enable effective and flexible data compression in GPU memory
  hierarchies.\comm{We show that \SADA can flexibly implement {\em
    multiple} different compression algorithms.}



\item We comprehensively evaluate the use of \SADA for data
  compression to alleviate the memory bandwidth bottleneck. Our
  evaluations across a wide variety applications from
  Mars~\cite{mars}, CUDA~\cite{sdk}, Lonestar~\cite{lonestar}, and
  Rodinia~\cite{rodinia} \green{benchmark} suites show that
  \SADA-based compression on average (1) reduces memory bandwidth
  by 2.1X, (2) improves performance by 41.7\%, and (3)
  reduces overall system energy by 22.2\%.

\item We discuss at least six other use cases of CABA that can improve
application performance and system management, showing that CABA is a primary
general framework for taking advantage of underutilized resources in modern GPU
engines. 

\end{itemize}

\subsection{Motivation: Bottlenecks in Resource Utilization}

\label{sec:motivation_caba} 

We observe that different bottlenecks and imbalances
during program execution leave resources unutilized within the GPU
cores. \green{We motivate our proposal, CABA,  by examining these
inefficiencies.} CABA
leverages these inefficiencies as an opportunity to perform useful work.

\textbf{Unutilized Compute Resources.} A GPU core employs fine-grained
multithreading~\cite{burtonsmith,cdc6600} of {\em warps}, i.e., groups of threads executing the
same instruction, to hide long memory and ALU operation latencies.
\comm{When a few warps are stalled due to these long-latency
  operations, the remaining warps are swapped in for execution,
  potentially hiding the performance penalties of the stalled warps.
}If the number of available warps is insufficient to cover these long
latencies, the core stalls or becomes idle. \green{To understand} the
key sources of inefficiency in GPU cores, we conduct an experiment
where we show the breakdown of the applications' execution time spent
on either useful work (\emph{Active Cycles}) or stalling due to one of
four reasons: \green{\emph{Compute, Memory, Data Dependence
    Stalls} and \emph{Idle Cycles}}.  We also vary the amount of
available off-chip memory bandwidth: (i) half (1/2xBW), (ii) equal to
(1xBW), and (iii) double (2xBW) the peak memory bandwidth of our
baseline GPU architecture. Section~\ref{sec:methodology_caba} details our baseline architecture and methodology.

\begin{figure*}[!h] \
\centering \includegraphics[width=1\textwidth]{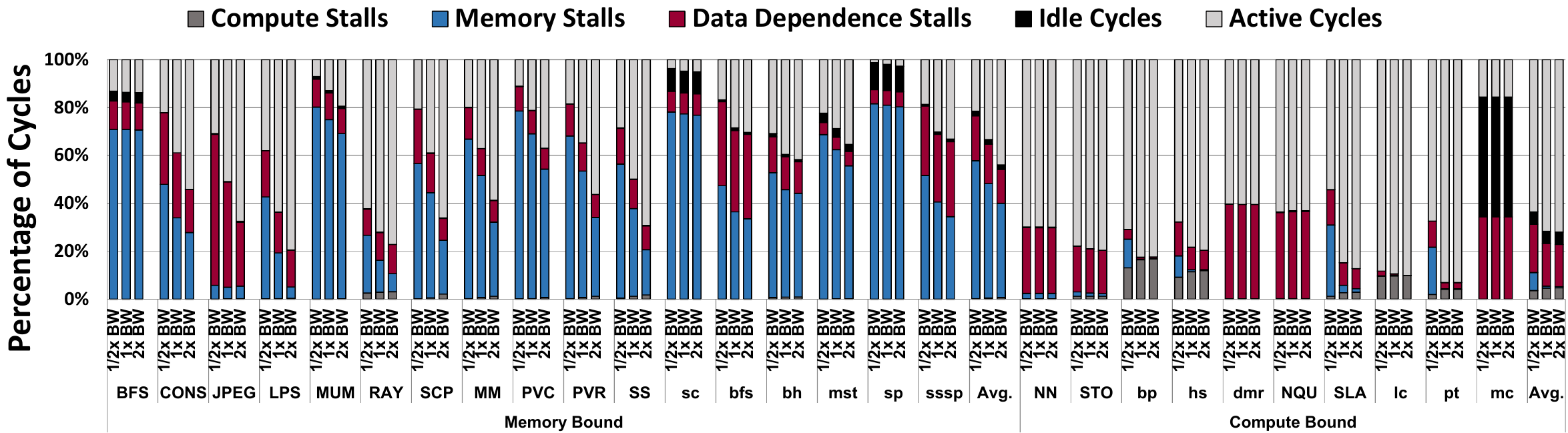}
\caption{\small{Breakdown of total issue cycles for 27 representative
CUDA applications. See Section~\ref{sec:methodology_caba} for methodology.}}
\label{fig:pipelinestalls2} \vspace{-0.0cm} \end{figure*}

Figure~\ref{fig:pipelinestalls2} shows the percentage of total issue
cycles, divided into five components (as described above).  The
first two components{\textemdash}\emph{Memory and Compute Stalls}{\textemdash}are attributed
to the main memory and \green{ALU-pipeline structural stalls}. These stalls are because of backed-up
pipelines due to oversubscribed resources that \comm{situation arises
  due to the oversubscription to corresponding shared resources and
}prevent warps from being issued to the respective pipelines.  The
third component (\emph{Data Dependence Stalls}) is due to data
\green{dependence} stalls. These stalls prevent warps from
issuing new instruction(s) when the previous instruction(s) from the
same warp are stalled on long-latency operations (usually memory load
operations).  In some applications (e.g., {\tt dmr}),
special-function-unit (SFU) ALU operations that may take tens of
cycles to finish are also the source of data dependence stalls.  The
fourth component, \emph{Idle Cycles}, refers to idle cycles when either all
the available warps are issued to the pipelines and not ready
to execute their next instruction or the instruction buffers are flushed due to a mispredicted branch.  All these components are
sources of inefficiency that cause the cores to be underutilized. The
last component, \emph{Active Cycles}, indicates the fraction of cycles
during which at least one warp was successfully issued to the
pipelines.

We make two observations from Figure~\ref{fig:pipelinestalls2}.
First, \emph{Compute, Memory}, and \emph{Data Dependence Stalls} are
the major sources of underutilization in many GPU applications.  We distinguish applications
based on their primary bottleneck as either \emph{Memory} or
\emph{Compute Bound}. \green{We observe that a majority of the applications in our workload pool (17 out of
27 studied) are \emph{Memory Bound}, and bottlenecked by the off-chip memory bandwidth.}

\green{Second, for the \emph{Memory Bound} applications, we observe
  that the \emph{Memory} and \emph{Data Dependence} stalls constitute
  a significant fraction (61\%) of the total issue cycles on our
  baseline GPU architecture (1xBW). This fraction goes down to 51\%
  when the peak memory bandwidth is doubled (2xBW), and increases
  significantly when the peak bandwidth is halved (1/2xBW), indicating
  that limited off-chip memory bandwidth is a critical performance
  bottleneck for \emph{Memory Bound} applications. Some applications,
  e.g., \emph{BFS}, are limited by the interconnect bandwidth.
  \green{In contrast}, the \emph{Compute Bound} applications are
  primarily bottlenecked by stalls in the ALU pipelines\comm{that
    constitute 11\% of the underutilized cycles}.  An increase or
  decrease in the off-chip bandwidth has little effect on the
  performance of these applications.}

\begin{figure}[!h] \centering 
\centering \includegraphics[width=0.69\textwidth]{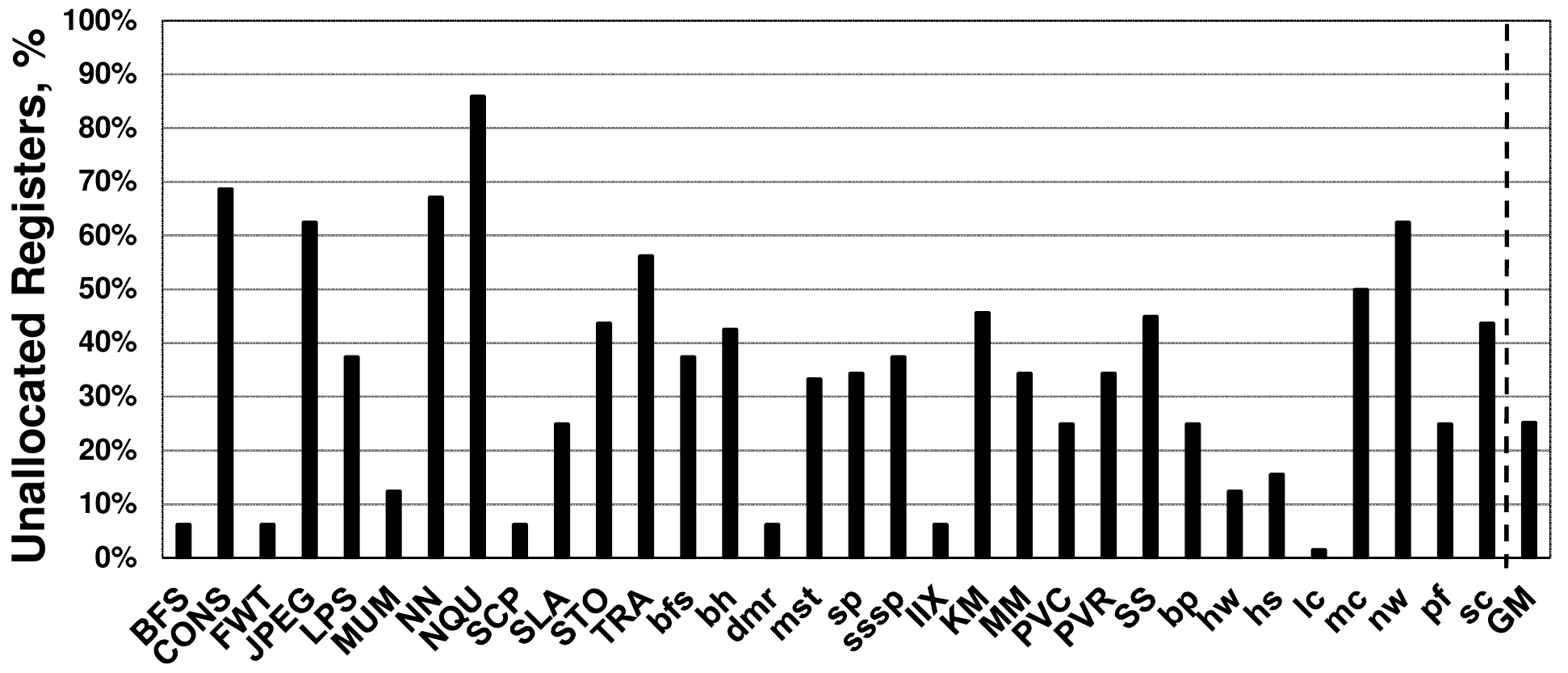}
\caption{Fraction of statically unallocated registers.} \label{fig:reg_util}
\end{figure}

\textbf{Unutilized On-chip Memory.} The \emph{occupancy} of any GPU Streaming
Multiprocessor (SM), i.e., the number of threads running concurrently,
is limited by a number of factors: (1) the available registers and
shared memory, (2) the hard limit on the number of threads and thread
blocks per core, (3) the number of thread blocks in the application
kernel. The limiting resource from the above, leaves the other resources
underutilized. This is because it is challenging, in practice, to achieve a
perfect balance in utilization of all of the above factors for different
workloads with varying characteristics. Very often, the factor determining the occupancy is the thread
or thread block limit imposed by the architecture. In this case, there
are many registers that are left unallocated to any thread
block. Also, the number of available registers may not be a multiple
of those required by each thread block. The remaining registers are
not enough to schedule an entire extra thread block, which leaves a
significant fraction of the register file and shared memory
unallocated and unutilized by the thread blocks.
Figure~\ref{fig:reg_util} shows the fraction of statically unallocated
registers in a 128KB register file (per SM) with a 1536 thread, 8
thread block occupancy limit, for different applications. We observe
that on average 24\% of the register file remains unallocated. This
phenomenon has previously been observed and analyzed in detail
in~\cite{unified-register, warped-register, spareregister-lakshminarayana-hpca14,
  energy-register, compiler-register}. We observe a similar trend with
the usage of shared memory (not graphed).

\textbf{Our Goal.} We aim to exploit the underutilization of compute
resources, registers and on-chip shared memory as an opportunity to
enable different optimizations to accelerate various bottlenecks in
GPU program execution.\comm{For example, unutilized computational
  resources could be used to perform efficient data
  compression/decompression that effectively converts the data
  communication problem into a computational one.} To achieve this goal, we
would like to enable efficient helper threading for GPUs 
to dynamically generate threads in hardware \green{that use} the available
on-chip resources for various purposes. In the next section, we present the
detailed design of our \SADA framework that enables the generation
and management of these threads.\comm{The aim of the framework is to
  simplify the application of data compression as well as other
  optimization techniques (e.g., memoization) to accelerate the
  execution of GPU architectures.}

\subsection{The \SADA Framework} \label{sec:idea} In order to understand the major
design choices behind the \SADA framework, we first present our major design
goals and describe the key challenges in applying helper threading to GPUs. We
then show the detailed design, hardware changes, and operation of \SADA.
Finally, we briefly describe potential applications of our proposed framework.
Section~\ref{sec:compression_caba} goes into a detailed design of one application
of the framework. 
 
\subsubsection{Goals and Challenges} 


The purpose of \SADA is to leverage underutilized GPU resources for useful
computation.  To this end, we need to efficiently execute subroutines that
perform optimizations to accelerate \comm{application bottlenecks}bottlenecks in
application execution.  The key difference between \SADA's \emph{assisted
execution} and regular execution is that \SADA must be \emph{low overhead} and,
therefore, helper threads need to be treated differently from regular threads.
The \emph{low overhead} goal imposes several key requirements in designing a
framework to enable helper threading.  
First, we should be able to easily manage helper threads{\textemdash}to enable,
trigger, and kill threads when required.  Second, helper threads need to be
flexible enough to adapt to the runtime behavior of the regular program.  Third,
a helper thread needs to be able to communicate with the original thread.
Finally, we need a flexible interface to specify new subroutines, with the
framework being generic enough to handle various optimizations.


With the above goals in mind, enabling helper threading in GPU architectures
introduces several new challenges.  First, execution on GPUs involves context
switching between hundreds of threads. These threads are handled at different
granularities in hardware and software. The programmer reasons about these
threads at the granularity of a thread block. \comm{Threads are scheduled onto
Streaming Multiprocessors (SMs) at the granularity of a block. }However, at any
point in time, the hardware executes only a small subset of the thread block,
i.e., a set of warps. Therefore, we need to define the \emph{abstraction
levels} for reasoning about and managing helper threads from the point of view
of the programmer, the hardware as well as the compiler/runtime. In addition,
each of the thousands of executing threads could simultaneously invoke an
associated helper thread subroutine.  To keep the management overhead low, we
need an efficient mechanism to handle helper threads at this magnitude.

Second, GPUs use fine-grained multithreading~\cite{cdc6600,burtonsmith} to
time multiplex the fixed number of compute units among the hundreds of threads.
Similarly, the on-chip memory resources (i.e., the register file and shared
memory) are statically partitioned between the different threads at compile
time. Helper threads require their own registers and compute cycles to
execute.  A straightforward approach would be to dedicate few registers and compute
units just for helper thread execution, but this option is both expensive and
wasteful. In fact, our primary motivation is to utilize \emph{existing idle
resources} for helper thread execution. In order to do this, we aim to enable
sharing of the existing resources between primary threads and helper threads at
low cost, while minimizing the interference to primary thread execution. In the
remainder of this section, we describe the design of our low-overhead \SADA
framework.

\subsubsection{Design of the \SADA Framework} \comm{\SADA enables helper threads, which we view as simple subroutines that
assist the primary program execution.} We choose to implement \SADA using a hardware/software co-design, as pure hardware or pure software approaches pose certain challenges that we describe below. There are two alternatives for a fully software-based
approach to helper threads.  The first alternative, treating each helper thread
as independent kernel code, has high overhead, since we are now treating the
helper threads as, essentially, regular threads. This would reduce the primary
thread occupancy in each SM (there is a hard limit on the number of
threads and blocks that an SM can support). It would also complicate the data communication
between the primary and helper threads, since no simple interface exists for
inter-kernel communication. The second alternative, embedding the
helper thread code within the primary thread kernel itself, offers little
flexibility in adapting to runtime requirements, since such helper threads cannot be triggered
or squashed independently of the primary thread. 

On the other hand, a pure
hardware solution would make register allocation for the assist warps and the
data communication between the helper threads and primary threads more difficult.
Registers are allocated to each thread block by the compiler and are then mapped to
the sections of the hardware register file at runtime. Mapping registers for
helper threads and enabling data communication between those registers and the primary thread
registers would be non-trivial.  Furthermore, a fully hardware approach would
make offering the programmer a flexible interface more challenging.

Hardware support enables simpler fine-grained management of helper threads,
aware of micro-architectural events and runtime program behavior.
Compiler/runtime support enables simpler context management for helper threads
and more flexible programmer interfaces. Thus, to get the best of both worlds, we propose a \emph{hardware/software cooperative approach}, where the hardware manages the scheduling and execution of helper thread subroutines, while the compiler
performs the allocation of shared resources (e.g., register file and shared
memory) for the helper threads and the programmer or the microarchitect provides the helper threads themselves. 

\textbf{Hardware-based management of threads.} To use the available
on-chip resources the same way that thread blocks do during program execution, we
dynamically insert sequences of instructions into the execution stream. We track
and manage these instructions at the granularity of a warp, and refer to them as
\emph{\textbf{Assist Warps}}. An \helperwarp is a set of
instructions issued into the core pipelines. Each instruction is executed in
lock-step across all the SIMT lanes, just like any regular instruction, with an
active mask to disable lanes as necessary. The \helperwarp does \emph{not} own a
separate context (e.g., registers, local memory), and instead shares both a
context and a warp ID with the regular warp that invoked it. In other words, each
assist warp is coupled with a \emph{parent warp}. In this sense, it is different
from a regular warp and does not reduce the number of threads that
can be scheduled on a single SM. Data sharing between the two warps becomes simpler, since the assist warps share the register file with the parent warp.
Ideally, an assist warp consumes resources and issue cycles that would otherwise
be idle. We describe the structures required to support hardware-based
management of assist warps in Section~\ref{sec:Components}. 

\textbf{Register file/shared memory allocation.} Each helper thread
subroutine requires a different number of registers depending on the
actions it performs. These registers have a short lifetime, with no
values being preserved between different invocations of an assist warp. To limit the register requirements for assist warps, we impose the restriction
that only one instance of each helper thread routine can be active for each
thread. All instances of the same helper thread for each parent thread use
the same registers, and the registers are allocated to the helper threads statically by the compiler. One of
the factors that determines the runtime SM occupancy is the number of registers required by a thread block (i.e, per-block register requirement). For each helper thread subroutine that is enabled, we add its register requirement
to the per-block register requirement, to ensure the availability of registers
for both the parent threads as well as every assist warp. The
registers that remain unallocated after allocation among the parent thread
blocks should suffice to support the assist warps. If not, register-heavy assist
warps may limit the parent thread block occupancy in SMs or increase the number
of register spills in the parent warps. Shared memory resources are partitioned in a similar manner and allocated to each assist warp as and if
needed.  

\textbf{Programmer/developer interface.} The assist warp subroutine can
be written in two ways. First, it can be supplied and annotated by the
programmer/developer using CUDA extensions with PTX instructions and then
compiled with regular program code. Second, the assist warp subroutines can be written by the microarchitect in the internal GPU instruction format. \comm{This approach enables the
microarchitect to take control at finer granularity, and }These helper thread
subroutines can then be enabled or disabled by the application programmer. This
approach is similar to that proposed in prior work (e.g.,~\cite{ssmt}). It
offers the advantage of potentially being highly optimized for energy and
performance while having flexibility in implementing optimizations
that are not trivial to map using existing GPU PTX instructions.  The
instructions for the helper thread subroutine are stored in an on-chip buffer
(described in Section~\ref{sec:Components}). 

Along with the helper thread subroutines, the programmer also provides: (1)
the \emph{priority} of the assist warps to enable the warp scheduler to
make informed decisions,  (2) the trigger conditions for each assist warp, and (3)
the live-in and live-out variables for data communication with the parent warps.

Assist warps can be scheduled with different priority levels in relation to
parent warps by the warp scheduler. Some assist warps may perform a function
that is required for correct execution of the program and are \emph{blocking}.
At this end of the spectrum, the \emph{high priority} assist warps are treated
by the scheduler as always taking higher precedence over the parent warp
execution. Assist warps
should be given a high priority only when they are \comm{on the critical path for
program execution and are }required for correctness. \emph{Low priority} assist
warps, on the other hand, are scheduled for execution only when computational
resources are available, i.e., during idle cycles. There is no guarantee that
these assist warps will execute or complete. 

The programmer also
provides the conditions or events that need to be satisfied for the deployment of
the assist warp. This includes a specific point within the original program
and/or a set of other microarchitectural events that could serve as a \emph{trigger}
for starting the execution of an assist warp.\comm{With assist warps that are pre-installed by the
microarchitect, the trigger events and priority are hard-coded for each assist
warp routine, and the programmer is provided with a knob to enable/disable a
specific assist warp routine.}

\subsubsection{Main Hardware Additions} \label{sec:Components}
\begin{figure}[h] \centering 
\includegraphics[width=0.69\textwidth]{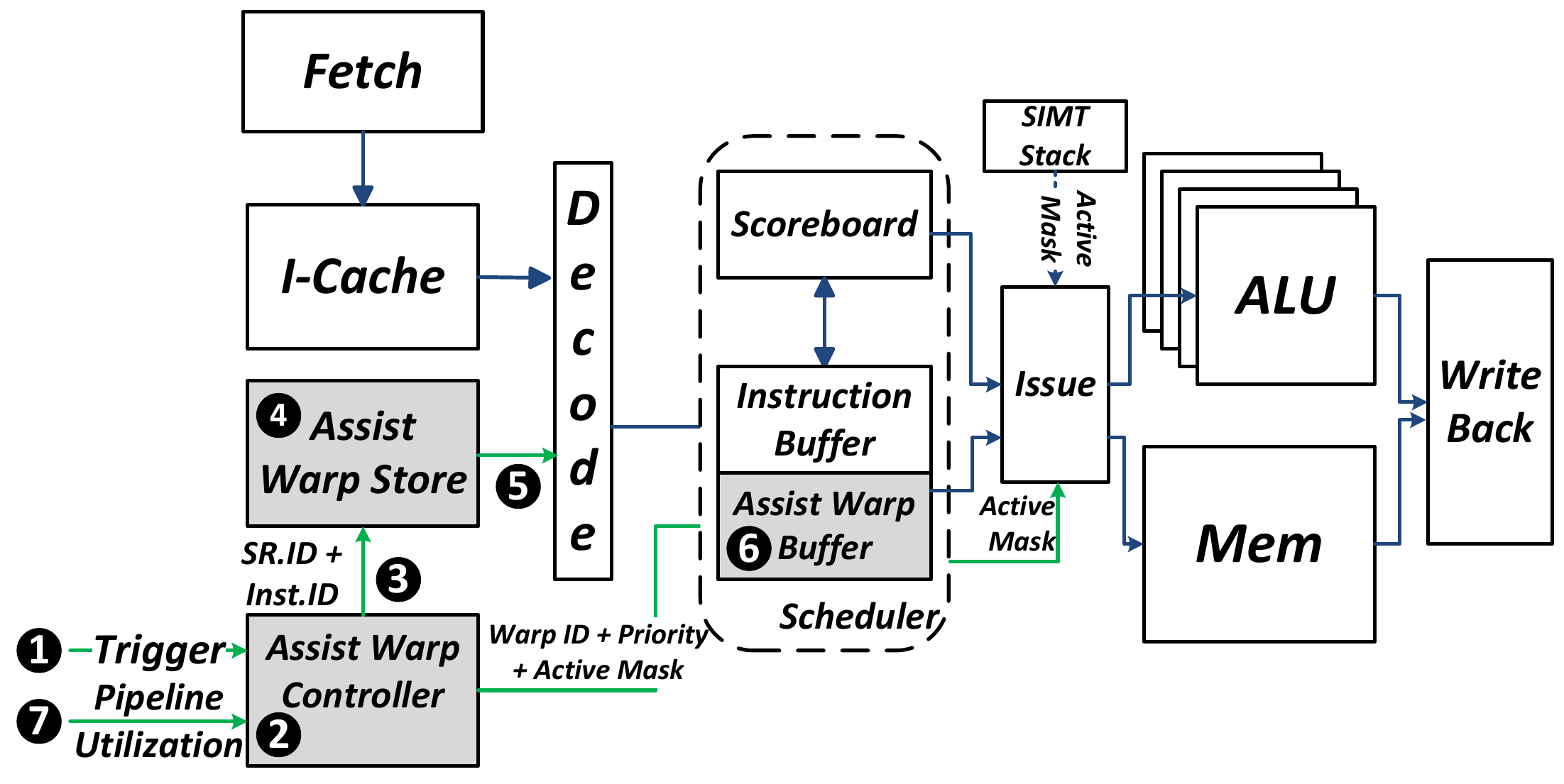} 
\caption{\SADA framework flow within a typical GPU pipeline~\cite{manual}. The
shaded blocks are the components introduced for the framework.}
\label{fig:framework} 
\end{figure}

Figure~\ref{fig:framework} shows a high-level block diagram of the GPU
pipeline~\cite{manual}.\comm{Instructions are fetched from the Instruction
cache, decoded and then buffered in the Instruction Buffer before a warp is
selected by the scheduler for issue into the ALU/memory pipelines. Accesses to
global memory also involve address generation and coalescing before requests are
send to the memory hierarchy.} To support assist warp execution, we add three new
components: (1) an Assist Warp Store to hold the assist warp
code, (2) an Assist Warp Controller to perform the deployment, tracking,
and management of assist warps, and (3) an Assist Warp Buffer to stage instructions from triggered assist
warps for execution.

\textbf{Assist Warp Store (AWS).} Different assist warp subroutines are possible
based on the purpose of the optimization. These code sequences for different
types of \helperwarps need to be stored on-chip. An on-chip storage structure
called the Assist Warp Store (\ding{205}) is preloaded with these instructions
before application execution. It is indexed using the subroutine index (SR.ID)
along with the instruction ID (Inst.ID).\comm{Indexing is based on the sequence
number. Special instructions are required to load the \helperwarp sequencer
before the program execution begins.}\comm{Note that an alternative
approach, is to simply contain the AWS as a part of the instruction cache. We, however, assume a separate structure to carry assist warp code.}

\textbf{Assist Warp Controller (AWC).} The AWC (\ding{203}) is responsible for the triggering, tracking, and management
of \helperwarp execution. It stores a mapping between
trigger events and a subroutine index in the AWS, as specified by the programmer. The AWC monitors for such events, and when they take place, triggers the fetch, decode and execution of instructions from the AWS for the respective assist warp. 

Deploying all the instructions within an assist warp, back-to-back, at the trigger point may require increased fetch/decode bandwidth and buffer space after decoding~\cite{ssmt2}. To avoid this, at each
cycle, only a few instructions from an assist warp, at most equal to the
available decode/issue bandwidth, are decoded and staged for execution. Within the AWC, we simply track the next instruction that needs to be executed for each assist warp and this is stored in the Assist Warp Table (AWT), as depicted in Figure~\ref{fig:AWT}. The AWT also tracks additional metadata required for assist warp management, which is described in more detail in Section~\ref{sec:mechanism_caba}.  

\textbf{Assist Warp Buffer (AWB).} Fetched and decoded instructions (\ding{203}) belonging to the assist warps that have been triggered need to
be buffered until the assist warp can be selected for issue by the scheduler. These instructions are then staged in the Assist
Warp Buffer (\ding{207}) along with their warp IDs. The AWB is contained within the \emph{instruction buffer (IB)}, which holds decoded instructions for the parent warps. The AWB makes use of the existing IB structures. The IB is typically partitioned among different warps executing in the SM. Since each assist warp is associated with a parent warp, the assist warp instructions are directly inserted into the \emph{same partition} within the IB as that of the parent warp. This simplifies warp scheduling, as the assist warp instructions can now be issued as if they were parent warp instructions with the same warp ID. In addition, using the existing partitions avoids the cost of separate dedicated instruction buffering for assist warps. We do, however,
provision a small additional partition with two entries within the IB, to hold
non-blocking \emph{low priority} assist warps that are scheduled only during idle
cycles. This additional partition allows the scheduler to distinguish \emph{low priority} assist warp instructions from the parent warp and \emph{high priority} assist warp instructions, which are given precedence during scheduling, allowing them to make progress. 

\subsubsection{The Mechanism}
\label{sec:mechanism_caba}

\begin{figure}[t] \centering 
\includegraphics[width=0.69\textwidth]{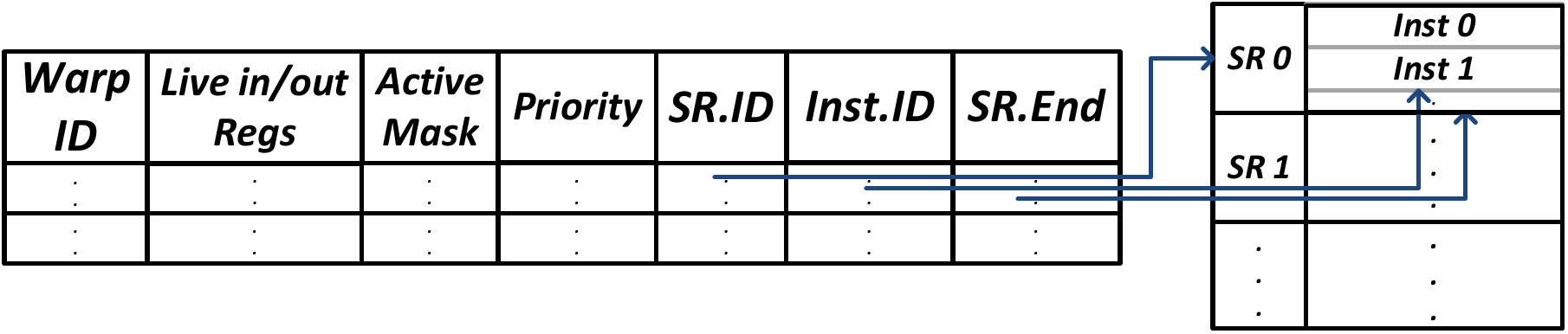} 
\caption{Fetch Logic: Assist Warp Table (contained in the AWC) and the Assist
Warp Store (AWS).} 
\label{fig:AWT} 
\end{figure}

\indent\textbf{Trigger and Deployment.} An \helperwarp is triggered (\ding{202}) by the AWC (\ding{203}) based on a specific
set of architectural events and/or a triggering instruction (e.g., a load
instruction).  When an \helperwarp is triggered, its specific
instance is placed into the Assist Warp Table (AWT) within the AWC
(Figure~\ref{fig:AWT}). Every cycle, the AWC selects an assist warp to deploy in
a round-robin fashion. The AWS is indexed (\ding{204}) based on the subroutine
ID (SR.ID){\textemdash}which selects the instruction sequence to be executed by the assist
warp, and the instruction ID (Inst.ID){\textemdash}which is a pointer to the next instruction
to be executed within the subroutine (Figure~\ref{fig:AWT}). The selected
instruction is entered (\ding{206}) into the AWB (\ding{207}) \comm{The AWC also generates
the priority of each \helperwarp to help the scheduler prioritize between the
\helperwarps and primary warps. }and, at this point, the instruction enters the active
pool with other active warps for scheduling. The Inst.ID for the assist warp is updated in the AWT to point to the next
instruction in the subroutine. When the end of the subroutine is reached, the
entry within the AWT is freed. 

\textbf{Execution.} Assist warp instructions, when selected for issue by the scheduler, are executed in much the same way as any other instructions. The scoreboard tracks the dependencies between instructions
within an assist warp in the same way as any warp, and instructions from different assist warps are interleaved in execution in order to hide latencies. \purple{We also provide an active mask (stored as a part of the AWT), which 
allows for statically disabling/enabling different lanes within a warp. This is
useful to provide flexibility in lock-step instruction execution when we do not need all threads within a warp to execute a specific assist warp subroutine.} 

\textbf{Dynamic Feedback and Throttling.} \Helperwarps, if not properly
controlled, may stall application execution. This can happen due to several
reasons.  First, \helperwarps take up issue cycles, and only a limited number of
instructions may be issued per clock cycle.  Second, \helperwarps require
structural resources: the ALU units and resources in the load-store pipelines
(if the assist warps consist of computational and memory instructions,
respectively). We may, hence, need to throttle assist warps
to ensure that their performance benefits outweigh the overhead. This requires mechanisms to appropriately balance and manage the
aggressiveness of assist warps at runtime. 

The overheads associated with \helperwarps can be controlled in different ways.
First, the programmer can statically specify the priority of the \helperwarp.
Depending on the criticality of the \helperwarps in making forward progress,
the assist warps can be issued either in idle cycles or with varying levels of priority in
relation to the parent warps.  For example, warps performing \emph{decompression} are given a
high priority whereas warps performing \emph{compression} are given a low
priority. Low priority assist warps are inserted into the dedicated
partition in the IB, and are scheduled only during idle cycles. This priority is
statically defined by the programmer. Second, the AWC can control the number of times
the \helperwarps are deployed into the AWB.\comm{In our framework applied to data compression, we
employ these options depending on the overhead associated with the
\helperwarps.}
The AWC monitors the utilization of the functional
units (\ding{208}) and idleness of the cores to decide when to throttle assist warp deployment. \comm{A simple
timeout mechanism is used by the AWC to periodically flush lower priority assist warps from
the Assist Warp Table if the AWC is unable to find idle cycles to issue them.} 

\textbf{Communication and Control.} An assist warp may need to communicate data
and status
with its parent warp. For example, memory addresses from the parent warp need to
be communicated to assist warps performing decompression or prefetching. The IDs
of the registers containing the live-in data for each assist warp are saved in the AWT
when an assist warp is triggered. Similarly, if an assist warp needs to report
results to its parent warp (e.g., in the case of memoization), the register
IDs are also stored in the AWT. When the assist warps execute,
\emph{MOVE} instructions are first executed to copy the live-in data from the
parent warp registers to the assist warp registers. Live-out
data is communicated to the parent warp in a similar fashion, at the end of
assist warp execution.

Assist warps may need to be \emph{killed} when they are not required (e.g., if the data
does not require decompression) or when they are no longer beneficial. In this case, the entries in the AWT and AWB are
simply flushed for the assist warp. 

\subsubsection{Applications of the \SADA Framework} We envision multiple
applications for the \SADA framework, e.g., data
compression~\cite{fvc,fpc,c-pack,bdi}, memoization~\cite{dynreuse, Arnau-memo,
danconnors}, data prefetching~\cite{fdp, stream1,stream2,stride1,stride2}. In Section~\ref{sec:compression_caba}, we provide a detailed case study of enabling data compression with
the framework, discussing various tradeoffs. We
believe \SADA can be useful for many other optimizations, and we discuss some of them
briefly in Section~\ref{sec:Applications}.

\subsection{A Case for CABA: Data Compression}
\label{sec:compression_caba}
Data compression is a technique that exploits the redundancy in the
applications' data to reduce capacity and bandwidth requirements for many modern
systems by saving and transmitting data in a more compact form.  Hardware-based
data compression has been explored in the context of on-chip
caches~\cite{fvc,fpc,c-pack,bdi,dcc,sc2-arelakis-isca14,zca-dusser-sc09,zvc,exploiting-pekhimenko-hpca15},
interconnect~\cite{noc-comp}, and main
memory~\cite{MXT,LinkCompression,MMCompression,lcp-micro,memzip,toggle-aware} as a
means to save storage capacity as well as memory bandwidth.  In modern GPUs,
memory bandwidth is a key limiter to system performance in many workloads
(Section~\ref{sec:motivation_caba}). As such, data compression is a promising
technique to help alleviate this bottleneck.  Compressing data enables less data
to be transferred from/to DRAM and the interconnect. 

In
bandwidth-constrained workloads, idle compute pipelines offer an opportunity to
employ \SADA to enable data compression in GPUs. We can use assist warps to (1)
decompress data, before loading it into the caches and registers, and (2)
compress data, before writing it back to memory. Since assist warps execute
instructions, \SADA offers some flexibility in the compression algorithms that
can be employed. Compression algorithms that can be mapped to the general GPU
execution model can be flexibly implemented with the \SADA framework. \comm{This
section describes how we can use the CABA framework to implement data
compression in GPUs.}  

\subsubsection{Mapping Compression Algorithms into Assist Warps} 

In order to employ \SADA to enable data compression, we need to map compression
algorithms into instructions that can be executed within the GPU cores. For a
compression algorithm to be amenable for implementation with \SADA, it ideally
needs to be (1) reasonably parallelizable and (2) simple (for low latency).
Decompressing data involves reading the encoding associated with each cache line
that defines how to decompress it, and then triggering the corresponding
decompression subroutine in CABA.\comm{The instruction storage in the AWS
constrains the number of possible encodings. } Compressing data, on the other
hand, involves testing different encodings and saving data in the compressed
format. 

We perform compression at the granularity of a cache line. The data needs to be
decompressed before it is used by any program thread.  In order to utilize the
full SIMD width of the GPU pipeline, we would like to decompress/compress all
the words in the cache line in parallel. With \SADA, helper thread routines are
managed at the warp granularity, enabling fine-grained triggering of assist
warps to perform compression/decompression when required. However, the SIMT
execution model in a GPU imposes some challenges: (1) threads within a warp
operate in lock-step, and (2) threads operate as independent entities, i.e.,
they do not easily communicate with each other. 

In this section, we discuss the architectural changes and algorithm adaptations
required to address these challenges and provide a detailed implementation and
evaluation of \emph{Data Compression} within the \SADA framework using the
\emph{Base-Delta-Immediate compression} algorithm~\cite{bdi}.
Section~\ref{sec:Other_Algos} discusses implementing other compression
algorithms.

\begin{figure}[!h] 
\includegraphics[scale=0.65]{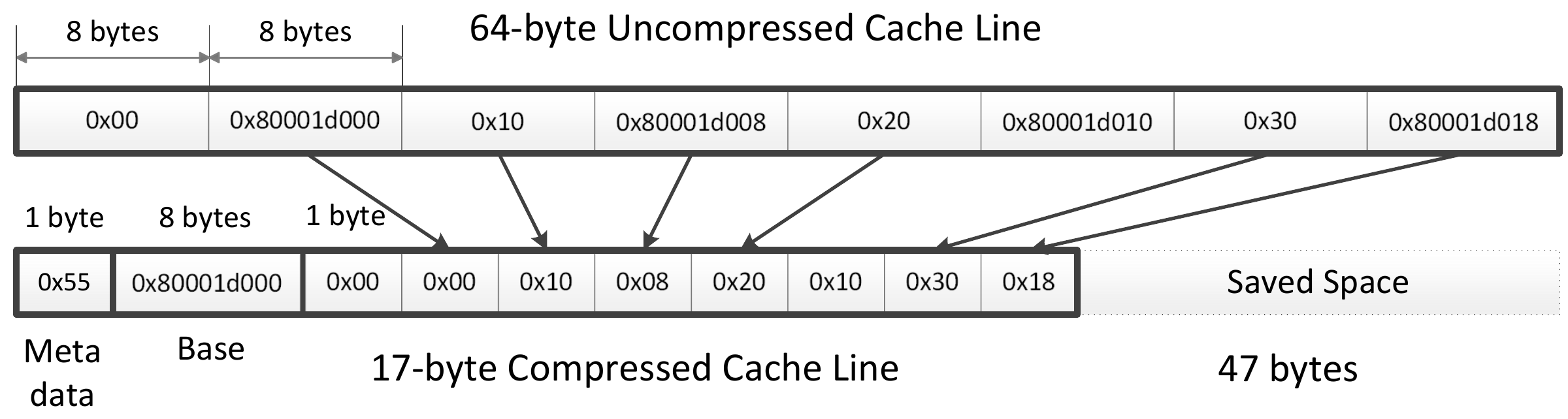} 
\caption{Cache line from \emph{PVC} compressed with BDI.} 
\label{fig:bdc-example2} \end{figure}

\textbf{Algorithm Overview.} \label{sec:algorithm} Base-Delta-Immediate
compression (BDI) is a simple compression algorithm that was originally proposed
in the context of caches~\cite{bdi}.  It is based on the observation that many
cache lines contain data with low dynamic range. BDI exploits this observation
to represent a cache line with low dynamic range using a common \emph{base} (or
multiple bases) and an array of \emph{deltas} (where a delta is the difference
of each value within
the cache line and the common base). Since the \emph{deltas} require fewer bytes
than the values themselves, the combined size after compression can be much
smaller. Figure~\ref{fig:bdc-example2} shows the compression of an example
64-byte cache line from the \emph{PageViewCount (PVC)} application using BDI.
As Figure~\ref{fig:bdc-example2} indicates, in this case, the cache line can be
represented using two bases (an 8-byte base value, $0x8001D000$, and an implicit
zero value base) and an array of eight 1-byte differences from these bases. As a
result, the entire cache line data can be represented using 17 bytes instead of
64 bytes (1-byte metadata, 8-byte base, and eight 1-byte deltas), saving 47
bytes of the originally used space.

Our example implementation of the BDI compression algorithm~\cite{bdi} views a
cache line as a set of fixed-size values i.e., 8 8-byte, 16 4-byte, or 32 2-byte
values for a 64-byte cache line.  For the size of the deltas, it considers three
options: 1, 2 and 4 bytes.  \comm{Note that all potential compressed sizes are
known statically after compression.} The key characteristic of BDI, which makes
it a desirable compression algorithm to use with the \SADA framework, is its
fast parallel decompression that can be efficiently mapped into instructions
that can be executed on GPU hardware. Decompression is simply a masked vector
addition of the deltas to the appropriate bases~\cite{bdi}.

\textbf{Mapping BDI to CABA.} In order to implement BDI with the \SADA
framework, we need to map the BDI compression/decompression algorithms into GPU
instruction subroutines (stored in the AWS and deployed as \helperwarps).

\textbf{Decompression.} To decompress the data compressed with BDI, we need a
simple addition of deltas to the appropriate bases. The CABA decompression
subroutine first loads the words within the compressed cache line into assist
warp registers, and then performs the base-delta additions in parallel, employing the
wide ALU pipeline.\footnote{Multiple instructions are required if the number of
deltas exceeds the width of the ALU pipeline. We use a 32-wide pipeline.} The
subroutine then writes back the uncompressed cache line to the cache. It skips
the addition for the lanes with an implicit base of zero by updating the active
lane mask\comm{manipulating the per-lane predicate registers~\cite{pguide}}
based on the cache line encoding.  We store a separate subroutine for each
possible BDI encoding that loads the appropriate bytes in the cache line as the
base and the deltas. \highlight{The high-level algorithm for decompression is
presented in Algorithm 1.}

\begin{algorithm} \caption{BDI: Decompression} \small{
\begin{algorithmic}[1] \State load \emph{base, deltas} \State
\emph{uncompressed\_data} $=$ \emph{base} $+$ \emph{deltas} \State store
\emph{uncompressed\_data} \end{algorithmic} } \end{algorithm} 

\textbf{Compression.} To compress data, the \SADA compression subroutine tests
several possible encodings (each representing a different size of base and
deltas) in order to achieve a high compression ratio. The first few bytes (2--8
depending on the encoding tested) of the cache line are always used as the base.
\purple{Each possible encoding is tested to check whether the cache line can be
successfully encoded with it. In order to perform compression at a warp
granularity, we need to check whether all of the words at every SIMD lane were
successfully compressed. In other words, if any one word cannot be compressed,
that encoding cannot be used across the warp. We can perform this check by
adding a global predicate register, which stores the logical AND of the per-lane
predicate registers.\comm{Each option is tested to check whether the line is
compressible using the specific encoding. The compressibility of all lanes for a
specific encoding, in parallel, is detected using a global predicate register.}
We observe that applications with homogeneous data structures can typically use the same
encoding for most of their cache lines~\cite{bdi}. We use this observation to reduce the
number of encodings we test to just one in many cases.} \comm{In BDI, the
number of instructions required to perform the compression depends on the
compressibility of the data.  If more encodings are tested, a better compression
ratio can be obtained at the cost of higher latency. The encoding complexity can
be decided based on the requirement at run time. If we just use one encoding
this would translate to approximately three operations. After compression, the
first few bytes (2--8 depending on the encoding used) of the cache line are
always used as the base.} All necessary operations are done in parallel using
the full width of the GPU SIMD pipeline.
\highlight{The high-level algorithm for compression is presented in Algorithm 2.}

\begin{algorithm} \caption{BDI: Compression} \small{
\begin{algorithmic}[1] \For{\emph{each base\_size}} \State load \emph{base,
values} \For{\emph{each delta\_size}} \State \emph{deltas} $=$ \emph{abs(values -
base)} \If{\emph{size(deltas) $<=$ delta\_size}} \State store \emph{base, deltas}
\State \textbf{exit} \EndIf \EndFor \EndFor \end{algorithmic} } \end{algorithm}

\subsubsection{Implementing Other Algorithms.} \label{sec:Other_Algos} The BDI
compression algorithm is naturally amenable towards implementation using assist
warps because of its data-parallel nature and simplicity.  The CABA framework
can also be used to realize other algorithms. The challenge in implementing
algorithms like FPC~\cite{fpc-tr} and C-Pack~\cite{c-pack}\comm{Our
technical report~\cite{caba-tr} and the original works~\cite{fpc-tr,c-pack}
provide more details on the specifics of these algorithms.}, which have
variable-length compressed words, is primarily in the placement of compressed
words within the compressed cache lines.  In BDI, the compressed words are in
\emph{fixed} locations within the cache line and, for each encoding, all the
compressed words are of the same size and can, therefore, be processed in
parallel. In contrast, C-Pack may employ multiple dictionary values as opposed
to just one base in BDI. In order to realize algorithms with \emph{variable
length words} and \emph{dictionary values} with assist warps, we leverage the
coalescing/address generation logic~\cite{coal1,coal2} already available in the
GPU cores. We make two minor modifications to these
algorithms~\cite{fpc-tr,c-pack} to adapt them for use with \SADA. First, similar
to prior works~\cite{c-pack,fpc-tr,MMCompression}, we observe that few encodings
are sufficient to capture almost all the data redundancy. In addition, the
impact of any loss in compressibility due to fewer encodings is minimal as the
benefits of bandwidth compression are only at multiples of a single DRAM burst
(e.g., 32B for GDDR5~\cite{GDDR5}). We exploit this to reduce the number of
supported encodings.  \comm{reduce the number of encodings supported, as we
observe that just a few encodings are sufficient to capture almost all of the
redundancy in data.\footnote{Prior works~\cite{c-pack,fpc-tr,MMCompression} made
a similar observation.}} Second, we place all the metadata containing the
compression encoding at the \emph{head} of the cache line to be able to determine how
to decompress the entire line \emph{upfront}. In the case of C-Pack, we place the
dictionary entries after the metadata.  

We note that it can be challenging to implement complex algorithms efficiently
with the simple computational logic available in GPU cores.  Fortunately, there
are already Special Function Units (SFUs)~\cite{sfu,sfu2} present in the GPU
SMs, used to perform efficient computations of elementary mathematical
functions.  SFUs could potentially be extended to implement primitives that
enable the fast iterative comparisons performed frequently in some compression
algorithms. This would enable more efficient execution of the described
algorithms, as well as implementation of more complex compression algorithms,
using \SADA. We leave the exploration of an SFU-based approach to future work. 

\highlight{We now present a detailed overview of mapping the FPC and C-PACK algorithms into
assist warps.}

\textbf{Implementing the FPC (Frequent Pattern Compression)
Algorithm.} For FPC, the cache line is treated as set of
fixed-size words and each word within the cache line is compressed into a simple
{prefix or encoding and a compressed word if it matches a set of frequent
patterns, e.g. narrow values, zeros or repeated bytes. The word is left
uncompressed if it does not fit any pattern. We refer the reader to the original
work~\cite{fpc-tr} for a more detailed description of the original algorithm.

The challenge in mapping assist warps to the FPC
decompression algorithm is in the serial sequence in which each word within a
cache line is decompressed. This is because in the original proposed version,
each compressed word can have a different size. To determine the location of a
specific compressed word, it is necessary to have decompressed the previous
word. We make some modifications to the algorithm in order to parallelize the decompression across
different lanes in the GPU cores. First, we move the word prefixes (metadata)
for each word to the front of the cache line, so we know \emph{upfront} how to
decompress the rest of the cache line. Unlike with BDI, each word within the
cache line has a different encoding and hence a different compressed word length
and encoding pattern. This is problematic as statically storing the sequence of
decompression instructions for every combination of patterns for all the words
in a cache line would require very large instruction storage. In order to
mitigate this, we break each cache line into a number of segments. Each segment
is compressed independently and all the words within each segment are compressed
using the \emph{same encoding} whereas different segments may have different
encodings. This creates a trade-off between
simplicity/parallelizability versus compressibility. Consistent with
previous works~\cite{fpc-tr}, we find that this doesn't significantly impact compressibility. 

\textbf{Decompression.} The high-level algorithm we use for decompression is presented
in Algorithm 3. Each segment within the compressed cache line is loaded in series. Each
of the segments is decompressed in parallel{\textemdash}this is possible because
all the compressed words within the segment have the same encoding. The
decompressed segment is then stored before moving onto the next segment. The
location of the next compressed segment is computed based on the size of the
previous segment.

 \begin{algorithm} \caption{FPC: Decompression} \small{ \begin{algorithmic}[1] \For{\emph{each segment}}
\State load \emph{compressed words} \State \emph{pattern specific decompression (sign
extension/zero value)} \State store \emph{decompressed words} \State
\emph{segment-base-address} $=$ \emph{segment-base-address} $+$ \emph{segment-size}
\EndFor \end{algorithmic}}  \end{algorithm} 

\textbf{Compression.} Similar to the BDI implementation, we loop through and test
different encodings for each segment. We also compute the address offset for each
segment at each iteration to store the compressed words in the appropriate
location in the compressed cache line. Algorithm 4 presents the high-level FPC
compression algorithm we use. 
 
\begin{algorithm} \caption{FPC: Compression} \small{ \begin{algorithmic}[1]
\State load \emph{words}
\For{\emph{each segment}}  \For{\emph{each encoding}}  \State \emph{test
encoding} \If{\emph{compressible}} \State \emph{segment-base-address} $=$
\emph{segment-base-address} $+$ \emph{segment-size} \State store \emph{compressed
words} \State \textbf{break} \EndIf \EndFor \EndFor \end{algorithmic} }
\end{algorithm}

\textbf {Implementing the C-Pack Algorithm.} C-Pack~\cite{c-pack} is a
dictionary based compression
algorithm where frequent "dictionary" values are saved at the beginning of the
cache line. The rest of the cache line contains encodings for each word which may
indicate zero values, narrow values, full or partial matches into the
dictionary or simply that the word is uncompressible. 

In our implementation, we reduce
the number of possible encodings to partial matches (only last byte mismatch),
full word match, zero value and zero extend (only last byte) and we limit the
number of dictionary values to 4. This enables fixed compressed word size within
the cache line. A fixed compressed word size enables compression and decompression
of different words within the cache line in parallel. If the number of required
dictionary values or uncompressed words exceeds 4, the line is left
decompressed. This is, as in BDI and FPC, a trade-off between simplicity and
compressibility. In our experiments, we find that it does not significantly
impact the compression ratio{\textemdash}primarily due the 32B minimum data size
and granularity of compression. 

\textbf{Decompression.} As described, to enable parallel decompression, we place the
encodings and dictionary values at the head of the line. We also limit the
number of encodings to enable quick
decompression. We implement C-Pack decompression as a series of instructions
(one per encoding used) to load all the registers with the appropriate
dictionary values. We define the active lane mask based on the encoding (similar
to the mechanism used in BDI) for each load instruction to ensure the correct
word is loaded into each lane's register. Algorithm 5 provides the high-level
algorithm for C-Pack decompression.  

\begin{algorithm} \caption{C-PACK: Decompression} \small{ \begin{algorithmic}[1] \State add \emph{base-address}
$+$ \emph{index-into-dictionary} \State load \emph{compressed words} \For{each encoding}
\State \emph{pattern specific decompression} \Comment {Mismatch byte load for zero
extend or partial match} \EndFor \State Store \emph{uncompressed words} \end{algorithmic} }
\end{algorithm}

\textbf{Compression.} Compressing data with C-Pack involves determining the
dictionary values that will be used to compress the rest of the line. In our
implementation, we serially add each word from the beginning of the cache
line to be a dictionary value if it was not already covered by a previous
dictionary value. For each dictionary value, we test whether the rest of the
words within the cache line is compressible. The next dictionary value is
determined using the predicate register to determine the next uncompressed
word, as in BDI. After four iterations (dictionary values), if all the
words within the line are not compressible, the cache line is left uncompressed.
Similar to BDI, the global predicate register is used to determine the
compressibility of all of the lanes after four or fewer iterations. Algorithm 6
provides the high-level algorithm for C-Pack compression. 

\begin{algorithm} \caption{C-PACK: Compression} \small{ \begin{algorithmic}[1]
\State load \emph{words} \For{each
dictionary value (including zero)} \Comment {To a maximum of four} \State \emph{test match/partial match}
\If{\emph{compressible}} \State Store \emph{encoding and mismatching
byte} \State
\textbf{break} \EndIf \EndFor \If{\emph{all lanes are compressible}} \State Store
\emph{compressed cache line} \EndIf \end{algorithmic} }
\end{algorithm} }

\subsubsection{Walkthrough of \SADA-based Compression} 
\begin{figure}[h] \centering 
\includegraphics[width=0.69\textwidth]{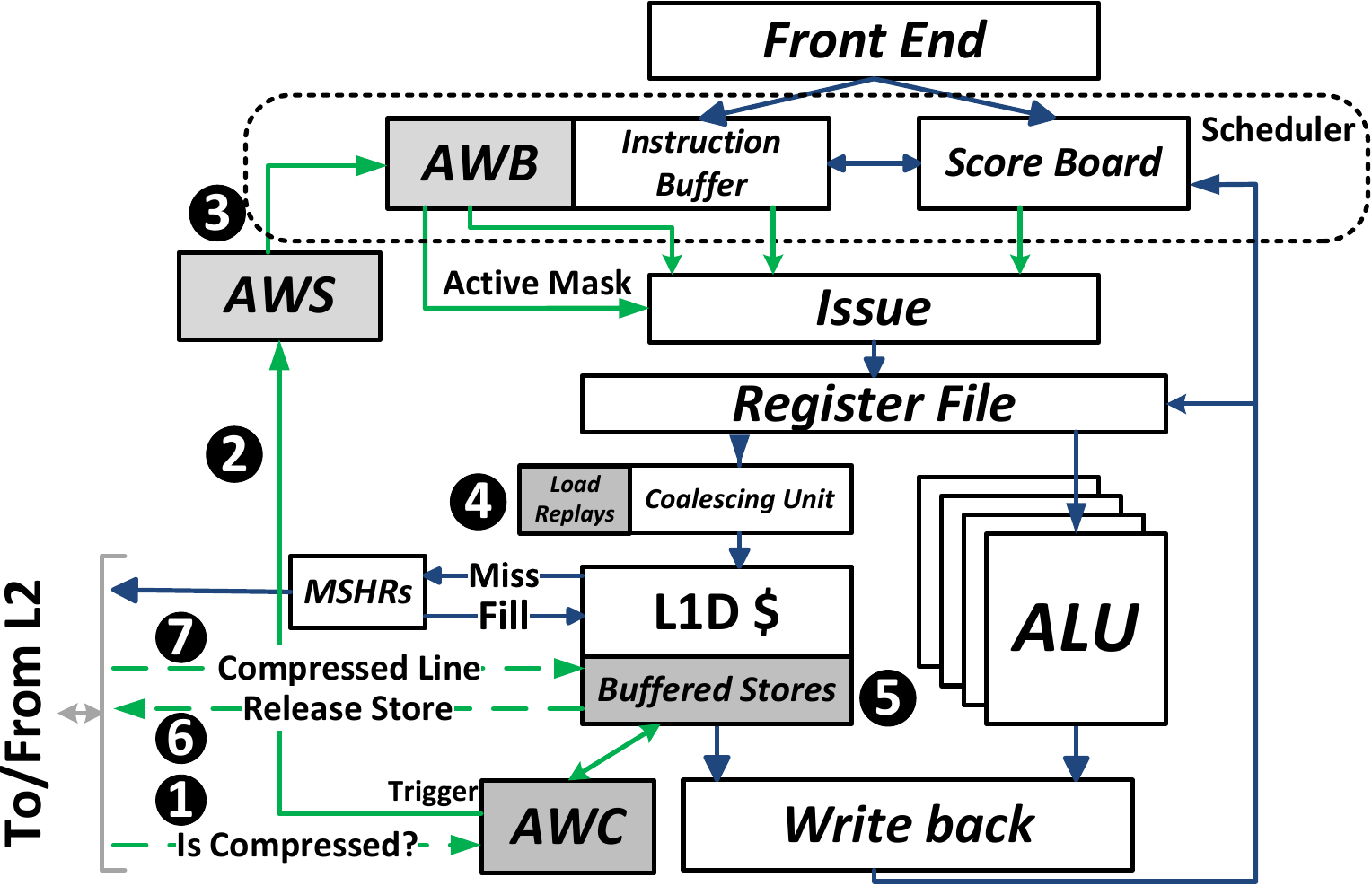}
\caption{Walkthrough of \SADA-based Compression.}
\label{fig:load} 
\end{figure}

We show the detailed
operation of \SADA-based compression and decompression mechanisms in
Figure~\ref{fig:load}. 
We assume a baseline GPU architecture with three levels in the memory hierarchy
-- two levels of caches (private L1s and a shared L2) and main memory.
Different levels can potentially store compressed data. In this section and in
our evaluations, we assume that only the L2 cache and main memory contain
compressed data.\comm{\footnote{Different memory spaces exist in GPU
architectures.  In this work we apply compression only to global memory.}} Note
that \comm{even though the data is stored in the compressed form in both main
memory and L2 cache,} there is no capacity benefit in the baseline mechanism as
compressed cache lines still occupy the full uncompressed slot, i.e., we only
evaluate the bandwidth-saving benefits of compression in GPUs.

\textbf{The Decompression Mechanism.} Load instructions that access
global memory data in the compressed form trigger the appropriate \helperwarp to
decompress the data before it is used. The subroutines to decompress data are
stored in the \emph{Assist Warp Store (AWS)}. The AWS is indexed by the
compression encoding at the head of the cache line and by a bit indicating
whether the instruction is a load (decompression is required) or a store
(compression is required). Each decompression assist warp is given \emph{high
priority} and, hence, stalls the progress of its parent warp until it completes
its execution. This ensures that the parent warp correctly gets the decompressed
value. 

\comm{The AWS feeds into the AWB where it  available and active for scheduling.} 

\textbf{L1 Access.} We store data in L1 in the uncompressed form.  An L1 hit
does not require an assist warp for decompression.
 
\textbf{L2/Memory Access.} Global memory data cached in L2/DRAM could
potentially be compressed. A bit indicating whether the cache line is compressed
is returned to the core along with the cache line (\ding{202}). If the data is uncompressed,
the line is inserted into the L1 cache and the writeback phase resumes normally.
If the data is compressed, the compressed cache line is inserted into the L1
cache. The encoding of the compressed cache line and the warp ID
are relayed to the Assist Warp Controller (AWC), which then triggers the AWS
(\ding{203}) to deploy the appropriate \helperwarp (\ding{204}) to decompress the line. During regular execution, the load information for each thread is
buffered in the coalescing/load-store unit~\cite{coal1,coal2} until all the data
is fetched. We continue to buffer this load information (\ding{205}) until the
line is decompressed. 

%
After the \SADA decompression subroutine ends execution, the original load that
triggered decompression is resumed (\ding{205}).

\comm{The registers are released at this point and regular execution is resumed
for the warp. }

\textbf{The Compression Mechanism.} \label{sec:comp} The \helperwarps to
perform compression are triggered by store instructions. When data is written to
a cache line (i.e., by a store), the cache line can be written back to main
memory either in the compressed or uncompressed form.  Compression is off the
critical path and the warps to perform compression can be scheduled when the
required resources are available. 

\comm{\begin{figure}[t] \centering 
\includegraphics[width=0.62\textwidth]{figures/Store2.pdf} \vspace{-0.2cm}
\caption{Walkthrough of the Compression Mechanism.} \label{fig:store}
\end{figure}}

\comm{The pending stores, however, need to be buffered until the corresponding
cache lines can be compressed. We consider two possible options for where to
buffer this data: (i) a special store buffer or (ii) a few dedicated sets within
the L1 cache (\ding{202}).\footnote{L1 caches are usually too small to fit the
working set of our applications, and hence the impact on performance of this
reduction in size is minimal.}}Pending stores are buffered in a few dedicated
sets within the L1 cache or in available shared memory (\ding{206}). In the case
of an overflow in this buffer space (\ding{206}), the stores are released to the
lower levels of the memory system in the uncompressed form (\ding{207}).  Upon
detecting the availability of resources to perform the data compression, the AWC
triggers the deployment of the \helperwarp that performs compression
(\ding{203}) into the AWB (\ding{204}), with \emph{low priority}. The scheduler
is then free to schedule the instructions from the compression subroutine. Since
compression is not on the critical path of execution, keeping such instructions
as low priority ensures that the main program is not unnecessarily delayed.  

\purple{\textbf{L1 Access.} On a hit in the L1 cache, the cache line is already
available in the uncompressed form. Depending on the availability of resources,
the cache line can be scheduled for compression or simply written to the L2 and
main memory uncompressed, when evicted.}  \comm{A write-evict policy is used in
the L1 (cite).}

\textbf{L2/Memory Access.} Data in memory is compressed at the granularity of a full cache line, but stores
can be at granularities smaller than the size of the cache line. This poses some
additional difficulty if the destination cache line for a store is already
compressed in main memory.  Partial writes into a compressed cache line would
require the cache line to be decompressed first, then updated with the new data,
and  written back to main memory. The common case{\textemdash}where the cache
line that is being written to is uncompressed initially{\textemdash}can be easily
handled. However, in the worst case, the cache line being partially written to
is already in the compressed form in memory. We now describe the mechanism to
handle both these cases.

Initially, to reduce the store latency, we assume that the cache line is uncompressed, and issue a store to the lower levels of the memory hierarchy, while buffering a copy in L1. If the cache line is found in L2/memory in the
uncompressed form (\ding{202}), the assumption was correct. The store then proceeds normally and the buffered stores are evicted from L1. If
the assumption is incorrect, the cache line
is
retrieved (\ding{208}) and decompressed before the store is retransmitted to the lower levels of the memory hierarchy.

\comm{If the cache line is found in L2/memory in the
uncompressed form (\ding{202}), the store proceeds normally. If the cache line
is already compressed, then it needs to be reloaded into the L1D (\ding{208})
and decompressed at the core before the store can be completed.  The store in
both these cases is buffered at the L1, but to reduce the store latency, is still transmitted to the lower
levels of the memory hierarchy assuming that the data is not compressed. If
this assumption is incorrect, we drop the store transaction and the cache line
is
retrieved, and then decompressed before the store is resumed. On the other
hand, if the cache line was correctly assumed to be uncompressed, a signal is
transmitted back to the core and the stores buffered at the L1 are
evicted.}

\textbf{Realizing Data Compression.} Supporting data compression requires
additional support from the main memory controller and the runtime system, as we
describe below. 
 
\textbf{Initial Setup and Profiling.} Data compression with \SADA
requires a one-time data setup before the data is transferred to the GPU.
\comm{Typically, the input data that the GPU kernels operate on is initialized
by the CPU, stored in CPU main memory, and then transferred to the GPU DRAM by
invoking a memory copy function (e.g., \emph{``CUDAMemCopy''} in
CUDA~\cite{pguide}). The data is transferred to GPU DRAM across the PCI Express
link. }We assume initial software-based data preparation where the input data is
stored in CPU memory in the compressed form with an appropriate compression
algorithm before transferring the data to GPU memory. Transferring data in the
compressed form can also reduce PCIe bandwidth usage.\footnote{This requires
changes to the DMA engine to recognize compressed lines.} 

Memory-bandwidth-limited GPU applications are the best candidates for employing
data compression using \SADA. The compiler (or the runtime profiler) is required
to identify those applications that are most likely to benefit from this
framework. For applications where memory bandwidth is not a bottleneck, data
compression is simply disabled.  \comm{Different kernels within an application
may also operate on different input data sets which may have varying levels of
compressibility. We use an initial sampling run to test the input data for
compressibility, and only compress those data sets which are most amenable. We
profile multiple different compression algorithms (e.g., BDI~\cite{bdi},
FPC~\cite{fpc}) to find the maximal opportunities for compression during the
profiling run.  }

\textbf{Memory Controller Changes.} \label{sec:mdcache} Data compression
reduces off-chip bandwidth requirements by transferring the same data in fewer
DRAM bursts. The memory controller (MC) needs to know whether the cache line
data is compressed and how many bursts (1--4 bursts in GDDR5~\cite{GDDR5}) are
needed to transfer the data from DRAM to the MC.  Similar to prior
work~\cite{GPUBandwidthCompression,lcp-micro}, we require metadata information
for every cache line that keeps track of how many bursts are needed to transfer
the data.\comm{For a maximum of 4 bursts, we need two bits of metadata for every
128B cache line.  Assuming a 4GB DRAM, we need $4GB / 128B * 2b = 8MB$ of
metadata space in DRAM.} Similar to prior work~\cite{GPUBandwidthCompression},
we simply reserve 8MB of GPU DRAM space for the metadata (\textasciitilde{0.2\%}
of all available memory).  Unfortunately, this simple design would require an
additional access for the metadata for every access to DRAM effectively doubling
the required bandwidth. To avoid this, a simple \emph{metadata (MD) cache} that
keeps frequently-accessed metadata on chip (near the MC) is required. Note that
this metadata cache is similar to other metadata storage and caches proposed for
various purposes in the memory controller,
e.g.,\cite{lcp-micro,page-overlays,enabling-meza-cal12,qureshi20,bus-energy,raidr}. Our
experiments show that a small 8 KB 4-way associative MD cache is sufficient to
provide a hit rate of 85\% on average (more than 99\% for many applications)
across all applications in our workload pool.\footnote{For applications where MD
cache miss rate is low, we observe that MD cache misses are usually also TLB
misses. Hence, most of the overhead of MD cache misses in these applications is
outweighed by the cost of page table lookups.} Hence, in the common case, a
second access to DRAM to fetch compression-related metadata can be avoided.

\subsection{Use Cases}
\label{sec:Applications}
\subsubsection{Memoization} Hardware memoization is a technique used to avoid
redundant computations by reusing the results of previous computations that have
the same or similar inputs. Prior work~\cite{Arnau-memo, memoing2, sage}
observed redundancy in inputs to data in GPU workloads. In applications limited
by available compute resources, memoization offers an opportunity to trade off
computation for storage, thereby enabling potentially higher energy efficiency and performance. In order to realize memoization in hardware, a look-up
table (LUT) is required to dynamically cache the results of computations as well
as the corresponding inputs. The granularity of computational reuse can be at
the level of fragments~\cite{Arnau-memo}, basic blocks,
functions~\cite{alvarez-reuse,block-reuse,alvarez-reuse2,dynreuse,danconnors}, or
long-latency instructions~\cite{memo-inst}.  The CABA framework is a natural
way to implement such an optimization. The availability of on-chip memory lends
itself for use as the LUT. In order to cache previous results in on-chip memory,
look-up tags (similar to those proposed in~\cite{unified-register}) are required
to index correct results. With applications tolerant of
approximate results (e.g., image processing, machine learning, fragment
rendering kernels), the computational inputs can be hashed to reduce the size of
the LUT.  Register values, texture/constant memory or global memory sections
that are not subject to change are potential inputs. \purple{An assist warp can be
employed to perform memoization in the following way: (1) compute the
hashed value for look-up at predefined trigger points, (2) use the
load/store pipeline to save these inputs in available shared memory, and (3) eliminate redundant computations by loading the
previously computed results in the case of a hit in the LUT.}

\subsubsection{Prefetching} Prefetching has been explored in the context of
GPUs~\cite{Meng,orchestrated-jog-isca13,owl-jog-asplos13,spareregister-lakshminarayana-hpca14,manythread-lee-micro10,Arnau,apogee-sethia-pact13} with the
goal of reducing effective memory latency. With memory-latency-bound applications, the load/store pipelines can be employed by the CABA
framework to perform opportunistic prefetching into GPU caches. The CABA framework can potentially enable the effective use of prefetching in GPUs due to several reasons: (1) Even simple prefetchers such as the
stream~\cite{stream1,stream2,fdp} or stride~\cite{stride1,stride2} prefetchers
are non-trivial to implement in GPUs since access patterns need to be tracked
and trained at the granularity of warps~\cite{apogee-sethia-pact13,manythread-lee-micro10}. CABA could enable
fine-grained book-keeping by using spare registers and assist warps to save
metadata for each warp. The computational units could then be used to
continuously compute strides in access patterns both within and across warps.
(2) It has been demonstrated that software prefetching and helper
threads~\cite{ht2,ht5, ht7, \comm{ht10,}ht12, ht9, ht7, ht1,spareregister-lakshminarayana-hpca14} are very
effective in performing prefetching for irregular access patterns. Assist warps
offer the hardware/software interface to implement application-specific
prefetching algorithms with varying degrees of complexity without the additional
cost of hardware implementation. (3) In bandwidth-constrained GPU systems,
uncontrolled prefetching could potentially flood the off-chip buses, delaying
demand requests. CABA can enable flexible prefetch throttling
(e.g.,~\cite{ebrahimi-micro09,ebrahimi-isca11,fdp}) by scheduling assist warps that perform prefetching, only when the memory pipelines are idle.  (4) Prefetching
with CABA entails using load or prefetch instructions, which not only
enables prefetching to the hardware-managed caches, but also simplifies usage of
unutilized shared memory or register file as prefetch buffers.

\subsubsection{Redundant Multithreading} Reliability of GPUs is a key concern,
especially today when they are popularly employed in many supercomputing
systems. Ensuring hardware protection with dedicated resources can be
expensive~\cite{characterizing-luo-dsn14}.
Redundant multithreading~\cite{gpu-rmt,rmt-alternatives,Qureshi-rmt} is an approach where redundant threads are used to
replicate program execution. The results are compared at different points in
execution to detect and potentially correct errors. The CABA framework can be
extended to redundantly execute portions of the original
program via the use of such approaches to increase the reliability
of GPU architectures.

\subsubsection{Speculative Precomputation} In CPUs, speculative
multithreading (~\cite{ht16, ht17, ht18}) has
been proposed to speculatively parallelize serial code and verify the
correctness later. Assist
warps can be employed in GPU architectures to speculatively
pre-execute sections of code during idle cycles to further improve parallelism
in the program execution. Applications tolerant to approximate results could
particularly be amenable towards this optimization~\cite{rfvp-yazdanbakhsh-taco16}.

\subsubsection{Handling Interrupts and Exceptions.} Current GPUs do not implement support for
interrupt handling except for some support for timer interrupts used
for application time-slicing~\cite{fermi}. CABA offers a natural mechanism for
associating architectural events with subroutines to be executed in
throughput-oriented architectures where thousands of threads could be active at
any given time. Interrupts and exceptions can be handled by special assist
warps, without requiring complex context switching or heavy-weight kernel
support. 
  
\subsubsection{Profiling and Instrumentation} 
Profiling and binary instrumentation tools like Pin~\cite{pin} and Valgrind~\cite{valgrind}
proved to be very useful for development, performance analysis and debugging
on modern CPU systems. At the same time, there is a lack~\footnote{With the 
exception of one recent work~\cite{gpuPin}.} of tools with
same/similar capabilities for modern GPUs. This significantly limits software development
and debugging for modern GPU systems. The CABA framework can potentially enable easy and efficient
development of such tools, as it is flexible enough to invoke user-defined
code on specific architectural
events (e.g., cache misses, control divergence).

\subsection{Methodology} \label{sec:methodology_caba}

We model the \SADA framework in GPGPU-Sim 3.2.1~\cite{GPGPUSim}.
Table~\ref{tab:meth} provides the major parameters of the simulated system. We
use GPUWattch~\cite{gpuwattch} to model GPU power and CACTI~\cite{cacti} to
evaluate the power/energy overhead  associated with the MD cache
(Section~\ref{sec:mdcache}) and the additional components (AWS and AWC) of the CABA framework.  We implement BDI~\cite{bdi} using the Synopsys Design
Compiler with 65nm library (to evaluate the energy overhead of
compression/decompression for the dedicated hardware design for comparison to
\SADA), and then use ITRS projections~\cite{ITRS} to scale our results to the
32nm technology node.

\begin{table}[h!] 
	\centering
	\begin{tabular}{ll} 
\toprule System Overview           &  15 SMs, 32 threads/warp,  6 memory channels\\ 
\cmidrule(rl){1-2} Shader Core Config &  1.4GHz, GTO scheduler~\cite{tor-micro12}, 2 schedulers/SM\\
\cmidrule(rl){1-2} Resources / SM     &  48 warps/SM, 32768 registers, 32KB Shared Memory \\ 
\cmidrule(rl){1-2} L1 Cache    &  16KB, 4-way associative, LRU replacement policy    \\ 
\cmidrule(rl){1-2} L2 Cache   &  768KB, 16-way associative, LRU replacement policy  \\ 
\cmidrule(rl){1-2} Interconnect   &  1 crossbar/direction (15 SMs, 6 MCs), 1.4GHz  \\ 
\cmidrule(rl){1-2} Memory Model  &  177.4GB/s BW, 6 GDDR5 Memory Controllers (MCs),\\ 
		& FR-FCFS scheduling, 16 banks/MC\comm{, 924 MHz} \\ 
\cmidrule(rl){1-2}GDDR5 Timing~\cite{GDDR5}  & $t_{CL}=12,:t_{RP}=12,:t_{RC}=40,:t_{RAS}=28,$\\
		&$t_{RCD}=12,:t_{RRD}=6:t_{CLDR}=5:t_{WR}=12$ \\
	\bottomrule \end{tabular}%
 \caption{Major parameters of the simulated systems.} 
\label{tab:meth}%
\end{table}%

\textbf{Evaluated Applications.} We use a number of CUDA applications derived
from CUDA SDK~\cite{sdk} (\emph{BFS, CONS, JPEG, LPS, MUM, RAY, SLA, TRA}),
Rodinia~\cite{rodinia} (\emph{hs, nw}), Mars~\cite{mars} (\emph{KM, MM, PVC, PVR, SS}) and
lonestar~\cite{lonestar} (\emph{bfs, bh, mst, sp, sssp}) suites.\comm{Note that
CUDA SDK implementation of breadth first search is different than that of
lonestar.}  We run all applications to completion or for 1 billion instructions
(whichever comes first).  \SADA-based data compression is beneficial mainly for
memory-bandwidth-limited applications.  In computation-resource limited applications,
data compression is not only unrewarding, but it can also cause significant
performance degradation due to the computational overheads associated with
assist warps. We rely on static profiling to identify memory-bandwidth-limited
applications and disable \SADA-based compression for the others.\comm{\footnote{The \SADA compression framework can
potentially adapt on-the-fly to different workloads and phases within a
workload.}} In our evaluation (Section~\ref{sec:Results}), we demonstrate
detailed results for applications that exhibit some compressibility in
memory bandwidth (at least 10\%). Applications without compressible data (e.g., sc, SCP) do not gain
any performance from the \SADA framework, and we verified that these
applications do not incur any performance degradation (because the assist warps
are \emph{not} triggered
for them).

\textbf{Evaluated Metrics.} We present Instruction per Cycle (\emph{IPC}) as the
primary performance metric. We also use \emph{average bandwidth
utilization}, defined as the fraction of total DRAM cycles that the DRAM data bus
is busy, and \emph{compression ratio}, defined as the ratio of the number of DRAM
bursts required to transfer data in the compressed vs. uncompressed form. As
reported in prior work~\cite{bdi}, we use decompression/compression latencies of 1/5
cycles for the hardware implementation of BDI.

\subsection{Results}
\label{sec:Results}
To evaluate the effectiveness of using \SADA to employ data compression, we
compare five different designs:
(i) \emph{Base} - the baseline system with no compression,
(ii) \emph{HW-BDI-Mem} - hardware-based \emph{memory bandwidth compression} with dedicated logic
(data is stored compressed in main memory but uncompressed in the last-level cache, similar to prior works~\cite{GPUBandwidthCompression,lcp-micro}),
(iii) \emph{HW-BDI} - hardware-based \emph{interconnect and memory bandwidth compression} 
(data is stored uncompressed only in the L1 cache)
(iv) \emph{CABA-BDI} - \SADAfull (\SADA) framework (Section~\ref{sec:idea}) with all associated overheads
of performing compression (for both interconnect and memory bandwidth),
(v) \emph{Ideal-BDI} - compression (for both interconnect and memory) with no
latency/power overheads for compression or decompression. This section provides our major results and analyses.\comm{All designs (unless stated otherwise)
  employ BDI compression algorithm~\cite{bdi}.}



\subsubsection{Effect on Performance and Bandwidth Utilization}

\begin{figure}[h!]
  \centering
  \includegraphics[width=0.69\textwidth]{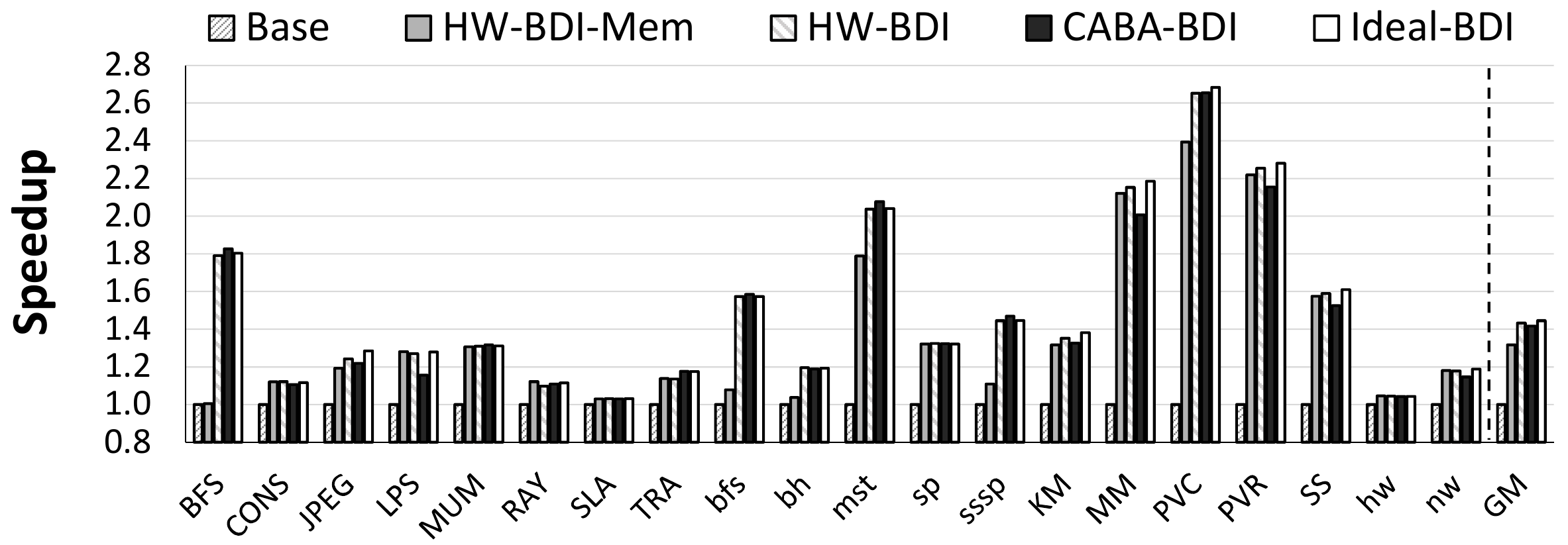}
  \caption{Normalized performance.}
  \label{fig:performance}
\end{figure}

\begin{figure}[h!]
  \centering
  \includegraphics[width=0.69\textwidth]{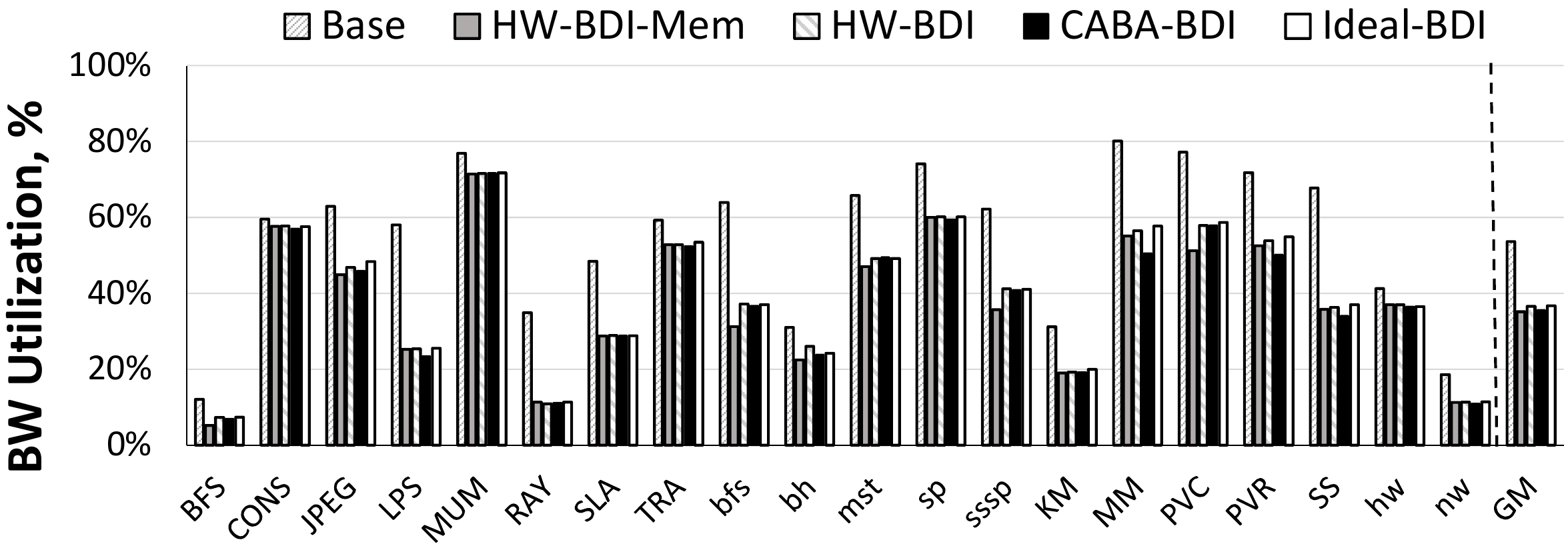}
  \caption{Memory bandwidth utilization.}
  \label{fig:bwutil}
\end{figure}

Figures~\ref{fig:performance} and~\ref{fig:bwutil} show, respectively,
the normalized performance (vs. \emph{Base}) and the memory bandwidth
utilization of the five designs.
We make three major observations.

First, all compressed designs are effective in providing high
performance improvement over the baseline. Our approach (CABA-BDI)
provides a 41.7\% average improvement, which is only 2.8\% less than
the ideal case (Ideal-BDI) with none of the overheads associated with
\SADA. CABA-BDI's performance is 9.9\% better than the
previous~\cite{GPUBandwidthCompression} hardware-based memory
bandwidth compression design (HW-BDI-Mem), and \emph{only} 1.6\% worse
than the purely hardware-based design (HW-BDI) that performs both
interconnect and memory bandwidth compression. We conclude that our
framework is effective at enabling the benefits of compression without
requiring specialized hardware compression and decompression logic.

Second, performance benefits, in many workloads, correlate with the reduction in memory bandwidth utilization. For a fixed amount of data,
compression reduces the bandwidth utilization, and, thus, increases
the effective available bandwidth. Figure~\ref{fig:bwutil} shows that
CABA-based compression 1) reduces the average memory bandwidth
utilization from 53.6\% to 35.6\% and 2) is effective at alleviating the
memory bandwidth bottleneck in most workloads.\comm{Applications where bandwidth demand is high and where
  data is highly compressible with BDI compression algorithm (see
  Figure~\ref{fig:ratio}) are the ones that usually benefit most. For
  example, all applications from the Mars suite, \emph{bfs},
  \emph{mst}, and \emph{ssst} from the lonestar suite exhibit
  significant reduction in bandwidth utilization due to high data
  compressibility (up to 3.9X compression ratio for \emph{PVC}).
  This, in turn, results in high performance improvements for these
  applications (e.g., 2.6X performance improvement in \emph{PVC})} In
some applications (e.g., \emph{bfs} and \emph{mst}), designs that
compress \emph{both} the on-chip interconnect and the memory
bandwidth, i.e. CABA-BDI and HW-BDI, perform better than the design
that compresses only the memory bandwidth (HW-BDI-Mem). Hence, \SADA
seamlessly enables the mitigation of the interconnect bandwidth
bottleneck as well, since data compression/decompression is flexibly
performed at the cores.

Third, for some applications, CABA-BDI performs slightly (within 3\%) better
 than Ideal-BDI and HW-BDI. The reason for this
counter-intuitive result is the effect of warp
oversubscription~\cite{tor-micro12,nmnl-pact13,Kayiran-micro2014,medic}. In these cases, too
many warps execute in parallel, polluting the last level
cache. CABA-BDI sometimes reduces pollution as a side effect of
performing more computation in assist warps, which slows down the
progress of the parent warps.


We conclude that the \SADA framework can effectively enable data
compression to reduce both on-chip interconnect and off-chip memory
bandwidth utilization, thereby improving the performance of modern
GPGPU applications.

%

\subsubsection{Effect on Energy}

\begin{figure}[h]
  \centering
  \includegraphics[width=0.69\textwidth]{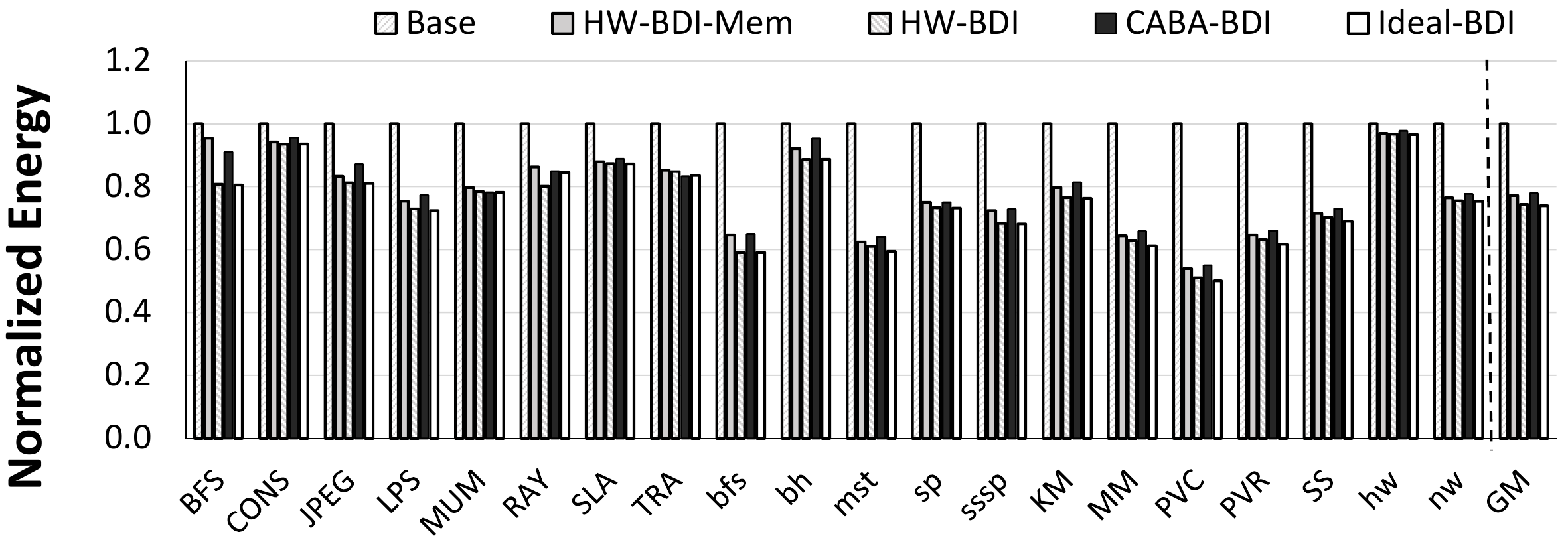}
  \caption{Normalized energy consumption.} 
  \label{fig:energy_caba}
\end{figure}

Compression decreases energy consumption in two ways: 1) by reducing
bus energy consumption, 2) by reducing execution time.
Figure~\ref{fig:energy_caba} shows the normalized energy consumption of the
five systems. We model the static and dynamic energy of the cores,
caches, DRAM, and all buses (both on-chip and off-chip), as well as
the energy overheads related to compression: metadata (MD) cache and
compression/decompression logic. We make two major observations.

First, CABA-BDI reduces energy consumption by as much as 22.2\% over
the baseline. This is especially noticeable for memory-bandwidth-limited
applications, e.g., \emph{PVC}, \emph{mst}.  This is a result of two
factors: (i) the reduction in the amount of data transferred between
the LLC and DRAM (as a result of which we observe a 29.5\% average reduction in DRAM power)
and (ii) the reduction in total execution time.  This observation
agrees with several prior works on bandwidth
compression~\cite{lcp-micro,memzip}.  We conclude that the \SADA
framework is capable of reducing the overall system energy, primarily
by decreasing the off-chip memory traffic.

Second, CABA-BDI's energy consumption is only 3.6\% more than that of
the HW-BDI design, which uses dedicated logic for memory bandwidth
compression. It is also only 4.0\% more than that of the Ideal-BDI
design, which has no compression-related overheads.\comm{The reduction
  in the execution time (due to less data sent over on-chip
  interconnect) usually alleviates the penalty of core-execution. It
  is also the primary cause for lower energy consumption with CABA-BDI
  in \emph{BFS} and \emph{RAY}.}  CABA-BDI consumes more energy
because it schedules and executes assist warps, utilizing on-chip
register files, memory and computation units, which is less
energy-efficient than using dedicated logic for compression. \purple{However, as
results indicate, this additional energy cost is small compared to the
performance gains of CABA (recall, 41.7\% over Base), and may be
amortized by using CABA for other purposes as well (see
Section~\ref{sec:Applications}).}

{\bf Power Consumption.} CABA-BDI increases the system power consumption by 2.9\%
over the baseline (not graphed), mainly due to the additional hardware
and higher utilization of the compute pipelines.\comm{However, the
  overheads incurred by the CABA-BDI mechanism are outweighed by the
  energy savings achieved for both off-chip and on-chip buses.}
However, the power overhead enables energy savings by reducing
bandwidth use and can be amortized across other uses of CABA
(Section~\ref{sec:Applications}).

\begin{figure}[h]
  \centering
  \includegraphics[width=0.69\textwidth]{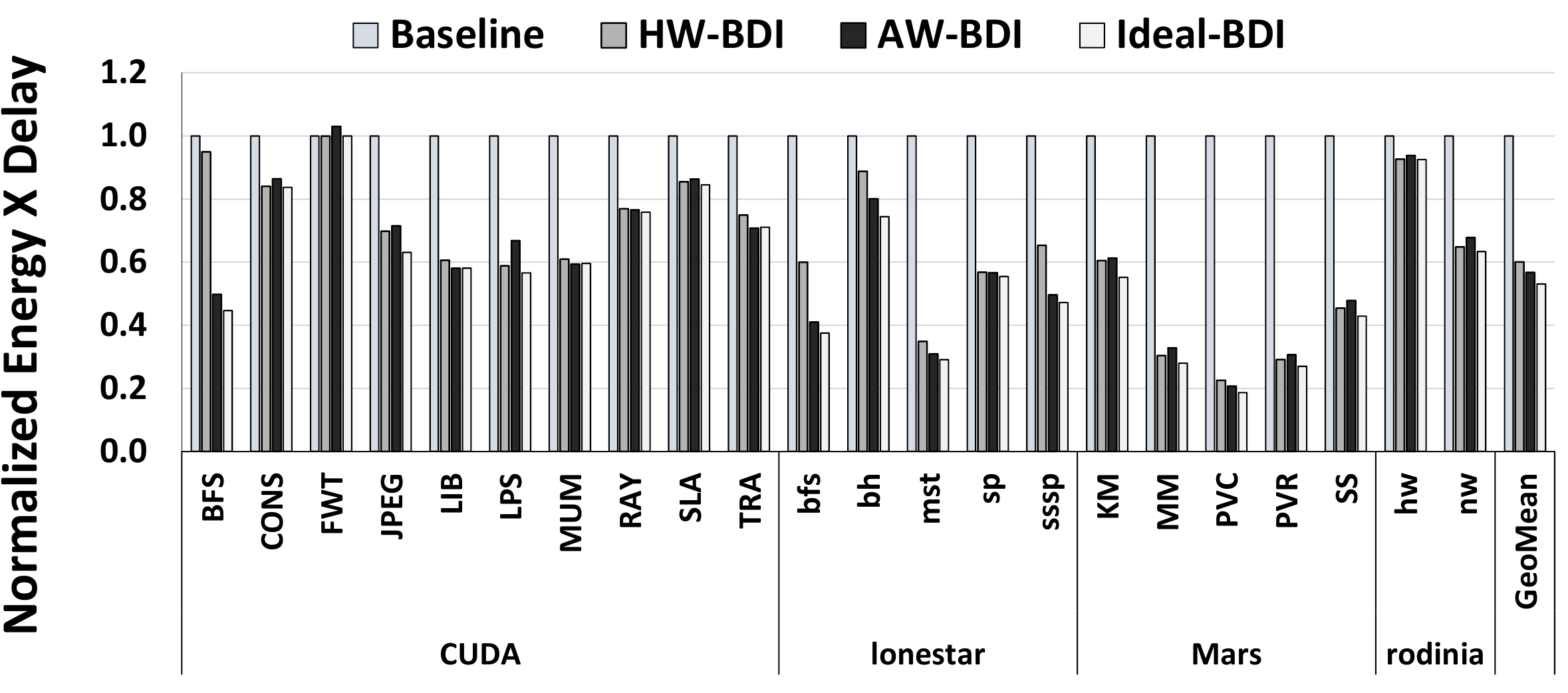}
  \caption{Energy-Delay product. }
  \label{fig:energy-perf}
\end{figure}

\highlight{ {\bf Energy-Delay product.} Figure~\ref{fig:energy-perf} shows the product of the
normalized energy consumption and normalized execution time for the evaluated
GPU workloads. This metric simultaneously captures two metrics of
interest{\textemdash}energy dissipation and execution delay (inverse of
performance). An optimal feature
would simultaneously incur low energy overhead while also reducing the execution
delay. This metric is useful in capturing the efficiencies of different
architectural designs and features which
may expend differing amounts of energy while producing the same performance
speedup or vice-versa. Hence, a lower Energy-Delay product is more desirable. We observe that
CABA-BDI has a 45\% lower Energy-Delay product than the baseline. This reduction
comes from energy savings from reduced data transfers as well as lower execution
time. On average, CABA-BDI is within only 4\% of Ideal-BDI which incurs none of the
energy and performance overheads of the CABA framework. }


\subsubsection{Effect of Enabling Different Compression Algorithms}
The \SADA framework is \emph{not limited 
to a single compression algorithm}, and can be effectively used to employ other
hardware-based compression algorithms (e.g., FPC~\cite{fpc} and C-Pack~\cite{c-pack}).
The effectiveness of other algorithms depends on two key factors:
(i) how efficiently the algorithm maps to GPU instructions, (ii) how
compressible the data is with the algorithm. 
We map the FPC and C-Pack algorithms to the \SADA framework and evaluate the
framework's efficacy.\comm{Our technical report~\cite{caba-tr} details how
these algorithms are mapped to \SADA.} 
\comm{We estimate that FPC requires eight add instructions to compute the address 
of consecutive compressed words within a cache line, C-Pack -- X instructions.}

\begin{figure}[!h]
  \centering
  \includegraphics[width=0.69\textwidth]{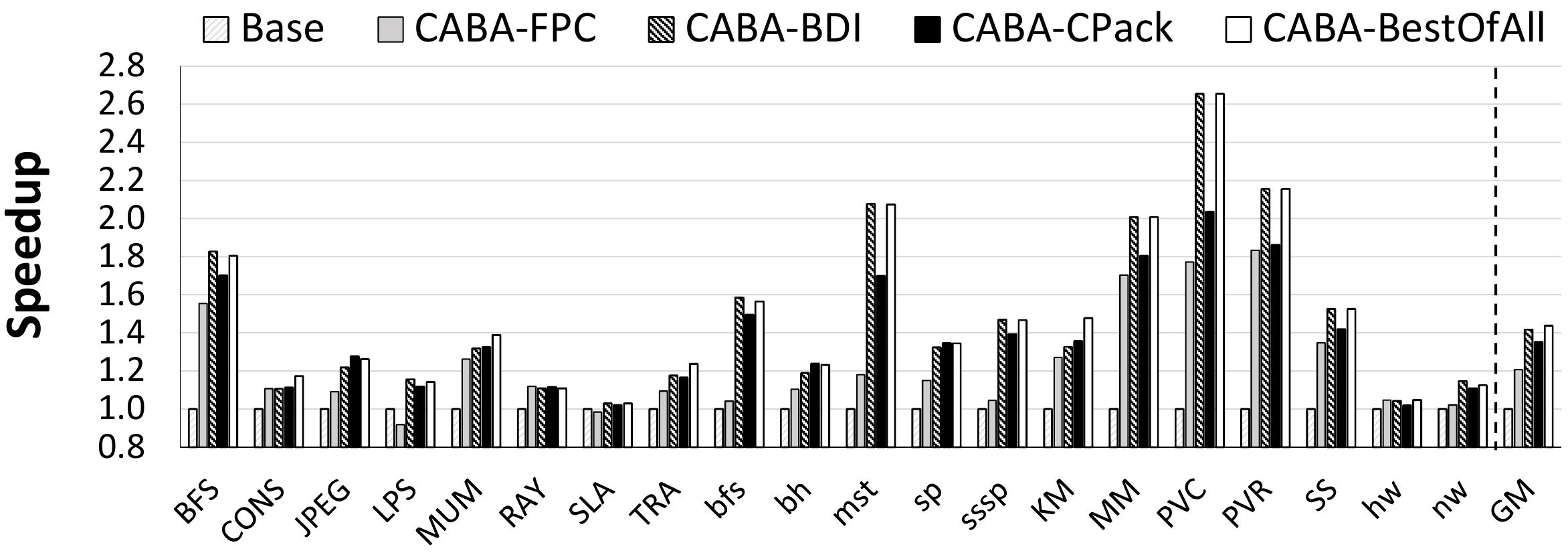}
  \caption{Speedup with different compression algorithms.}
  \label{fig:fpc}
\end{figure}

Figure~\ref{fig:fpc} shows the normalized speedup with four
versions of our design: \emph{CABA-FPC}, \emph{CABA-BDI},  \emph{CABA-C-Pack},
and \emph{CABA-BestOfAll}  
with the FPC, BDI, C-Pack compression algorithms. CABA-BestOfAll is an idealized design that selects and uses the best of all three algorithms in terms of compression ratio for \emph{each cache line}, assuming no selection overhead. 
We make three major observations.

\begin{figure}[h]
  \centering
  \includegraphics[width=0.69\textwidth]{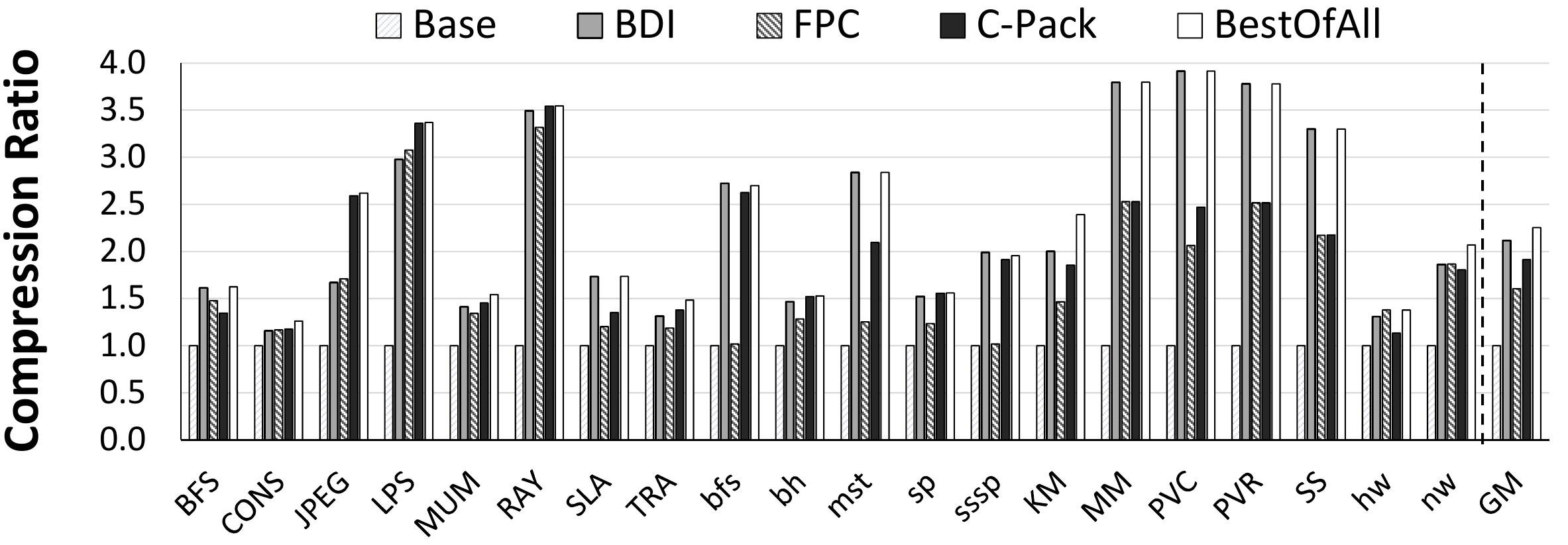}
  \caption{Compression ratio of algorithms with \SADA.}
  \label{fig:ratio}
\end{figure}

First, \SADA significantly improves performance with any compression algorithm (20.7\% with FPC, 35.2\% with C-Pack). Similar to CABA-BDI,
the applications that benefit the most are those that are both memory-bandwidth-sensitive (Figure~\ref{fig:bwutil})
and compressible (Figure~\ref{fig:ratio}). 
We conclude that our proposed framework, \SADA, is general and flexible enough
to successfully enable different compression
algorithms.

Second, applications benefit differently from each algorithm. 
For example, \emph{LPS, JPEG, MUM, nw} have higher compression ratios with FPC or C-Pack, 
whereas \emph{MM, PVC, PVR} compress better with BDI. 
This motivates the necessity of having \emph{flexible data compression} with different
algorithms within the same system. Implementing multiple compression
algorithms completely in hardware is expensive as it adds significant area
overhead, whereas \SADA can flexibly enable the use of different algorithms via
its general assist warp framework.

\comm{Third, the performance benefits with CABA-FPC are significantly lower than that of CABA-BDI.
There are two main reasons for this. First, as Figure~\ref{fig:ratio} shows, the compression
ratio with FPC is significantly lower than that of BDI
(2.1X vs. 1.6X correspondingly).\footnote{Our analysis shows that for our applications
the data patterns that can be compressed with BDI (e.g., array of pointers, pixels), but cannot
be compressed with FPC, are quite frequent. This leads to significantly higher compression ratios with BDI than
with FPC.} Second, the overhead of decompression with
FPC is higher than with BDI). This leads to a higher performance penalty when
applying CABA-FPC than CABA-BDI. In one application (\emph{LPS}), this overhead causes a degradation (e.g., 8.2\%)
in performance. In contrast, CABA-C-Pack, is much closer in performance to CABA-BDI, due to its competitive
compression ratio (1.9X vs. 2.1X with BDI).}

Third, the design with the best of three compression algorithms, CABA-BestOfAll,
can sometimes improve performance more than each
 individual design with just one compression algorithm (e.g., for \emph{MUM} and \emph{KM}). This 
happens because even within an application, different cache lines compress better with different algorithms. At the same time, different compression related overheads of different algorithms can cause one to have higher performance than another even though the latter may have a higher compression ratio. For example, CABA-BDI provides higher performance on \emph{LPS} than CABA-FPC, even though BDI has a lower compression ratio than FPC for \emph{LPS}, because BDI's compression/decompression latencies are much lower than FPC's. Hence, a mechanism
that selects the best compression algorithm based on \emph{both} compression ratio and the relative cost of compression/decompression is desirable to get the best of multiple compression algorithms.
The \SADA framework can flexibly enable the implementation of such a mechanism,
whose design we leave for future work. 

\subsubsection{Sensitivity to Peak Main Memory Bandwidth}
As described in Section~\ref{sec:motivation_caba}, main memory (off-chip) bandwidth is a major
bottleneck in GPU applications. In order to confirm that \SADA works for different designs
with varying amounts of available memory bandwidth, we conduct an experiment where CABA-BDI is used in three systems with 0.5X, 1X and 2X amount of bandwidth
of the baseline.

\begin{figure}[h!]
  \centering
  \includegraphics[width=0.69\textwidth]{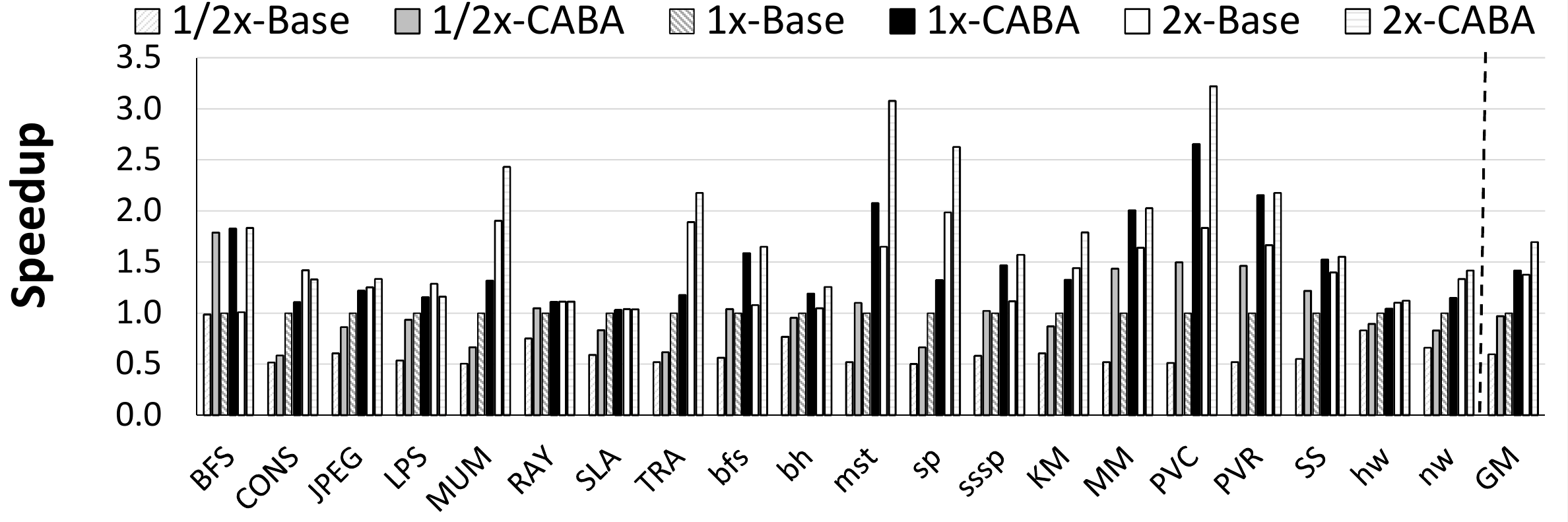}
  \caption{Sensitivity of \SADA to memory bandwidth.}
  \label{fig:bw}
\end{figure}

Figure~\ref{fig:bw} shows the results of this experiment. We observe that, as
expected, each \SADA design
(\emph{*-CABA}) significantly outperforms the corresponding baseline designs with the same
amount of bandwidth. The performance improvement of \SADA is often equivalent to
the doubling the off-chip memory bandwidth. We conclude that \SADA-based bandwidth compression, on average, offers almost
all the performance benefit of doubling the available off-chip
bandwidth with only modest complexity to support \helperwarps.  
 
\comm{
\subsubsection{Sensitivity to Decompression Latency}
Figure~\ref{fig:latency} shows the sensitivity of the performance to the decompression 
latency (varied from 0 in AW-Ideal to 16 instructions) assuming compression ratio achieved with BDI algorithm. 
As expected, the performance drops as we increase the latency of the decompression (that is on the critical
path of the execution). With 16 instructions for decompression the assist warp approach loses more than half of its
performance benefit with several applications degrading more than 20\% over the baseline. 
These results show some practical boundaries on how
complex can the decompression algorithm be (when it is mapped onto the GPU) to be efficient with our design.
 
\begin{figure}[h!]
  \centering
   \vspace{-0.2cm}
  \includegraphics[width=0.49\textwidth]{figures/LatencySensitivity4.pdf}
   \vspace{-0.6cm}
  \caption{Effect on performance with different decompression latency: 0 (AW-Ideal), 4, 8, and 16 instructions. }
  \label{fig:latency}
  \vspace{-0.2cm}
\end{figure}
}
\comm{\subsubsection{Sensitivity to the Block/Atom Size}
\todo{Nandita}{This is an experiment where we vary cache line size: 64, 128, 256, and atom (chunk) size: 8,16,32}

\subsubsection{Sensitivity to Available Bandwidth}
Vary the bandwidth to see the effect on performance (and maybe energy).

\subsubsection{Sensitivity to Cache Size}
Vary L1 and L2 sizes to show that mostly bandwidth matters.

\subsubsection{Sensitivity to SM Number}
Vary the number of SMs to see the effect.}

\subsubsection{Selective Cache Compression with \SADA}
\label{sec:caches}
 In addition to reducing bandwidth consumption, data compression can also
increase the \emph{effective capacity} of on-chip caches. While compressed caches can be
beneficial{\textemdash}as higher effective cache capacity leads to lower
miss rates{\textemdash}supporting cache compression requires several changes in the cache
design~\cite{fpc,c-pack,bdi,dcc,exploiting-pekhimenko-hpca15}. \comm{In this work, we do not aim to propose a new
compressed cache design, but investigate the capacity benefits of
compression in the L1/L2 caches based on the design originally proposed for
FPC~\cite{fpc}. }

\begin{figure}[!h] \centering 
\includegraphics[width=0.69\textwidth]{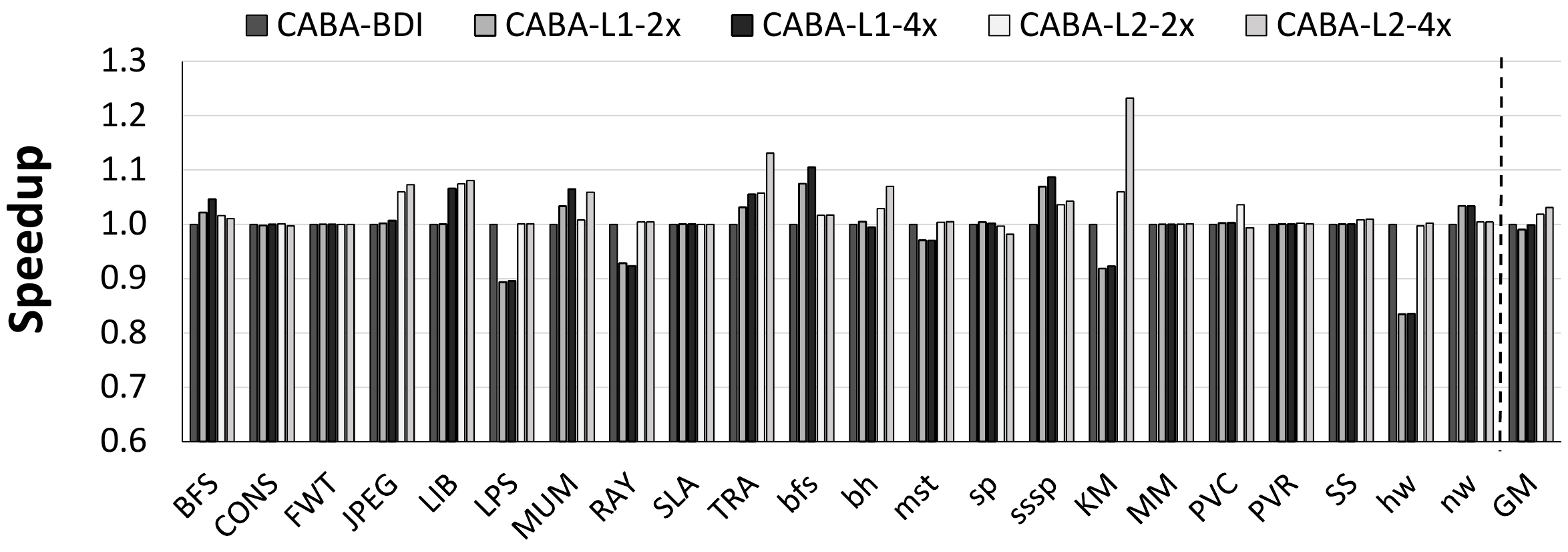}
\caption{Speedup of cache compression with \SADA.}
\label{fig:comp-caches} 
\end{figure}

 Figure~\ref{fig:comp-caches} shows the effect of four cache compression
designs using CABA-BDI (applied to both L1 and L2 caches
with 2x or 4x the number of tags of the baseline\footnote{The number of tags limits the effective compressed cache size~\cite{fpc,bdi}.}) on performance.  We make two major observations.
First, several applications from our workload pool are not only bandwidth
sensitive, but also cache capacity sensitive. For example, \emph{bfs} and \emph{sssp}
significantly benefit from L1 cache compression, while \emph{TRA} and
\emph{KM} benefit from L2 compression.  Second, L1 cache compression can severely
degrade the performance of some applications, e.g., \emph{hw} and \emph{LPS}.
The reason for this is the overhead of decompression, which can be
especially high for L1 caches as they are accessed very frequently.  This
overhead can be easily avoided by disabling compression at any level of the
memory hierarchy. 

\subsubsection{Other Optimizations}
We also consider several other optimizations of the \SADA framework for data
compression:
(i) avoiding
the overhead of decompression in L2 by storing data in the uncompressed form and 
(ii) optimized load of \emph{only useful} data.
\begin{figure}[!h] 
\centering 
\includegraphics[width=0.69\textwidth]{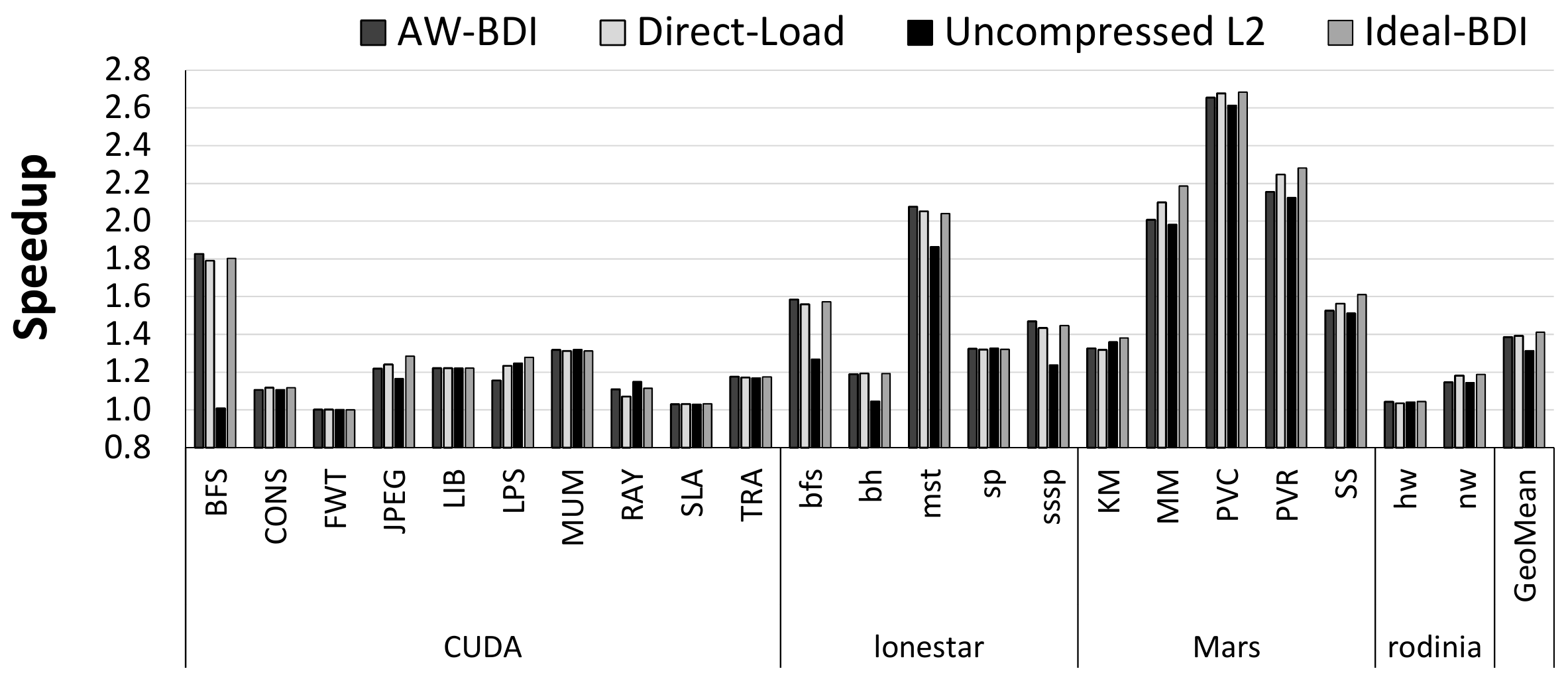}
\caption{Effect of different optimizations (Uncompressed data
in L2 and Direct Load) on applications' performance.} 
\label{fig:optimizations} 
\end{figure} 

\textbf{Uncompressed L2.}
\comm{This optimization provides another tradeoff between the savings in the on-chip traffic (when data in L2 is compressed -- default
option),
or savings in the additional latency of decompression (when data in L2 is uncompressed). 
Several applications in our workload pool (e.g., \emph{RAY}) benefit from
storing data uncompressed as these applications have higher hit rates in the L2
cache.}
The \SADA framework allows us to store compressed data selectively at different levels of the memory hierarchy. \comm{For example, \emph{either one or both}
levels of cache can hold uncompressed data with compression only at main memory
or \emph{all levels of memory} could hold compressed data. }We consider an
optimization where we avoid the overhead of decompressing data in L2 by storing
data in uncompressed form. This provides another
tradeoff between the savings in on-chip traffic (when data in L2 is
compressed -- default option), and savings in decompression latency (when data
in L2 is uncompressed). Figure~\ref{fig:optimizations} depicts the performance
benefits from this optimization. Several applications in our
workload pool (e.g., \emph{RAY}) benefit from storing data uncompressed as
these applications have high hit rates in the L2 cache. \purple{We conclude that
offering the choice of enabling or disabling compression at different levels of
the memory hierarchy can provide higher levels of the software stack (e.g.,
applications, compilers, runtime system, system software) with an additional performance knob.} \comm{\SADA framework provides application developers an additional
per-application performance knob with on-demand data compression
at different levels in the memory hierarchy.}

\textbf{Uncoalesced requests.} 
Accesses by scalar threads from the same
warp are coalesced into fewer memory transactions~\cite{pguide}. 
If the requests from different threads within a
warp span two or more cache lines, multiple lines have to be retrieved and
decompressed before the warp can proceed its execution. 
Uncoalesced requests can significantly increase the number of \helperwarps that need to be executed.
An alternative to decompressing each cache line (when only a few bytes from each line
may be required), is to enhance the coalescing unit to supply only the correct
\emph{deltas} from within each compressed cache line. 
The logic that maps bytes within a cache line to the appropriate registers will need to be
enhanced to take into account the encoding of the compressed line to determine
the size of the \emph{base} and the \emph{deltas}. 
As a result, we do not decompress the entire cache lines and only extract the
data that is needed. In this case, the cache line is not
inserted into the L1D cache in the uncompressed form, and hence every line needs to
be decompressed even if it is found in the L1D cache.\footnote{This
optimization also benefits cache lines that might \emph{not} have many uncoalesced
accesses, but have poor data reuse in the L1D.} \emph{Direct-Load} in
Figure~\ref{fig:optimizations} 
depicts the performance impact from this optimization. The overall
performance improvement is 2.5\% on average across all applications (as high as
4.6\% for \emph{MM}). 


\subsection{Related Work}
To our knowledge, this work is the first to (1) propose a flexible and
general cross-layer abstraction and framework for employing idle GPU resources for useful
computation that can aid regular program execution, and (2) use the
general concept of {\em helper threading} to perform memory and
interconnect bandwidth compression. We demonstrate the benefits of our
new framework by using it to implement multiple compression algorithms
on a throughput-oriented GPU architecture. We briefly discuss related
works in helper threading and bandwidth compression.

\textbf{Helper Threading.} Previous
  works~\cite{ssmt,ssmt2,ht2,ht5\comm{,ht0},ht20,ht12,ht0,assisted-execution,
    assisted-execution-04,ht11,ht19,ht25,ht4,ht7,ht9,ht1} demonstrated
  the use of \emph{helper threads} in the context of Simultaneous
  Multithreading (SMT) and multi-core processors, primarily to speed
  up single-thread execution by using idle SMT contexts or idle
  cores in CPUs. These works typically use helper threads (generated
  by the software, the hardware, or cooperatively) to pre-compute
  useful information that aids the execution of the primary thread
  (e.g., by prefetching, branch outcome pre-computation, and cache
  management). No previous work discussed the use of helper threads
  for memory/interconnect bandwidth compression or cache compression.

While our work was inspired by these prior studies of helper threading
in latency-oriented architectures (CPUs), developing a framework for
helper threading (or {\em assist warps}) in throughput-oriented
architectures (GPUs) enables new opportunities and poses new
challenges, both due to the massive parallelism and resources present
in a throughput-oriented architecture.  Our CABA framework exploits these new
opportunities and addresses these new challenges, including (1)
low-cost management of dozens of assist warps that could be running
concurrently with regular program warps, (2) means of state/context
management and scheduling for assist warps to maximize effectiveness
and minimize interference, and (3) different possible applications of
the concept of assist warps in a throughput-oriented architecture.

\comm{
In these prior works, helper threads are either generated with the
help of the compiler (e.g.,~\cite{ht11, ht19, ht25}) or completely in
hardware (e.g.,~\cite{ht4}) and used to perform optimizations like
prefetching in ~\cite{ssmt, ht2,ht5, ht7, ht10, ht12, ht9, ht7, ht1,
  assisted-execution-04}, branch prediction~\cite{ssmt}, cache
management~\cite{ssmt} or speculative
multi-threading~\cite{ht15,ht16,ht17,ht18}.
}

In the GPU domain, CudaDMA~\cite{cudadma} is a recent proposal that
aims to ease programmability by decoupling execution and memory
transfers with specialized DMA warps. This work does not provide a
general and flexible hardware-based framework for using GPU cores to run warps that aid the main program.


\textbf{Compression.} Several prior
works~\cite{LinkCompression,CompressionPrefetching,GPUBandwidthCompression,
  lcp-micro,memzip, sc2-arelakis-isca14, exploiting-pekhimenko-hpca15} study memory and cache compression with
several different compression
algorithms~\cite{fpc,bdi,c-pack,sc2-arelakis-isca14,zvc,fvc}, in the context of CPUs
or GPUs. Our work is the first to demonstrate how one can adapt some
of these algorithms for use in a general helper threading framework
for GPUs. As such, compression/decompression using our new framework
is more flexible since it does not require a specialized hardware
implementation for any algorithm and instead utilizes the existing GPU
core resources to perform compression and decompression. Finally, assist warps
are
applicable beyond compression and can be used for other purposes.

\subsection{Summary} 

This work makes a case for the \SADAfull (\SADA) framework, which
employs \emph{assist warps} to alleviate different
bottlenecks in GPU execution. \SADA is based on the key observation
that various imbalances and bottlenecks in GPU execution leave on-chip
resources, i.e., computational units, register files and on-chip
memory, underutilized. CABA takes advantage of these idle resources and employs
them to perform useful work that can aid the execution of the main program and
the system. 

We provide a detailed design and analysis of
how \SADA can be used to perform flexible data compression in GPUs to
mitigate the memory bandwidth bottleneck.  Our extensive evaluations
across a variety of workloads and system configurations show that the
use of \SADA for memory compression significantly improves system
performance (by 41.7\% on average on a set of bandwidth-sensitive GPU
applications) by reducing the memory bandwidth requirements of both the
on-chip and off-chip buses.\comm{We also show that \SADA can flexibly
implement multiple different compression algorithms, which have
different effectiveness for different applications or data
patterns.} 

We conclude that CABA is a general substrate that can
alleviate the memory bandwidth bottleneck in modern GPU systems by
enabling flexible implementations of data compression algorithms.  We
believe \SADA is a general framework that can have a wide set of use
cases to mitigate many different system bottlenecks in
throughput-oriented architectures, and we hope that future work
explores both new uses of \SADA and more efficient implementations of
it.

%% file: sections/misc.tex
\section{Conclusions and Future Work}
\subsection{Future Work}
\label{sec:futurework}
This dissertation opens new avenues for research. In this section, we
describe several such research directions in which the ideas and approaches in this
thesis can be extended to address other challenges in programmability,
portability, and efficiency in various systems. 
\subsubsection{Enabling rapid and fine-grain code adaptation at runtime, driven
by hardware} 
As we increasingly rely on clouds and other virtualized environments, co-running
applications, unexpected contention, and \emph{lack of visibility} into
available hardware resources make software optimization and dynamic
recompilation limited in effectiveness. The hardware today cannot easily help
address this challenge since the existing hardware-software contract requires
that
hardware rigidly execute the application as defined by software.

With the abstractions proposed in this dissertation, we cannot adapt the
code itself, but only the underlying system and hardware based on the program
properties. Future work would involve enabling the system/hardware to dynamically change
computation depending on availability of resources and runtime program
behavior. The idea is to have the application only convey higher-level
functionality, and then enable a \emph{codesigned} hardware-software system to
dynamically change the implementation as the program executes. The benefit of
enabling such capability in hardware is greater efficiency in adapting software
and more fine-grain visibility into dynamic hardware state and application
behavior. For example, the program describes a potentially sparse computation
(e.g., sparse matrix-vector multiply). The hardware then dynamically elides
computation when it detects zero inputs. Other examples include changing graph
traversal algorithms to maximize data locality at runtime or altering the
implementation of a forward pass in each neural network layer, based on resource
availability and contention. 
The challenge is in designing a \emph{general} hardware-software system that
enables many such runtime optimizations and integrates flexibly with frameworks
such as Halide~\cite{halide}, TensorFlow~\cite{tensorflow}, and
Spark~\cite{rdd-zaharia-nsdi12}. 

The approach would be to design a clearly defined hardware-software abstraction
that expresses what computation can be changed based on runtime information
(resource availability/bottleneck). This abstraction should integrate well into
common building-block operations of important applications to obtain generality. 
Another approach is to determine how to effectively abstract and
communicate fine-grained dynamic hardware state, bottlenecks, and contention, to
enable software frameworks, databases, and other software systems to dynamically
adapt applications accordingly. 


\subsubsection{Widening the scope of cross-layer and full stack optimizations} 
This dissertation has demonstrated the significant \emph{performance,
portability,} and \emph{productivity} benefits of cross-layer abstractions that
are carefully architected to enable \emph{full-system coordination and
communication} to achieve these goals, from the programming model, compiler and
OS, to each hardware component (cache, memory, storage, and so on). 

Future work would investigate enabling full-system coordination and communication
to achieve other challenges such as security, quality-of-service (QoS), meeting
service-level objectives (SLOs), and reliability. The challenge is in designing
cross-layer abstractions and interfaces that enable very disparate components
(e.g., cache, storage, OS thread scheduler) to communicate and coordinate, both
horizontally and vertically in the computing stack, to meet the same goal

The first steps would involve looking into enhancing the existing compute stack
to enable these holistic designs and cross-layer optimizations. This will
involve determining how to express application-level requirements for these
goals, how to design low-overhead interfaces to communicate these requirements
to the system and each hardware component, and then enhance the system
accordingly to achieve the desired goal. Insights from this thesis may be
directly applicable to solutions here. 

The long-term research goal is to research \emph{clean-slate approaches} to
building systems with modularized components  that are designed to provide
full-system guarantees for performance isolation, predictability, reliability,
and security. The idea is to compose the overall system from smaller modules
where each module or \emph{smallest unit} is designed to provide some guarantee
of (for example) strict performance isolation. Similarly the interfaces between
module should preserve the guarantees for the overall system. Initial research
questions: What are the semantics that define a  module? What are the semantics
that define interactions between modules? 

\subsubsection{Integrating new technologies, compute paradigms, and specialized
systems} 
Future systems will incorporate a diverse set of technologies and specialized
chips, that will rapidly evolve. This poses many new challenges across the
computing stack
in the \emph{integration} of new technologies, such as memristors, persistant
memory,
and optical devices, and
new paradigms, such as quantum computing, reconfigurable fabrics,
application-specific hardware, and processing-in-memory substrates. 
Below are extensions to this thesis, relating to different aspects of this
problem. 

 \textbf{Enabling applications to automatically leverage new systems and
 architectures (\emph{a software approach}).}
 Automatically generating high-performance code for any new  hardware
 accelerator/specialization today, without fully rewriting applications, is a
 challenging task. Software libraries have been demonstrated to be very
 inefficient and are not general. Automatic code generating frameworks and
 optimizing compilers (e.g.,~\cite{hpvm,tvm,delite}) are  promising approaches
 to target a changing set of architectures, without rewriting application code.
 However, such tool chains use a common intermediate representation (IR) to
 summarize the application and then use \emph{specialized backends} that
 optimize the IR according to the characteristics of the new architecture and
 produce target code. Integration of new architectures or even new
 instructions/functionality in CPUs/GPUs (e.g., to add instructions to leverage
 a processing-in-memory accelerator) into these systems takes significant time,
 effort, and cross-layer expertise. 

 The goal is to develop frameworks to enable automatic generation of
 high performance code for specialized/new hardware, without rewriting existing
 applications. This would significant reduce the effort required to evaluate and
 deploy new hardware innovations including new architectures, substrates, or
 finer granularity hardware primitives (e.g., a faster in-memory operation). 

The first steps would be to design an abstraction that enables easy and flexible
 description of the performance characteristics, constraints, and the
 \emph{semantics} of the interface to hardware, i.e., the functionality
 implemented by the instruction-set architecture (ISA) or new primitive. Next,
 develop tools that enable automatic integration into existing compiler IRs to
 generate backends based on the description. Existing IRs should also be
 enhanced to capture more semantic content if required. 

 A longer-term goal is to develop hardware-software frameworks that enable flexible evaluation and design
 space exploration of how to abstract new hardware technologies and substrates
 in terms of the overall programmability (how many applications can leverage the
 technology), portability (how easily we can enhance the architecture without
 changing the interface), and performance (efficiency of the new architecture).

 \textbf{Enabling software-transparent hardware specialization and
 reconfiguration (\emph{a hardware approach}).} 
 Today, even within general-purpose cores, architects are turning to different
 forms of hardware customization and reconfigurability for important computation
 as a means to drive improvements in performance and energy efficiency. Examples
 include Tensor Cores within GPUs to speed up tensor operations in machine
 learning, accelerating data intensive computation using processing-in-memory
 technologies, or reconfigurable fabrics (e.g., CGRAs) to accelerate important
 computations. 

 The goal is to enable seamless integration of specialization and
 reconfigurability into general-purposes architectures, \emph{transparently} to
 the  software stack. This would enable specialized designs to continuously
 evolve without creating new primitives/instructions each time (and hence no
 recompilation/rewriting is required). This addresses the critical portability
 and compatibility challenges associated with these approaches. For example,
 this would enable seamless addition of specialized/reconfigurable hardware
 support within general-purpose cores for: sparse and irregular computations
 (e.g, graph processing), managed languages (e.g., support for object-based
 programming, garbage collection), critical computations (e.g, numerical loops
 in machine learning) and frequent operations (e.g, queries in databases), among
 very many possibilities. Abstractions proposed in this dissertation, such as
 Expressive Memory, are not sufficiently rich to include specialized
 \emph{computation} as opposed to just memory access. 

 The first steps would be to enhance general-purpose architectures to \emph{(1)}
 allow flexible customization and reconfiguration of different components such
 as the compute units, memory hierarchy, and coherence protocols; and \emph{(2)}
 enable seamless transition between general-purpose and specialized computation,
 and enable safe reconfiguration of hardware components at runtime. The next
 steps would be 
 to design a rich (and future-proof) programming abstraction that captures
 sufficient application information to enable this flexible customization and
 runtime reconfiguration.

\subsection{Conclusions}
In this dissertation, we observed that the \emph{interfaces} and abstractions
between the layers of the computing stack{\textemdash}specifically, the
hardware-software interface{\textemdash}significantly constrain the ability of
the hardware architecture to intelligently and efficiently manage key resources
in CPUs and GPUs.
This leads to challenges in programmability and portability, as the application
software is forced to do much of the heavy lifting in optimizing the code for
performance. It also leaves significant performance on the table, as the
application has
little access to key resources and architectural mechanisms in hardware, and
little visibility into available resources in the presence of virtualization or
co-running applications. 

We proposed rich low-overhead cross-layer abstractions that communicate
higher-level program information from the application to the underlying system
software and hardware architecture. These abstractions enable a wide range of
hardware-software cooperative mechanisms to optimize for performance by managing
critical resources more efficiently and intelligently. More efficient resource
management at the hardware and system-level 
makes performance
less sensitive to how well an application is optimized for the underlying
architecture. 
This reduces the burden on the application developer and makes performance
more portable across architecture generations. 
We demonstrated how such cross-layer abstractions can be designed
to be \emph{general}, enabling a wide range of hardware-software mechanisms,
and \emph{practical}, requiring only low overhead additions to existing systems
and interfaces. Using 4 different contexts in CPUs and GPUs, we validate the
thesis: \emph{a rich low-overhead cross-layer interface that
communicates higher-level application information to hardware enables many
hardware-software cooperative mechanisms that significantly
improve performance, portability, and programmability.}

First, we proposed Expressive Memory, a rich cross-layer interface in CPUs to communicate
higher-level semantics of data structures and their access semantics 
to the operating system and hardware. We demonstrated its effectiveness in
improving the portability of memory system optimizations and in enabling a wide
range of cross-layer optimizations to improve memory system performance. Second,
we introduced the Locality Descriptor, a cross-layer abstraction in GPUs that
enables expressing and exploiting data locality in GPU programs. We demonstrated
significant performance benefits by enabling the hardware to leverage
knowledge of data locality properties, while reducing programming effort when
optimizing for data locality. Third, we proposed Zorua, a framework that
decouples GPU programming models from hardware resource management. We
demonstrated how Zorua enhances programmability, portability, and performance,
by enabling hardware to dynamically manage resources based on the program
requirements. Finally, we introduced the Assist Warp abstraction in GPUs to
effectively leverage idle memory and compute bandwidth to perform useful work. 
We hope that the ideas, analyses, and techniques presented in this dissertation
can be extended to address challenges in hardware-software codesign and the
design of cross-layer interfaces in future computer systems. 

%% file: sections/appendix.tex
\begin{appendices}
\section{Other Works of the Author}
In addition to the works presented in this thesis, I have also contributed to
several other research works done in collaboration with students and professors
at CMU and ETH. In this section, I briefly overview these works.

Toggle-aware compression~\cite{toggle-pekhimenko-hpca16} tackles the additional bit flips in wires when
data is communicated in a compressed form. Data compression, on the one hand,
significantly decreases the amount of data transmitted and hence improves
performance and reduces memory bandwidth consumption. On the other hand,
however, compression increases the overall \emph{entropy} in data, and hence
causes more \emph{bit flips} in the wires when data is transmitted across the
processor interconnect in the form of flits. This increase in bit flips leads to
increased power consumption. This work is the first to identify this challenge
in GPUs and quantify its effects. We propose simple techniques to alleviate
this phenomenon and reduce energy consumption.

ChargeCache~\cite{chargecache} tackles the long access latencies in modern DRAM chips. In
this work, we leverage the key observation that a recently-accessed DRAM row has
more charge, and hence can be accessed with reduced timing parameters.
ChargeCache tracks recently-accessed rows in each bank, and any subsequent
access to a recently-accessed row is handled by the memory controller with
reduced DRAM parameters. Due to the temporal locality in row accesses seen in
most workloads, ChargeCache is able to significantly reduce DRAM access
latency.

Copy-Row DRAM (CROW)~\cite{crow} is a flexible in-DRAM substrate that can be used to
improve the latency, energy efficiency, and reliability of DRAM chips. This
substrate allows duplicating select rows in DRAM that contain
frequently-accessed data or are most sensitive to refresh or reliability issues.
The rows that contain duplicate data can be activated simultaneously, which
reduces the latency
of access. The rows reserved for duplicate data can also be used to map weak
cells in DRAM to reduce refresh overheads and improve reliability.  

SoftMC~\cite{softmc} is the first FPGA-based testing platform that can be used to
control and test memory modules designed for the commonly-used DDR (Double Data
Rate) interface. SoftMC provides an intuitive high-level programming interface
to the user, while also providing the flexibility to control and issue low-level
memory commands and implement a range of mechanisms and tests. We demonstrate
its capability with two example use cases: characterizing retention time of DRAM
cells and validating recently-proposed mechanisms that access recently-refreshed
or recently-accessed rows faster than other DRAM rows.

TOM~\cite{tom-hsieh-isca16} is a programmer-transparent mechanism in GPUs to address two
critical challenges in enabling processing-in-memory (PIM) via the logic layer in
3D-stacked memory architectures: First, its unclear which code should be
offloaded
to the PIM cores and, second, it's challenging to map data to the same memory
stack as the PIM core that will operate on it. TOM uses a compiler-based
technique to automatically offload GPU code to the PIM logic with simple
cost-benefit analysis. TOM further uses a software-hardware cooperative
mechanism to predict which memory page will be accessed by the offloaded code
and ensure that the page is placed in the same memory stack. 

IMPICA~\cite{impica} is an in-memory pointer-chasing accelerator that leverages the
logic layer in 3D-stacked memories. IMPICA enables parallelism in serial
pointer-chasing with address-access decoupling. It also addresses the
significant challenges of virtual-to-physical address translation by using a
region-based page table near the PIM core. In this work, we identify the key
challenges in enabling processing in memory for irregular pointer-chasing
applications and design IMPICA to effectively address these challenges and
significantly accelerate pointer-chasing.  

Gaia~\cite{gaia} is a geo-distributed machine learning system that employs an
efficient communication mechanism over wide-area networks (WANs). Gaia decouples the
communication within a data center from the communication between data centers,
enabling different communication and consistency models for each. In this work,
we present a new machine learning synchronization model, Approximate Synchronous
Parallel (ASP) for WAN communication that dynamically eliminates insignificant
communication between data centers. At the same time, Gaia is generic and
flexible enough to run a wide range of machine learning algorithms. 

SMASH~\cite{smash} is a hardware-software cooperative mechanism that enables
highly-efficient indexing and storage of sparse matrices. The key idea of SMASH
is to explicitly enable the hardware to recognize and exploit sparsity in data.
To this end, we devised a novel software encoding based on a hierarchy of
bitmaps that can be used to
efficiently compress any sparse matrix and, at the same time, can be directly
interpreted by the hardware to enable highly-efficient indexing. SMASH enables
significant speedups for sparse matrix computation by eliminating the expensive
pointer-chasing operations required in state-of-the-art sparse matrix compression schemes.

\end{appendices}